\documentclass[10pt]{iopart}

\usepackage{iopams}
\usepackage{graphicx}
\usepackage{color}
\usepackage{ulem}
\usepackage{cite}

\usepackage{xcolor}
\definecolor{trueblue}{rgb}{0.0, 0.45, 0.81}
\definecolor{crimsonglory}{rgb}{0.75, 0.0, 0.2}

\usepackage{hyperref}
\hypersetup{colorlinks,linkcolor={trueblue},citecolor={crimsonglory},urlcolor={crimsonglory}}

\renewcommand\contentsname{}

\begin{document}

\review[Electronic and optical properties of strained graphene]{Electronic and optical properties of strained graphene and other strained 2D materials: a review}

\author{Gerardo G. Naumis$^1$, Salvador Barraza-Lopez$^2$, Maurice Oliva-Leyva$^3$, and Humberto Terrones$^4$}

\address{$^1$ Depto. de Sistemas Complejos, Instituto de F\'{i}sica, Universidad Nacional Aut\'{o}noma de
M\'{e}xico, Apdo. Postal 20-364, Mexico City 01000, Mexico}

\address{$^2$ Department of Physics, University of Arkansas, Fayetteville, AR 72701, USA}

\address{$^3$ Instituto de Investigaciones en Materiales, Universidad Nacional Aut\'{o}noma de
M\'{e}xico, Apdo. Postal 70-360, Mexico City 04510, Mexico.}

\address{$^4$ Department of Physics, Applied Physics, and Astronomy. Rensselaer Polytechnic Institute,
Troy, New York, NY 12180, USA}


\ead{naumis@fisica.unam.mx}
\date{\today}


\begin{abstract}
This review presents the state of the art in strain and ripple-induced effects on the electronic
and optical properties of graphene. It starts by providing the crystallographic description of
mechanical deformations, as well as the diffraction pattern for different kinds of representative
deformation fields. Then, the focus turns to the unique elastic properties of graphene, and to how strain is produced. Thereafter, various theoretical approaches used to study the electronic properties of strained graphene are examined, discussing the advantages of each. These approaches provide a platform to describe exotic properties, such as a fractal spectrum related with quasicrystals, a mixed Dirac–Schr\"odinger behavior, emergent gravity, topological insulator states, in molecular graphene and other 2D discrete lattices. The physical consequences of strain on the optical properties are reviewed next, with a focus on the Raman spectrum. At the same time, recent advances to tune the optical conductivity of graphene by strain engineering are given, which open new paths in device applications. Finally, a brief review of strain effects in multilayered graphene and other promising 2D materials like silicene and materials based on other group-IV elements, phosphorene, dichalcogenide- and monochalcogenide-monolayers is presented, with a brief discussion of interplays among strain, thermal effects, and illumination in the latter material family.
\end{abstract}

\vspace{2pc}
%

\noindent{\submitto{\RPP}}
%
%
\ioptwocol
\tableofcontents

\section{Introduction}

Graphene was the first truly two dimensional (2D) crystal ever discovered \cite{Novoselov2004,Novoselov2005}. A key ingredient
for its discovery was the observation that it becomes visible in an optical microscope if placed on a SiO$_2$ substrate with a carefully chosen thickness \cite{Roddaro2007}. Eventually, this one-atom thick carbon membrane turned out to have the highest known electrical and thermal conductivity \cite{Balandin2008}, as well as the highest stiffness and strength. It supports very high strain prior to mechanical failure \cite{Lee2008}, can be strained well beyond the linear regime, and bent and wrinkled \cite{Lee2008}. In many applications, graphene lays or grows on a substrate. Due to the mismatch between the graphene and substrate lattice parameters, atoms move to reduce their energy, producing a certain amount of strain too.

There has been an ever increasing interest in using strain and the soft-properties of graphene to control its physical properties \cite{Vozmediano2010,Novoselov2012,Bissett2014,Roldan2015,BarrazaLopez2015,Galiotis2015,Jing2015,Amorim2016,Deng2016,Meunier2016}, and the word \textit{straintronics} \cite{Pereira2009a} appropriately describes this aim. Understanding how strain affects graphene's electronic and optical properties is of paramount importance, but the interest is not only of a technological nature. Strained graphene is a playground for new Physics, from exotic topological quantum phases, to analogies with other fields (from Quantum Electrodynamics to Quantum Gravity), and even connects with traditional arts like origami and kirigami. Furthermore, strain in graphene invites exploration of similar effects on 2D materials like hexagonal boron nitride (hBN), transition metal dichalcogenide NbSe$_2$, TaS$_2$, MoS$_2$ monolayers, silicene, monochalcogenide monolayers, among many others \cite{Novoselov2012, Roldan2015}, and on their multi-layers/superlattices \cite{Novoselov2012}.

Some examples of the effect of strain are: (i) the experimental observation of Dirac cone replicas on substrates \cite{Pletikosi2009,Yankowitz2012} with halved group velocity \cite{Ortix2012} and sometimes accompanied by a reversal of the Hall effect \cite{Yankowitz2012,Ponomarenko2013}, (ii) the first atomic observation of the Quantum Hall effect fractal spectrum \cite{Dean2013,Ponomarenko2013} first predicted in 1976 and known as the Hofstadter butterfly \cite{Hofstadter1976}, (iii) in-plane  graphene/hBN  heterostructures with controlled domain sizes \cite{Zheng2013} as a potential pathway to two-dimensional electronic devices \cite{NaumisTerrones2009}. There are proposals for (iv) spin-polarization switches via strain \cite{Diniz2016}, (v) piezoelectricity by Li  doping \cite{Ong2012}, (vi) giant pseudomagnetic fields \cite{Levy2010} and (vii) giant two-dimensional band-piezoelectric effect by biaxial-strain \cite{Xiaomu2015}. Other 2D materials show potential for piezoelectricity \cite{Carvalho,LiYang,Hennig,Droth2016}, display interesting connections with  glass networks \cite{Thorpe2012,Mehboudi2016} and glass constraint (rigidity) theory \cite{Naumis2015Frontiers}, and can be strained in a peculiar manner when illuminated \cite{newest}.

The bibliography in the field continues to grow. Other reviews serve as a guide through this information forest \cite{Vozmediano2010,Bissett2014,Roldan2015,BarrazaLopez2015,Galiotis2015,Jing2015,Amorim2016,Deng2016,Meunier2016}, each with its own focus. The goal of the present review is to provide a didactic and basic platform to understand in simple terms representative models and results, with enough physical significance to get a grasp of the main effects of strain, the emerging consensus on the field, and its new directions. It will be shown that many of the effects of strain are already present in the simplest case, the isotropic expansion.
From this realization, one is able to understand more complex situations, like the appearance of a mixed Dirac-Schr\"odinger behavior \cite{Pereira2009a,Montambaux2009a,Montambaux2009b,Montambaux2012,Montambaux2013,Roman2015a}, and pseudomagnetic fields in the Dirac equation \cite{Vozmediano2010}.

Our second focus is to cover several theoretical approaches to treat strain in graphene, not all of them covered in existing reviews. The tight-binding approach, the Dirac equation with pseudoelectromagnetic fields, a perspective from discrete differential geometry, and results from \textit{ab initio}, density-functional theory (DFT) will be considered here. Each of such approaches leads to comparisons with experimental results and has its own
virtues. For example, the pseudomagnetic field approach is excellent to provide a link with quantum electrodynamics but
is better suited for studying smooth spatially varying strain. If short-wavelength strain is present, like in graphene
grown on a substrate, the tight-binding approach is better suited. The approach based on discrete differential geometry has the advantage of laying out the theory directly onto the atomic lattice. DFT helps to design and improve theories and can provide direct connections to important experimental questions too.

Many of the methods and concepts used to treat strain in graphene are also applicable to other 2D materials. With this idea in mind, a short mini-review of other strained 2D materials is given with the aim of helping readers to discover and identify new areas of research in which the methods learned for graphene are well suited for further use.

With previous considerations in mind, each Section of this review was written to be as independent
as possible from the others. At the same time, the aim is to offer an integrated body of knowledge, with a logical and pedagogical structure.

The layout of this work is as follows. In Section \ref{Description} deformations in graphene are  described using crystallography and elasticity theory. Section
\ref{Deformations} is devoted to study the mechanical properties of graphene to explain how different types of deformations are produced. In Section \ref{Electronics}, the electronic properties of strained graphene are discussed, providing
different theoretical approaches as well as numerical and experimental results. Optical properties are reviewed in Section \ref{Optical}, while Sections \ref{Multilayered} and \ref{Other2D} provide an overview of multilayered graphene and other 2D materials different from graphene within a unifying context of strain. Conclusions and an outlook of the field are presented afterwards.


\section{Description of pristine and deformed graphene} \label{Description}

This section deals with unstrained graphene, but it provides tools for the description of strain too.

\subsection{Unstrained graphene: crystal structure, reciprocal lattice and diffraction}
\begin{figure}
\centering
\includegraphics[width=\linewidth]{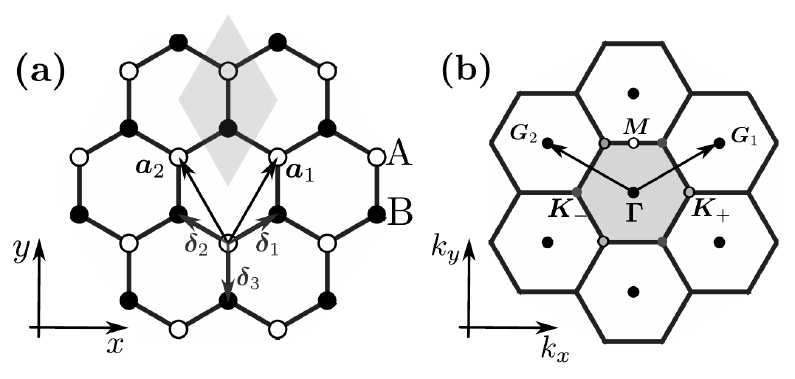}
\caption {(a) Graphene lattice showing the unit cell (shaded), the lattice vectors $\mathbf{a}_1$ and $\mathbf{a}_2$, and first-neighbor vectors $\boldsymbol{\delta}_1$, $\boldsymbol{\delta}_2$ and $\boldsymbol{\delta}_3$. The bipartite sublattices A and B are also shown, as well as the definition of the reference system used in this work. (b) First Billouin zone (shaded) of unstrained graphene
showing the high symmetry points.  The Fermi level
and the Dirac points coincide with the inequivalent high symmetry points $\boldsymbol{K}_{+}$ and $\boldsymbol{K}_{-}$.
This is no longer true for strained graphene.}
\label{RedGrafeno}
\end{figure}

Consider the honeycomb lattice in figure \ref{RedGrafeno}(a). The honeycomb lattice
is not a Bravais lattice but a lattice with a basis, since there are two environments for carbon atoms that are usually denoted as $A$ and $B$ sublattices and shown with open and closed circles in figure ~\ref{RedGrafeno}(a). Notice that atoms in the $A$ sublattice only have first neighbors belonging in the $B$ lattice and \textit{viceversa}. Such subdivision
means that the lattice is bipartite, and many of the electronic and optical properties depend on this general observation \cite{CastroNeto2009}.
Non-periodic bipartite lattices
like the  quasiperiodic Penrose lattice \cite{Naumis1994},
the random binary alloy in a square lattice \cite{Naumis2002}, or vacancies in
graphene \cite{Naumis2007,BarriosVargas2011a,BarriosVargas2013} share some electronic features with pristine graphene, like
zero-energy confined state modes or pseudomobility energy edges \cite{Naumis1994}.

The $x$ axis in figure \ref{RedGrafeno}(a) defines the \textit{zigzag} direction, and each sublattice (say the $A$ sublattice) is a 2D Bravais triangular lattice with lattice vectors:
\begin{equation}\label{VB}
\boldsymbol{a}_{1}=\frac{a}{2}(\sqrt{3},3), \ \ \ \ \ \ \boldsymbol{a}_{2}=\frac{a}{2}(-\sqrt{3},3),
\end{equation}
where $a = 1.42$ ${\mbox{\AA}}$ is the distance between carbon atoms \cite{CastroNeto2009}. The $B$ sublattice is obtained by a shift of atoms belonging to the $A$ sublattice by  $\boldsymbol{\delta}_{1}$. However, it is customary to define a triad of vectors:
\begin{equation}\label{PV}
\boldsymbol{\delta}_{1}=\frac{a}{2}(\sqrt{3},1), \ \ \ \
\boldsymbol{\delta}_{2}=\frac{a}{2}(-\sqrt{3},1),\ \ \ \
\boldsymbol{\delta}_{3}=a(0,-1),
\end{equation}
that point out to the first neighbours of $A$. $\boldsymbol{\delta}_{2}$ and $\boldsymbol{\delta}_{3}$ are the first images  of $\boldsymbol{\delta}_{1}$
under the symmetry group of the Bravais lattice, and are equivalent to the first images of $\boldsymbol{\delta}_1$ in a trigonal kaleidoscope.
Such images are obtained from the Bravais lattice and $\boldsymbol{\delta}_1$ as follows:
\begin{equation}
 \boldsymbol{\delta}_{2}=\boldsymbol{\delta}_{1}+\boldsymbol{a}_{2}-\boldsymbol{a}_{1},\ \ \ \
 \boldsymbol{\delta}_{3}=\boldsymbol{\delta}_{1}-\boldsymbol{a}_{1}.
 \label{kaleidoscope}
\end{equation}

The corresponding reciprocal lattice,
seen in figure \ref{RedGrafeno}(b), has the following reciprocal lattice vectors:
\begin{equation}
\boldsymbol{G}_{1}=\frac{2\pi}{3a}(\sqrt{3},1), \ \ \ \ \ \ \boldsymbol{G}_{2}=\frac{2\pi}{3a}(-\sqrt{3},1).
\end{equation}

As shown in figure \ref{RedGrafeno}(b), the first Brillouin zone (1BZ) is built from the Wigner-Seitz construction, resulting in an hexagon with two inequivalent high-symmetry points   $\boldsymbol{K}_{\pm}=(\pm 4\pi/(3\sqrt{3}a),0)$ \cite{CastroNeto2009} that are labeled $K$ and $K'$ in other works in the field. Notice that the form of the 1BZ is a property of the Bravais lattice.

An important feature that has a crucial impact on the optical and electronic properties of graphene
is the fact that points $\boldsymbol{K}_{+}$ and $\boldsymbol{K}_{-}$ in figure \ref{RedGrafeno}(b) correspond to the intersection of diffraction
Bragg lines (instead of Bragg planes, as it is the case for 3D, bulk materials). In figure \ref{RedGrafeno}(b), a Bragg (Voronoi) line (corresponding to the 1BZ boundary) bisects
a reciprocal lattice vector $\boldsymbol{G}$ at right angles \cite{Ashcroft}. In graphene,
if $\boldsymbol{k}$ is a wavevector in reciprocal space, the diffraction lines closer to
the $\Gamma$ point are described by the Laue conditions \cite{Ashcroft}:
\begin{equation}
 2 \boldsymbol{k} \cdot \boldsymbol{G}_1=\pm |\boldsymbol{G}_1|^{2},
 \label{Laue1}
 \end{equation}
 \begin{equation}
 2 \boldsymbol{k} \cdot \boldsymbol{G}_2=\pm |\boldsymbol{G}_2|^{2},
  \label{Laue2}
\end{equation}
and:
\begin{equation}
 2 \boldsymbol{k} \cdot ( \boldsymbol{G}_1+\boldsymbol{G}_2)=\pm | \boldsymbol{G}_1+\boldsymbol{G}_2|^{2},
  \label{Laue3}
\end{equation}
where $\boldsymbol{K}_{+}$ and $\boldsymbol{K}_{-}$ are at intersections of pairs of straight lines given by equations (\ref{Laue1}), (\ref{Laue2}) or  (\ref{Laue3}).

Diffraction impacts the electronic properties through the generation
of stationary waves and van Hove singularities in the density of electronic states (DOS) \cite{Ashcroft} (both related through an integral over
isoenergetic level curves), making it worthwhile to
calculate the diffraction properties of the lattice. The diffraction pattern is given by the norm of the Fourier transform of atomic positions,
multiplied by the atomic form factor  (also known as the structure factor) \cite{Ashcroft}.

\begin{figure}
\centering
\includegraphics[width=0.9\linewidth]{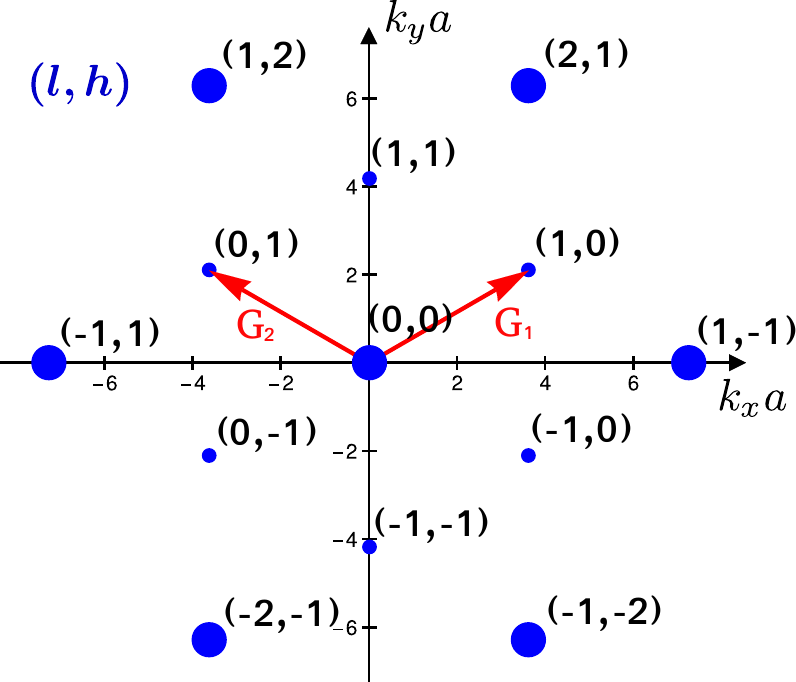}
\caption {Theoretical computation of graphene's diffraction pattern. The position of each diffraction spot was computed from equation  (\ref{PattersonGraphene})
while the intensity (here depicted by spot size) is obtained from equation (\ref{StructureFactorGraphene}) for the Miller indexes $l$ and $h$ that label each spot.}
\label{RedDiffraction}
\end{figure}

To obtain the diffraction pattern, assume that the electronic density results in an scattering potential  $V(\boldsymbol{r})$ which
can be described as delta functions with a weight $V_0$ centered at carbon atoms:
\begin{equation}\label{diff_8}
 V(\boldsymbol{r})=V_0 \sum_{l} \delta(\boldsymbol{r}-\boldsymbol{r}_l),
\end{equation}
where $\boldsymbol{r}_l$ are positions of carbon atoms.
The Fourier transform of this potential, denoted by $\widetilde{V}(\boldsymbol{k})$, is obtained by integrating over the entire area $S$:
\begin{equation}
  \widetilde{V}(\boldsymbol{k})=\int_{S} V(\boldsymbol{r}) e^{i\boldsymbol{k} \cdot \boldsymbol{r}} dS.
\end{equation}
For further reference, this transform will be labeled $\widetilde{V}_{gp}(\boldsymbol{k})$, which can be written as:
\begin{equation}
  \widetilde{V}_{gp}(\boldsymbol{k})= \sum_{\boldsymbol{G}}  V_0[1+ e^{i\boldsymbol{k} \cdot \boldsymbol{\delta}_1}]\delta(\boldsymbol{k}-\boldsymbol{G}),
  \label{FTgraphene}
\end{equation}
with $\boldsymbol{G}=l\boldsymbol{G}_1+h\boldsymbol{G}_2$ and $l,h$ integers. The norm of this transform is (see figure \ref{RedDiffraction}):
\begin{equation}
 |\widetilde{V}_{gp}(\boldsymbol{k})|^{2}= \sum_{\boldsymbol{G}} [4V_0^{2} \cos^{2} (\boldsymbol{k} \cdot \boldsymbol{\delta}_1/2)]\delta(\boldsymbol{k}-\boldsymbol{G}).
 \label{PattersonGraphene}
\end{equation}
The delta term in previous equation indicates the location of diffraction spots in the reciprocal lattice, while the term in brackets is the spot amplitude, known as the structure factor that will be denoted by $F(l,h)$.
For graphene, the diffraction spots form a triangular lattice, and therein, the amplitudes are determined by the structure factor:
\begin{equation}
 F_{gp}(l,h)=4V_0^{2} \cos^{2} ([l \boldsymbol{\delta}_1 \cdot \boldsymbol{G}_1+h\boldsymbol{\delta}_1 \cdot \boldsymbol{G}_2 ]/2),
\end{equation}
or:
\begin{equation}
 F_{gp}(l,h)=4V_0^{2} \cos^{2} \left(\frac{\pi}{3}(2l-h)\right).
 \label{StructureFactorGraphene}
\end{equation}

Previous equation predicts intensities $4V_0^{2}$ or $V_0^{2}$ for the diffraction peaks (rods, due to the two-dimensional nature of the lattice). Equation (\ref{StructureFactorGraphene}) is in  good
agreement with  electron diffraction results \cite{Meyer2007}. However, the experimental shape
and widths of the peaks show deviations from the standard diffraction behavior presented here when graphene
is not flat, as there is a superposition of diffraction rods with slightly different orientations \cite{Meyer2007}.
In bilayer graphene, the rods present further variations in intensity as the crystal is tilted, thus providing a means to
distinguish between monolayer and multilayer graphene \cite{Meyer2007}.

\subsection{Description of deformed graphene}\label{DescriptionStrain}

A mechanical deformation displaces atoms away from their original positions on the crystalline structure. Within Cauchy-Born assumption, the new positions are usually described by
a displacement field $\bi{u}(\bi{r})$ where $\bi{r}$ represents a material point of unstrained graphene \cite{Amorim2016}. After being deformed, the new
position of the atom is:
\begin{equation}
\bi{r}^{\prime}=\bi{r}+\bi{u}(\bi{r}).
\end{equation}
Two kinds of problems arise in elasticity \cite{Malvern,Lai}. One is to find $\bi{u}(\bi{r})$ given some prescribed forces, usually at the boundaries.
The other is to find the forces once  $\bi{u}(\bi{r})$ is given.

The symmetric strain tensor:
\begin{equation}
 \bar{\epsilon}(\bi{r})_{ij}=(\partial_{j}u_{i}(\bi{r})+\partial_{i}u_{j}(\bi{r}))/2,
\end{equation}
(where $i,j=x,y$) measures deformations, discarding rigid-body movements such as rotations or translations \cite{Landau}. On the other hand, the antisymmetric strain tensor measures rigid-body displacements  \cite{Landau}:
\begin{equation}
 \bar{\omega}(\bi{r})_{ij}=(\partial_{j}u_{i}(\bi{r})-\partial_{i}u_{j}(\bi{r}))/2.
\end{equation}

Stresses acting on the system are described by a tensor $\bar{\bi{s}}(\bi{r})$ whose components $\bar{s}(\bi{r})_{ij}=F_{i}/A_{j}$ are
forces along direction $i$ applied over a surface whose normal is along direction $j$.

In linear elasticity, stress is related to $\bi{u}(\bi{r})$
by means of the-so called elasticity (or constitutive) equations, a generalization of Hooke's law for crystalline bodies that applies for small strain fields \cite{Malvern,Landau}:
\begin{equation}
 \bar{s}(\bi{r})_{ij}=\bar{C}_{ijls}\bar{\epsilon}(\bi{r})_{ls},
\end{equation}
where $\bar{C}_{ijls}$ is a rank-$4$ tensor containing the elastic constants of the material (and sometimes written as a rank-2 tensor in Voigt notation). A full discussion on the elastic properties of graphene and how to produce different kinds of deformations will be provided in Section \ref{Deformations}.

First-neighbour vectors  $\boldsymbol{\delta}_{n}$ become space-dependent when strain is applied. Within Cauchy-Born approximation, they are given by \cite{Kitt2013}:
\begin{equation}\label{ew}
\boldsymbol{\delta}_{n}^{\prime} (\boldsymbol{r}) \approx \boldsymbol{\delta}_{n} + (\boldsymbol{\delta}_{n}\cdot \nabla) \bi{u}(\bi{r}) =(\bar{\bi{I}} + \nabla \bi{u}(\bi{r}))\cdot \boldsymbol{\delta}_{n}
\label{Deltan},
\end{equation}
where $\bar{\bi{I}}$ is the $2\times2$
identity matrix, and $\nabla \bi{u}(\bi{r})$ is the Jacobian of the displacement field, known as the displacement gradient tensor whose components are:
\begin{equation}
[\nabla \bi{u}(\bi{r})]_{ij}=\bar{\epsilon}(\bi{r})_{ij}+ \bar{\omega}(\bi{r})_{ij}
\label{Jacobian}.
\end{equation}

Actual deformations do not involve rigid-body rotations, i.e. $\bar{\boldsymbol{\omega}}(\bi{r})=0$. This way, equations (\ref{Deltan}) and  (\ref{Jacobian}) yield \cite{Kitt2013}:
\begin{equation}
\boldsymbol{\delta}_n^{\prime}=(\bar{\bi{I}}+\bar{\boldsymbol{\epsilon}}(\bi{r}))\cdot\boldsymbol{\delta}_n\quad(n=1,2,3)
\label{DeltaGeneral}.
\end{equation}

Some particular cases of strain will be next discussed to illustrate their consequences on the lattice and on diffraction; this discussion will be useful when discussing electronic and optical properties later on.

One of the most illustrative cases of strain is the uniform case, for which the positions of carbon atoms at location $\boldsymbol{r}$ are transformed by:
\begin{equation}
\boldsymbol{u}(\boldsymbol{r})=\bar{\boldsymbol{\epsilon}}\cdot\boldsymbol{r},
\end{equation}
where $\bar{\boldsymbol{\epsilon}}$ is a uniform strain
tensor, i.e. its components are position-independent:
\begin{equation}
 \bar{\boldsymbol{\epsilon}}=\left(\begin{array}{cc}
\epsilon_{Z} & \gamma_S \\
\gamma_S & \epsilon_{A}
\end{array} \right)
\label{EpsilonGraphene}.
\end{equation}
In previous expression, the space-independent parameters $\epsilon_A$ and $\epsilon_Z$ denote the uniaxial strain
applied along the zigzag and armchair directions, and $\gamma_S$ is the
shear strain, respectively.

\begin{figure}
\centering
\includegraphics[width=\linewidth]{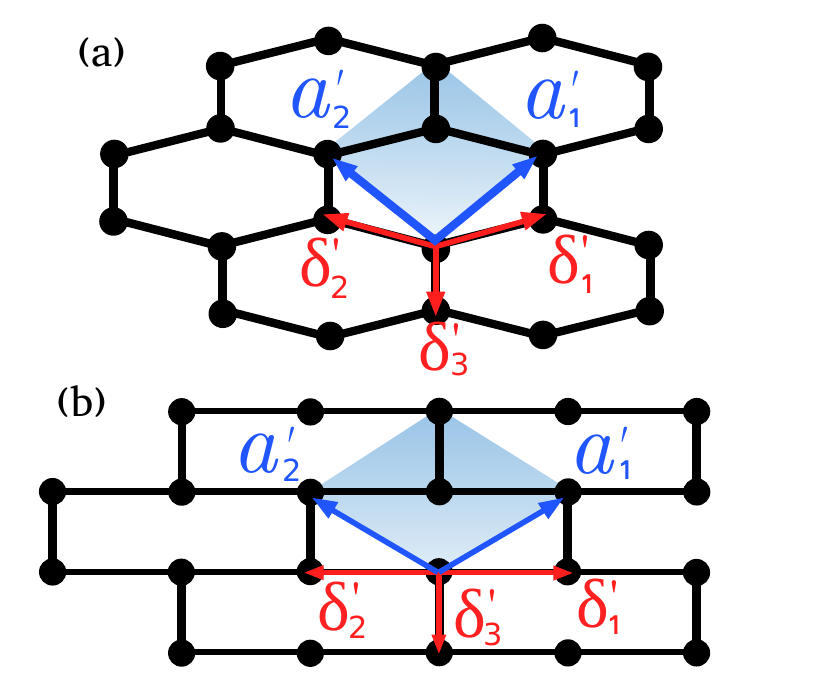}
\caption {Two examples of deformed lattices in which the deformed unit vectors $\boldsymbol{a}_1^{\prime}$, $\boldsymbol{a}_2^{\prime}$ and
the first-neighbour vectors $\boldsymbol{\delta}_1^{\prime},\boldsymbol{\delta}_2^{\prime},\boldsymbol{\delta}_3^{\prime}$ are indicated. The corresponding
unit cell is indicated by the shadowed area. In panel (a), the graphene lattice is slightly
distorted, while panel (b) shows the limiting case of a brick-wall lattice. Networks (a) and (b)
share the same topology.}
\label{Fig_Deformedlattices}
\end{figure}

With this field, new atomic positions are:
\begin{equation}
\boldsymbol{r}^{\prime}=(\bar{\bi{I}} + \bar{\boldsymbol{\epsilon}})\cdot\boldsymbol{r},
\end{equation}
where $\bar{\bi{I}}$  is the $2\times2$ identity matrix. According to figure \ref{Fig_Deformedlattices},
 lattice vectors are transformed as:
\begin{equation}
\boldsymbol{a}_i^{\prime}=(\bar{\bi{I}}+\bar{\boldsymbol{\epsilon}})\cdot\boldsymbol{a}_i \quad (i=1,2)
\label{newunitary},
\end{equation}
and using equation (\ref{DeltaGeneral}), the set of transformed space-independent first-neighbour vectors turn into:
\begin{equation}
\boldsymbol{\delta}_n^{\prime}=(\bar{\bi{I}}+\bar{\boldsymbol{\epsilon}})\cdot\boldsymbol{\delta}_n\quad(n=1,2,3)
\label{DeltaUniform}.
\end{equation}

The functional dependence of $\boldsymbol{\delta}_n^{\prime}$ on $\boldsymbol{r}$ was dropped out in equation (\ref{DeltaUniform}), as all first-neighbour vectors are deformed by the same amount under a uniform strain.

Equations (\ref{newunitary}) and (\ref{DeltaUniform}) indicate that strain changes the Bravais lattice
through $\boldsymbol{a}_i^{\prime}$ and the graphene space group as well, due to the
modification of decoration vectors $\boldsymbol{\delta}_n^{\prime}$.

Equation (\ref{DeltaUniform}) implies that Cauchy-Born rule applies, which will be the case for graphene under uniform biaxial strain. However, for crystals like graphene that are characterized by a lattice with a
basis, the strained nearest-neighbor vectors are transformed following a sublattice-dependent rule due to the additional degrees of freedom introduced by the basis atoms. According to Midtvedt \textit{et al.} \cite{Midtvedt2016}, equation (\ref{DeltaUniform}) should be generalized into $\boldsymbol{\delta}_{n}^{\prime}=(\bar{\bi{I}}+\bar{\boldsymbol{\epsilon}})\cdot\boldsymbol{\delta}_{n}+
\boldsymbol{\Delta}$, where the vector $\boldsymbol{\Delta}$ is sublattice-dependent. This point is also emphasized by Zhou and Wang \cite{Zhou} when they indicate that the deformation field for the two atoms within the blue-shaded unit cell follow stresses that must pull them in \textit{ opposite} vertical directions in bringing a honeycomb lattice (figure \ref{Fig_Deformedlattices}(a)) onto a brick lattice (figure \ref{Fig_Deformedlattices}(b)). The pull in opposite directions for the two atoms in the unit cell is a clear example of Cauchy-Born rule violation \cite{Ericksen2008}, which occurs on lattices with a basis such as graphene, and has not been mentioned in any review of strain in graphene thus far.

\begin{figure}
\centering
\includegraphics[width=\linewidth]{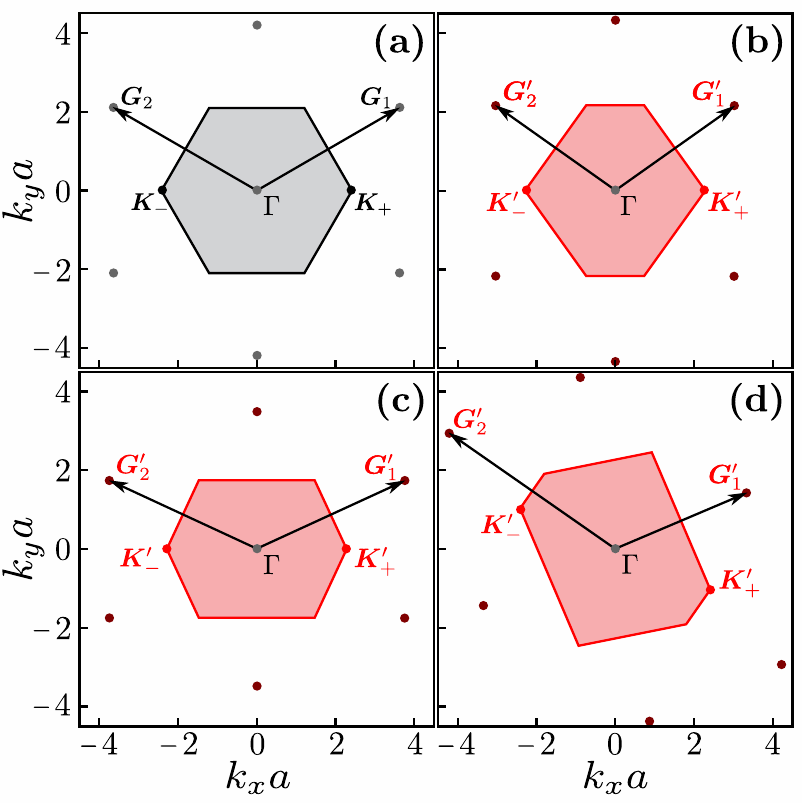}
\caption {1BZ and reciprocal lattice vectors for representative cases of uniform strain: (a) unstrained graphene, (b) uniaxial  strain along the zigzag-direction (zigzag strain) with $\bar{\epsilon}_{xx}\equiv\epsilon_Z=0.2$, $\bar{\epsilon}_{yy}\equiv\epsilon_A=-\nu\bar{\epsilon}_{xx}$, $\bar{\epsilon}_{xy}\equiv\gamma_S=0$, (c) uniaxial armchair strain with ${\epsilon}_A=0.2$, ${\epsilon}_{Z}=-\nu\bar{\epsilon}_{yy}$, $\gamma_{S}=0$ and (d) shear strain with ${\epsilon}_{Z}={\epsilon}_{A}=0$ and ${\gamma}_{S}=0.2$. (The Poisson ratio $\nu$ was set to 0.165.)}
\label{NewBZ}
\end{figure}

The reciprocal lattice is deformed by strain too: using the deformed cell unit vectors given by equation~(\ref{newunitary}), the corresponding new reciprocal lattice vectors are given by:
\begin{equation}
\boldsymbol{G}_i^{\prime}=(\bar{\bi{I}}+\bar{\boldsymbol{\epsilon}})^{-1}\cdot\boldsymbol{G}_i \quad (i=1,2).
\label{UniformGvectors}
\end{equation}
Figure \ref{NewBZ} shows reciprocal lattice vectors for some representative cases of uniform strain.

Following equations (\ref{Laue1}), (\ref{Laue2}) and (\ref{Laue3}), the new high-symmetry points in the corners of the 1BZ of the uniformly strained reciprocal lattice are also obtained by the Wigner-Seitz construction of the primitive cell (figure \ref{strainedlattice}(b)), resulting in the following positions \cite{Gomez2016}:
\begin{equation}
  \boldsymbol{K}_{+}^{\prime}=\bar{\bi{M}_1}^{-1}\cdot\boldsymbol{C}_{+} \quad \\
  \quad \boldsymbol{K}_{-}^\prime=\bar{\bi{M}_2}^{-1}\cdot\boldsymbol{C}_{-}
 \label{K},
 \end{equation}
with:
\begin{equation}
 \bar{\bi{M}_i}=\left(\begin{array}{cc}
(G_i^\prime)_x & (G_i^\prime)_y \\
(G_1^\prime)_x+(G_2^\prime)_x & (G_2^\prime)_y+(G_2^\prime)_y
\end{array} \right),
\end{equation}
 and:
\begin{equation}
 \boldsymbol{C}_{\pm}=\frac{1}{2}\left(\begin{array}{c}
\pm |\boldsymbol{G}_{i}^\prime|^2  \\
\mp |\boldsymbol{G}_1^\prime+\boldsymbol{G}_2^\prime|^{2}
\end{array} \right).
\end{equation}
Here, $(G_i^\prime)_x$ and $(G_i^\prime)_y$ are the $x$ and $y$
components of the deformed reciprocal vectors  $\boldsymbol{G}_i^{\prime}$
for $i=1,2$.

Figure~\ref{NewBZ} presents the shape of the 1BZ under different types of uniform strain, including the corresponding high-symmetry points. To first order in the strain tensor, it can
be demonstrated that high symmetry point positions are given by \cite{Pereira2009a}:
\begin{equation}
 \boldsymbol{K}_{\pm}^{\prime}=\pm \frac{4\pi}{3\sqrt{3}a}(1-{\epsilon}_{Z}/2 -{\epsilon}_{A}/2, -2{\gamma}_{S}).
\end{equation}

Finally, the diffraction pattern of the uniform deformed lattice is a scaling of equation (\ref{FTgraphene}):
\begin{equation}
 |\widetilde{V}_{gpu}(\boldsymbol{k})|^{2}=\sum_{\boldsymbol{G}^{\prime}} 4V_0^{2} \cos^{2} \left(\frac{ \boldsymbol{k} \cdot \boldsymbol{\delta}_1^{\prime}}{2}\right)\delta(\boldsymbol{k}-\boldsymbol{G}^{\prime}),
\end{equation}
where  $\boldsymbol{G}^{\prime}$ is a reciprocal lattice vector modified by strain. This creates diffraction peaks at the following locations in reciprocal space:
$\boldsymbol{G}^{\prime}=l\boldsymbol{G}_1^{\prime}+h\boldsymbol{G}_2^{\prime}$,
having the same amplitudes as in undeformed graphene, since 
$\boldsymbol{G}_{i}^{\prime}\cdot \boldsymbol{\delta}_{1}^{\prime}=\boldsymbol{G}_{i}\cdot \boldsymbol{\delta}_{1}$.

Another important case that will be discussed now is that of a periodic strain field that arises when graphene lays over a substrate like Ir, Fe or hBN. In this case, the
resulting structure forms a superlattice that is usually described as a modulated crystal.

A periodic strain field can be expressed as a Fourier series:
\begin{equation}
\boldsymbol{u}(\boldsymbol{r})=\sum_{ \Delta \bf{g}} \widetilde{\boldsymbol{u}}(\Delta \boldsymbol{g}) e^{i \Delta \bf{g} \cdot \bf{r}},
\end{equation}
using an expansion over certain wavevectors $\Delta \boldsymbol{g}$, where $\widetilde{\boldsymbol{u}}(\Delta \boldsymbol{g})$ is the amplitude of each wavevector.
The question here is how to choose the vectors $\Delta \boldsymbol{g}$: extracting $\Delta \boldsymbol{g}$ from experimental data can be a quite
complex task due to the presence of beatings \cite{Artaud2016}.

From a theoretical point of view, periodicity holds if two linearly independent vectors $\boldsymbol{T}_1$ and  $\boldsymbol{T}_2$ exist such that:
\begin{equation}
\boldsymbol{u}(\boldsymbol{r})=\boldsymbol{u}(\boldsymbol{r}+\boldsymbol{T}_1)=\boldsymbol{u}(\boldsymbol{r}+\boldsymbol{T}_2).
\end{equation}
In the above case, two reciprocal lattice basis vectors $\Delta \boldsymbol{g}_1$ and $\Delta \boldsymbol{g}_2$ can be found from the standard orthogonality relation:
\begin{equation}\label{orthogonalitysupercell}
\Delta \boldsymbol{g}_{i}\cdot \boldsymbol{T}_{j}=2\pi\delta_{i,j},
\end{equation}
where $i,j=1,2$. Then $\Delta \boldsymbol{g}=\Delta \boldsymbol{g}(s_1,s_2)\equiv s_1 \Delta \boldsymbol{g}_1+s_2 \Delta \boldsymbol{g}_2$
for $s_1$ and $s_2$ integers.

The resulting modulated crystal is said to be commensurate when the vectors $\boldsymbol{T}_1$ and  $\boldsymbol{T}_2$  fall into the graphene lattice:
\begin{eqnarray}
 \boldsymbol{T}_1=n_1\boldsymbol{a}_1+n_2\boldsymbol{a}_2,\nonumber\\
 \boldsymbol{T}_2=m_1\boldsymbol{a}_1+m_2\boldsymbol{a}_2,
 \label{Supercondition}
\end{eqnarray}
for some integers $n_1$, $n_2$, $m_1$ and $m_2$.   Commensurability (expressed in equation (\ref{Supercondition})) implies that
a supercell can be defined. In building that supercell, one chooses values for $n_1$, $n_2$, $m_1$ and $m_2$ to employ the smallest number of unit cells of both graphene and its supporting substrate \cite{Artaud2016}. (The system behaves as a quasicrystal when no integer solutions for equation (\ref{Supercondition}) exist, since in that case it has at least two incommensurate lengths \cite{Janot}. Quasicrystals are crystals with classical forbidden symmetries, that are described using more reciprocal lattice vectors than the dimensionality of the physical space \cite{Janot}.)

In the periodic and modulated supercell, the reciprocal lattice and the diffraction pattern can be found as follows. Lattice positions under a periodic deformation obey:
\begin{equation}
 \widetilde{V}(\boldsymbol{r})=V_0 \sum_{l} \delta(\boldsymbol{r}-\boldsymbol{r}_l-\boldsymbol{u}(\boldsymbol{r}_l)),
\end{equation}
(compare to equation (\ref{diff_8})) which can also be expressed as:
\begin{equation}
  \widetilde{V}(\boldsymbol{r})=V_0 \sum_{\boldsymbol{r_l}}  e^{i \bf{k} \cdot (\bf{r_l}+\boldsymbol{u}(\boldsymbol{r}_l))}.
\end{equation}\label{eq37}
Periodicity of the strain field permits expanding equation (\ref{eq37}) as a Fourier series:
\begin{equation}
  e^{i \bf{k} \cdot \boldsymbol{u}(\boldsymbol{r}_l)}=\sum_{\Delta \bf{g}} \widetilde{U}_{\boldsymbol{k}}(\Delta \boldsymbol{g}) e^{-i \Delta \bf{g} \cdot \bf{r}_l},
  \label{expu}
\end{equation}
where $\Delta \boldsymbol{g}=s_1\Delta \boldsymbol{g}_1+s_2\Delta \boldsymbol{g}_2$ are the wavevectors determined from the periodicity of $\boldsymbol{u}(\boldsymbol{r}_l)$, while
$\widetilde{U}(\boldsymbol{g})$ are the coefficients of the expansion, found by projecting $e^{i \bf{k} \cdot \boldsymbol{u}(\boldsymbol{r}_l)}$
into the kernel of the transformation:
\begin{equation}
 \widetilde{U}_{\boldsymbol{k}}(\Delta \boldsymbol{g})=\sum_{\boldsymbol{r}_l}e^{i \bf{k} \cdot \boldsymbol{u}(\boldsymbol{r}_l)}e^{i \Delta \bf{g} \cdot \bf{r}_l}.
\end{equation}

Using equation (\ref{expu}), the Fourier transform becomes:
\begin{equation}
 \widetilde{V}(\boldsymbol{k})=V_0 \sum_{\Delta \boldsymbol{g},\boldsymbol{r_l}} \widetilde{U}_{\boldsymbol{k}}(\Delta \boldsymbol{g}) e^{i( \bf{k}- \Delta \bf{g})\cdot \bf{r_l}}.
\end{equation}

 One can also use  equation  (\ref{FTgraphene}) to rewrite the preceding equation as:
\begin{equation}
 \widetilde{V}(\boldsymbol{k})= \sum_{\Delta \boldsymbol{g}} \widetilde{U}_{\boldsymbol{k}}(\Delta \boldsymbol{g})\widetilde{V}_{gp}(\boldsymbol{k}-\Delta \boldsymbol{g}) ,
\label{FinalConvolution}
 \end{equation}
which is a convolution of two functions, the Fourier transform of the graphene structure and the modulation Fourier transform, a result well known for modulated crystals or quasicrystals \cite{Janot,Naumis1998,NaumisThorpe1999,NaumisHierarchy2005}. Equation (\ref{FinalConvolution}) can also be obtained using projections of a higher dimensional lattice \cite{Janot,NaumisAragon2003,Janssen2004} within the cut and projection method \cite{Janot,Macia2006}.

\begin{figure}
\centering
\includegraphics[width=0.9\linewidth]{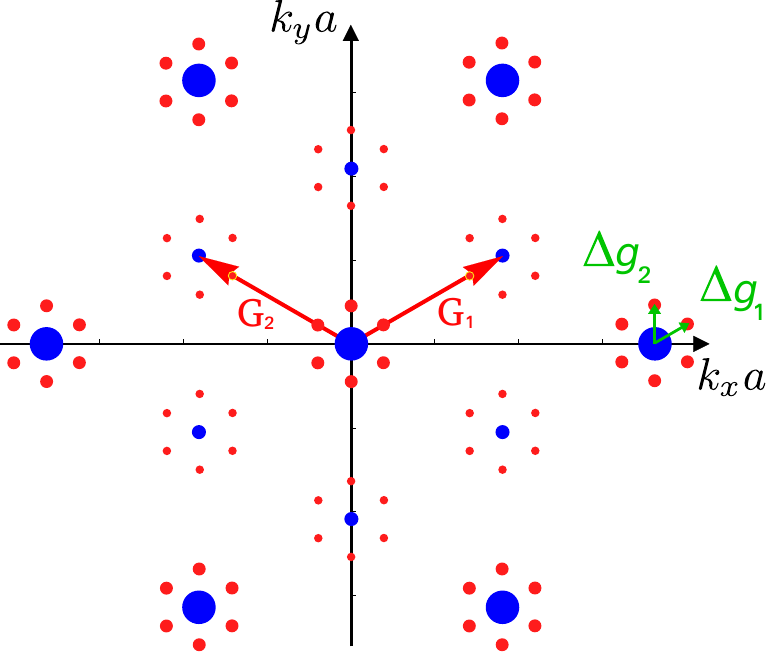}
\caption{Theoretical diffraction pattern for periodically strained graphene, as obtained from equation (\ref{ModulatedDiffraction}).
On each graphene's diffraction spot (indicated in blue), phason satellites
indicated in red appear due to the periodic modulation. The original reciprocal lattice vectors of graphene, and the new set of reciprocal superlattice vectors are indicated in capital (red) and lowercase (green) letters. Notice that here only first harmonics phason satellites are displayed, i.e. $\Delta \boldsymbol{g}=s_1\Delta \boldsymbol{g}_1+s_2\Delta \boldsymbol{g}_2$ for $s_1=\pm1$ and $s_2=\pm1$.
(Spot sizes represent relative intensity.)}
\label{FigModulatedDiffraction}
\end{figure}

Using equations (\ref{PattersonGraphene}) and (\ref{FinalConvolution}), the diffraction pattern turns out to be:
\begin{eqnarray}
  |\widetilde{V}(\boldsymbol{k})|^{2}= \sum_{\boldsymbol{G},\Delta \boldsymbol{g}}4V_0^{2} \cos^{2}\left(\frac{\boldsymbol{k}\cdot\delta_{1}}{2} \right)\nonumber\\
\qquad \times
 |\widetilde{U}_{\boldsymbol{k}}(\Delta \boldsymbol{g})|^{2}\delta \left(\boldsymbol{k}-[\boldsymbol{G}+\Delta \boldsymbol{g}]\right),
\label{ModulatedDiffraction}
 \end{eqnarray}
i.e. the diffraction pattern --shown in figure \ref{FigModulatedDiffraction}-- contains the graphene's spots displayed in figure \ref{RedDiffraction} (for $\Delta \boldsymbol{g}=0$), plus extra spots at positions $\boldsymbol{G}+\Delta \boldsymbol{g}$. The generic
features of figure \ref{FigModulatedDiffraction} do coincide with X-ray or LEEDS diffraction experiments of graphene on substrates (e.g. \cite{Diaye2008,Wang2016}). Given that the period of the superlattice tends to be much larger that graphene's lattice parameter  ($|\Delta \boldsymbol{g}|<<|\boldsymbol{G}|$),
the new diffraction spots in figure \ref{FigModulatedDiffraction} appear as satellites of the original graphene lattice.
The relative peak intensity is obtained by squaring the amplitudes given in equations (\ref{StructureFactorGraphene}) and (\ref{ModulatedDiffraction}):
\begin{eqnarray}
 |\widetilde{V}(\boldsymbol{\boldsymbol{G}+\Delta \boldsymbol{g}})|^{2}=4V_0^{2}
 |\widetilde{U}_{ \boldsymbol{G}+\Delta \boldsymbol{g}    }(\Delta \boldsymbol{g})|^{2}\nonumber\\
\qquad\times \cos^{2}\left(\frac{\pi}{3}(2l-h)+\frac{(l+h)}{2}  \boldsymbol{\delta}_1\cdot \Delta \boldsymbol{g}\right).
\label{ModulatedFlh}
 \end{eqnarray}
 
This interference effect is akin to the beats appearing in acoustics when two waves of nearly similar frequencies are superposed.

If the modulation and the graphene periods are commensurate in the sense of equation (\ref{Supercondition}), satellite peaks will eventually coincide with points of the reciprocal lattice of graphene. But when commensurability does not hold, the diffraction pattern is densely filled with spots in a self-similar (i.e. fractal) way \cite{Janot}. Satellites are
related to new degrees of freedom known as phasons \cite{Lubensky1985,Janot,NaumisThorpe1999,NaumisHierarchy2005} that are Goldstone modes associated with the extra  broken-symmetries provided by the modulation, and have very different dynamics when contrasted with more usual  Goldstone modes like phonons \cite{Lubensky1985,Janssen2004}.

This Section ends with a  brief discussion of random strain fields. When $\bi{u}(\bi{r})$ is a random quantity with a given distribution, the description is similar to a crystalline lattice with noise due to thermal fluctuations. For example, the diffraction pattern has peaks at the
same position as in unstrained graphene, but with a certain width resulting from the convolution of the dispersion distribution. The width is determined by the mean square value of $\bi{u}(\bi{r})$, as it happens with the Debye-Waller factor \cite{Ashcroft}.

\section{Deformations in graphene}\label{Deformations}

Other reviews provide an already exhaustive presentation of this topic \cite{Galiotis2015,Colin15,Liu2015,Akinwande2016}. For that reason, the mechanical properties of graphene are briefly reviewed here. First, a description of graphene's elastic properties is made, to then describe how to produce deformations.

\subsection{Elastic coefficients: Experimental and theoretical characterization}\label{elasticcoeffs}

Graphene is the thinnest elastic membrane in nature, with an exceptional stress-strain behaviour, including the highest stiffness and strength ever measured \cite{Lee2008,Bunch2008}. At the same time, it 
can be easily bent to get complex folded structures \cite{Bao2009,Kim2011} and can withstand elastic deformations of up to $25\,\%$ \cite{Lee2008}, that are much larger than in any other known material. Owing to these outstanding mechanical properties, graphene is an ideal candidate for nanomechanical systems \cite{Bunch2007,Wang2014,Mathew2016} and flexible electronic devices \cite{Kim2015,Jang16}.

Many of the mechanical properties of graphene can be understood from continuum mechanics \cite{Reddy2006,Scarpa2010,Wei2009,Colombo2011,Lindahl2012}. Within the theory of linear elasticity for two-dimensional (2D) membranes \cite{Landau1986,Nelson2004}, the elastic energy density  $\mathcal{U}$ (energy per unit area) of a strained graphene membrane is given by:
\begin{equation}\label{EED}
2 \mathcal{U}=\frac{\mathcal{E}}{1+\nu}\Tr(\bar{\boldsymbol{\epsilon}}^{2}) + \frac{\mathcal{E}\nu}{1-\nu^{2}}(\Tr\bar{\boldsymbol{\epsilon}})^{2},
\end{equation}
where $\bar{\boldsymbol{\epsilon}}$ is the rank-two strain tensor, $\mathcal{E}$ is the Young's modulus and $\nu$ is the Poisson's ratio (that was introduced in figure \ref{NewBZ}). The stress-strain relation for in-plane deformations is obtained from equation (\ref{EED}), by setting $\bar{\bi{s}}=\partial\mathcal{U}/\partial\bar{\boldsymbol{\epsilon}}$, and keeping in mind that the strain tensor $\bar{\boldsymbol{\epsilon}}(\bi{r})$ may be position-dependent.

\begin{figure}
\centering
\includegraphics[width=\linewidth]{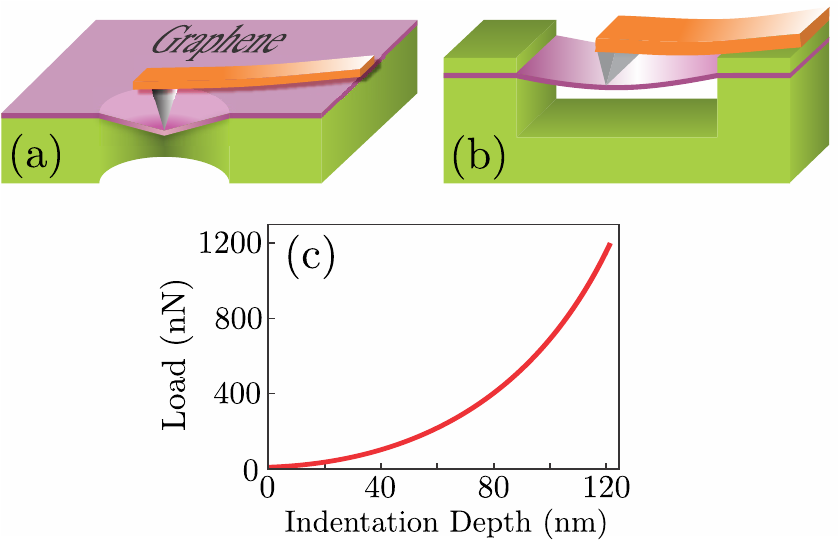}
\caption {Schematics of an atomic force microscopy nanoindentation experiments of suspended graphene membranes for (a) circular and (b) ribbon configurations. (c) Typical force-displacement curve of an indented graphene membrane.}
\label{Fig_Indentation}
\end{figure}

From an experimental point of view, the characterization of the elastic coefficients $\mathcal{E}$ and $\nu$ is challenging owing to difficulties related to imposing a measurable uniform stress, and to technical issues in handling membranes of monoatomic thickness.  Lee \textit{et al.} \cite{Lee2008} reported the value of Young's modulus of graphene as obtained from nanoindentation experiments. As depicted in figure \ref{Fig_Indentation}(a), graphene was suspended over a circular cavity and indented by the tip of an atomic force microscope. The experimental force-displacement relation was subsequently fitted to the Schwering-type equation \cite{Begley2004,Komaragiri2005}:
\begin{equation}\label{Schwering}
F=\pi s_{0}\delta + \frac{\mathcal{E}}{r^{2}}\delta^{3},
\end{equation}
where $F$ is the applied force, $\delta$ is the indentation at the central point, $r$ is the radius of the drum and $s_{0}$ is the  pre-tension accumulated in the membrane during the preparation procedure. Typically, pre-tension values are small: $0.07-1.00$ N/m. Using a least-square fitting of equation (\ref{Schwering}) to the force-displacement curves, the Young's modulus was determined to be $\mathcal{E}=340 \pm 50\,\mbox{N/m}$ \cite{Lee2008}.
If the thickness of graphene is assumed to be $d=0.335\,\mbox{nm}$, the derived $\mathcal{E}$ corresponds to an ultrahigh 3D Young's modulus of $\mathcal{E}^{3D}=\mathcal{E}/d=1.0 \pm 0.1\,\mbox{TPa}$, which is close to the in-plane Young's modulus of bulk graphite $(1.02 \pm 0.03\,\mbox{TPa})$.

Lee and coworkers estimated the breaking strength of graphene as $42 \pm 4$ N/m, and discovered that the elastic response is highly nonlinear for strains above $10\,\%$.

Subsequently, these experiments were interpreted by Cadelano \textit{et al.} \cite{Cadelano2009} within a generalized nonlinear stress-strain relation for graphene that incorporates cubic terms in strain to equation (\ref{EED}). In addition to the Young's modulus and the Possion's ratio, this approach also estimated three nonlinear elastic coefficients
from atomistic simulations: nonlinear features play a crucial role in determining graphene's properties upon significant load and up to its mechanical failure.

Other nanoindentation works \cite{Huang2011,Lee2013,Lopez2015} report a Young's modulus $\mathcal{E}$ similar to the one found in \cite{Lee2008}. In particular, Huang \textit{et al.} \cite{Huang2011} performed nanoindentation experiments with a wedge-shaped tip on graphene ribbons having a geometry similar to the one depicted in figure \ref{Fig_Indentation}(b). On the basis of geometry, the force-displacement relation $F(\delta)$ was approximated to a suspended bridge model as follows \cite{Herbert2011}:
\begin{equation}\label{Herbert}
F=\frac{8 w s_{0}}{l}\delta + \frac{8 w \mathcal{E}}{l^{3}}\delta^{3},
\end{equation}
where $w$ and $l$ are the width and the length of the graphene ribbon, respectively. The Young's modulus was estimated to be $\mathcal{E}=335 \pm 20\,\mbox{N/m}$ from equation (\ref{Herbert}) \cite{Huang2011}. Even though atomic force microscope nanoindentation is the most employed method to characterize the elastic properties of graphene, other experimental techniques have also been utilized for this purpose, yielding similar results \cite{Bunch2008,Wong2010,Politano2015}.

On the experimental side, the Poisson's ratio of graphene $\nu$ has not been reported from a direct measurement, but Politano \textit{et al.} \cite{Politano2015} estimated $\nu=0.19$ for graphene grown on metal substrates. Computational estimates for $\nu$ are listed next.

The typically quoted value is $\nu=0.16$, which corresponds to the Poisson's ratio for graphite in the basal plane. However, a wide distribution of theoretical values for $\nu$ exists in the literature, ranging from $0.1$ to $0.4$. Atomistic Monte Carlo simulations show $\nu$ to be small ($\approx0.1$) in a broad temperature interval, being $0.12$ at room temperature \cite{Zakharchenko2009}. $\nu$ can be negative at higher temperature ($T\geq1700\,\mbox{K}$), so that graphene becomes an auxetic material. To emphasize the scatter on the estimates of $\nu$, an \textit{ab initio} calculation reveals an isotropic in-plane elastic response of graphene at small strains with a Poisson's ratio of $0.186$ which becomes anisotropic for large strain \cite{Liu2007}. Molecular dynamics (MD) simulations show a Poisson's ratio of $0.21$ for graphene, and the ratio significantly depends on the size and chirality in the case of graphene nanoribbons, with a larger value in the armchair direction than in the zigzag direction.

The bending rigidity $\kappa$ is another important elastic parameter used to characterize the performance of nanoelectromechanical graphene devices. Within the theory of elasticity for thin plates, the bending rigidity is determined to be $\kappa=\mathcal{E}_{3D}d^{3}/12(1-\nu^{2})$. Evaluating this expression with the parameters of graphite ($\mathcal{E}_{3D}\approx1\,\mbox{TPa},\ d\approx 0.34\,\mbox{nm},\ \nu\approx0.16$) yields $\kappa\approx20\,\mbox{eV}$, which is an order of magnitude larger than the value estimated for graphene ($\kappa\approx1.2\,\mbox{eV}$, as extracted from graphite's phonon spectrum).

\textit{Ab initio} calculations
predict $\kappa\approx1.46\,\mbox{eV}$, and an analytical estimation based on empirical potentials gives $\kappa\approx1.4\,\mbox{eV}$. By using a bond-orbital model, Zhang \textit{et al.} \cite{Zhang2011} demonstrated that the breakdown of the plate phenomenology for a graphene monolayer is due to the decoupling of bending and tensional deformations. At the same time, they reported a precise expression for the bending rigidity $\kappa$ of $n$-layered graphene for $n>2$, and argue that the thin-plate treatment can be applied to multilayer graphene in situations involving out-of-plane deformations and no layer-sliding.

\subsection{Strain patterns and methods to produce them}

Strain can either arise naturally or be produced in a controlled way in graphene. Strain can be produced by bending or elongating the graphene/substrate lattice, by lattice mismatch, and/or by thermal expansion mismatches between graphene and its supporting substrate. Compressive strain usually gives way to out-of-plane corrugations, resulting in the formation of
ripples and wrinkles \cite{Deng2016}. Several reviews cover these scenarios \cite{Roldan2015,Deng2016,Zhang2015}, but two topics are next discussed given their relevance in subsequent Sections.

\subsubsection{Graphene on substrates: moir\'e superlattices.}

When on a substrate, graphene experiences strain due to surface corrugations or to a lattice/rotational mismatch with respect to the supporting substrate \cite{Artaud2016,Ponomarenko2013}. Using lattice and rotational mismatch, uniaxial periodic strain \cite{BaiKeKe2014} and two-dimensional structures known as moir\'e patterns \cite{Artaud2016} can be produced. The structural mismatch between graphene and its support results in a superlattice over a distance know as the moir\'e period,
which usually ranges from $1$ to $20\,\mbox{nm}$ \cite{BaiKeKe2014,Artaud2016}: as it was discussed in Section \ref{DescriptionStrain}, a new length scale is thus introduced by the superlattice.

Moir\'e superlattices provide a powerful strategy
to engineer electronic and optical properties when the interaction between graphene and the substrate is weak and no covalent bonds are formed \cite{Artaud2016,Diaye2008,Fabien2015}.

\begin{figure}
\centering
\includegraphics[width=\linewidth]{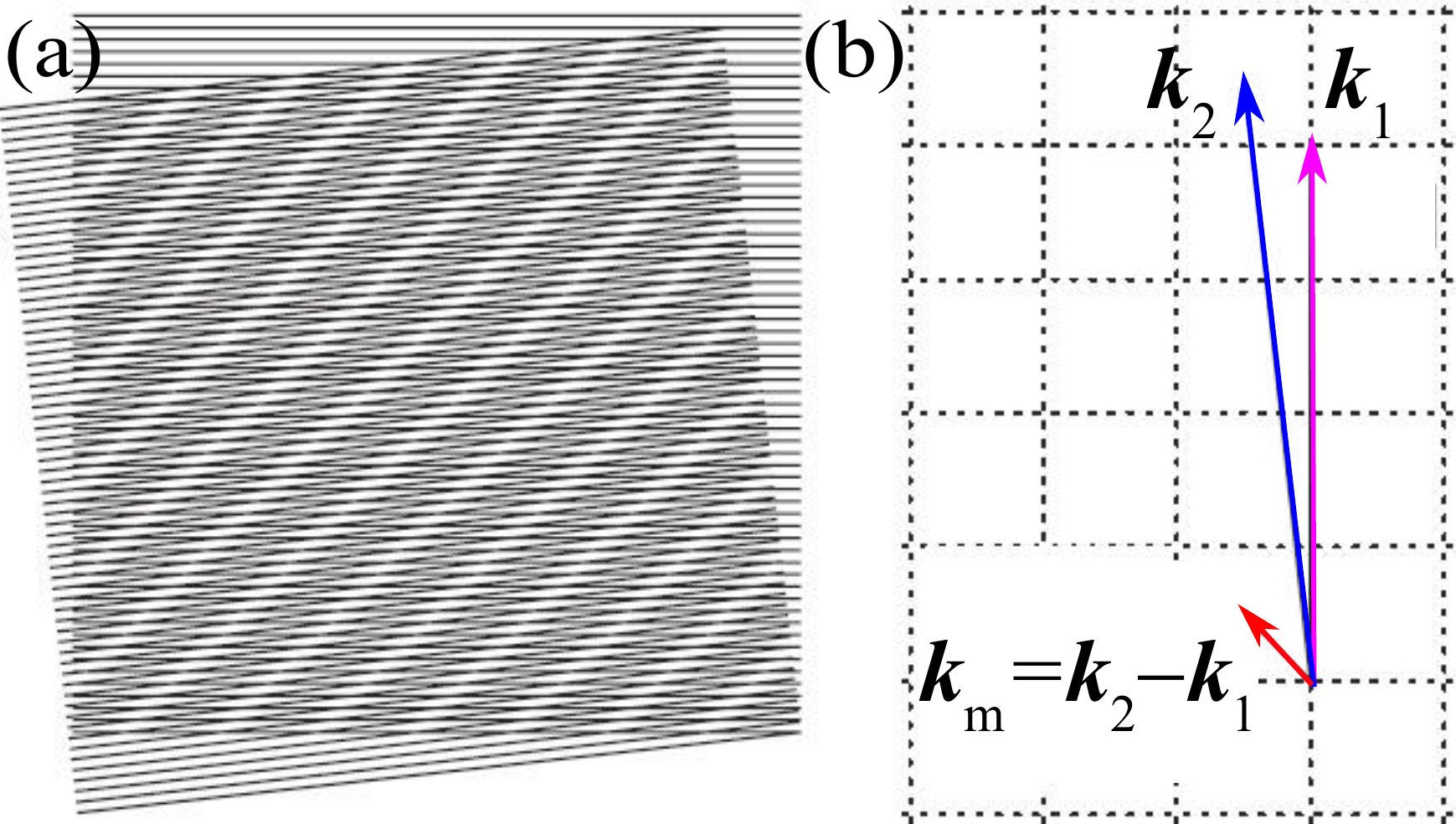}
\caption { (a) Two superposed patterns with a small lattice parameter and angular differences lead to a moir\'e superlattice, seen here as a striped diagonal pattern with a larger periodicity than the original lattices. (b) Wave vectors associated with the superposed lattices of (a). The wavevector of the resulting moir\'e is given by
the difference between the original wavevectors. Adapted from \cite{Diaye2008} with permission.}
\label{moireexample}
\end{figure}

In figure \ref{moireexample}(a), a moir\'e pattern is created by the superposition of two lattices with a small lattice and rotational mismatch. This leads to the observed striped diagonal pattern that has a larger spatial period and is analogous to the beating phenomena observed when two sound-waves of slightly different frequencies are superposed. As shown in figure~\ref{moireexample}(b), the wavevector associated with this lattice, denoted by $\boldsymbol{k}_m$,
is given by $\boldsymbol{k}_m=\boldsymbol{k}_{2}-\boldsymbol{k}_{1}$, where $\boldsymbol{k}_{1}$ and
$\boldsymbol{k}_{2}$ are wavevectors corresponding to each of the superposing lattices \cite{Diaye2008}. These wavevectors are perpendicular to the stripes seen in figure \ref{moireexample}(a), and their magnitudes are proportional to the inverse of the lattice parameter. There are many practical issues to take into account in graphene over substrates, and the description of the moir\'e pattern can become quite complex \cite{Artaud2016}.

\begin{figure}
\centering
\includegraphics[width=\linewidth]{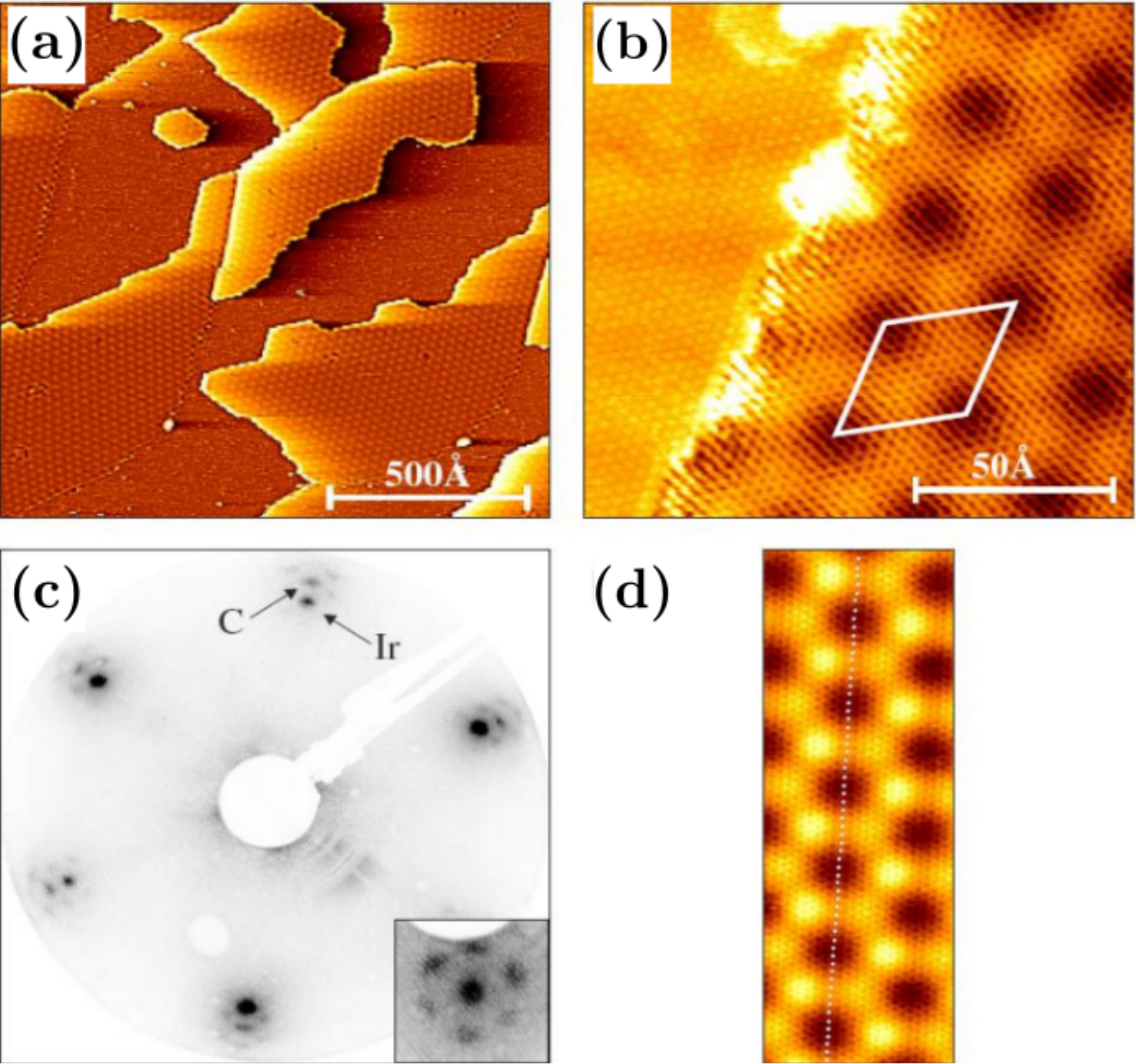}
\caption {(a) Graphene on Ir$(111)$ as seen on an electronic microscope. (b) A moir\'e pattern with a periodicity of $25.3\,\mbox{\AA}$ can be seen on a graphene flake
attached to a Ir step: the unit cell of the superlattice is represented by a rhombus. (c) Low energy electron diffraction pattern (LEED) showing two main spots corresponding to graphene (C) and Ir$(111)$ (Ir). Smaller satellites reflect the moir\'e periodicity. These spots arise from reciprocal lattice vectors of the supercell, which depend upon the difference between the graphene and Ir$(111)$ reciprocal lattice vectors.  (d) A strip of the superstructure, where a carbon row is shown in white. Reproduced from  \cite{Diaye2008} with permission.}
\label{moiregrapheneIr}
\end{figure}

To make a point about the alluded complexity, consider graphene on Ir$(111)$ as obtained \textit{via} the pyrolytic cleavage of ethylene \cite{Diaye2008} (see \cite{Hattab2012} too). Figure \ref{moiregrapheneIr}  presents the resulting superstructure, as obtained from scanning tunneling microscopy (STM) and low energy electron diffraction (LEEDS) \cite{Diaye2008}. There is a slight lattice mismatch between the graphene
and Ir$(111)$ lattices, resulting in a moir\'e pattern with a repeat distance of $2.53\,\mbox{nm}$ \cite{Diaye2008}.
Further refinement indicates that this is an excellent approximation (although the actual structure actually comprises three beatings instead of two \cite{Artaud2016}).

Figure \ref{moiregrapheneIr}(b) shows the moir\'e superlattice within a white rhombus. As explained in Section \ref{DescriptionStrain}, the diffraction pattern of a superlattice is given by a decoration of  graphene's diffraction pattern with phason satellites, and figure \ref{moiregrapheneIr}(c) --a diffraction pattern of graphene on top of  Ir$(111)$, obtained from the  LEEDS experiment-- represents experimental realization of figure \ref{FigModulatedDiffraction}. The satellites that originate due to the periodic modulation are
clearly seen and reveal the superlattice periodicity; they are spaced according to the wavevectors $\boldsymbol{k}_m$, which have a much more smaller norm than the substrate and graphene's reciprocal lattices.
Rotations over the substrate provide a way to tailor the periodicity of the superlattice further \cite{Ponomarenko2013}.

Graphene suffers four geometrical transformations as it is twisted, strained and sheared with respect to its substrate:  an isotropic rescaling, a directional rescaling in two directions, and a rotation. In mechanical terms, the isotropic rescaling produces biaxial strain, while the directional rescaling leads to uniaxial strain. For graphene on Ir$(111)$, the biaxial compression is estimated as $\epsilon \approx -0.29\,\%$ and the uniaxial compression as $\epsilon \approx -0.41\,\%$ \cite{Artaud2016}.

\subsubsection{Ripples and bending.}\label{RipplesDeformation}

Ripples and bending are vertical displacements of the atoms from a plane that are usually
described by a height $z(\boldsymbol{r})$ as a function of atomic position.
Although ripples and bending are usually treated as different from strain, they are coupled to strain by geometry.
Graphene is a soft-material in the sense that it wrinkles, it can be folded, and it
is even possible to do origami with it \cite{Pereira2009b,Kamien2014,Blees2015,Grosso2015,Castlee2016}. In fact, it is difficult to grow perfectly flat graphene \cite{Meyer2007, meyersolid07, stolyarova07, Vinogradov12}.
Graphene exhibits a high asymmetry
in tensile versus compressive strain: while the carbon-carbon bond length can be increased up to $25\,\%$ the carbon bond is almost incompressible, as  compressive stress rather induces out-of-plane deformations.

It has been observed that growing graphene on an anisotropic substrate produces one-dimensional periodic ripples \cite{BaiKeKe2014}, and
MD simulations are helpful to clarify how ripples and strain are intertwined
in suspended graphene sheets \cite{Sloan2013,Barraza2013,Monteverde2015}.

Two types of ripples originate to relax external strain in suspended graphene.  The first is a one-dimensional
ripple that is orthogonal to the strain front, if external strain is applied along a single direction.
The second type has a two-dimensional sinusoidal shape, and emerges when strain is applied  along two orthogonal directions simultaneously \cite{Monteverde2015}.

\begin{figure}
\centering
\includegraphics[width=\linewidth]{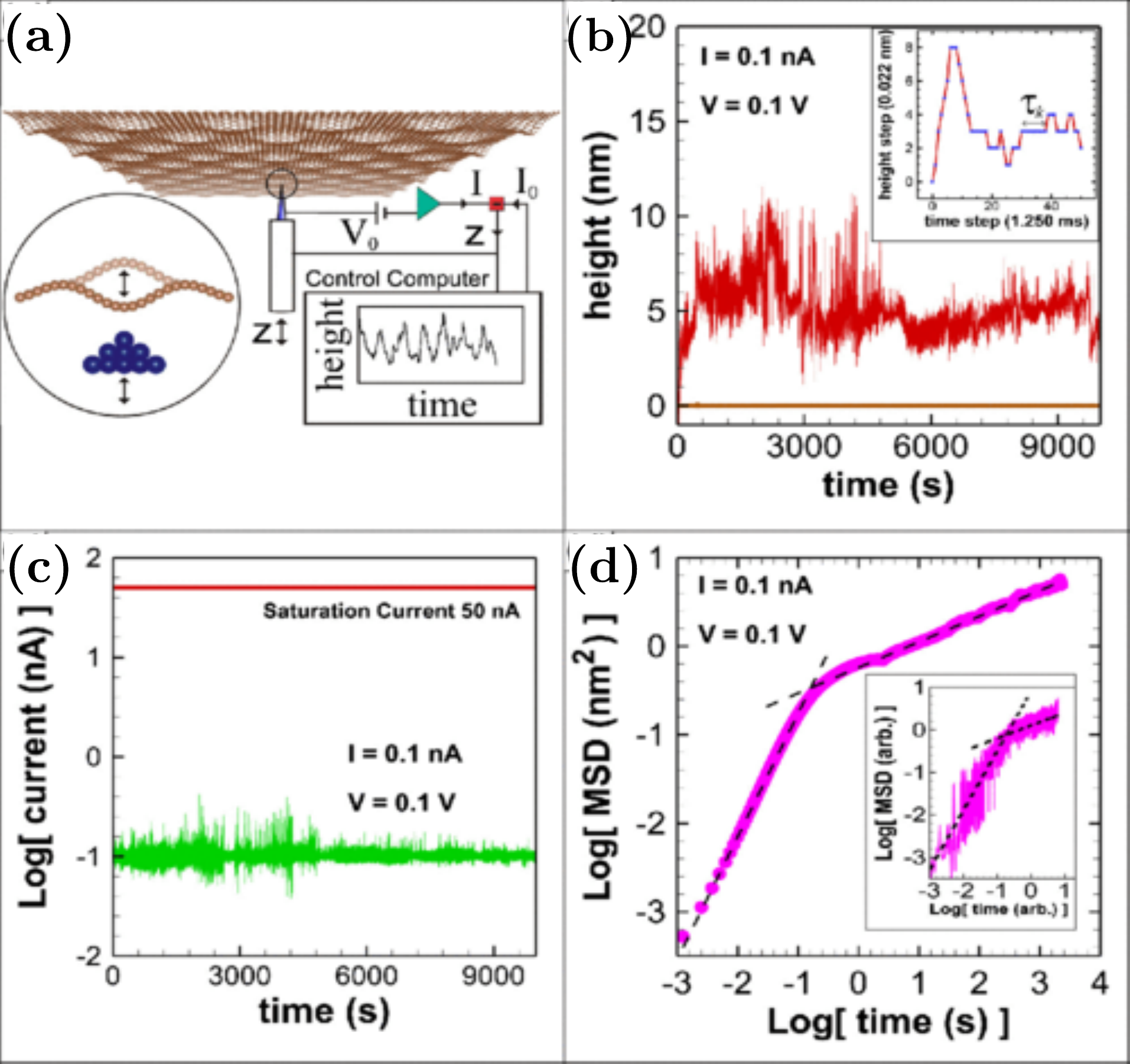}
\caption {Height dynamics of freestanding graphene: (a) Experimental setup using a scanning tunnel microscope with an inset showing
a zoom of the microscope tip and the graphene membrane vibrations. (b) Time trace of membrane height (above) and from a rigid sample (below). The inset shows a zoom of the same trace height. (c) Tunneling current as a function of time. (d) Mean-squared displacement  of membrane height as a function of time.
Dashed lines are fits with the result of a simulation using exponential wait times and Cauchy jump lengths.
Reproduced from \cite{Ackerman2016} with permission. Copyrighted by the American Physical Society.}
\label{Freestanding}
\end{figure}

The typical height of the ripples is between $5\,\mbox{\AA}$ and $10\,\mbox{\AA}$ \cite{Sloan2013,Barraza2013,Monteverde2015,Ackerman2016}, with a  wavelength of around 10 nm. However, the out-of-plane dynamics of freestanding samples display complexity. A recent experiment with a scanning tunneling microscope (STM) measured  the vertical motion of  graphene \cite{Ackerman2016}. The set-up of the experiment as well as the reported measurements of height $z(\boldsymbol{r})$, the tunnel current and the mean quadratic displacement
of the membrane are shown in figure \ref{Freestanding}. The dynamics exhibits rare long-scale excursions reflected in the anomalous mean-squared displacements and
Cauchy-Lorentz power law jump distributions \cite{Ackerman2016}. This random quivering of graphene
membranes has been proposed to generate electricity for nanomachines.

\section{Electronic properties}\label{Electronics}

The electronic quality of a material is determined from its charge carrier mobility $\mu$, that must be complemented by the charge concentration $n$ in graphene, since charge carriers can be tuned continuously between electrons and holes by
 external electric fields \cite{Geim2007}.

For typical carrier concentrations of $n \approx 10^{11}\,\mbox{cm}^{-2}$, values of $\mu$ exceeding  $1.0\times10^{5}\,\mbox{cm}^{2}\mbox{V}^{-1}\mbox{s}^{-1}$ (at room temperature) and $1.0\times10^{6}\,\mbox{cm}^{2}\mbox{V}^{-1}\mbox{s}^{-1}$ (at liquid-helium temperatures) have been measured on suspended graphene \cite{Bolotin2008,Castro2009}.  These suspended devices are extremely fragile and difficult to anneal  \cite{Mayorov2011}. Flexural modes, which are out-of-plane membrane vibrations, produce most of the electronic scattering once extrinsic defects are removed \cite{Castro2009}. In suspended graphene, a significant amount of strain is needed to suppress flexural mode scattering \cite{Castro2009}. A very good compromise between $n$, $\mu$, and the ease of building the experimental set-up
is achieved by using hBN as the substrate \cite{Mayorov2011}, in which $n \approx 10^{11}\,\mbox{cm}^{-2}$ with a reported value of $\mu$ equal to $1.0\times10^{5}\,\mbox{cm}^{2}\mbox{V}^{-1}\mbox{s}^{-1}$.

As a comparison,
some undoped (intrinsic) semiconductors like InSb exhibit a value of  $\mu$ as high as $7.7\times10^{4}\,\mbox{cm}^{2}\mbox{V}^{-1}\mbox{s}^{-1}$ at room-temperature \cite{Geim2007}, and typical doped semiconductors like n-Ge can reach $5.0\times10^{3}\,\mbox{cm}^{2}\mbox{V}^{-1}\mbox{s}^{-1}$.
When high quality graphene obtained by mechanical cleavage on top of an oxidized Si wafer is used, this extreme electronic quality  translates into a mean free path  $l=(h/e)\mu (n/\pi)^{1/2}$
of the order of $100$ nm for $n \approx 10^{12}\,\mbox{cm}^{-2}$, where $h$ is Planck's constant and $e$ the electron charge \cite{Mayorov2011}. The same extreme electronic quality is behind the observation of ballistic behavior and quantum Hall effects at room temperature \cite{Geim2007}.

The spread on the magnitude of $\mu$ has been attributed to graphene and substrate quality, and to experimental setups \cite{Castro2009}. In other words, imperfections
due to wrinkles, edges, flexural scattering and strain affect the electronic properties.  Therefore, it is important to keep in mind that even suspended graphene has a certain amount of strain that must be taken into account.  In spite of this, it is now accepted that most of the electronic and optical properties of pristine graphene (excluding strong electron-electron correlations) are well described by a two-band tight-binding Hamiltonian defined in a honeycomb lattice \cite{CastroNeto2009,DasSarma2011,Roche08,DresselhausBook}. For low-energy excitations, this approach leads to an effective Dirac Hamiltonian in reciprocal space.

There are several paths to calculate and understand the effects of strain on electronic properties, most of them based in methods
used for studying pristine graphene \cite{DasSarma2011,Roche08}. One approach is to use Density Functional theory (DFT). The second common approach is the use of a modified tight-binding Hamiltonian (TBH), solved by numerical diagonalization or analytical calculations. The third path is to approximate the TBH near the Dirac points, leading
to an effective Dirac equation where pseudomagnetic fields appear. The last approach is used whenever the strain field varies smoothly in space.

All approaches mentioned in previous paragraph will be reviewed here. In particular, analytical solvable strain fields that provide useful tools to understand the main consequences of strain will be presented. Eventually, a comparison can be made between the different methods for such solvable cases.
Under this perspective, one of the most important and instructive cases is that of uniform strain. It served to quantify gaps as a function of strain \cite{Pereira2009a}, and to clarify some early issues in the derivation of the effective Dirac Hamiltonian  for strain \cite{Oliva2013}.

The effects of slowly-spatially varying strain \cite{Volovik2015,Oliva2015a}  are:
\begin{itemize}
 \item A shift of the Dirac point. The shift is given by a pseudomagnetic vector potential.
 \item A change in the metric of the lattice and a consequent modification of the metric of reciprocal space.
 \item The energy scale is changed due to the modification of bond lengths.
 \item The Fermi velocity becomes anisotropic.
 \item A pseudoelectric (deformation potential) field appears.
\end{itemize}

The Dirac approximation is strongly modified for strain variations within the scale given by carbon-carbon bonds, leading to:
\begin{itemize}
\item Shift, merging and reproduction of Dirac points.
\item The creation of energy gaps.
 \item The reciprocal lattice can even loose its meaning. In some cases it can be replaced by a superlattice, like the moir\'e superlattice. A procedure akin to the magnetic lattice concept can be introduced.
 \item Electronic spatial localization.
 \end{itemize}

Strain produces effects that go beyond a spatial variation of the Fermi  velocity. In both slow and fast spatially varying strain, the Dirac points will not coincide with the high symmetry points of the distorted reciprocal lattice \cite{Oliva2013} (whenever it is possible to define it). Previous two lists do not include the effects of uncorrelated, local random strain, which
are better described and understood by local impurity fields that will be discussed later.

 Strain can be local or have a long-range nature, as well as being random or correlated. For example, local strain can be produced by the tip of a scanning microscope \cite{Salvador2012,Ackerman2016}, while long-range correlated strain occurs over periodic substrates. Given that
strain breaks the symmetry of the honeycomb lattice, it is possible to classify the specific broken symmetries in some cases \cite{Evers2008}, and include terms in the Hamiltonians with the specific symmetry consideration \cite{Amorim2016}. Such a path is not followed here to avoid overlap with other reviews. Instead, representative cases of strains/ripples will be discussed.

These simple models cover the following measurable effects: a gap phase diagram \cite{Hasegawa,Pereira2009a}, a fractal spectrum \cite{Naumis2014,Roman2015a,Roman2015b}, anisotropic Fermi velocity \cite{deJuan2012,Oliva2013,Oliva2015a}, mixed Dirac-Schr\"odinger behavior \cite{Pereira2009a,Montambaux2009a,Montambaux2009b,Montambaux2012,Montambaux2013,Roman2015a}, pseudomagnetic and pseudoelectric fields \cite{Vozmediano2010,Sloan2013,Barraza2013,Volovik2014,Pacheco2014,Oliva2015a}, Landau levels \cite{Guinea2010,Guinea2010b}, non-trivial topological modes, and optical dicroism \cite{Pereira2010,Oliva2015b}.

The electronic properties of unstrained graphene and some of the general effects of disorder will be reviewed in the next Subsection, giving way to different approaches to treat strained graphene in remaining Subsections.

\subsection{Electronic properties of  pristine and disordered graphene}\label{ElectronicGraphene}

One of the most fruitful approaches to study the electronic properties of graphene is the tight-binding (TB) approximation based on $\pi$-electrons \cite{Meunier2016}. Within this method, the
contributions to the electronic behavior from the three valence electrons belonging in the $\sigma$ carbon orbitals are neglected, leading to
the following Hamiltonian matrix model in which only $\pi-$orbitals are considered \cite{CastroNeto2009}:
\begin{equation}
\boldsymbol{H}_0=-t_{0} \sum_{\boldsymbol{r}}\sum_{n=1}^{3}  a_{\boldsymbol{r}}^{\dag}
b_{\boldsymbol{r}+\boldsymbol{\delta}_{n}}+H.c.,
\label{TB0}
\end{equation}
where $\boldsymbol{r}$ runs over all $A$ sites of the Bravais lattice, and the hopping integral
(also known as the transfer integral)  $t_{0} \approx$ 2.7 eV is obtained by fitting to experimental (ARPES) or numerical data \cite{CastroNeto2009}.
$a_{\boldsymbol{r}}^{\dag}$ and $b_{\boldsymbol{r}+\boldsymbol{\delta}_{n}}$ are creation and
annihilation electron operators on the $A$ sublattice (at position $\boldsymbol{r}$) and the $B$ sublattice (at
position $\boldsymbol{r}+\boldsymbol{\delta}_{n}$), respectively. This electronic model describes two electrons in a honeycomb
lattice with a first nearest neighbour interaction.


Second nearest neighbours can be included by using
a second transfer integral $t_0^{sn}\approx 0.68$ eV which adds extra terms to
equation (\ref{TB0}). It is possible to reproduce the
energy dispersion in the whole Brillouin zone using a TB Hamiltonian that includes up to third nearest neighbours \cite{Reich2002,Foa2014}. Such corrections  play an important
role for disordered \cite{BarriosVargas2011a,BarriosVargas2013,BarriosVargas2012} 
and for excitations with energies at least 1 eV away from the Fermi level that are no longer considered low-energy in pristine graphene.

The bipartite nature of the lattice has many important consequences for electronic properties. As indicated before,
other bipartite lattices with quasiperiodic order \cite{Naumis1994} or disorder \cite{Naumis2002,Naumis2007,BarriosVargas2011a,BarriosVargas2013}
share some features with graphene, like a symmetric spectrum,  zero-energy confined state modes or  pseudomobility energy edges \cite{Naumis1994,Naumis2002}.

Returning to the first-neighbour model with a single $\pi-$orbital per site, equation (\ref{TB0}),
one can reduce the Hamiltonian to a  $2\times2$ matrix (because the lattice only contains two-non equivalent sites) by a Fourier transform:
\begin{equation}
a_{\boldsymbol{r}}^{\dag}=\frac{1}{\sqrt{N}}\sum_{\boldsymbol{k}}e^{i\boldsymbol{k}\cdot\boldsymbol{r}}a_{\boldsymbol{k}}^{\dag},
\end{equation}
where the wavevector $\boldsymbol{k}$ is introduced. Using a similar transformation for $b_{\boldsymbol{r}+\boldsymbol{\delta}_{n}}$, one gets
the Hamiltonian:
\begin{equation}
\boldsymbol{H}_0=- t_{0}\sum_{\boldsymbol{k}}\sum_{n=1}^{3} e^{-i\boldsymbol{k}\cdot\boldsymbol{\delta}_n} a_{\boldsymbol{k}}^{\dag}b_{\boldsymbol{k}} + H.c.
\label{TB0K}
\end{equation}
The corresponding Schr\"odinger equation is an effective $2\times2$ Hamiltonian matrix $H(\boldsymbol{k})$,
acting on a wavevector  with components $(a_{\boldsymbol{k}},b_{\boldsymbol{k}})$ and eigenvalues $E(\boldsymbol{k})$:
\begin{equation}
 \left( \begin{array}{cc}
0 &  H_{AB}(\boldsymbol{k})\\
H_{AB}^{*}(\boldsymbol{k})  & 0  \\
\end{array} \right)
\left( \begin{array}{c} a_{\boldsymbol{k}} \\ b_{\boldsymbol{k}} \end{array} \right) =
E(\boldsymbol{k})\left( \begin{array}{c} a_{\boldsymbol{k}} \\ b_{\boldsymbol{k}} \end{array} \right),
\label{Hab}
\end{equation}
where $H_{AB}(\boldsymbol{k})=-t_{0}f(\boldsymbol{k})$
and $f(\boldsymbol{k})$ is the following complex function:
\begin{equation}
f(\boldsymbol{k})=\sum_{n=1}^{3} e^{-i\boldsymbol{k}\cdot\boldsymbol{\delta}_n}.
\label{finplane}
\end{equation}
Graphene's energy dispersion is found from equation (\ref{Hab}):
\begin{equation}
 E(\boldsymbol{k})=\pm t_{0} |\sum_{n=1}^{3} e^{-i\boldsymbol{k}\cdot\boldsymbol{\delta}_n}|
 \label{E0}.
 \end{equation}

\begin{figure}
\centering
\includegraphics[width=\linewidth]{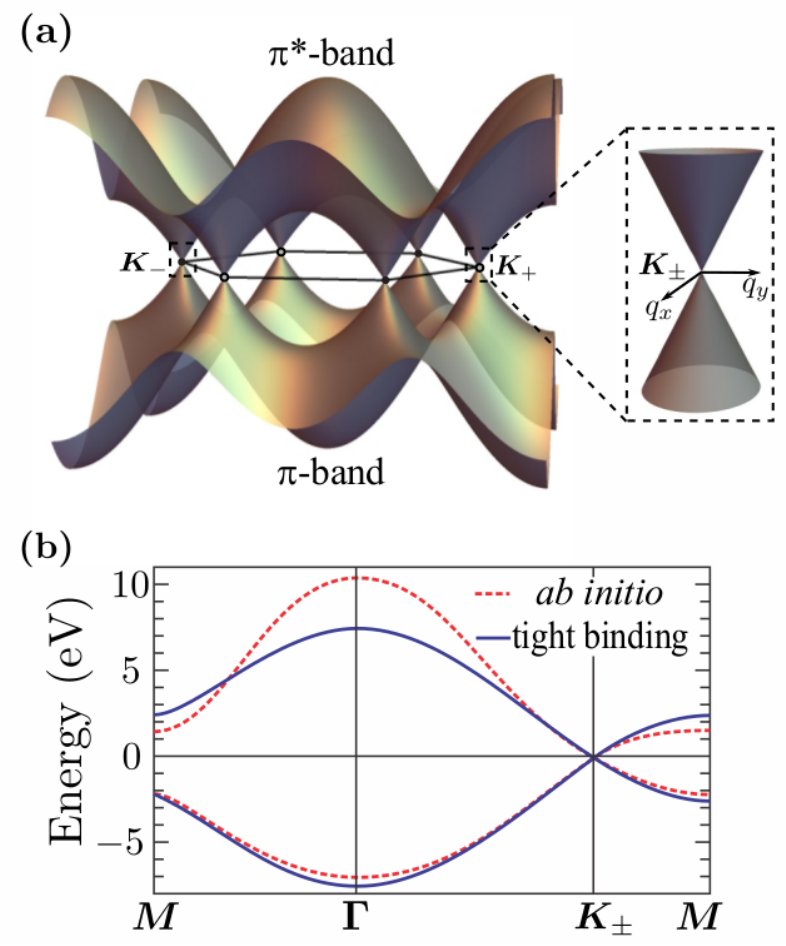}
\caption {Comparison between tight-binding and \textit{ab initio} (DFT)
electronic band structures for pristine graphene. (a) Energy dispersion obtained from equation (\ref{E0}). A zoom-in at the Fermi
energy showing a cone is displayed as well. The vertices of the cones touch at the Dirac point
at $\boldsymbol{K}_{\pm}$. (b) Band structure along a high-symmetry path in the Brillouin zone. The
dotted line is obtained from equation (\ref{E0}), and the solid line is from a DFT calculation that includes $\sigma$ and $\pi$ orbitals.}
\label{TBDFT}
\end{figure}

The surface $E(\boldsymbol{k})$ obtained from equation (\ref{E0}), as well as a transversal cut over a high-symmetry path in reciprocal space are presented in figure \ref{TBDFT}(a). A comparison with a full DFT calculation is also shown in figure \ref{TBDFT}(b). The agreement
near the Fermi energy is excellent \cite{Meunier2016}, but one must remember that the magnitude of the Fermi velocity is underestimated in DFT. Without charge pumping by external electric fields,
 the orbitals are half-filled and thus the Fermi energy ($E_F$) lies at $E=0$. Equation (\ref{E0}) leads to an effective Dirac equation (see Section \ref{Dirac}) and displays conical dispersions near $E=0$ called Dirac cones. The condition $E=0$ leads to a pair of special $\boldsymbol{k}$ points labeled by $\boldsymbol{K}^D$
 for which $E(\boldsymbol{K}^{D})=0$.

  $\boldsymbol{K}^{D}$ happens to coincide with $\boldsymbol{K}_{\pm}$ for pristine graphene. Although there is some confusion in the literature about this point \cite{Oliva2013,Oliva2015a}, this is no longer the case for strained graphene \cite{Li2010,Oliva2013,Meunier2016}. The existence of two inequivalent Dirac points with the same energy
 leads to the concept of a valley \cite{CastroNeto2009} that is key for electronic and optical properties.
 In particular, effects arising from structural disorder depend upon their feasibility of producing intra- or inter-valley
 scattering \cite{Katsnelson2008}.

Equation (\ref{E0}) can be written without reference to vectors $\boldsymbol{\delta}_n$: using equation (\ref{kaleidoscope}), equation (\ref{E0}) transforms into:
 \begin{equation}
 E(\boldsymbol{k})=\pm t_{0} |1+e^{i\boldsymbol{k}\cdot\boldsymbol{a}_1}+e^{i\boldsymbol{k}\cdot(\boldsymbol{a}_1-\boldsymbol{a}_2)}|
 \label{E0a}.
 \end{equation}
In other words, as long as the transfer integral is the same for all bonds, the energy dispersion only depends on the structure of the Bravais lattice. Furthermore, equation (\ref{E0a}) indicates the existence of a peculiar phase-difference in the wavefunction among neighbors in the same bipartitie lattice at the Dirac point where $E(\boldsymbol{K}^{D})=0$. From equation (\ref{E0a}), this can only be satisfied when there is a phase difference of $2\pi/3$  between second-nearest neighbours, that belong to the same bipartite sublattice. The amplitude must be zero in the other sublattice. Moreover, each bipartite sublattice is made from triangles, implying that states near the tip of the Dirac cone have a certain amount of frustration, in the sense that phase differences can not be equal to $\pi$ between consecutive identical sites.

  Equation (\ref{E0}) yields the Dirac cone:  for crystal momentum $\mathbf{q}$ near the Dirac point such that $\boldsymbol{k}=\boldsymbol{K}^{D}+\boldsymbol{q}$ (figure \ref{TBDFT}) one has:
  \begin{equation}
   E(\boldsymbol{k})=E(\boldsymbol{q})=\pm \hbar v_{F}|\boldsymbol{q}|,
  \end{equation}
where $v_{F}$ is the Fermi velocity:
\begin{equation}
 v_{F}=\frac{3t_{0}a}{2\hbar}.
\end{equation}

This leads to a linear DOS:
\begin{equation}\label{DOS0}
\rho_{0}(E)=\frac{2|E|}{\pi \hbar^{2} v_{F}^{2}},
\label{rho}
\end{equation}
and to the following carrier density:
\begin{equation}
 n_{0}(E)=\mbox{sgn}(E)\frac{2|E|^{2}}{\pi \hbar^{2} v_{F}^{2}}.
\label{carrier}
\end{equation}

Dirac cones are topologically protected
and thus robust to second nearest neighbour interaction \cite{Hatsugai2011}.
For small graphene sheets, the linear behavior
of the dispersion can change due to edge-related effects. For example,
a gap can be opened in graphene nanoribbons depending on the
edge type as well as on the number of hexagons along
the nanoribbon width \cite{Roche08}, making this a useful effect to design
electronic devices \cite{NaumisTerrones2009}.

\begin{figure}
\centering
\includegraphics[width=\linewidth]{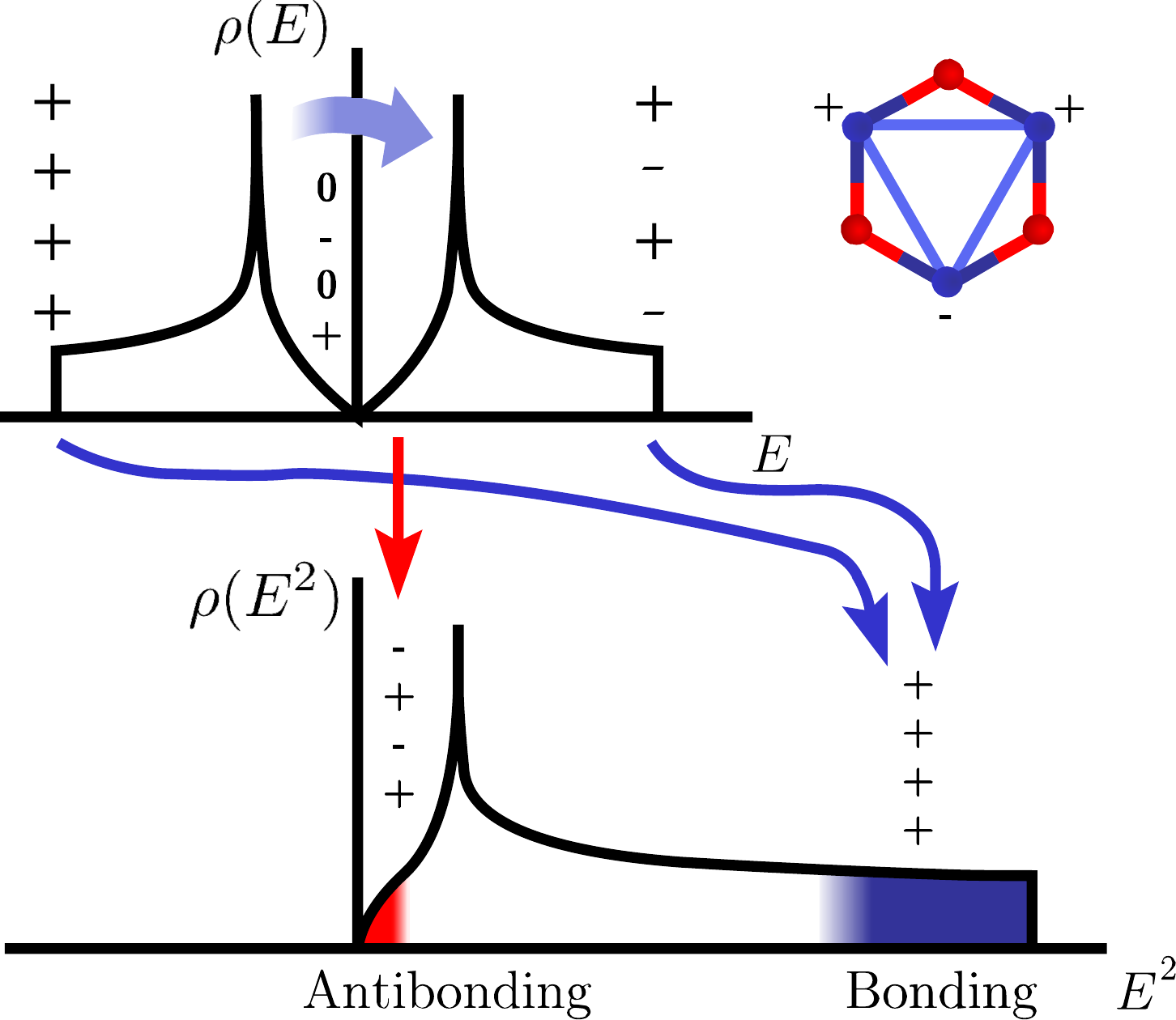}
\caption{Sketch of the Hamiltonian eigenvalue renormalization from a graphene hexagonal lattice into a triangular one
by the transformation $H^{2}$: the
graphene's density of states $\rho(E)$ is transformed into $\rho(E^{2})$, resulting in a folding around $E=0$ that is indicated by arrows.
Band edges, central states, and phase differences among sites are represented by $\pm$ signs. Central states at $E=0$ have a zero amplitude in one
sublattice \cite{BarriosVargas2013}. When one of the sublattices is renormalized, states near $E=0$
result in edge band states with an antibonding nature in a triangular lattice \cite{Naumis2007,BarriosVargas2012},
as indicated in the triangle that appears inside the hexagon.
Due to frustration, states are pushed to higher energies, leading to van Hove degeneracies seen as a peaks.}
\label{Renormalization}
\end{figure}

Less known and important for the introduction of disorder and spin, is the
physical reason behind the cone appearance: a wavefunction frustration
in the underlying triangular lattice \cite{Naumis2007,BarriosVargas2012}.
Indeed, the square of the Hamiltonian in equation (\ref{TB0K}) ($\mathbf{H}_0^2$) is diagonal by virtue of equation (\ref{Hab}):
\begin{equation}
 \left( \begin{array}{cc}
H_{AB}^{2}(\boldsymbol{k}) &  0\\
0   & H_{AB}^{2}(\boldsymbol{k})  \\
\end{array} \right)
\left( \begin{array}{c} a_{\boldsymbol{k}} \\ b_{\boldsymbol{k}} \end{array} \right) =
E^{2}(\boldsymbol{k})\left( \begin{array}{c} a_{\boldsymbol{k}} \\ b_{\boldsymbol{k}} \end{array} \right),
\label{Hab2}
\end{equation}
implying that the components of the wavefunction on the $A$ and $B$ sublattices are decoupled. Thus $H^{2}$
describes a triangular lattice, and the squaring of $H$ renormalizes one of the bipartite sublattices \cite{Naumis2007,BarriosVargas2012} with an spectrum folded around $E=0$ that is illustrated in figure \ref{Renormalization}.

As explained before, states near $E=0$ need to be close to an
antibonding nature in a triangular lattice. This produces frustration, since wavefunctions can not have a phase difference of $\pi$ between all neighboring sites in a triangular lattice.
In the absence of disorder, some states lower their energy by having  phase
differences close to $2\pi/3$ as $E\rightarrow 0$.
Frustration implies that many states are pushed away to higher energies, thus producing a van Hove singularity at energy $E^{2}=t_0^{2}$.
This leads to a simple picture of graphene's spectrum from the underlying triangular sublattice.

The wavefunction frustration-driven picture is summarized as follows:
 \begin{itemize}
  \item Band edges in graphene are obtained from the maximum of $E^{2}(\boldsymbol{k})$, associated with the diffraction spots at $\Gamma$ points, i.e.  for $\boldsymbol{k}=l \boldsymbol{G}_{1}+h\boldsymbol{G}_{2}$  with $l$ and $h$ integers. Here $\nabla_{\boldsymbol{k}} E^{2}(\boldsymbol{k})=0$.
  \item The Dirac points in graphene correspond to the minimums of the function $E^{2}(\boldsymbol{k})$. Here $\nabla_{\boldsymbol{k}} E^{2}(\boldsymbol{k})=0$ too (in graphene, the operator $\nabla_{\boldsymbol{k}} E$ is not defined at the Dirac cone tip). Dirac points coincide with the high-symmetry points $K_{\pm}$, a result expected from the diffraction properties because two Bragg lines intersect therein. Diffraction leads to stationary waves, i.e. to a vanishing group velocity in the triangular lattice.
  \item Since  $E^{2}(\boldsymbol{k})$ is a periodic bounded function, there must be a third singularity \cite{Naumis2016}. This
  corresponds to the van Hove singularity  at $E^{2}(\boldsymbol{k})=  t_{0}^{2}$. The singularity is a saddle
  point of $E^{2}(\boldsymbol{k})$.
 \end{itemize}

When very strong impurities or vacancies are added,
the wavefunction has more amplitude in regions
of lower frustration that have a decreasing exponential probability with size, leading
to a kind of Lifshitz tail \cite{Naumis2007,Abergel2010}. As a result, a pseudomobility
edge appears near the Dirac cones \cite{Naumis2007,Abergel2010}, as confirmed in ARPES experiments of graphene doped with hydrogen impurities \cite{Bostwick2009}. The corresponding
wavefunctions have an interesting multifractal behavior \cite{BarriosVargas2012}.
In a similar way, resonant states appear near the Fermi energy when uncorrelated impurities are added  \cite{Abergel2010,BarriosVargas2011b,Martinazzo2010,BarriosVargas2011b}.

Zero energy modes appear due to disorder or to the presence of boundary modes that are associated
with topological properties. These modes decouple from the renormalization and are related with highly degenerate modes with the property that the sum of wavefunction amplitudes must add to zero for the neighbors of any site in the lattice \cite{Naumis2002,BarriosVargas2013}.

In other bipartite lattices such as random binary alloys, zero-energy modes are strictly localized and confined \cite{Kirkpatrick1972,Naumis2002}. In quasiperiodic
lattices, zero energy modes form beautiful fractal nodal lines carrying up to $10\,\%$ of the spectral weight \cite{Kohmoto1986}.
For doped graphene, the number of states was obtained by using a sum over moments and disordered
configurations \cite{BarriosVargas2013}. Such modes are especially important for magnetic properties.

The effects of disorder can be classified by the kind of symmetry they break \cite{Evers2008}.
Different types of randomness realize  all possible ten  symmetry  classes  of
Dirac  Hamiltonians \cite{Bernard2002}, while symmetry considerations lead to different kinds
of extra terms \cite{Amorim2016} to the unperturbed Hamiltonian, equation (\ref{TB0}).  As a rule of thumb,
the range of the potential determines if intra- or inter-valley
scattering is allowed \cite{Katsnelson2008,Tikhonenko2009,Abergel2010}. Inter-valley scattering, associated with long range potentials,
allows access to the chirality degree of freedom allowing localization \cite{Katsnelson2008,Abergel2010}, which is otherwise evaded for short-range potentials. Here, chirality is given as the phase difference of the wave-function
projections onto the A and B sublattices \cite{Naumis2007}, and it arises from the projection of pseudospin in the momentum direction \cite{Foa2014}.

The main reason for antilocalization is the absence of backscattering: to reverse
the trajectory of an electron, one must change its momentum from $\boldsymbol{p}$ to $\boldsymbol{-p}$, implying a change
of valley. But $\boldsymbol{p}$ is coupled with chirality, which is not changed by short-range potentials \cite{Foa2014}.
However, backscattering, as created by group-I impurities \cite{Bostwick2009} or vacancies \cite{Naumis2007}, also depends upon the energetic range of the disorder \cite{Foa2014}. Further details are given in works by Foa-Torres, Roche and Charlier \cite{Foa2014}, and Katnelson \cite{Katsnelson2012}.

\subsection{Tight-binding approach to strain}

As indicated in Section \ref{DescriptionStrain}, atomic positions change from
$\boldsymbol{r}$ to $\boldsymbol{r}^{\prime}=\boldsymbol{r}+\boldsymbol{u}(\boldsymbol{r})$ in strained graphene.
As a result, distances between atoms change, modifying the hopping parameter along the way.
The TB Hamiltonian for strained graphene is obtained by replacing the original lattice positions and hopping parameter in equation (\ref{TB0K}) with those resulting from the structural distortion \cite{CastroNeto2009}:
\begin{equation}
\boldsymbol{H}=-\sum_{\boldsymbol{r}^{\prime},n} t_{\boldsymbol{r}^{\prime},\boldsymbol{\delta}_{n}^{\prime}(\boldsymbol{r}) }a_{\boldsymbol{r}^{\prime}}^{\dag}
b_{\boldsymbol{r}^{\prime}+\boldsymbol{\delta}_{n}^{\prime}(\boldsymbol{r}) }+H.c.,
\label{TB}
\end{equation}
where $\boldsymbol{r}^{\prime}$ runs over all sites of the deformed honeycomb lattice and the hopping integral
$t_{\boldsymbol{r}^{\prime},\boldsymbol{\delta}_{n}^{\prime}(\boldsymbol{r}) }$
varies due to the modification of carbon-carbon distances. The operators
$a_{\boldsymbol{r}^{\prime}}^{\dag}$ and $b_{\boldsymbol{r}^{\prime}+\boldsymbol{\delta}_{n}^{\prime}(\boldsymbol{r}) }$ correspond to creating and
annihilating electrons on the $A$ sublattice (at position $\boldsymbol{r}^{\prime}$) and $B$
sublattice (at position $\boldsymbol{r}^{\prime}+\boldsymbol{\delta}_{n}^{\prime}(\boldsymbol{r})$).

An important feature of equation (\ref{TB}) is the modification of the first-neighbour vectors  $\boldsymbol{\delta}_{n}$ into a space-dependent
set of first-neighbour vectors $\boldsymbol{\delta}_{n}^{\prime}(\boldsymbol{r})$ whose values depend upon $\nabla \boldsymbol{u(\boldsymbol{r})}$
as indicated in equation (\ref{Deltan}). The lack of this lattice correction in earlier versions of the TB Hamiltonian produced some controversies concerning the contribution of this term to local pseudomagnetic fields in the Dirac approach. Eventually, it has been found that lattice corrections do not contribute to the pseudomagnetic field \cite{Kitt2013,Sloan2013,Barraza2013,deJuan2013}, although they are important for the TB  Hamiltonian in circumstances where the mean-field Dirac approach is not valid (see Sections \ref{Dirac} and \ref{secdiscrete}).

\begin{figure*}[t]
\centering
\includegraphics[width=\textwidth]{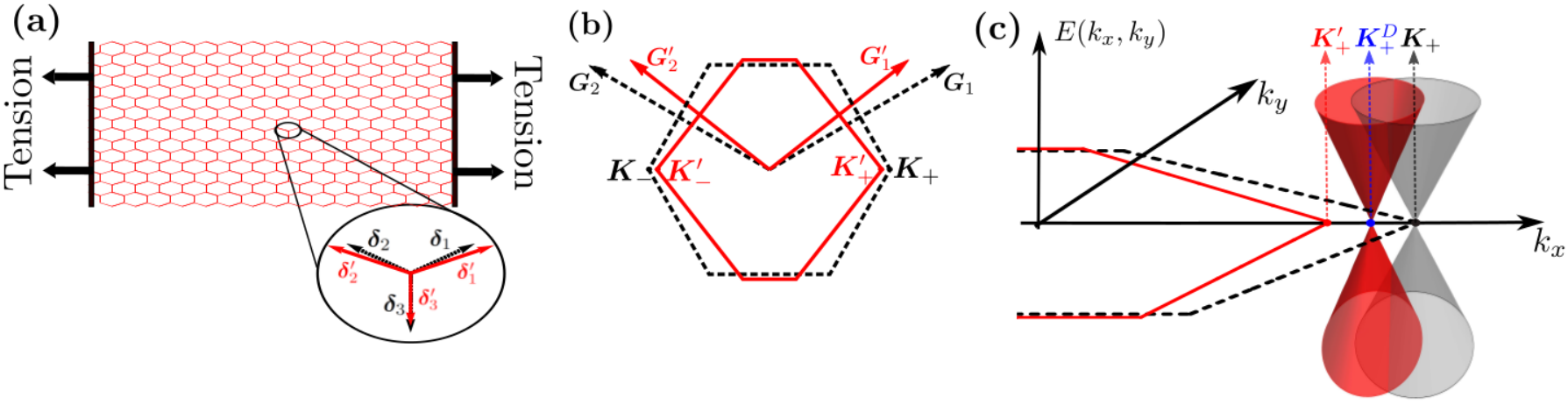}
\caption{(a) Uniformly-strained graphene lattice showing  the strained vectors $\boldsymbol{\delta}_{i}^{\prime}$ --that point to the neighbors of $A-$sites-- and their unstrained counterparts $\boldsymbol{\delta}_i$. (b)  Corresponding strained reciprocal lattice, showing the strained and unstrained reciprocal vectors $\boldsymbol{G}^{\prime}$ and $\boldsymbol{G}$, as well as the  strained and unstrained high-symmetry points $\boldsymbol{K}_{\pm}^{\prime}$ and $\boldsymbol{K}_{\pm}$. (c) Energy dispersion, showing how the distortion  of the reciprocal lattice transforms the original Dirac cone into a distorted one with a directional dependent Fermi velocity. The  cone vertex is also translated to a new point $\boldsymbol{K}^{D}$ which does not coincide with  $\boldsymbol{K}_{\pm}^{\prime}$. The vector displacement from the original position is directly given by the pseudomagnetic vector potential. Adapted from \cite{Oliva2015a} with permission.}
\label{strainedlattice}
\end{figure*}

Displacements that keep distances between
neighbours constant can still modify angles between atomic bonds with a minuscule energy cost. Such kind of (nearly isometric) deformation is
known as a floppy-mode \cite{Thorpe1985,Flores2010} and it provides a
link to the study of electronic properties and network topological constraints \cite{Naumis2015Frontiers} known in glasses as the Phillips-Thorpe rigidity theory \cite{Phillips1979,Thorpe1985,Flores2010}.

To complete the model, modifications of the TB parameters are needed. Usually, this is given by an estimation of overlap changes for $\pi$-orbitals as a function of the carbon-carbon distance \cite{Suzuura2002,KimNeto2008,Ribeiro}:
\begin{equation}
t_{\boldsymbol{r}^{\prime},\boldsymbol{\delta}_{n}^{\prime}(\boldsymbol{r})}=t_{0}\exp[-\beta(|\boldsymbol{\delta}_{n}^{\prime}(\boldsymbol{r})|/a - 1)],
\end{equation}
where $\beta$ is the electron Gr\"uneisen  parameter \cite{Katsnelson2012}:
\begin{equation}\label{eqGruinesen}
\beta=-\frac{\partial \ln t}{\partial \ln a},
\end{equation}
estimated for graphene to lie in the interval $\beta \approx 2-3$. This parameter is usually found by
analyzing the $G-$mode Raman scattering peak of strained graphene \cite{Mohiuddin2009,Ding2010}, or from
\textit{ab initio} calculations \cite{Cheng2011}. The Gr\"uneisen  parameter is a measure of the phonon mode softening rate or
hardening. It is a fundamental quantity to understand strain effects as well as thermomechanical properties.

Once the Hamiltonian is written, it can be diagonalized either by numerical calculations or (in some cases) analytically. Some examples of these procedures are presented in the
following Section, as follows:
\begin{enumerate}
	\item Uniform strain field. Here the strain field is independent of the position (but it can depend on sublattice).
	\item Isotropic expansion. This is a particular case of the uniform strain field.
	\item Uniaxial non-uniform strain field. There is a direction in which strain is either zero or constant, while its magnitude is arbitrary in the perpendicular direction.
	\item Arbitrary periodic strain fields due to substrates or ripples.
\end{enumerate}
Finally, the reader is reminded that equation (\ref{TB}) and the derivations following from this model do not take hopping terms to second
and third nearest neighbours into account.

\subsection{Uniformly strained graphene}\label{UniformStrain}

In this case, the strain displacement field and the strain tensor $\bar{\boldsymbol{\epsilon}}$ are space-independent,
as explained in Section \ref{DescriptionStrain}. A uniformly strained lattice preserves periodicity, and thus the dynamics can be solved for in reciprocal space.

In Section \ref{DescriptionStrain}, it was demonstrated that the deformed cell unit vectors are given by equation~(\ref{newunitary}),
leading to the new set of first-neighbour vectors $\boldsymbol{\delta_n}'$ indicated
in figure \ref{strainedlattice}(a), and that the corresponding new reciprocal-lattice vectors are given by equation (\ref{UniformGvectors}) (figure \ref{strainedlattice}(b)).

The uniformly-strained lattice is still periodic, so a Fourier transformation of operators
$a_{\boldsymbol{r}^{\prime}}^{\dag}$ and $b_{\boldsymbol{r}^{\prime}+\boldsymbol{\delta}_{n}^{\prime}}$ can be employed to determine the electron dynamics.
Using equation (\ref{TB}), the  Hamiltonian for the strained lattice is \cite{Oliva2013}:
\begin{equation}
\boldsymbol{H}=-\sum_{\boldsymbol{k}}\sum_{n=1}^{3} t_{n} e^{-i\boldsymbol{k}\cdot(\bar{\bi{I}} +
\bar{\boldsymbol{\epsilon}})\cdot\boldsymbol{\delta}_{n}} a_{\boldsymbol{k}}^{\dag}
b_{\boldsymbol{k}} + H.c.,
\label{k-H}
\end{equation}
where $t_{n}$ ($n=1,2,3$) is the hopping integral between each of the three nearest neighbors of a site:
\begin{equation}
t_{n}=t_0\exp[-\beta(|(\bar{\bi{I}}+\bar{\boldsymbol{\epsilon}})\cdot\boldsymbol{\delta}_{n}|/a - 1)]
\label{texp}.
\end{equation}

Equation~(\ref{k-H}) can be written as a $2\times2$ matrix, whose eigenvalues provide the energy-momentum dispersion for graphene under uniform strain:
\begin{equation}\label{DR68}
 E(\boldsymbol{k})=\pm |\sum_{n=1}^{3} t_{n} e^{-i\boldsymbol{k}\cdot(\bar{\bi{I}} +
\bar{\boldsymbol{\epsilon}})\cdot\boldsymbol{\delta}_{n}} |,
\end{equation}
and which can be simplified to give:
\begin{equation}
 E(\boldsymbol{k})=\pm\sqrt{\gamma+g(\boldsymbol{k})},
  \label{Ekxky}
\end{equation}
where $\gamma={t_1}^2+{t_2}^2+{t_3}^2$ and,
\begin{eqnarray}\label{eq66}
 g(\boldsymbol{k})=\sum_{n=1}^3\sum_{s>n}^3 2t_n t_ s\cos\left[\boldsymbol{k}\cdot(\bar{\bi{I}}+
 \bar{\boldsymbol{\epsilon}})\cdot(\boldsymbol{\delta}_n-\boldsymbol{\delta}_s)\right],
\end{eqnarray}
equation (\ref{eq66}) is valid for any anisotropic honeycomb lattice \cite{Oliva2016a}.

\begin{figure}
\centering
\includegraphics[width=\linewidth]{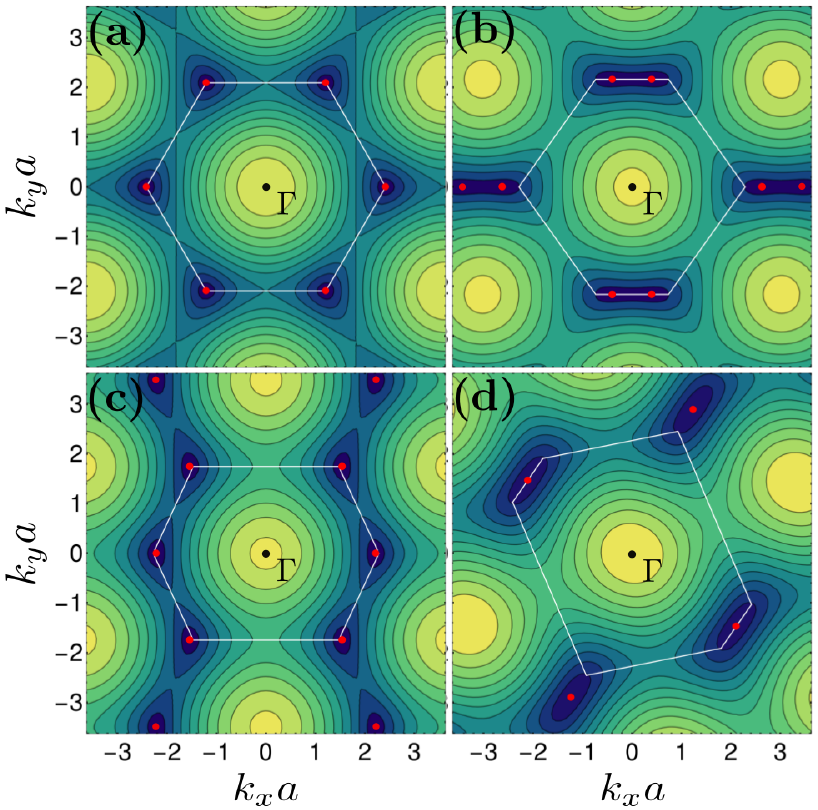}
\caption{Energy dispersion obtained from equation (\ref{Ekxky}). Dark colors are close to zero energy, and light color implies a higher energy isovalue. The  1BZ indicated by white lines is identical to figure \ref{NewBZ}. The Dirac points, where the Fermi (zero) energy lies in, are indicated by red dots. (a) Unstrained graphene.
(b) Uniaxial zigzag strain with $\epsilon_{Z}=0.2$, $\epsilon_{A}=-\nu\epsilon_{xx}$, and $\gamma_{S}=0$. (c) Uniaxial armchair strain with $\epsilon_{Z}=0.2$, $\epsilon_{A}=-\nu\epsilon_{Z}$, and $\gamma_{S}=0$. (d) Shear strain ($\epsilon_{A}=\epsilon_{Z}=0$ and $\gamma_{S}=0.2$).
Pairs of Dirac cones merge onto one in this latter case, as the reciprocal lattice turns into a square.
For all strains, Dirac points $\boldsymbol{K}^{D}$ do not have the same position as $\boldsymbol{K}_{+}^\prime$ and $\boldsymbol{K}_{-}^\prime$, which are the vertices of the polygons.}
\label{contour1}
\end{figure}

Typical contour plots resulting from the energy dispersion given by equation (\ref{Ekxky}) are shown in figure~\ref{contour1}(a) for unstrained graphene, and in figures \ref{contour1}(b), \ref{contour1}(c) and \ref{contour1}(d) for other representative types of strain. The high symmetry points of the strained reciprocal lattice and the 1BZ obtained from equation (\ref{K}) are shown as white lines in figure \ref{contour1} as well. The Dirac points $\boldsymbol{K}^D$ are found by solving $E(\boldsymbol{K}^D)=E_{F}=0$ and shown by red dots in figure~\ref{contour1}. The Dirac points do not coincide with the high symmetry points $\boldsymbol{K}_{-}^{\prime}$ and  $\boldsymbol{K}_{+}^{\prime}$ of the strained reciprocal lattice: the condition  $E(\boldsymbol{K}^D)=0$ leads to k-points that are different from the high-symmetry points of the reciprocal space given by
equation (\ref{K}). Dirac points coincide with the $\boldsymbol{K}_{-}^{\prime}$ and  $\boldsymbol{K}_{+}^{\prime}$ high symmetry points only in unstrained graphene. Dirac points are shifted by strain and their actual location needs to be found from the dispersion relation explicitly.

This crucial point has not been taken into account in  several papers available on the literature,  although it has been observed in DFT calculations (\cite{Naumov2011,Kerszberg2015} and Section \ref{DFT_H}),  in relativistic field theory approaches \cite{Volovik2015,Zubkov2015}, and in TB calculations \cite{Oliva2013,Oliva2015a,Goerbig2008}, and is analyzed in subsections \ref{Dirac} in more detail.

Up to first order in the displacement field, the Dirac point is located at \cite{Oliva2015a}:
\begin{equation}\label{NewKD}
\boldsymbol{K}_{\pm}^D\approx(\boldsymbol{I}-\bar{\boldsymbol{\epsilon}})\cdot\boldsymbol{K}_{\pm} \pm \boldsymbol{A}_s,
\end{equation}
where the choice of $\pm 1$ labels the valley, and $\boldsymbol{A}_s$ turns out to be a vector potential whose rotational induces a pseudomagnetic field (Section \ref{Dirac}), and is given by:
\begin{equation}\label{VP}
\bi{A}_s=\left(\frac{\beta}{2a}({\epsilon}_{Z}-{\epsilon}_{A}), -\frac{\beta}{a}{\gamma}_{S}\right).
\end{equation}
The pseudomagnetic potential will be studied in more depth later on.

\begin{figure*}[t]
\centering
\includegraphics[width=0.95\textwidth]{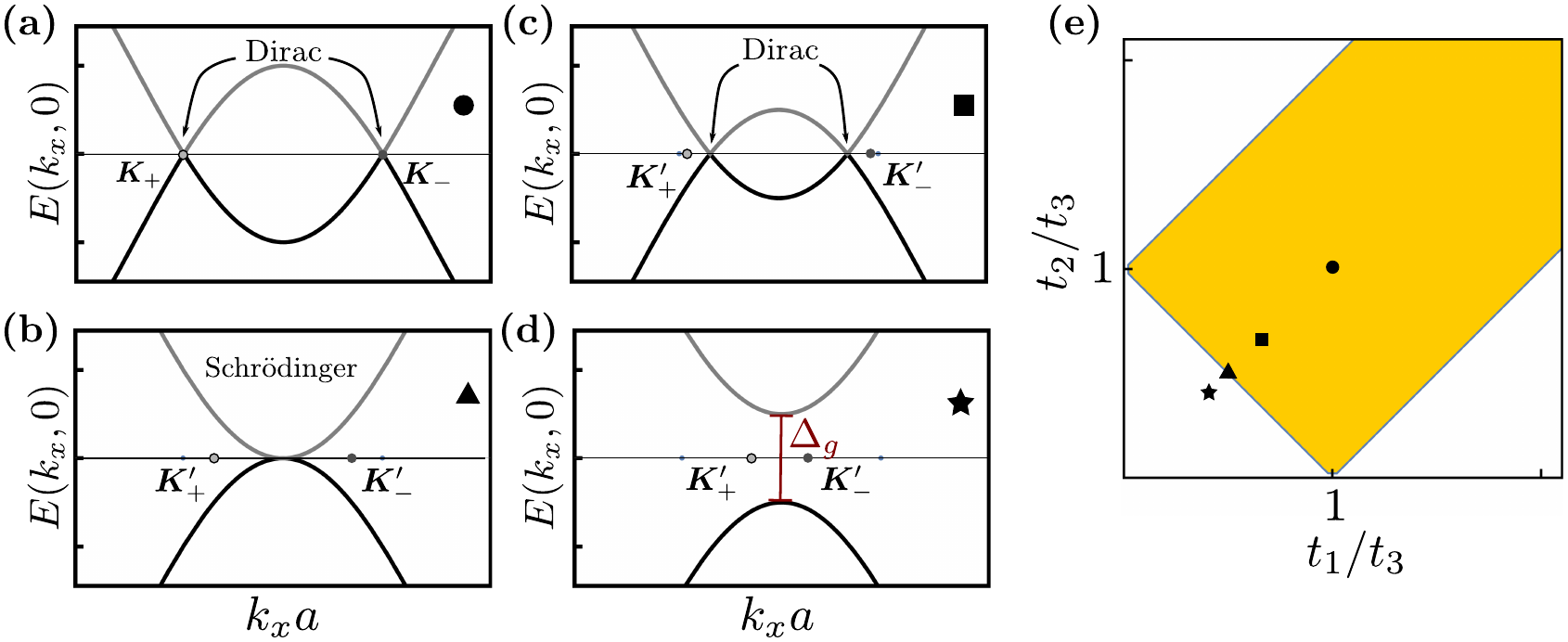}
	\caption{Energy dispersion $E(k_{x},0)$ and  (a) $t_2/t_3=1$, (b) $t_2/t_3=1.5$, (c) $t_2/t_3=2$, and (d) $t_2/t_3=2.5$. The colored area in panel (e) is the region of a gapless electronic spectra. The four symbols marked in (e) signal the coordinates in parameter space corresponding to subplots (a), (b), (c) and (d).}
	\label{Fig_Hasegawa}
\end{figure*}

The relationship between high-symmetry points and Dirac points is also shown in figure~\ref{strainedlattice}(c), which depicts the deformation of the Dirac cone \cite{Oliva2013}. As shown in Section \ref{Dirac}, this produces an angle-dependent Fermi velocity as the energy dispersion becomes elliptical.

This deformation has two contributions \cite{Oliva2013}: a change of reciprocal space, and the modification in the hopping term $t_0$. Both effects can be analyzed separately if
the Gr\"{u}neisen parameter is set to zero ($\beta=0$ in equation (\ref{eqGruinesen})), corresponding to a lattice with unchanged transfer integrals ($t_0$) and the same connectivity of pristine graphene, and a rescaled reciprocal space.

As an example of this phenomena, one can strain the honeycomb lattice and turn it into a brick wall lattice. Since the lattice connectivity is the same and $\beta=0$, both Hamiltonians are the same. It follows that the energy eigenvalues are the same for the brick wall and honeycomb. However, the reciprocal lattice of the brick wall is different from the honeycomb, and its low-energy dispersion is a deformed cone.

A puzzling point is yet to be addressed in the literature. According to previous developments, energetic effects --which depend on the parameter $\beta$-- and geometric effects --which depend on the position of
$\boldsymbol{K}_{\pm}$-- predict a separation of $\bi{K}^D$ from $\boldsymbol{K}_{\pm}$. However, geometrical effects are captured in diffraction and, as explained in Section \ref{Description}, diffraction leads to singularities in the DOS. Therefore, it is difficult to imagine a situation in which nothing happens to the electron dispersion at the intersection of Bragg lines at $\boldsymbol{K}_{\pm}$, while a singularity is displaced to $\boldsymbol{K}_{\pm}$.

This suggests a missing interplay between energetic and geometric effects, and several paths are available to solve this dilemma. One path requires to assume an electronic stabilization of the structure, like in the Humme-Rothery phases
or in the Peierls instability. Another path invokes the breakdown of Cauchy-Born on lattices with a basis, to modify the parameter $\beta$ \cite{Midtvedt2016}. In any case, the shift of Dirac points can be dramatic, as displayed in figures \ref{contour1} and \ref{Fig_Hasegawa}.

In figure \ref{Fig_Hasegawa}, energy-momentum dispersions $E({k}_x,{k}_y)$ of special significance are shown for $k_y=0$, by setting $t_1=t_2=t_0$ and using four values of $t_2/t_3$.

First, the dispersion for the undistorted lattice ($t_2/t_3=1$) is displayed in figure \ref{Fig_Hasegawa}(a). The Dirac points
separate from the high symmetry points when $t_2/t_3=2$ in figure \ref{Fig_Hasegawa}(c). As seen in figures \ref{contour1} and  \ref{Fig_Hasegawa}(b), Dirac points merge for deformations larger than $20\,\%$. For this critical stress, the dispersion relation is linear along  the $y$ direction (relativistic, Dirac behavior)
and quadratic along the $x$ direction \cite{Pereira2009a,Montambaux2009a,Montambaux2009b,Montambaux2012,Montambaux2013,Roman2015a}  (non-relativistic, Schr\"{o}dinger behavior). As seen in figure \ref{Fig_Hasegawa}(d), a gap $\Delta_g$ appears as the hopping anisotropy continues to increase.

In general, the spectrum remains gapless as long as the Hasegawa triangular inequalities are satisfied \cite{Hasegawa}:
\begin{equation}\label{DH}
\left\vert \frac{\vert t_{1}\vert}{\vert t_{3}\vert} -1 \right\vert \leq \frac{\vert t_{2}\vert}{\vert t_{3}\vert} \leq \left\vert\frac{\vert t_{1}\vert}{\vert t_{3}\vert} +1 \right\vert,
\end{equation}
which correspond to the coloured area in figure~\ref{Fig_Hasegawa}(e). Each of the four representative key points in the evolution of $E({k}_x,{k}_y)$ seen
in figures~\ref{Fig_Hasegawa}(a), \ref{Fig_Hasegawa}(b), \ref{Fig_Hasegawa}(c), and \ref{Fig_Hasegawa}(d) are represented in figure~\ref{Fig_Hasegawa}(e), which shows that a gap opens past a critical strain when ${t_{2}}/{t_{3}}\ge2$.

The Hasegawa inequalities have important predictive power. For example, one sees in figure~\ref{Fig_Hasegawa}(e) that the point $(\frac{t_{1}}{t_{3}},\frac{t_{2}}{t_{3}})=(1,1)$ --corresponding to the dispersion seen in figure~\ref{Fig_Hasegawa}(a)-- is surrounded by an appreciable shaded area.
This means that the gap needs a certain threshold strain to open, that in turn depends on the direction of the applied deformation.

To understand previous point, note that deformations in the
armchair direction keep the ratio between $t_1$ and  $t_2$ fixed while $t_3$ changes. Thus, armchair strain changes hoppings along the diagonal line $t_{1}/t_3=t_{2}/t_3$ in figure~\ref{Fig_Hasegawa}(e), and one learns that tensile strain along the armchair direction (armchair strain) will never lead to a gap opening from this figure.

For strain in the zig-zag direction, the argument is reverted,
and the system moves in the direction  $t_{1}/t_3=-t_{2}/t_3$. In this direction, the shaded ribbon-like area has its narrowest width and a gap opens.

Pereira and Castro-Neto used the Hasegawa inequalities to show that a uniform uniaxial strain in the zigzag direction opens a gap once the elongation reaches $23\,\%$. In contrast, a uniform uniaxial strain in the armchair direction is not able to open a gap \cite{Meunier2016,Pereira2009a}.

\begin{figure}[t]
\centering
\includegraphics[width=0.75\linewidth]{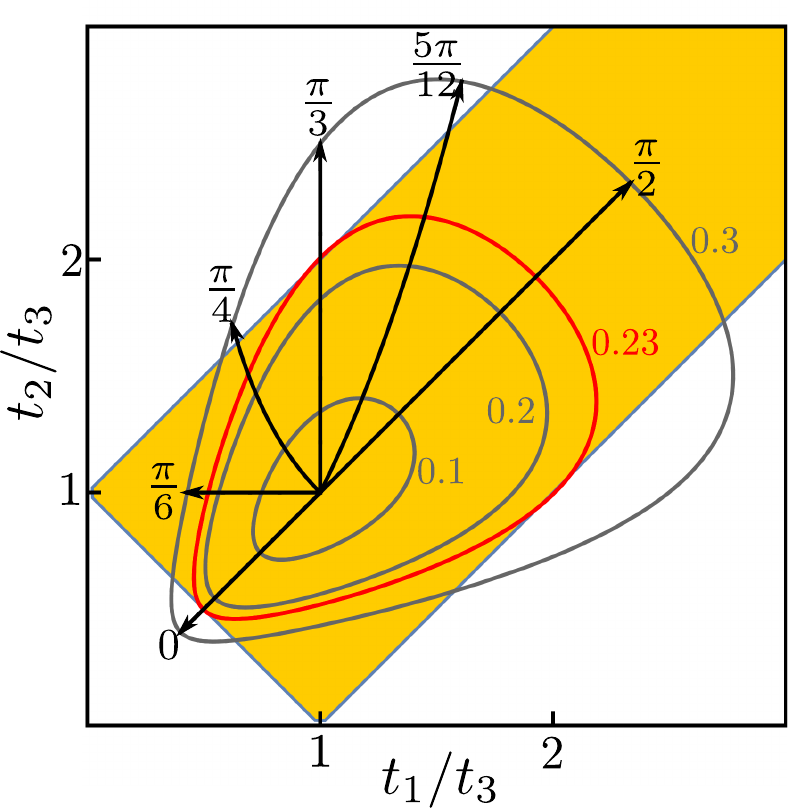}
\caption{Isostrain curves ($\epsilon=0.1,0.2,0.23,0.3$) for the direction of the applied strain indicated by arrows. The angles are measured from the zig-zag direction.  The corresponding values of $t_1/t_3$ and $t_2/t_3$ are also indicated and compared with the Hasewaga region. The critical (red) isostrain curve indicates the minimal strain ($23\,\%$) necessary to open a gap. No gap is open for tensile strain applied along the armchair direction. Inspired in a plot presented in \cite{Pereira2009a}.}
\label{Fig_PolarGap}
\end{figure}

Figure~\ref{Fig_PolarGap} presents polar plots of isostrain curves  for different angles of the applied strain indicated by arrows \cite{Pereira2009a}. The corresponding values of $t_1/t_3$ and $t_2/t_3$ are also indicated and compared with the Hasewaga region. The red isostrain curve indicates the minimal strain ($23\,\%$) needed to open an electronic band gap.

\begin{figure}
\centering
\includegraphics[width=\linewidth]{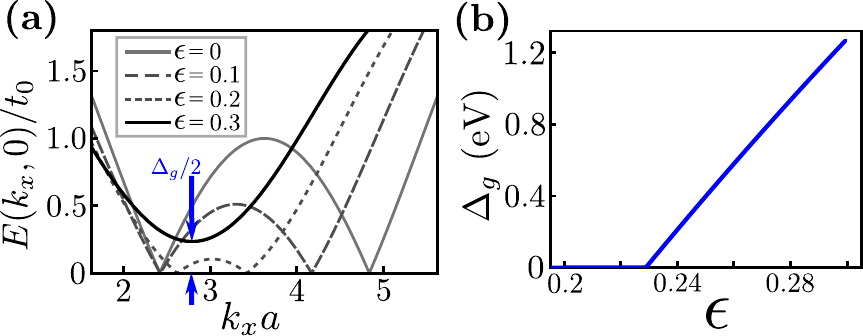}
\caption{(a) Energy dispersion for $k_y=0$ and different values of
$k_x$ using a zigzag strain. Observe the gap opening indicated by an arrow, as well as the parabolic band shape. (b) Gap size as a function of strain in the zigzag direction. The gap opens linearly once the critical strain of $23\,\%$ is attained. Inspired on a plot presented in \cite{Pereira2009a}.}\label{Fig_Gap}
\end{figure}

This physics is re-emphasized in figure~\ref{Fig_Gap}(a), which presents a plot of the energy dispersion for $k_y=0$ and different values of $k_x$ using a zigzag strain, leading to a parabolic band dispersion once the gap opens. Figure ~\ref{Fig_Gap}(b) presents the resulting evolution of the energy bandgap, which increases linearly once the critical strain of $23\,\%$ is attained.

The gap opening has been confirmed by \textit{ab initio} simulations \cite{Choi2010}.  Ni \textit{et al.}~confirmed the possibility of a gap opening \cite{Ni2008}, although the first reported minimal tensile strain for a gap opening was about $0.8\,\%$. Later on, the same group  revised to $26\,\%$ their estimate of the  minimal tensile strain for gap opening \cite{Ni2009}, in good agreement with the tight-binding calculation. The problem in the first estimation by Ni \textit{et~al.} \cite{Ni2009} was, precisely, the overlooking of the displacement of Dirac points from the high-symmetry points.

\begin{figure*}[t]
\centering
\includegraphics[width=0.75\textwidth]{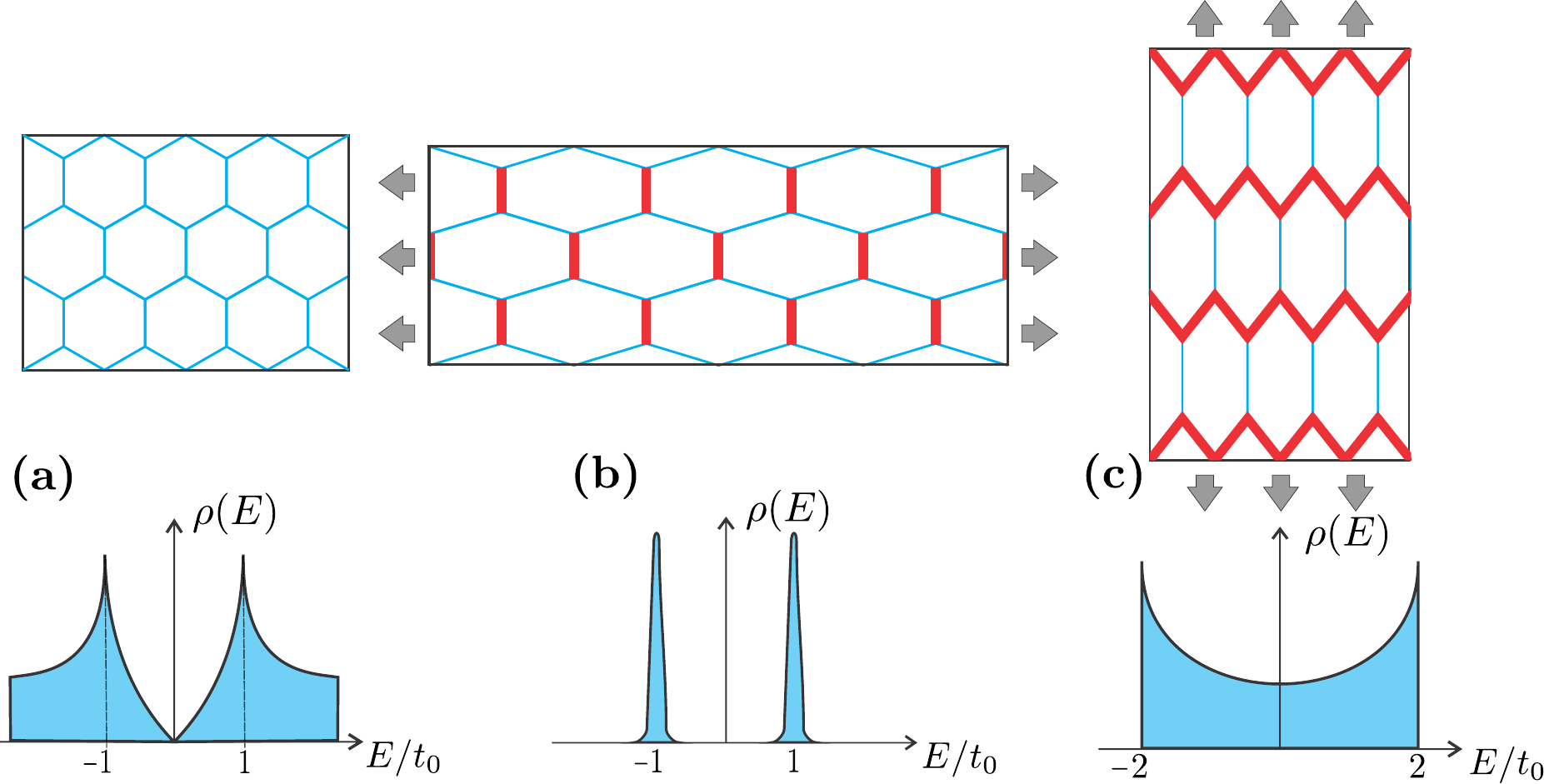}
\caption {A comparison of uniform strain in two limits. (a) Unstrained graphene. (b) Large strain applied along the zigzag direction. (c) Large strain along the armchair direction. The corresponding DOS is indicated in the lower row. In (b), the lattice turns into weak perturbed dimers that are indicated in red. In turn, the DOS corresponds to a system of dimers, a gapped two-level Hamiltonian with multiply-degenerate energy levels at $E=\pm t_0$ (the van Hove singularity in unstrained graphene). In (c), strain preserves chains of bonds indicated in red, and the interaction between chains goes to zero. As a result, the spectrum is made from weakly perturbed degenerate linear chains, with a DOS determined from a linear chain. No gap can be generated in this case, and these linear chains resemble a quasi-1D system known as a Luttinger liquid \cite{Fradkin}.}
\label{StrainDOS}
\end{figure*}

Figure \ref{StrainDOS} is aimed to understand why zigzag and armchair strain have such a different effect on the opening of a band gap. Apply a huge tensile strain in the zigzag direction and note that, while $t_3$ keeps its value due to the invariance $\boldsymbol{\delta}_3'=\boldsymbol{\delta}_3$, both $t_1$ and $t_2$ go to zero. As seen in figure \ref{StrainDOS}(b), the system is nearly dimerized in this limit, keeping bonds only along the $y$ direction. In that case,
there are only two eigenvalues of the Hamiltonian ($E=\pm t_0$), each with degeneracy $N/2$ where $N$ is the number of sites. The location of these states is precisely the same of the van Hove singularity seen on the DOS for pure graphene (figure~\ref{Renormalization}). A gap opens, given the finite spacing among these van Hove singularities.

But as illustrated in figure \ref{StrainDOS}(c), the situation is different for armchair strain. Here, bonds along the zigzag
direction are almost preserved while the interaction between zigzag chains goes to zero with increasing strain, becoming isolated chains at this limit \cite{Ando2002b}. The DOS of these linear conducting chains appears in figure \ref{StrainDOS}(c), in which a semiconducting gap never opens. This limit brings the system close to a quasi-1D system known as a Luttinger liquid \cite{Fradkin}. Several systems share this behavior (for example chalcogenide compounds, stripe phases of the copper oxide high-T$_{c}$ superconductors, carbon nanotubes,
two-dimensional gases in large magnetic fields, etc.), in which several exotic properties appear \cite{Fradkin}.

The mean electron velocity, as given by the group velocity of the wave packet, evolves with strain:
\begin{equation}
 \boldsymbol{v}(\boldsymbol{k})=\frac{1}{\hbar}\nabla_{\boldsymbol{k}}E(\boldsymbol{k}),
 \label{vxvy}
\end{equation}
where $\nabla_{\boldsymbol{k}}$ is the gradient operator in $\boldsymbol{k}$-space.

For electrons in graphene under uniform strain, one substitutes equation (\ref{Ekxky}) into equation (\ref{vxvy}) to obtain:
\begin{equation}
 \boldsymbol{v}(\boldsymbol{k})= \pm \frac{1}{2\hbar E(\boldsymbol{k})}\nabla_{\boldsymbol{k}}g(\boldsymbol{k}).
 \label{vectorv}
\end{equation}

This way, the components of  $\boldsymbol{v}(\boldsymbol{k})$ are given by \cite{Gomez2016}:
\begin{eqnarray}
v_l(\boldsymbol{k})=\pm\frac{1}{\hbar}\sum_{n=1}^3\sum_{s>n}^3[(1+\epsilon_{ll})(\delta_n^l-\delta_s^l)+\epsilon_{lm}(\delta_n^m-\delta_s^m)]\nonumber\\
\qquad\times t_{n} t_{s}\frac{\sin[\boldsymbol{k}\cdot(\bar{\bi{I}}+\bar{\boldsymbol{\epsilon}})\cdot(\boldsymbol{\delta}_n-\boldsymbol{\delta}_s)]}{\sqrt{\gamma+g(\boldsymbol{k})}},
\end{eqnarray}
where $l,m=\{x,y\}$, $l\ne m$, and $\delta_s^l$ denotes the $l$-component of the vector $\boldsymbol{\delta}_s$.

It is possible to calculate the electronic and optical conductivity from equation (\ref{vxvy}). Some groups have studied the electronic conductivity using uniform uniaxial strain for graphene nanoribbons
coupled to metallic leads in heterojunctions using the TB method and the Landauer formalism \cite{Wang2015}. It becomes easier to apply the effective Dirac equation formalism in some cases (see Section \ref{Optical}).

\subsubsection{Simplest strain: the isotropic expansion.}\label{IUS}

The significance of the isotropic expansion resides in
being a limiting case for checking
the consistency of any strain theory. As it will be  discussed  within the effective Dirac Hamiltonian approach in Section~\ref{Dirac}, such test leads to the discovery of problems in some formulations \cite{Oliva2013,Volovik2015,Oliva2015a}.

Isotropic biaxial strain depends
upon the single parameter $\epsilon_{Z}=\epsilon_A=\epsilon=(a'-a)/a$ and $\gamma_S=0$. This way, the lattice parameter $a$ rescales to $a'$:
\begin{equation}
 a\rightarrow a'=(1+\epsilon)a,
\end{equation}
given that the strain tensor is diagonal and $\bar{\boldsymbol{\epsilon}}=\epsilon \bar{\bi{I}}$.
As a result of a rescaled lattice constant, all hopping parameters (equation (\ref{texp})) scale to
the same value $t_0'$:
\begin{equation}
 t_0\rightarrow t_0'=t_1=t_2=t_3\approx (1-\beta \epsilon)t_0,
\end{equation}
which will rescale the energy scale (equation (\ref{rescale})) but leaves the symmetries of the Hamiltonian unchanged.

The symmetry of the reciprocal lattice is preserved under an isotropic expansion, but the reciprocal lattice vectors change to $\boldsymbol{G}_i'=\boldsymbol{G}_i/(1+\epsilon)$ because an expansion in real space shrinks reciprocal space.

The Fermi velocity  is rescaled the following way:
\begin{equation}\label{resFV}
 v_F^{\prime}=3t_0'a'/2\hbar \approx (1-\beta \epsilon +\epsilon)v_F,
 \label{eq:vFprime}
\end{equation}
while eigenvalues related to the original ones ($E_{0}$) by:
\begin{equation}\label{rescale}
 E=\frac{t_0^{\prime}}{t_{0}}E_{0}\approx (1-\beta \epsilon) E_{0},
\end{equation}
and the dispersion relation has a rescaling of the energy and of reciprocal space as follows:
\begin{equation}
E(\boldsymbol{k})= \pm \sqrt{\gamma + g(\boldsymbol{k})},
\label{Eisotropic}
\end{equation}
where $\gamma=3(1-\beta \epsilon)^{2}t_0^{2}$ and
\begin{eqnarray}
 g(\boldsymbol{k})=(1-\beta \epsilon)^{2}t_0^{2}\times\nonumber\\
 4\cos \left( \frac{3}{2}k_xa'\right) \cos \left( \frac{\sqrt{3}}{2}k_ya'\right)+2\cos \left( \sqrt{3}k_ya'\right).
\end{eqnarray}

Given that the form of the energy dispersion is preserved, the DOS
can be calculated from equation (\ref{rho}) by replacing $v_{F}$ with $v_F^{\prime}$:
\begin{equation}
 \rho (E)\approx [1+2\epsilon(\beta-1)]\rho_0(E),
 \label{eq:rho}
\end{equation}
where $\rho_0(E)$ is the DOS of unstrained graphene. The carrier concentration is obtained from $n_0(E)$ using equation (\ref{carrier}):
\begin{equation}
 n(E)\approx [1+2\epsilon(\beta-1)]n_0(E).
 \label{eq:carrier}
\end{equation}
The rescaling of $v_F$ and  $n(E)$ is in excellent agreement with DFT simulations \cite{Choi2010}. Such rescaling provides an estimation of $\beta$:  when comparing equation (\ref{eq:vFprime}) with DFT simulations \cite{Choi2010} of $v_F^{\prime}/v_F$, one gets that $\beta=2.4$.

In conclusion, isotropic strain rescales the DOS and $n(E)$ scales linearly on $2(\beta-1)\epsilon$. This explains the similar scaling of group velocities, currents and conductivities for strain without a strong spatial gradient. Such strain can be locally replaced by an average strain, equations (\ref{eq:rho}) and (\ref{eq:carrier}). The phenomena described in this Subsection has consequences on strain-charge coupling, enhanced or decreased chemical reactivity, modulation of optical properties, among others.

\subsection{Non-uniform uniaxial strain}\label{Uniaxialstrain}

\begin{figure}
\centering
\includegraphics[width=\linewidth]{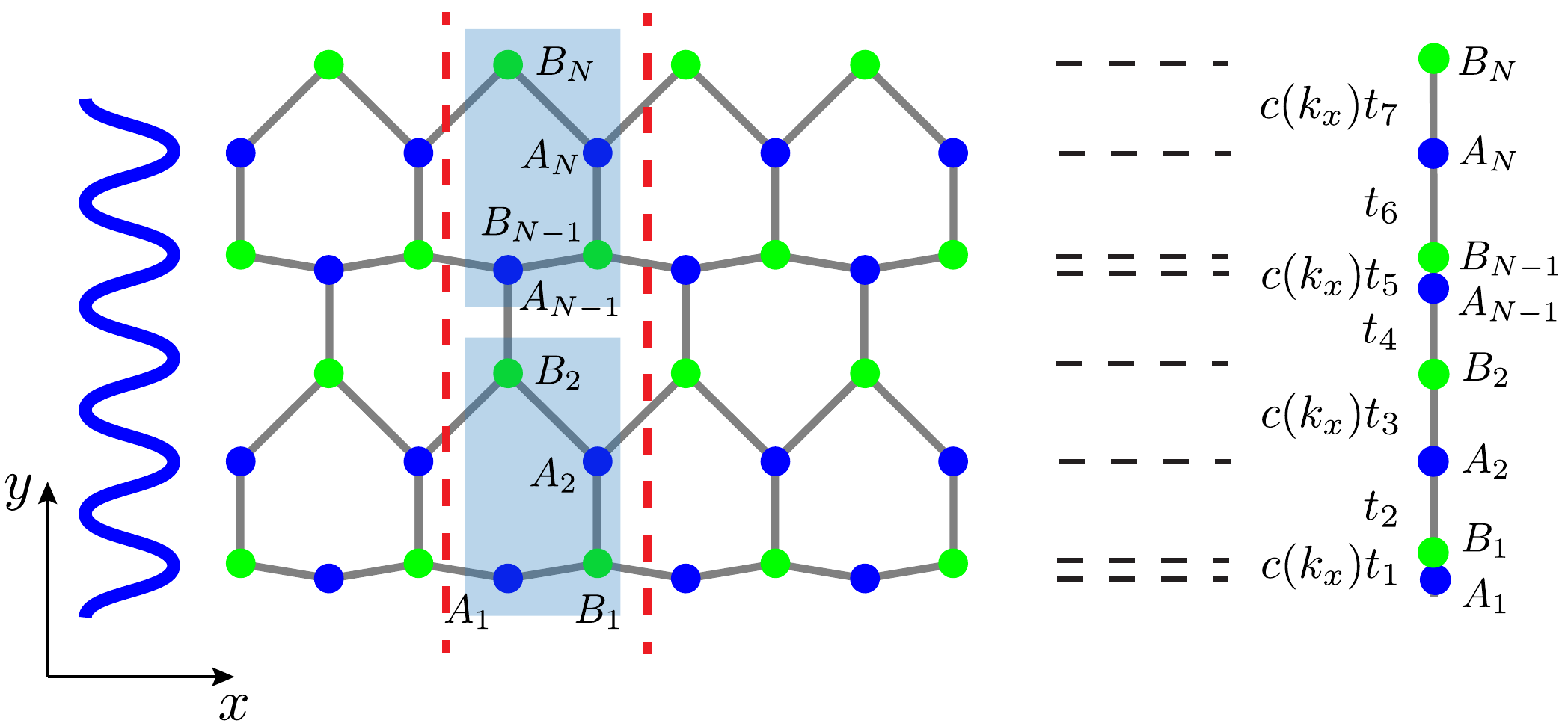}
\caption {Mapping of the graphene TB Hamiltonian with uniaxial strain along the armchair ($y$) direction into a
modulated chain described by equation ($\ref{HZZ}$). Strain is represented by the wavy line to the left. Vertical dotted lines represent the unit cell boundaries along the $x$ direction. Inequivalent positions of carbon atoms inside the unit cell are indicated by
$A_j$ and $B_j$, where $A$ and $B$ denotes the corresponding bipartite lattice and  $j$ labels the position along the $y-$direction.
The resulting map is a chain joined by effective bonds, and an effective dimerization occurs when $c(k_x)=0$. (The five-period wavy pattern to the left and the modulated chain do not possess an identical period, as only two periods of the modulated chain are shown.)}
\label{ZZmap}
\end{figure}

The first objective of this Subsection is to introduce the effects due to non-uniform uniaxial strain as
viewed from a tight-binding perspective. Although one can use numerical diagonalization
to solve the Hamiltonian given by equation (\ref{TB0K}), it is instructive to enhance our understanding by seeking analytical solutions. This allows to map the system to well-known one-dimensional equations which can be solved to enhance our physical insight. Furthermore, this type of strain can be produced experimentally
\cite{BaiKeKe2014}, and it can also be considered an approximation for the case of graphene
on a crystalline substrate with a relative rotational fault. The approach lends to an interesting comparison with
the effective Dirac equation, and points the way towards using the supercell formalism
for other types of strain. As sketched out in figure~\ref{ZZmap}, one can take advantage of uniaxial strain by noting that its Hamiltonian
 maps into one-dimensional effective chains \cite{Roman2015a,Roche08,Naumis2014,Roman2015b}.

Consider a nanoribbon in which a general uniaxial strain field $\boldsymbol{u}(\boldsymbol{r})=(0,u(y))$  is applied. For such a lattice  periodically strained along the armchair direction, the symmetry along the unstrained direction-- chosen  as the $x$ direction-- is not broken, so that the following solution to the Schr\"odinger equation can be proposed: $\Psi(\boldsymbol{r}')=\exp(ik_x x)\psi(y)$. The resulting Hamiltonian thus depends upon $k_x$ and is labeled $H(k_x)$. As seen in figure~\ref{ZZmap}, the non-strained Hamiltonian
can be built from a cell with orthogonal lattice vectors and four sites \cite{Roche08}.

Focusing attention to the right of figure~\ref{ZZmap}, one notes that all sites within the periodic supercell become inequivalent along the $y$ direction when strain is applied. One labels them
as $A_j$ and $B_j$, where $j$ is an index for the position in the path, and $A$ or $B$ labels the original bipartite lattice in the absence of strain.

The resulting Hamiltonian is that of a one-dimensional
modulated chain \cite{Naumis2014}:
\begin{eqnarray}
H(k_x)=  \sum_{j=1}^{N-1} \left[ t_{j+1} a_{j+1}^{\dag} b_{j}+
 c(k_x)t_{j} a_{j}^{\dag} b_{j} \right]+H.c.,
\label{HZZ}
\end{eqnarray}
where $a_j$, $a_{j}^{\dag}$ and $b_j$, $b_{j}^{\dag}$ are the annihilation and creation operators in the $A$ and $B$ sublattices, respectively,
and $N$ is the number of sites in the $A$ sublattice along the periodic path. For odd $j$, the effective bonds are defined through:
\begin{equation}
t_{j}=t_0 \exp\left[-\beta (u(y_{B_j})-u(y_{A_j}))/a\right],
\label{tsm}
\end{equation}
and for even $j$ as:
\begin{equation}
 t_{j}=t_0 \exp\left[-\beta (u(y_{A_{j+1}})-u(y_{B_j}))/2a\right].
 \label{tsmeven}
\end{equation}

The factor $c(k_x)$ contains the phase in the $x-$direction:
\begin{equation}
 c(k_x)=2\cos(\sqrt{3}k_xa/2).
\end{equation}

An interesting situation arises for $k_x=2/\sqrt{3}a$, where $c(k_x)=0$ and the chain decouples
into dimers \cite{Naumis2014}. For unstrained graphene, the dimers produce a massive degeneracy
leading to van Hove singularities at $E=\pm 3t_0$. For strained graphene, the degeneracy can be completely or partially removed \cite{Naumis2014}. This has
interesting consequences for electronic localization as well as for the topological properties of edge modes,
as will be discussed for the case of periodic strain.

\begin{figure}
\centering
\includegraphics[width=\linewidth]{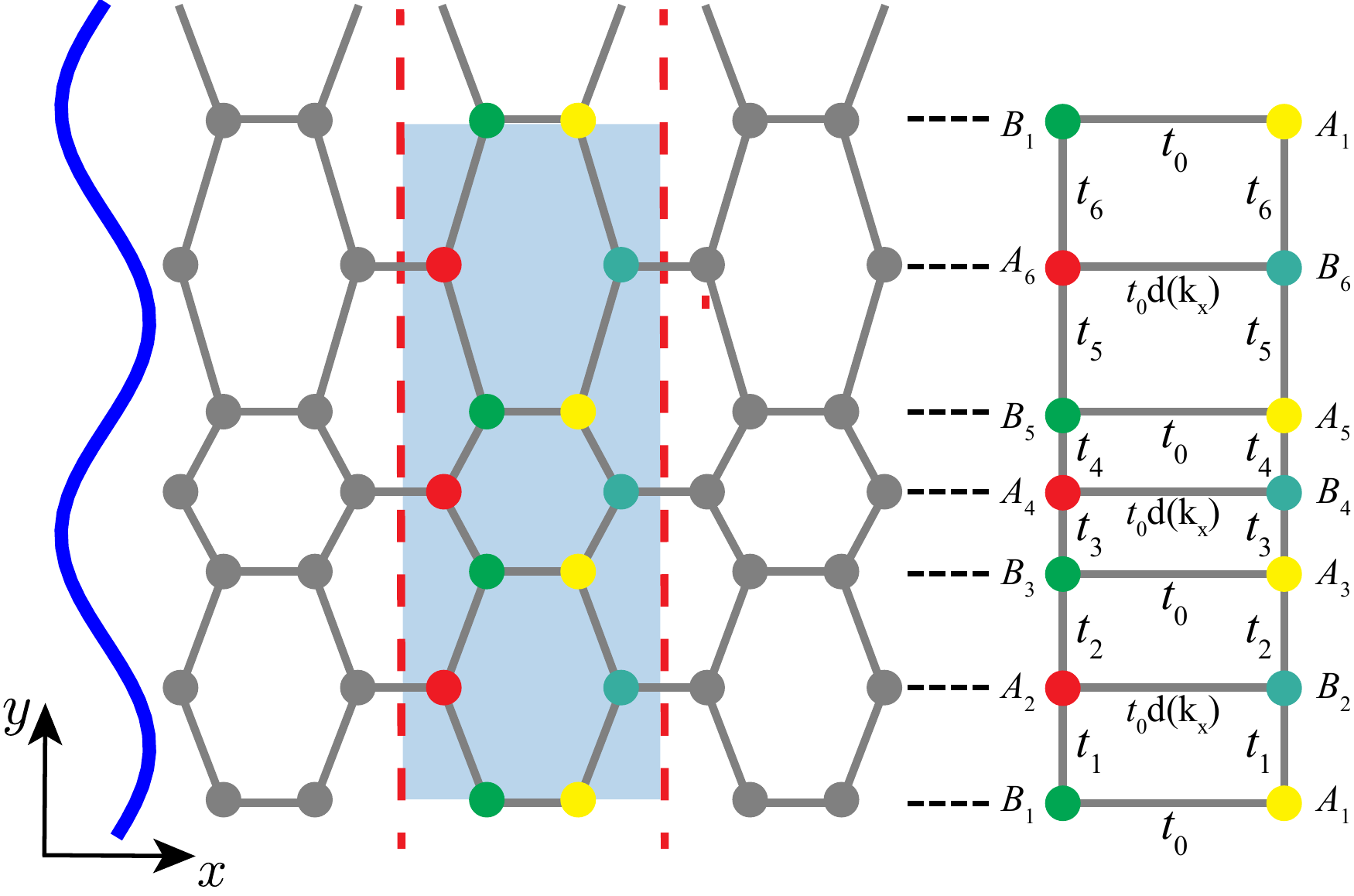}
\caption{Mapping of the graphene TB Hamiltonian for uniaxial strain along the zigzag ($y$) direction into a ladder Hamiltonian
described by equation ($\ref{HARM}$). Strain is represented by the wavy-line to the left.
The vertical dotted line represents the cell boundary along the $x$ direction, while the horizontal dotted lines are projections of  atomic positions in the $y$ direction.
Inequivalent positions of carbon atoms are indicated by colors; they map onto the two parallel chains (an effective ladder) shown to the right. Each atom is labeled $A_j$ or $B_j$ according to
their bipartite lattice, and $j$ is the position in the ladder.}
\label{ArmchairMap}
\end{figure}

The process becomes slightly different when periodic strain is applied along the zigzag direction \cite{Roman2015a}. As sketched out in figure \ref{ArmchairMap}, the resulting map
describes two modulated chains (i.e. a ladder) coupled by bonds of strength
$t_0$ and $t_0d(k_x)$, where $t_{j}$ are the values
of the transfer integrals along the chains in the $y$ direction.
The sites are best labeled $A_j$ or $B_j$, with $j$ the position along the chains.

The resulting Hamiltonian is \cite{Roman2015a}:
\begin{eqnarray}
H(k_x)&=&\sum_{j=1}^{N}
t_{0}\left[d(k_x)a_{2j}^{\dag}b_{2j}+a_{2j+1}^{\dag}b_{2j+1}\right]\nonumber\\
&& + \sum_jt_{j}a_j^{\dag}b_{j+1}+H.c.
\label{HARM}
\end{eqnarray}
The values of $t_j$ are:
\begin{equation}
 t_j=t_0 \exp{\left[-\beta \left(l_{j}/a-1\right)\right]},
 \label{effectiveodd}
\end{equation}
where $l_{j}$ are bond lengths:
\begin{eqnarray}
l_{j}= \nonumber\\
\left(\left(\delta^{x}_{s+2}\right)^2+
\left[\delta^{y}_{s+2}+u_y\left(y_{B_{j+1}}\right)-u_y\left(y_{A_{j}}\right)\right]^2\right)^{1/2},
\end{eqnarray}
and $s=0,-1$. $\delta_{s+2}^x$ and $\delta_{s+2}^y$ denote the $x$ and $y$ components of each of the
vectors $\boldsymbol{\delta}_1$ and $\boldsymbol{\delta}_2$. For armchair nanoribbons, the phase factor $d(k_x)$
is given by:
\begin{equation}
 d(k_x)=e^{ik_xa}.
\end{equation}

Using these effective mappings, the spectrum can be found by several methods. Since an
effective potential appears, one can expect different effects depending on the form of the
potential. For example, if the effective potential is periodic, as it is the case for graphene on a
substrate, a result akin to the Harper equation is found. This equation arises in the problem
of a constant magnetic field on a lattice. However, the effective equation for periodic
strained graphene is not exactly equal to the Harper equation. This result and others
concerning the spectrum and localization will be analyzed in  Section \ref{PeriodicStrain}.

\subsection{Ripples and bending}\label{Ripples}

As explained in Section \ref{DescriptionStrain}, (out of plane) ripples can appear when strain is applied, and this effect is commonly seen in MD simulations at finite temperature \cite{Fasolino,Zhang2011,Dumitrica}. 

\begin{figure}
\centering
\includegraphics[width=\linewidth]{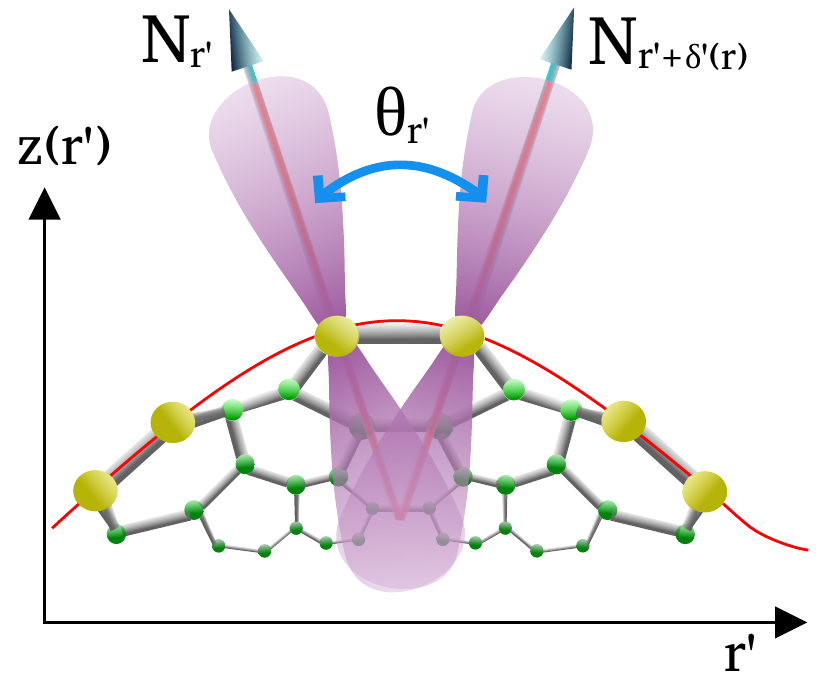}
\caption {Overlap of $\pi-$orbitals in rippled graphene. The red curve indicates a cut thorough the graphene membrane.
Local normal vectors to the surface and the corresponding angle between normals are indicated for a pair of first-neighbour atoms.}
\label{Orbitals}
\end{figure}

An important consideration in performing TB calculations for ripples is that $\pi$-orbitals change their overlap when the local normal to the graphene plane ($\boldsymbol{N_{\boldsymbol{r}}}$)
becomes a function of position due to curvature (see figure \ref{Orbitals}). Let $\theta_{\boldsymbol{r}'}$ determine
the relative orientation of a carbon  atom in the new strained position $\boldsymbol{r}'$. This angle
depends on the local curvature of the layer. The effect of the relative change of orientation for $\pi-$orbitals and
the inter-atomic distances changes has been described
in \cite{KimNeto2008,CastroNeto2009,Guinea2010}.
In this case, the transfer integral becomes  \cite{KimNeto2008,CastroNeto2009}:
\begin{eqnarray}
t_{\boldsymbol{r}',\boldsymbol{r}'+\boldsymbol{\delta}_{n}^\prime (\boldsymbol{r})}&=& t_0 \left[1+\alpha\left(1-\boldsymbol{N}_{\boldsymbol{r}'}\cdot
\boldsymbol{N}_{\boldsymbol{r}'+\boldsymbol{\delta}_{n}^{\prime} (\boldsymbol{r})}\right)\right]\nonumber\\
&&\times\exp{\left[-\beta(l_{\boldsymbol{r}',\boldsymbol{r}'+\boldsymbol{\delta}_{n}^{\prime} (\boldsymbol{r})}/a-1)\right]},
\end{eqnarray}
where $\boldsymbol{N}_{\boldsymbol{r}'}$ is the unit vector normal to the surface in the site $\boldsymbol{r}'$ in figure \ref{Orbitals}:
\begin{equation}
\boldsymbol{N}_{\boldsymbol{r}'}=\frac{\hat{\boldsymbol{z}}-\nabla z}{\sqrt{1+(\nabla z)^2}}.
\label{N}
\end{equation}
Here, $\nabla=(\partial_x,\partial_y)$ is the 2D gradient operator, and $\hat{\boldsymbol{z}}$ is the unit vector in the direction perpendicular to the plane and $z$ is the out-of-plane displacement field.
$l_{\boldsymbol{r}',\boldsymbol{r}'+\boldsymbol{\delta}_{n}^{\prime} (\boldsymbol{r})}$ is the interatomic distance between two neighboring
sites after a ripple or bent is applied. Here, $\alpha\approx0.4$
is a constant which couples the hopping term to a relative change of orientation between neighboring $\pi$-orbitals. This is the simplest approach, as other orbitals can be hybridized for
sharp ripples, thus resulting in a more complex Hamiltonian \cite{Huertas2006,Manes2013}.

It is useful to study uniaxial ripples to understand general features of curved graphene \cite{Monteverde2015}. For example, consider zigzag terminated graphene, where ripples are independent of the $x-$coordinate. In such scenario, the Hamiltonian given by equation (\ref{HZZ}) can be used with $t_j$ replaced by:
\begin{equation}
t_j = t_0
\left[1+\alpha\left(1-\boldsymbol{N}_{j+1}\cdot\boldsymbol{N}_j\right)\right]
\exp{\left[-\beta(l_{j}/a-1)\right]},
\label{tripple}
\end{equation}
where $\boldsymbol{N}_j=\boldsymbol{N}(y_j)$ is defined in equation (\ref{N}), and:
\begin{equation}
l_{j}=\sqrt{a^{2}+\left[z\left(y_{j+1}\right)-z\left(y_j\right)\right]^2}.
\end{equation}
Here:
\begin{equation}
y_{j}/a=\frac{1}{4}\left\{3j+\frac{1}{2}\left[1-(-1)^j\right]\right\},
\end{equation}
is the position of carbon atoms in unrippled graphene and $j=1$,$2$,...,$N$ labels the sites, as  displayed in
 figure \ref{ZZmap}.
Similar expressions exist for armchair-terminated graphene \cite{Roman2015b}.
For small amplitude and long wavelength ripples, the model resembles graphene under planar strain.
Due to the breakdown of sublattice symmetry, gaps open for wavelengths of the order of the lattice parameter \cite{Roman2015b}. Strained folded graphene nanoribbons have been proposed to assist and enhance electronic transport \cite{Carrillo2016}.

\subsection{Graphene on substrates: superlattices}\label{PeriodicStrain}

As it was explained in Section \ref{DescriptionStrain}, strain can also be produced \textit{via} a substrate, which leads to the creation of superlattices. Bloch's theorem
can thus be applied using the Fourier components of the resulting potential, as it was done in
Section \ref{DescriptionStrain} for the diffraction pattern. However, quasi-periodic behavior is to be expected when the lattice parameters are incommensurate.

The nature of the electronic properties  depends upon the interaction strength between graphene and the type of binding to the supporting substrate. If the interaction is weak, mainly of a
van der Waals type, the substrate does not modify the properties of graphene substantially. This is the case for graphene on hBN \cite{Dean2010,Kindermann2012,Amet2012}, or on the carbon face of SiC \cite{Haas2007}, which
 share similarities with electrons in lattices under external magnetic fields. Thus small gaps appear,
as well as Dirac cone replicas and different kinds of localization. Such systems
are very interesting for the study of exotic topological quantum phases.
Many of these properties can be understood from the perspective of a periodic uniaxial strain,
in which defining one new lattice parameter is required, as discussed in the following Subsection.

Metals as Fe, Pt and Re form partial covalent
bonds with graphene, leading to a strong interaction and a potential loss of a Dirac-cone dispersion. As the interaction becomes strong, graphene becomes rippled with amplitudes ranging between $0.03\,\mbox{\AA}$ and $1.6\,\mbox{\AA}$.

\subsubsection{Periodic uniaxial strain and ripples: the simplest superlattice.}

Consider armchair terminated nanoribbons with an oscillating uniaxial strain of the type:
\begin{equation}
u(y)/a=
\frac{\lambda}{\sqrt{3}\beta}\cos{\left[\frac{4\pi\sigma}{\sqrt{3}}\left(y/a-\sqrt{3}/4\right)\right]},
\label{uy}
\end{equation}
where $\lambda$ is a parameter that controls the amplitude of the strain and $\sigma$ is the
spatial frequency, proportional to the inverse of the strain wavelength (the constants for this strain field have been chosen to simplify the final equation).  Using a linear approximation for $t_{\boldsymbol{r}'\!,n}$, equation (\ref{effectiveodd}) becomes:
\begin{equation}
t_j=t_0\left[1+\lambda\sin{(\pi\sigma)}\sin{(2\pi\sigma j)} \right].
\label{tarmperiodic}
\end{equation}
The hopping parameter given by equation (\ref{tarmperiodic}) for graphene under uniaxial strain differs from
the Harper case by the coupling between left and right chains displayed in figure \ref{ArmchairMap}, as the graphene Hamiltonian at hand is actually made from two coupled Harper chains.

The functional dependence of $t_{j}$ is identical with that appearing in the Harper chain, which is the one-dimensional
effective equation resulting from the problem of an electron in a constant magnetic
field \cite{Hofstadter1976} and where $\sigma$ is the ratio between the magnetic flux and
the flux quantum. The energy spectrum for that problem is the  Hofstadter butterfly,
 a fractal \cite{Hofstadter1976} with interesting
topological quantum phases \cite{Naumis2016}.

For  rational $\sigma$ (such that $\sigma=P/Q$ with $P$ and $Q$ integers and the lattices are commensurate),
the system is periodic with period $Q$ for even $Q$  (or $2Q$ for odd $Q$). Bloch's theorem can be applied, resulting in
a Hamiltonian matrix of size $Q\times Q$ for even $Q$  (or $2Q\times 2Q$ for odd $Q$).

For commensurate strain (fractional $\sigma$), the relation among applied strain and system size can be understood from the perspective of superlattices: the strain modulation generates a new lattice involving $Q$ sites in the
strained direction, creating in turn a reciprocal lattice vector $\Delta \boldsymbol{g}$ that defines a new 1BZ. The effective potential can thus be written as a Fourier series, and the spectrum is found following details given in Section \ref{DescriptionStrain}.

As in the original Harper equation, the resulting Hamiltonian is no longer periodic when $\sigma \neq P/Q$ (with $P$ and $Q$ integers), being instead quasiperiodic due to incommensurability of the wavelength of the strain field and the lattice parameter \cite{Janot}. Inconmmensurability leads to fractal properties \cite{Macia2006}. Furthermore,
perturbation theory can not be used to solve for the spectra and eigenvectors due to the small divisor problem \cite{Macia2006}. Although for irrational $\sigma$ this procedure is no longer valid, it is still possible to study a sequence of rational approximants \cite{Macia2006}.

\begin{figure}
\centering
\includegraphics[width=\linewidth]{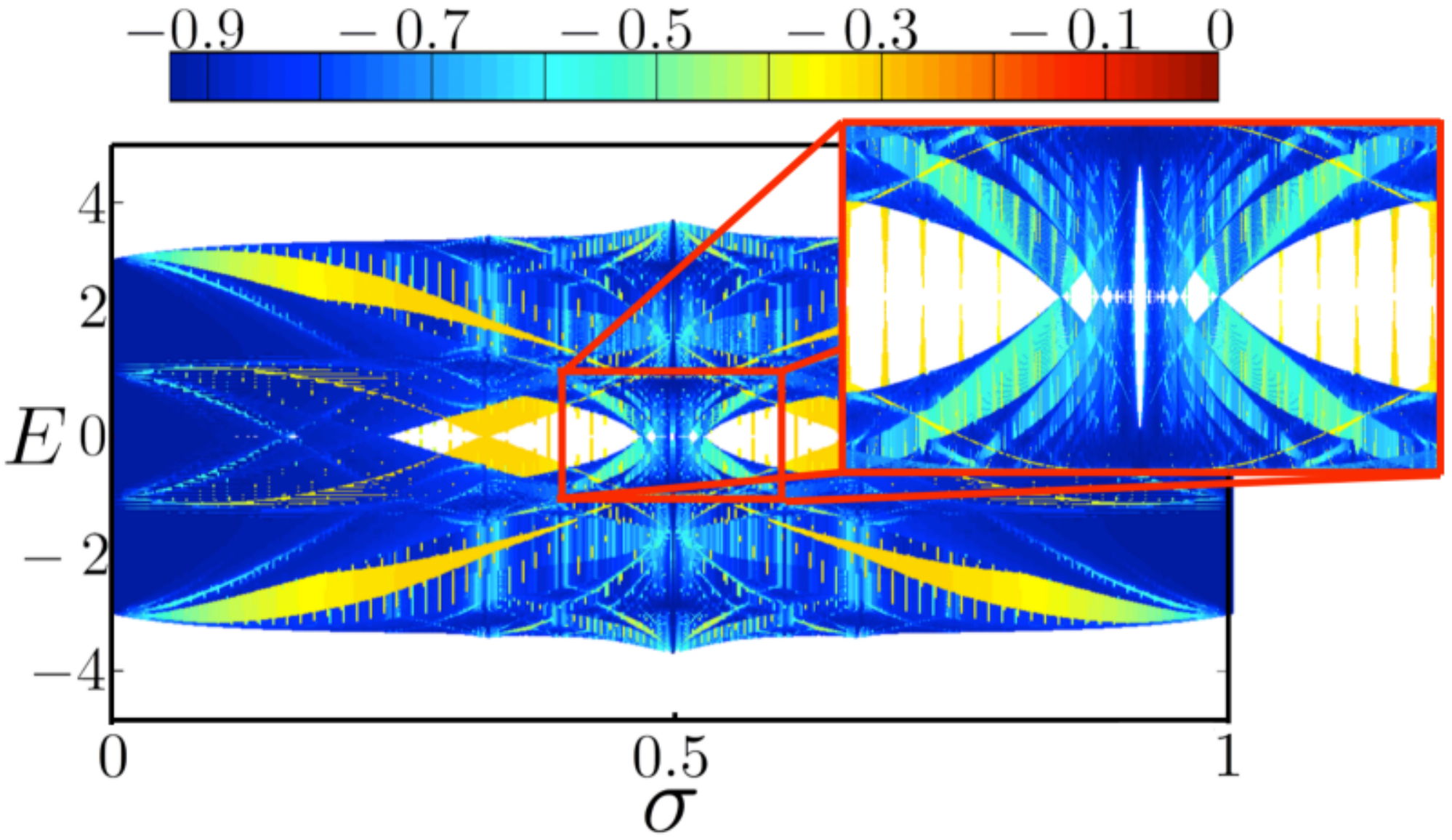}
\caption { Spectrum for uniaxial periodically strained armchair-terminated graphene. Colours represent the different localization
degrees of electron wavefunctions defined in terms of $\alpha(E)$, equation~(\ref{alpha(E)}). Blue-colored points
are extended states while red-colored points indicate states that tend to be localized. The spectrum contains features akin to the Hoftsadter butterfly
fractal. The inset shows a zoom-in around $\sigma=1/2$. Notice the gap opening for $\sigma=1/2$, when the strain wavelength is of the order of the lattice parameter. Such gap opens due to the bipartite symmetry breaking. In contrast, the long wavelength limit, i.e. $\sigma \rightarrow 0$, the spectrum is gapless with extended states. DOS cuts along
constant $\sigma$ lines are presented in figure~\ref{ArmchairDOS}. Reproduced from \cite{Roman2015a} with permission. Copyrighted by the American Physical Society.}
\label{ZigZagSpectrum}
\end{figure}

As it happens with
Harper equation, the energy spectrum depends on the magnitude of $\sigma$. As an example, figure \ref{ZigZagSpectrum} shows the spectrum
as a function of $\sigma$ for a typical value of $\lambda$. The spectrum has a
complex nature, with gaps at the Fermi level for some values of $\sigma$. The localization of states
is also very complex. To show this, colours in figure \ref{ZigZagSpectrum} represent the localization of eigenfunctions,
evaluated through the parameter $\alpha(E)$ that is defined as:
\begin{equation}\label{alpha(E)}
 \alpha(E)=\frac{\ln \sum_{j=1}^{N} |\psi_{j}(E)|^{4} }{\ln N},
\end{equation}
where $\psi_{j}(E)$ is the wavefunction at site $j$ with energy eigenvalue $E$.

Blue-colored states in figure~\ref{ZigZagSpectrum} represent extended states occurring as $\sigma \rightarrow 0$ and $\sigma \rightarrow 1$.
When $\sigma \rightarrow 0$, the strain wavelength is significantly larger than graphene's lattice
parameter. As $\sigma \rightarrow 1/2$, the situation is more complicated, since
states are extended for some rational values of $\sigma$, but they can be surrounded by more localized states that were coloured yellow in the figure.

\begin{figure}
\centering
\includegraphics[width=\linewidth]{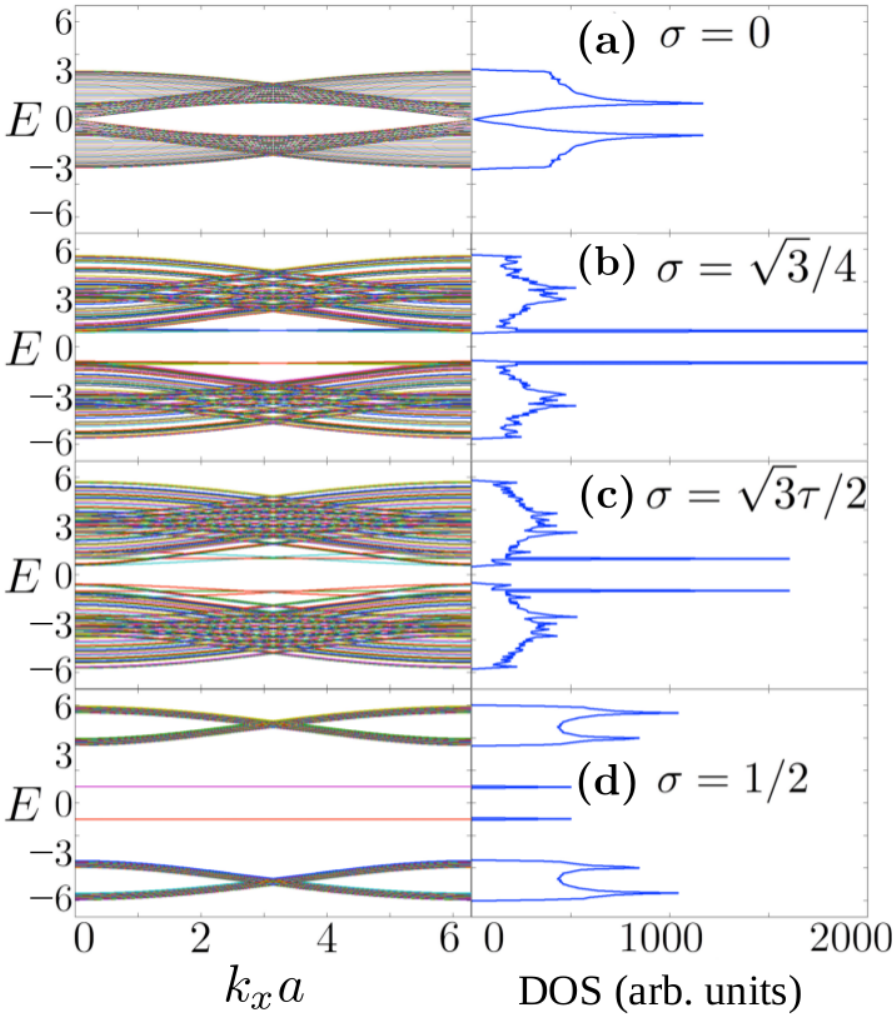}
\caption { Energy bands as a function of $k_{x}a$ for strained armchair terminated graphene with values of $\sigma$ as indicated. The resulting DOS appearing at the left shows van Hove singularities and band gaps.}
\label{ArmchairDOS}
\end{figure}

There are interesting localization properties for irrational $\sigma$  \cite{Hofstadter1976,Janot,Satija2013}. A powerful way
to study these properties is through the trace map that is defined
as a product of transfer matrices \cite{NaumisTraceMap1999}.
As seen in figure \ref{ArmchairDOS}, there are bandgap openings and closings as well as localization as a function of $k_x$.
Within the same figure, several extreme large peaks are observed in the DOS, especially for irrational values of $\sigma$.
These peaks have a counterpart in carbon nanotubes \cite{Nemec2006}, and many of these results are also
obtained for periodic ripples in graphene nanoribbons \cite{Roman2015b}.

A fundamental observation from figures \ref{ZigZagSpectrum} and \ref{ArmchairDOS} is that no gap opens for long-wavelength strain ($\sigma \rightarrow 0$). When the strain modulation has a wavelength comparable to the lattice parameter,
an asymmetry between A and B sites arises and a gap opens. In this case a tagged field, or in approximate way a reduction from the group $C_{6v}$ to
the $C_{3v}$ group, takes place. As it will be discussed later, the breaking of the sublattice symmetry creates an effective mass as well \cite{Barraza2013,Roman2015b,Manes2013,Skomski2014}.

\subsubsection{More general types of superlattices.}

The features discussed for periodic uniaxial strain are observed in more general cases. An isotropic
expansion induces a renormalization of the Fermi velocity, while biaxial strain has several consequences
in the band structure, including narrow gap openings that can be obfuscated due to the metallic nature of the substrate.

Superlattices induce a superperiod, and one can use
such periodicity to solve a tight-binding equation in the 1BZ of the superlattice,
as it was done when discussing uniaxial periodic strain before. TB parameters will depend on strain and ripples produced by the substrate and on the particular interaction with the substrate \cite{Jeil2015}.

Details of effects on electronic structure from interactions between graphene and its supporting substrate are still work in progress. For example, there are several theoretical models to understand graphene on hBN \cite{Ortix2012,Jeil2015}. The electronic dispersion strongly depends on the chosen parameters and on whether or not the perturbation is inversion-symmetric.   Electronic many-body exchange interactions seem to also play a role \cite{Jeil2015}. Thus, the opening of a gap in graphene on hBN is a highly debated issue, since some studies affirm the existence of a gap opening  \cite{Jeil2015} while others rule it out \cite{Ortix2012,Ponomarenko2013}. Recently, careful  ARPES measurements have shown gaps of $100$ meV and $160$ meV, depending on the sampled valley ($K_+$ or $K_-$), indicating the existence of a strong inversion-symmetry-breaking potential \cite{Eryin2016}. Local defects in the substrate can induce exotic phases like a broken chiral symmetry due to Kekul\'e ordering too \cite{Gutierrez2016}.

In spite of this, the following three general features occur for graphene on a weakly-interacting supporting substrate: the creation of new Dirac cones (second or third generation Dirac cones) around the original ones, the opening of minigaps, and Landau levels. The generation of second and third generation Dirac cones is not surprising from the results deduced
 in Subsection \ref{DescriptionStrain} concerning diffraction on modulated structures. Following such ideas, this Section provides a framework to understand gap openings and Dirac cone replicas, and reviews relevant experimental results.

To do so, one considers the effect of periodic modulations, assuming that the interaction between the substrate and
graphene is weak. Strain and ripples modify the creation/annihilation operators and the TB parameters of the Hamiltonian defined in equation (\ref{TB}). In addition, the substrate creates a space-dependent on-site potential that must be included into equation (\ref{TB}), leading to a more general Hamiltonian denoted by $\boldsymbol{H}_{Sl}$:
\begin{eqnarray}
\boldsymbol{H}_{Sl}=
\boldsymbol{H}+\sum_{\boldsymbol{r}^{\prime}\in A} V_{A}(\boldsymbol{r}^{\prime})a_{\boldsymbol{r}^{\prime}}^{\dag}
a_{\boldsymbol{r}^{\prime}}+\sum_{\boldsymbol{r}^{\prime} \in B} V_{B}(\boldsymbol{r}^{\prime})b_{\boldsymbol{r}^{\prime}}^{\dag}
b_{\boldsymbol{r}^{\prime}},
\label{TBSubstrate1}
\end{eqnarray}
where $V_{A}(\boldsymbol{r}^{\prime})$ and $V_{B}(\boldsymbol{r}^{\prime})$ are local on-site potential energies at $A$ and $B$ sublattices that are created by the interaction of graphene with the supporting substrate. For hBN, the B and N sites have different interactions with graphenes so that
$V_{A}(\boldsymbol{r}^{\prime})$ and $V_{B}(\boldsymbol{r}^{\prime})$  are different, even under the assumption
that hBN matches the graphene lattice  exactly.
Under such approximation of matching lattices, $V_{A}(\boldsymbol{r}^{\prime})$ and $V_{B}(\boldsymbol{r}^{\prime})$ are space independent,
and the Hamiltonian is reduced to:
\begin{eqnarray}
\boldsymbol{H}_{Sl}=\boldsymbol{H}_0+\sum_{\boldsymbol{r}\in A} V_{A}a_{\boldsymbol{r}}^{\dag}
a_{\boldsymbol{r}}+\sum_{\boldsymbol{r} \in B} V_{B}b_{\boldsymbol{r}}^{\dag}
b_{\boldsymbol{r}}.
\label{TBSubstrate2}
\end{eqnarray}

Previous expression is a tagged field akin to the one obtained under uniaxial strain for small strain wavelength ($\sigma \rightarrow1/2$). This way, equation (\ref{Hab}) for pristine
graphene is converted into:
\begin{equation}
 \left( \begin{array}{cc}
V_{A} &  H_{AB}(\boldsymbol{k})\\
H_{AB}^{*}(\boldsymbol{k})  & V_{B}  \\
\end{array} \right)
\left( \begin{array}{c} a_{\boldsymbol{k}} \\ b_{\boldsymbol{k}} \end{array} \right) =
E(\boldsymbol{k})\left( \begin{array}{c} a_{\boldsymbol{k}} \\ b_{\boldsymbol{k}} \end{array} \right).
\label{HabTagged}
\end{equation}
The new eigenvalues are:
\begin{equation}
 E(\boldsymbol{k})=\bar{V}\pm  \sqrt{\bar{V}^{2}+E_{0}^{2}(\boldsymbol{k})-V_A V_B},
\end{equation}
where the average potential is defined as $\bar{V}=(V_A+V_B)/2$ and  $E_{0}^{2}(\boldsymbol{k})$ is the energy dispersion for
pristine graphene. Since $E_{0}^{2}(\boldsymbol{K}_{\pm})=0$, a gap of the following magnitude opens:
\begin{equation}
 \Delta=|V_A -V_B|.
\end{equation}

The  diagonal terms in equation (\ref{HabTagged}) can be expressed as an average potential plus a mass term
in the effective Dirac Hamiltonian \cite{Barraza2013,Pacheco2014}, as can be readily seen by expanding $E_{0}^{2}(\boldsymbol{k})$ around $\boldsymbol{K}_{\pm}$. This argument explains  why short-wavelength strain fields open small gaps qualitatively.

More generally, $V_{A}(\boldsymbol{r}^{\prime})$ and $V_{B}(\boldsymbol{r}^{\prime})$ have a period
determined by the superlattice. Then, one follows the procedure detailed in Section \ref{DescriptionStrain} to obtain a Fourier expansion of the potentials, operators and tight-binding parameters of the supercell Hamiltonian, equation (\ref{TBSubstrate2}).  The effects of Fourier components on the electronic properties can be analyzed from second-order perturbation theory \cite{Ortix2012,Jeil2015}. However, and as indicated before, there is no consensus about which terms to include. Therefore, only the basic mechanisms behind the formation of Dirac cones replicas from a TB approach will be discussed here.

All operators are Fourier-decomposed first. For example, the two sets of annihilation and creation operators are written using two wavevectors $\boldsymbol{k_1}$ and $\boldsymbol{k_1+k}$:
 \begin{equation}
  a_{\boldsymbol{r}^{\prime}}^{\dag}= \frac{1}{\sqrt{N}}\sum_{\boldsymbol{k_1}}e^{i\boldsymbol{[(k_1+k)}\cdot(\boldsymbol{r}+\boldsymbol{u}(\boldsymbol{r}))]}a_{\boldsymbol{k}+\boldsymbol{q}}^{\dag},
  \end{equation}
  and:
  \begin{equation}
  b_{\boldsymbol{r}^{\prime}+\boldsymbol{\delta}_{n}^{\prime}(\boldsymbol{r})} \approx \frac{1}{\sqrt{N}}\sum_{\boldsymbol{k}}e^{-i\boldsymbol{k}\cdot (\boldsymbol{r}+\boldsymbol{u}(\boldsymbol{r}))}e^{-i\boldsymbol{k}\cdot \boldsymbol{\delta}_{n}^{\prime}(\boldsymbol{r})}b_{\boldsymbol{k}}.
  \label{bperiodic}
  \end{equation}

The periodicity of $\boldsymbol{u}(\boldsymbol{r})$ is accounted for by introducing a Fourier expansion in the reciprocal space superlattice vectors $\Delta \boldsymbol{g}(s_1,s_2)$:
\begin{equation}
e^{-i\boldsymbol{k_1}\cdot \boldsymbol{u}(\boldsymbol{r})}=\sum_{\Delta \boldsymbol{g}}e^{i \Delta \boldsymbol{g} \cdot \boldsymbol{r}} \tilde{V}_{\boldsymbol{k}_1}( \Delta \boldsymbol{g}),
\end{equation}
where $\tilde{V}_{\boldsymbol{k}_1}(\Delta \boldsymbol{g})$ are the expansion coefficients. In equation (\ref{bperiodic}), the strained first-neighbour
vectors (lattice corrections) $\boldsymbol{\delta}_{n}^{\prime}(\boldsymbol{r})$ are given by equation (\ref{Deltan}), i.e. they transform via the tensor operation $(\bar{\bi{I}}+\nabla \boldsymbol{u}(\boldsymbol{r}))$. To gain a better understanding, assume that $\boldsymbol{u}(\boldsymbol{r})$ has a long wavelength character.  Then one can
neglect the gradient of the field, resulting in the approximation $\boldsymbol{\delta}_{n}^{\prime}(\boldsymbol{r})=\boldsymbol{\delta}_{n}$. This result coincides with a graphene lattice under uniform strain due to the gentle existing strain gradient, and with calculations performed without inclusion of
lattice corrections \cite{Kitt2013}. Under this assumption, bond length variations are negligible and $t_{\boldsymbol{r}^{\prime},\boldsymbol{r}^{\prime}+\boldsymbol{\delta}_{n}^{\prime}(\boldsymbol{r})} \approx t_0$. Short-wavelength strain is neglected for simplicity as well ($V_{A}(\boldsymbol{r}^{\prime})=V_{B}(\boldsymbol{r}^{\prime})=0$).

\begin{figure*}[t]
\includegraphics[width=\textwidth]{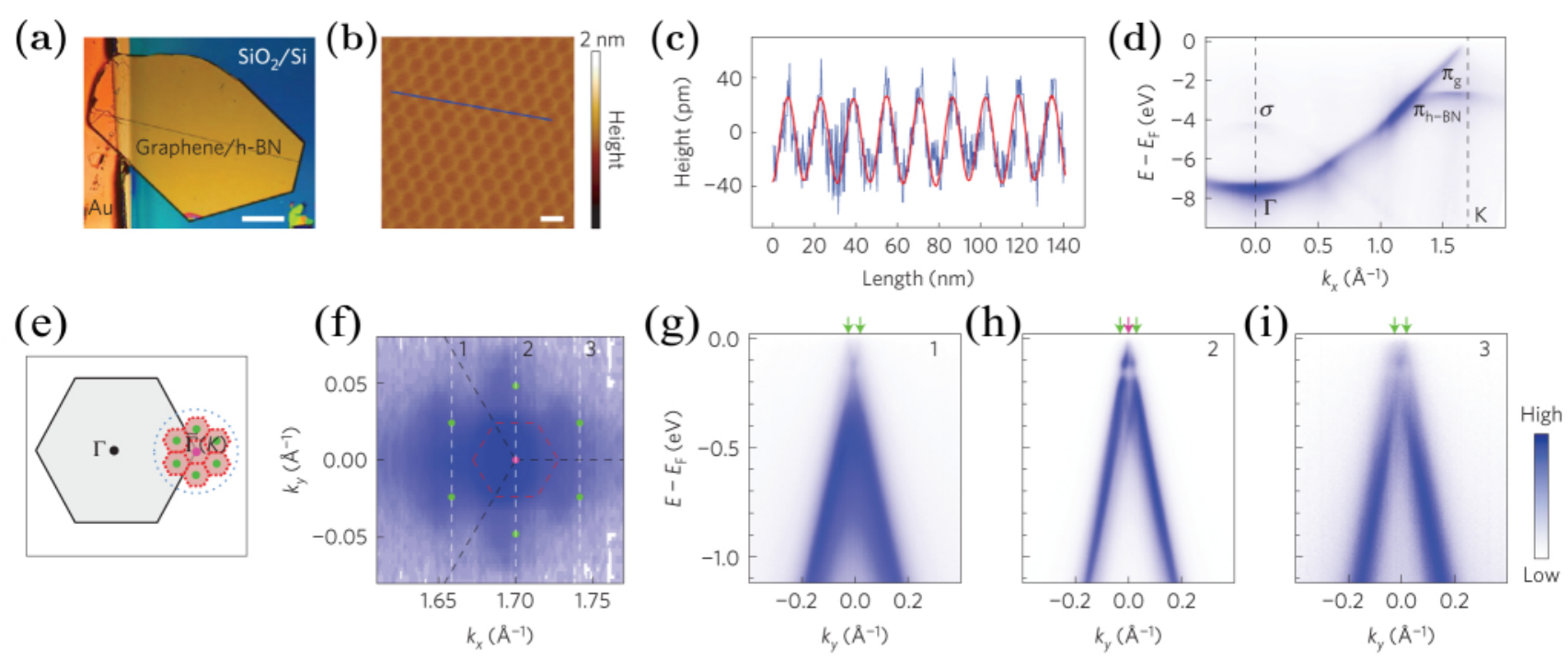}
\caption {Dirac cone replicas and band gap opening in an ARPES spectrum of graphene on hBN. (a) Optical image of the sample (scale bar is 100 $\mu$m). (b) Atomic force microscope image of the moir\'e pattern (scale bar is $20$ nm). (c) Height-profile along the blue line shown in (b). Raw data is shown in blue and the red curve is obtained from a high-pass filter. (d) Dispersion along a $\Gamma-K_+$ path in the 1BZ, showing the $\sigma$ and $\pi$ bands of graphene and hBN. (e) Schematics of graphene's 1BZ (gray hexagon)  and the phason satellites indicated in green, corresponding to  2BZ centers. The 2BZs areas are shadowed with pink and indicated with pink-dotted lines. The
$\Gamma$ and $\bar{\Gamma}(K_+)$ points of graphene are also indicated. (f) Constant energy map at $E_F$. Green dots indicate Dirac cone replicas observed at the six nearest 2BZ centers, while the red dot is the original Dirac cone. Graphene's 1BZ is indicated by broken black lines, while the superlattice BZ is indicated with red broken lines.  (g), (h) and (i) are dispersions along cuts 1, 2 and 3 that are indicated in panel (f) by white broken lines. The Dirac point is indicated with a pink arrow, while Dirac point replicas are emphasized in green arrows. Reprinted from \cite{Eryin2016} by permission from Macmillan Publishers Ltd. Copyright (2016).}
\label{ARPES_graphenehBN}
\end{figure*}

When these previous assumptions hold, the Hamiltonian given by equation (\ref{TBSubstrate1}) turns into:
\begin{equation}
\boldsymbol{H}_{Sl}(\boldsymbol{k})=\sum_{\Delta \boldsymbol{g},\boldsymbol{k}} \tilde{V}_{-\Delta {\boldsymbol{g}}}(\Delta \boldsymbol{g}) E_{0}(\boldsymbol{k})a_{\boldsymbol{k}-\Delta\boldsymbol{g}}^{\dag}b_{\boldsymbol{k}}.
\label{ModulatedH}
\end{equation}

For the first-generation high-symmetry points, obtained by setting $\boldsymbol{k}=\boldsymbol{K}_{\pm}$, one has $\boldsymbol{H}_{Sl}(\boldsymbol{k})=0$  since $E_0(\boldsymbol{K}_{\pm})=0$. This indicates that $\boldsymbol{K}_{\pm}$ are still Dirac points on first-generation Dirac cones.  When considering values of crystal momentum that are close to phason satellites of $\boldsymbol{K}_{\pm}$ (i.e. $\boldsymbol{k}=\boldsymbol{K}_{\pm}+\Delta \boldsymbol{g}_1+\boldsymbol{q}$), where  $\boldsymbol{q}$ is a small wavevector, equation
(\ref{ModulatedH}) reads as follows:
\begin{eqnarray}
\boldsymbol{H}_{Sl}(\boldsymbol{k})=\sum_{\Delta \boldsymbol{g},\boldsymbol{k}} \tilde{V}_{-\Delta {\boldsymbol{g}}}(\Delta \boldsymbol{g}) E(\boldsymbol{K}_{\pm}+\Delta \boldsymbol{g}_1+\boldsymbol{q})\nonumber\\
\times a_{\boldsymbol{K}_{\pm}+\Delta \boldsymbol{g}_1+\boldsymbol{q}-\Delta\boldsymbol{g}}^{\dag}b_{\boldsymbol{K}_{\pm}+\Delta \boldsymbol{g}_1+\boldsymbol{q}}.
\label{ModulatedTips}
\end{eqnarray}

$E(\boldsymbol{K}_{\pm}+\Delta \boldsymbol{g}_1+\boldsymbol{q})=0$ for $\boldsymbol{q}=-\Delta \boldsymbol{g}_1$, leading to a Dirac point at momentum $\boldsymbol{K}_{\pm}+\Delta \boldsymbol{g}_1$ corresponding to the satellite. Higher-order satellites will produce the same effect, leading to higher-order Dirac replicas.

Transfer integrals in the TB Hamiltonian are not severely affected in the absence of strong strain gradients. As a result, the Hamiltonian matrix has the same eigenvalues as in graphene, but the metric of reciprocal space is changed by a folding of the original energy dispersion into the second Brillouin zone (2BZ); a generic effect of small periodic modulations in the limits of the Brillouin zone \cite{Ashcroft}. Extra Fourier components enter the convolution when the derivative $\nabla \boldsymbol{u}(\boldsymbol{r})$ is not neglected.

Dirac cone replicas appear in the diffraction spots as a result of graphene's phason satellites. Considering the squared Hamiltonian introduced in Section \ref{ElectronicGraphene}, the electronic
dispersion has a minimum at intersections of Bragg lines at the boundaries of the 2BZ. As a result, each new phason
satellite produces a new minima in $\nabla_{\boldsymbol{k}} E^{2}(\boldsymbol(k))$ due to a vanishing group velocity. This minima has a parabolic dependence in the squared triangular lattice, since it corresponds to the lower band edge (Section  \ref{ElectronicGraphene}). The lowermost band turns into a Dirac cone when a ``square root'' operation is applied to recover graphene. In other words, the extra diffraction spots produce singularities in the electron dispersion that enter as standing waves in the supercell structure. The effect of a non-trivial atomic basis in monoatomic unit cells is well known for hexagonal-close packed (hcp) structures, where one talks about Jones zones due to the vanishing of the structure factor associated with the hexagonal top and bottom faces of the unitary cell prism \cite{Ashcroft}.

To conclude, graphene can be highly strained on hBN, leading to Dirac cone replicas \cite{Kindermann2012,Amet2012,Summerfield2016,Jarillo2011}. Such replicas can be understood by including strain in
a TB approximation \cite{Ortix2012,Jeil2015}, or in terms of a pseudomagnetic field in the Dirac equation \cite{SanJose2014}.

Figure \ref{ARPES_graphenehBN} presents an ARPES measurement of the electron dispersion of graphene on hBN  \cite{Eryin2016}.
The ARPES technique extracts the electron energy with respect to the Fermi level ($E-E_{F}$) as a function of the component of electron momentum in the plane of the sample,
denoted by $k_{||}$.

Second-generation Dirac cones (SDCs) are clearly seen in figure \ref{ARPES_graphenehBN}(f), \ref{ARPES_graphenehBN}(g), \ref{ARPES_graphenehBN}(h) and \ref{ARPES_graphenehBN}(i). The experiment is quite challenging as the separation of the original Dirac cone and the cloned
Dirac cones is extremely small, of the order of the reciprocal superlattice vector $\Delta \boldsymbol{g} \approx 0.05 A^{-1}$, requiring samples of extreme quality directly grown by remote plasma-enhanced chemical vapor deposition \cite{Eryin2016}.

The ARPES study reveals SDCs at the corners of the superlattice Brillouin zone, ocurring only at one of the two valleys \cite{Eryin2016} unambiguously. Gaps of approximately $100$ meV and approximately $160$ meV are
observed at the SDCs and the original graphene Dirac cone  \cite{Eryin2016}, and imply a
strong inversion-symmetry-breaking potential for graphene on hBN.

Other important features of graphene on hBN are Landau levels and a fractal spectrum in ultraclean graphene
on hBN subjected to electrostatic and magnetic fields \cite{Ponomarenko2013}.
The measured spectrum was a repeating butterfly-shaped motif, known as Hofstadter's butterfly \cite{Hofstadter1976}. This spectrum
has also been found in bilayer graphene \cite{Wang2012}, as it will be discussed in detail in
Section \ref{Multilayered}. The main reason for the fractal spectrum is the non-commensurability
of graphene-sustrate lattices, as in the case of an uniaxial strained lattice. Interestingly, there are second- and third-generation Dirac points,
leading to pronounced peaks in resistivity \cite{Ponomarenko2013}. Small lattice incommensurability prevents the
opening of gaps seen in perfectly lattice-matched graphene on hBN, and leads to a renormalized Dirac dispersion with a
trigonal warping \cite{Ortix2012,Kindermann2012}.

\begin{figure}
\centering
\includegraphics[width=\linewidth]{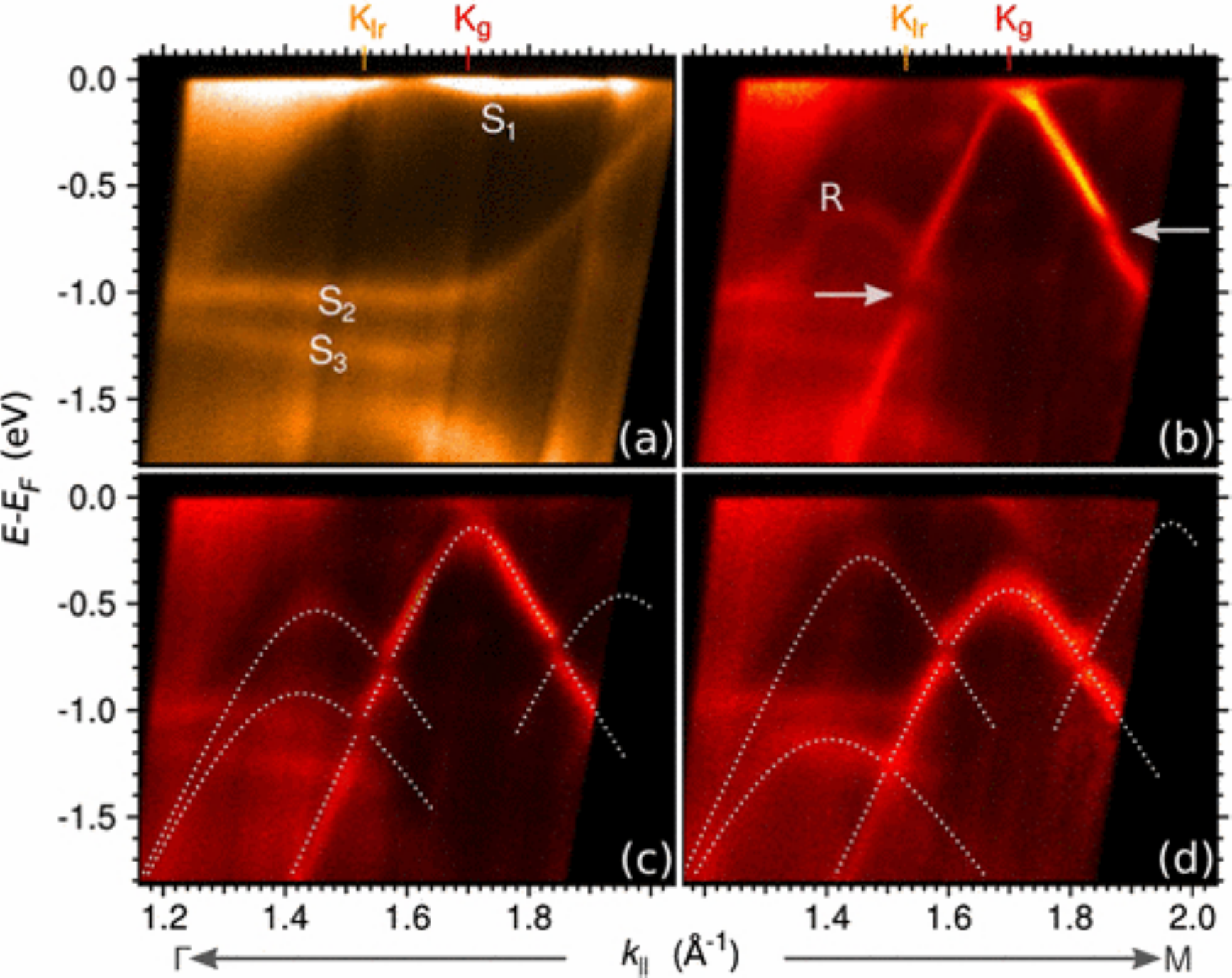}
\caption {ARPES spectrum of graphene on Ir$(111)$. The $K_+$ points of iridium ($K_{Ir}$) and graphene ($K_{g}$) are indicated for reference. (a) Clean Ir$(111)$ on an azimuthal angle of $0.5^{\circ}\pm 0.1^{\circ}$.
Three surface states are indicated by $S_{1}$, $S_{2}$, $S_{3}$.
(b) ARPES for Ir$(111)$ covered by graphene using the same angle. Horizontal arrows
denote minigaps in the primary Dirac cone.
A replica band is indicated by the letter R.  (c) and (d): Spectra
 for azimuthal angles $1.4^{\circ}\pm 0.1^{\circ}$
and $3.0^{\circ}\pm 0.1^{\circ}$, respectively. Dotted lines are results from tight-binding
calculations on a superlattice. Reproduced from \cite{Pletikosi2009} with permission. Copyrighted by the American Physical Society.}
\label{grapheneIr}
\end{figure}

Graphene over Ir$(111)$ displays effects induced by interaction with the substrate too. Figure \ref{grapheneIr} shows the ARPES spectra at different azimuthal angles around the $\Gamma-K-M$ direction \cite{Pletikosi2009}. For comparison, figure \ref{grapheneIr}(a) shows the spectrum
of Ir$(111)$ without graphene, while figures \ref{grapheneIr}(b), \ref{grapheneIr}(c) and \ref{grapheneIr}(d) display the spectrum with graphene included, along three different azimuthal angles.

The main effects
observed are  minigaps that are indicated by arrows, and replica bands, including with a Dirac cone replica \cite{Pletikosi2009}.
The measured gap width is between $0.1$ and $0.2$ eV. The Dirac cone of graphene is not hybridized with Ir,
indicating a weak interaction with the substrate \cite{Pletikosi2009}.

These minibands and gaps can be reproduced in a tight-binding calculation that includes up to three nearest neighbours, as shown with dotted
lines in figure \ref{grapheneIr} \cite{Pletikosi2009}. However, there is an important issue here. In principle, the reciprocal
vector of the moir\'e supercell implies a long wavelength periodic potential, since the moir\'e reciprocal lattice vector
$\Delta \boldsymbol{g}=\boldsymbol{G}_{gr}-\boldsymbol{G}_{Ir}$
is smaller than $\boldsymbol{G}_{gr}$ or $\boldsymbol{G}_{Ir}$, which are the original reciprocal vectors of graphene and Ir$(111)$ respectively.

As stated in previous Sections, and as seen in figure \ref{ZigZagSpectrum},  a long wavelength periodic potential is not able to
open a gap. Gap opening requires breaking the symmetry of the bipartite lattices within each unit cell \cite{Skomski2014}, as discussed
for the case of uniaxial strain before (in terms of the low-energy Dirac approximation, a band gap requires the existence of an effective mass).
This must indeed be the case for graphene on Ir$(111)$, where atoms in the two bipartite lattices are locally (and globally) inequivalent \cite{Artaud2016}.
From the above considerations, the strain field must have contribution from short and long wave-length periodic strain.

A band gap of $0.26$ eV has been experimentally determined for graphene on SiC \cite{Zhou2007}, and also
attributed to the bipartite lattice symmetry breaking \cite{Zhou2008}. This band gap, nevertheless, has also been assigned to many body interactions \cite{Bostwick2007}. A wider
band gap of  $0.5 1$ eV, has been observed for graphene on MgO $(111)$ and attributed to the bipartite lattice symmetry breaking \cite{Kong2010,Gaddam2011}.
Another interesting feature of strained generated superlattices is the possibility of  angle-dependent bandgap engineering \cite{Cervantes2016}, as well as spin-dependent transport and polarization \cite{Sattari2016}.

\subsection{Topological phases}\label{Topologicalphases}

The study of topological modes due to strain is becoming an emerging field \cite{Amorim2016}. These
are quantized edge modes that are robust against disorder because they are topologically protected \cite{Liang2011}. Topological effects on carrier transport were
first studied when magnetic fields were applied to two-dimensional electron gases confined at the interface of two semiconductors within the context
of the quantum Hall effect (QHE) \cite{Hofstadter1976}. Eventually, it was
recognized that topological modes were quantum phases with the peculiarity of not  breaking structural symmetries,
unlike the common cases seen in all previously known phase transitions \cite{Liang2011}.
Since the systems do not break structural symmetries, they are described using topological invariants instead of
structural order parameters. For unstrained graphene, most of the QHE topological aspects
are well understood now \cite{Hatsugai2011}.
Since  strain can be treated as a pseudomagnetic field, it is natural to expect topologically protected properties in
deformed graphene.

One of the first tasks to observe the topological QHE modes on graphene was to design a strain field leading to a constant pseudomagnetic field \cite{Guinea2010}. Such proposal was recently
refined to program extreme pseudomagnetic fields by uniaxial stretch  of
nanoribbons  designed with varying width \cite{ZhuShuze2015} and on ``molecular graphene'' \cite{Manoharan,Manoharan2013} (more on this in Section \ref{secdiscrete}). It is possible to generate Landau states using this profile, which is one of the QHE hallmarks.
In fact, as explained in Section \ref{Multilayered}, the first experimental observation of the QHE fractal spectrum and topological modes has been
observed in bilayer graphene on a hBN substrate.
The experimental tuning of Dirac states by strain in the topological insulator Bi$_2$Se$_3$ is expected to have an important
impact in the field as well \cite{Liu2014}, and several proposals concerning this idea were published before such experimental achievement took place. As an example,
a strain-based graphene electronic device was proposed to observe
a zero-field topological quantum phase transition between the time-reversal-symmetry-broken quantum spin Hall (QSH) and
quantum anomalous Hall states. The main feature of such device is the absence of an actual magnetic field \cite{Guassi2015}.

Uniaxial strain in nanoribbons is next employed to show how non-trivial topological properties  arise. Consider the Hamiltonian for uniaxial strain along
the zigzag direction given by equation (\ref{HZZ}), with hopping parameters given by equations (\ref{tsm}) and (\ref{tsmeven}). Furthermore, consider a strain field $\boldsymbol{u} (y)=(0,\cos(2\pi\Lambda y))$. For a wavelength  such that $\Lambda=1/(2a)$,  $t_j$ takes
only two values. By performing a Fourier transform of
equation (\ref{HZZ}) using:
\begin{equation}
a_j=\frac{1}{\sqrt{N/2}}\sum_{k_y}e^{-ik_y(j)3/2}a_{k_y},
\label{fouriera}
\end{equation}
and:
\begin{equation}
b_j=\frac{1}{\sqrt{N/2}}\sum_{k_y}e^{-ik_y(j)3/2}b_{k_y},
\label{fourierb}
\end{equation}
the Hamiltonian becomes:
\begin{equation}\label{EcSSH}
H(\boldsymbol{k})=h_x(\boldsymbol{k})\sigma_x+h_y(\boldsymbol{k})\sigma_y,
\end{equation}
where $\sigma_x$ and $\sigma_y$ are the  $x$ and $y$ Pauli matrices,
\begin{eqnarray}
h_x(\boldsymbol{k})=\nonumber\\
2(1-\lambda)\cos(\sqrt{3}k_x/2)+(1+\lambda/2)\cos(3k_y/2),
\end{eqnarray}
and:
\begin{equation}
h_y(\boldsymbol{k})=(1+\lambda/2)\sin(3k_y/2).
\end{equation}

Equation (\ref{EcSSH}) is the Su-Schrieffer-Heeger model for polyethylene \cite{Su1979}, in which  non-trivial topological
phases appear once a gap is opened \cite{Su1979,Pershoguba2012} for amplitudes  $\lambda>1/2$. For the gapless system ($\lambda<1/2$) there are topological
modes still, corresponding to flat bands that join Dirac points, known as Fermi arcs \cite{Roman2015a,DasSarma2013} that  appear in Weyl semimetals \cite{Volovik2011,Volovik2013}. The invariants used for the gapped spectra are ill-defined for such topological states with $\lambda<1/2$,
and one needs to use a new invariant \cite{Volovik2011,Volovik2013}.

As Landau levels are known to appear in monolayer and bilayer graphene due to the substrate interaction \cite{Ponomarenko2013}, the topology associated with constant magnetic fields is expected  \cite{Fradkin,Volovik2013}, including the fractal Chern-beating phenomena \cite{Satija2013} and topological collisions at van Hove singularities \cite{Naumis2016}. Finally, time-dependent strain is able to generate interesting topological properties as well \cite{Iadecola2013,Delplace2013,Majgraphene2015,RomanTaboada2017}.

\subsection{Continuum models: effective Dirac equation}\label{Dirac}

As seen in Section \ref{ElectronicGraphene}, $\pi$-electrons in pristine graphene have a linear dispersion relation in the low-energy regime ($\vert E\vert\lesssim 1$ eV) near the corners of the 1BZ \cite{Wallace1947}. This dispersion can be described in terms of a $2\times2$ Hamiltonian that is obtained by expanding the tight-binding Hamiltonian in momentum space around the $\bi{K}_{+}$ or the $\bi{K}_{-}$ point (see equations (\ref{Laue1}--\ref{Laue3}) and figure~\ref{RedGrafeno}). The low-energy Hamiltonian is obtained by means of the replacement $\bi{k}=\bi{K}_{\pm} + \bi{q}$. Such subsequent expansion up to first order in $\bi{q}$ around $\bi{K}_{+}$ gives \cite{Slonczewski1958,Semedoff1984}:
\begin{equation}\label{HDirac}
H_{ps}=\hbar v_{F}
\left(
\begin{array}{cc}
0 & q_{x}-iq_{y}\\
q_{x}+iq_{y} & 0
\end{array}
\right) =\hbar v_{F}\boldsymbol{\sigma}\cdot\boldsymbol{q},
\end{equation}
where $\boldsymbol{\sigma}=(\sigma_{x},\sigma_{y})$ is a vector whose components are Pauli matrices. Analogous expansion around $\bi{K}_{-}$ gives $H_{ps}=\hbar v_{F}\boldsymbol{\sigma}^{*}\cdot\boldsymbol{q}$, with $\boldsymbol{\sigma}^{*}=(-\sigma_{x},\sigma_{y})$. Making the replacement $\bi{q}\rightarrow -i\hbar\nabla$ in correspondence to the $\bi{k}\cdot\bi{p}$ or effective mass approximation, the Hamiltonian in equation (\ref{HDirac}) is a  two-dimensional equivalent of the Dirac Hamiltonian for massless fermions \cite{Bjorken}.

However, in contrast to the relativistic problem, the role of the velocity of light $c$ is played by the Fermi velocity $v_{F}\approx c/300$, and the two-component description given by Pauli matrices operates on the sublattice degree of freedom instead of the real spin, hence the term {\it pseudospin} that is highlighted as $H_{ps}$. {\it Pseudospin up} is another way to call sublattice (site) $A$ and {\it pseudospin down} labels site $B$.

The low-energy charge carries in graphene are relativistic fermions, giving rise to a number of unprecedented phenomena in Condensed Matter Physics, such as an anomalous quantum Hall effect, Klein tunneling, Zitterbewegung, a ``minimum'' conductivity of $\sim4e^{2}/h$ even when the carrier concentration tends to zero,
a universal optical transmittance expressed in terms of the fine-structure constant, among others \cite{Katsnelson2007,Cooper2012}.

\subsubsection{Uniform strain.} The procedure to obtain the effective Dirac Hamiltonian for graphene under a uniform strain is similar to that used for pristine, undeformed graphene. One starts with a TB Hamiltonian in momentum space for strained graphene:
\begin{equation}\label{G}
H =-\sum_{n=1}^{3}t_{n}
\left(\begin{array}{cc}
0 & e^{-i\bi{k}\cdot(\bar{\bi{I}} +
\bar{\boldsymbol{\epsilon}})\cdot\boldsymbol{\delta}_{n}}\\
e^{i\bi{k}\cdot(\bar{\bi{I}} +
\bar{\boldsymbol{\epsilon}})\cdot\boldsymbol{\delta}_{n}} & 0
\end{array}\right),
\end{equation}
and considers momenta close to the Dirac points to capture the strain-induced anisotropy appropriately \cite{Oliva2013,Volovik2014,Oliva2015a}. Here, one needs to find the new Dirac point positions explicitly from the dispersion relation, equation (\ref{DR68}). As mentioned in Section \ref{UniformStrain}, the Dirac points $\bi{K}_{\pm}^D$ are given by equations (\ref{NewKD}) and (\ref{VP}) to first order in strain tensor. Then, expanding equation (\ref{G}) around the Dirac points by means of the substitution $\bi{k}=\bi{K}_{\pm}^D+\bi{q}$, one finds that \cite{Oliva2013,Volovik2014,Oliva2015a,Pellegrino2011}:
\begin{equation}\label{NewH}
H_{ps}=\hbar
v_{F}\boldsymbol{\sigma}\cdot(\bar{\bi{I}}+\bar{\boldsymbol{\epsilon}}-\beta\bar{\boldsymbol{\epsilon}})\cdot\bi{q},
\end{equation}
is the effective Dirac Hamiltonian for uniformly strained graphene. It can be immediately verified that, for $\bar{\boldsymbol{\epsilon}}=0$, the Hamiltonian (\ref{NewH}) reduces to the Hamiltonian (\ref{HDirac}) of unstrained graphene.  At the same time, the Hamiltonian in equation (\ref{NewH}) is a particular case of the generalized effective Dirac Hamiltonian reported in \cite{Oliva2016a} for a honeycomb lattice with weak anisotropy in the hopping parameters, whereas an extension up to second in the strain tensor $\bar{\boldsymbol{\epsilon}}$ is carried out in \cite{Oliva2017}.

Some remarks on the effective Hamiltonian, equation (\ref{NewH}), follow. First, it is independent of the choice of reference system. Second, it has two contributions due to strain: a $\beta$-independent term, $\hbar
v_{F}\boldsymbol{\sigma}\cdot\bar{\boldsymbol{\epsilon}}\cdot\bi{q}$, which is purely
geometric, and a $\beta$-dependent term
$-\hbar v_{F}\beta\boldsymbol{\sigma}\cdot\bar{\boldsymbol{\epsilon}}\cdot\bi{q}$ arising from strain-induced changes in the hopping parameters. For graphene, both contributions have the
same order of magnitude. Third, one identifies a Fermi
velocity tensor from equation~(\ref{NewH}) that is given by  \cite{Oliva2013,Oliva2015a}:
\begin{equation}\label{VecKD}
\bar{\bi{v}}=
v_{F}(\bar{\bi{I}}+\bar{\boldsymbol{\epsilon}}-\beta\bar{\boldsymbol{\epsilon}}),
\end{equation}
whose tensorial character is due to the elliptic shape of the isoenergetic contours around $\bi{K}^{D}$. The principal axes of $\bar{\bi{v}}$ are collinear with the principal axes of $\bar{\boldsymbol{\epsilon}}$. For uniaxial strain, for example, the eigenvalues of $\bar{\bi{v}}$ are \cite{Pereira2009a}:
\begin{equation}
v_{\parallel}=v_{F}(1-\tilde{\beta}\epsilon),\ \ \ \ v_{\perp}=v_{F}(1+\tilde{\beta}\nu\epsilon),
\end{equation}
where $\tilde{\beta}=\beta-1$, $\nu$ is the Poisson's ratio, $\epsilon$ is the strain and $v_{\parallel}(v_{\perp})$ is the Fermi velocity parallel (perpendicular) to the direction of the applied strain. In \cite{Betancur2015}, an accurate and alternative approach is carried out to estimate $v_{\parallel}$ and $v_{\perp}$ from a fitting of the $\pi$-bands obtained with a DFT calculation for graphene under uniaxial strains along the zigzag and armchair directions. Morevoer, it is worth mentioning that an analysis beyond first order in $\bar{\boldsymbol{\epsilon}}$ reveals that the principal axes of the Fermi velocity tensor are only collinear with the principal axes of $\bar{\boldsymbol{\epsilon}}\left(\theta \right)$ if the stretching is along the zigzag or armchair directions \cite{Oliva2017}.

Consider an isotropic expansion ($\bar{\boldsymbol{\epsilon}}=\epsilon\bar{\bi{I}}$) as a consistency test. It was demonstrated in Section \ref{IUS} that the Fermi velocity of graphene is given by $v_{F}(1+\epsilon-\beta\epsilon)$ to first order in $\epsilon$ (see equation (\ref{resFV})). Therefore, any expression reported as the Fermi velocity tensor for strained graphene must yield $v_{F}(1+\epsilon-\beta\epsilon)$ when evaluated for $\bar{\boldsymbol{\epsilon}}=\epsilon\bar{\bi{I}}$. When this simple test is not fulfilled \cite{deJuan2012,Ramezani2013,Shallcross2016b}, the most probably cause will be an expansion around points of the reciprocal space which are not the true Dirac points.

One can use the Hamiltonian (\ref{NewH}) to evaluate other quantities up to the first order in strain. For example, the local density of states (LDOS) is given by \cite{deJuan2013,Oliva2014,Oliva2014C}:
\begin{equation}\label{DOSSG}
\rho(E)\approx\rho_{0}(E)\Bigl(1-\tilde{\beta}\mbox{Tr}(\bar{\boldsymbol{\epsilon}})\Bigr),
\end{equation}
where $\rho_{0}(E)$ is the DOS of unstrained graphene given by equation (\ref{DOS0}) and $\mbox{Tr}(\bar{\boldsymbol{\epsilon}})$ denotes the trace of $\bar{\boldsymbol{\epsilon}}$. Since $\tilde{\beta}>0$, the effect of strain on the LDOS depends on the sign of $\Tr(\bar{\bepsilon})$, which can be written as
$\Tr(\bar{\bepsilon})=S/S_{0}-1$, being $S$ ($S_{0}$) the area of the strained (unstrained) graphene sample. Thus, for an expanded sample ($S/S_{0}>1$) the effect is a decrease of the LDOS, whereas the effect is an increase for a compressed sample ($S/S_{0}<1$). The LDOS does not change for a shear strain ($S/S_{0}=\Tr(\bar{\bepsilon})=0$). These results permit understanding scanning tunnelling spectroscopy (STS) measurements which are sensitive to the LDOS \cite{Jang2014}.

Anisotropic Dirac quasiparticles  described by the effective Hamiltonian $H=\hbar(v_{x}\sigma_{x}q_{x}+v_{y}\sigma_{y}q_{y})$ have Landau levels given by $E_{n}=\sqrt{v_{x}v_{y}/v_{F}^{2}}E_{n}^{(0)}$ \cite{Goerbig2008}, where $E_{n}^{(0)}=\mbox{sgn}(n)\sqrt{2e\hbar Bn}$ are the conventional relativistic Landau levels \cite{Goerbig2011}. This result can be extrapolated to strained graphene as described by equation (\ref{NewH}) after diagonalizing the Fermi velocity tensor (equation (\ref{VecKD})). After algebraic manipulations to first order in the strain tensor, the Landau levels of uniformly strained graphene are given by:
\begin{equation}
E_{n}=E_{n}^{(0)}\left(1-\frac{\tilde{\beta}}{2}\mbox{Tr}(\bar{\boldsymbol{\epsilon}})\right).
\end{equation}

Last expression explains corrections of the Landau levels in the presence of a position-dependent Fermi velocity \cite{Storz2016}, which is interpreted as emergent gravity within an approach based on general relativity  \cite{Khaidukov2016}.

\begin{figure*}
\centering
\includegraphics[width=\textwidth]{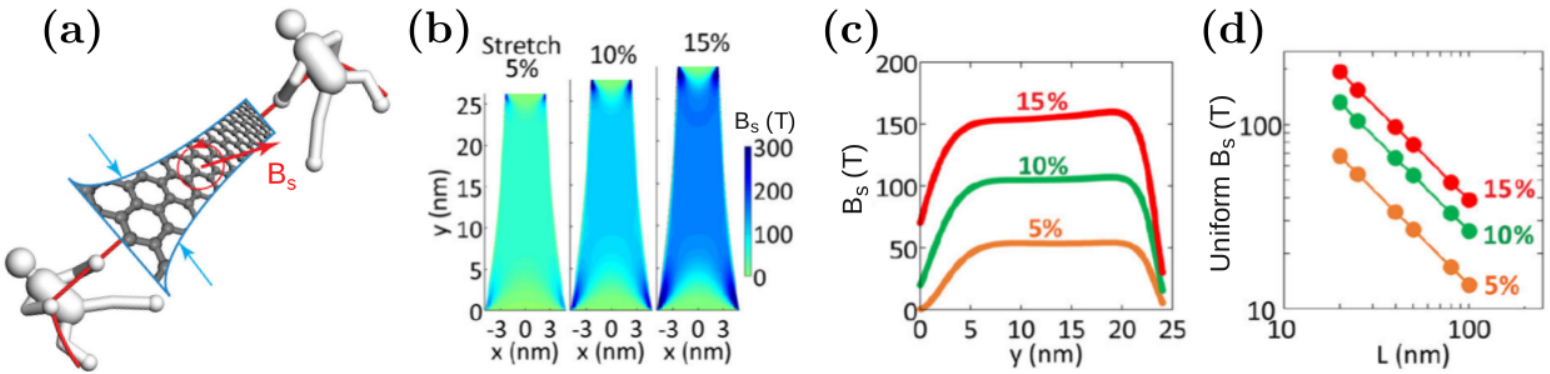}
\caption {(a) Schematics of a graphene strip of varying width under a uniaxial pulling, leading to a uniform pseudomagnetic field $\bi{B}_{s}$. (b) Resulting pseudomagnetic fields in the graphene nanoribbon shown in (a) under a uniaxial stretch of $5\,\%$, $10\,\%$, and $15\,\%$, respectively. (c) Intensity of the pseudomagnetic field as the function of position along the centerline of the graphene ribbon for various applied stretches. (d) Intensity of the pseudomagnetic field is shown to be linearly proportional to the applied uniaxial stretch and inversely proportional to the length of the graphene ribbon $L$. Adapted from \cite{ZhuShuze2015} with permission.}
\label{Fig_PLL}
\end{figure*}

\subsubsection{Nonuniform strain: Gauge fields and position-dependent Fermi velocity.}
If the spatial variation of strain is small on the interatomic scale, one expands the Hamiltonian obtained for uniform strain, equation (\ref{NewH}), around the true Dirac point in the momentum space, and goes to real space by the replacement \cite{Volovik2014,Oliva2015a}:
\begin{equation}\label{rule}
{v}_{mn}q_{l}\rightarrow
{v}_{mn}(\bi{r})\left(-i\frac{\partial}{\partial r_{l}}-K_{l}^{D}(\bi{r})\right) -
\frac{i}{2}\frac{\partial{v}_{mn}(\bi{r})}{\partial
r_{l}},
\end{equation}
where the last term assures hermiticity of the ensuing Hamiltonian \cite{deJuan2013,deJuan2012}.
This transformation makes strain space-dependent $\bar{\boldsymbol{\epsilon}}=\bar{\boldsymbol{\epsilon}}(\bi{r})$ \cite{Zhang1994,Linnik2012}. Thus, the Dirac points and the Fermi velocity (both functions of $\bar{\boldsymbol{\epsilon}}$ for uniform strain) now become functions of position $\bi{r}$, as denoted by $\bi{K}_{\pm}^D(\bi{r})$ and $\bar{\bi{v}}(\bi{r})$, respectively. As a consequence, $\bi{K}_{\pm}^D(\bi{r})$ can be interpreted as a pseudovector potential (gauge field), which yields an alternative physical picture for the strain-induced pseudomagnetic field \cite{Yang2011}. The rotational of the gauge field $\bi{K}^{D}(\bi{r})$ gives \cite{Oliva2015a}:
\begin{eqnarray}\label{pseudoB}
B_{s}&=&\nabla\times\bi{K}_{\pm}^D(\bi{r})\nonumber\\
&=&\nabla\times(\bar{\boldsymbol{\epsilon}}(\bi{r}) - \bar{\boldsymbol{\omega}}(\bi{r}))\cdot\bi{K}_{\pm} \pm\nabla\times\bi{A}_{s}(\bi{r})\nonumber\\
&=&\pm(\partial_{x}A_{y}-\partial_{y}A_{x})\nonumber\\
&=&\mp\frac{\beta}{2a}(2\partial_{x}{\epsilon}_{xy}(\bi{r})+\partial_{y}{\epsilon}_{xx}(\bi{r})-\partial_{y}{\epsilon}_{yy}(\bi{r})),
\end{eqnarray}
which is perpendicular to the graphene sample and has opposite signs for different valleys due to the preserved time-reversal symmetry under mechanical deformations.

The inclusion of the local rotation tensor $\bar{\boldsymbol{\omega}}(\bi{r})$ through equation (\ref{ew}) can be employed to demonstrate the lack of $\bi{K}_{\pm}$-dependent pseudovector potentials \cite{Kitt2012,Kitt2013} (another way is presented within the context of the discrete approach \cite{Sloan2013,Barraza2013} in Section \ref{secdiscrete}). Equation (\ref{pseudoB}) is the expression of the strain-induced pseudomagnetic field $\bi{B}_{s}$ that appears in early derivations \cite{Suzuura2002,Sasaki2008} where, unlike the approach presented here, the pseudovector potential comes from an expansion of the strained TB Hamiltonian around the fixed points $\bi{K}_{\pm}$. The emergence of the pseudomagnetic field due to nonuniform strain has also been considered from a symmetry analysis \cite{Manes2013,Cabra2013}, and from the context of a quantum field theory in curved spaces \cite{Yang2015,Iorio2015,Arias2015,Pavel2017}.

Replacing equation (\ref{rule}) into the Hamiltonian given by equation (\ref{NewH}), the effective Dirac Hamiltonian for graphene under a
nonuniform strain can be written as \cite{Volovik2014,Oliva2015a}:
\begin{equation}\label{HG}
H_{ps}=-i\hbar\boldsymbol{\sigma}\cdot\bar{\bi{v}}(\bi{r})\cdot\nabla - \hbar
v_{F}\boldsymbol{\sigma}\cdot\bi{A}_{s}(\bi{r}) - \hbar
v_{F}\boldsymbol{\sigma}\cdot\boldsymbol{\Gamma}_{s}(\bi{r}),
\end{equation}
where the position-dependent Fermi velocity tensor
$\bar{\boldsymbol{v}}(\bi{r})$ is given by:
\begin{equation}\label{GV}
\bar{\bi{v}}(\bi{r})=
v_{F}\bigl(\bar{\boldsymbol{I}}+\bar{\boldsymbol{\epsilon}}(\bi{r})-\beta\bar{\boldsymbol{\epsilon}}(\bi{r})\bigr),
\end{equation}
and the $l$-component of the corresponding complex vector field $\boldsymbol{\Gamma}_{s}(\bi{r})$ is:
\begin{equation}\label{Gamma}
\Gamma_{s,l}=\frac{i}{2
v_{F}}\partial_{j}{v}_{lj}(\bi{r})=\frac{i(1-\beta)}{2}\partial_{j}{\epsilon}_{lj}(\bi{r}),
\end{equation}
with an implicit sum over repeated indices. The complex gauge field $\boldsymbol{\Gamma}_{s}$ is due to a position-dependent Fermi velocity, and its presence guarantees the hermiticity of the Hamiltonian, equation (\ref{HG}). Unlike $\bi{A}_{s}$, $\boldsymbol{\Gamma}_{s}$ is a purely imaginary so that it cannot be interpreted as a gauge field. Therefore, it does not give rise to pseudomagnetism \cite{deJuan2013}. At present, experimental signatures of such complex gauge field $\boldsymbol{\Gamma}_{s}$ remain as open questions (see Section \ref{secdiscrete} for a discussion of hermiticity within the context of space-dependent pseudospin Hamiltonians).

Besides graphene, the procedure described in previous paragraph has been recently employed to obtain the effective low-energy Hamiltonian of 3D Dirac semimetals \cite{Zubkov2015} and certain class of 3D Weyl semimetals \cite{Cortijo2016} under mechanical deformations. In general, this approach corresponds to the scheme of emergence of gauge fields and gravity in the vicinity of the Dirac, Weyl or Majorana points in the energy spectrum \cite{Froggatt1991,Volovik2003,Horava2005}.

Nowadays, transport signatures of pseudomagnetic fields are investigated intensively \cite{Falko2013,Roy2013,Cosma2014,Roy2014,Bahamon2015,Burgos2015,Stegmann2016,Venderbos2016,Kauppila2016,Yeh2016,Mikkel2016b,Mikkel2016c}.
These strain-derived fields permit observing Landau quantization and a pseudo-quantum Hall effects under zero external magnetic fields. Early proposals of  such strain distributions \cite{Guinea2010,Guinea2010b} are technically challenging because they require complex stress fields applied at the boundaries of graphene sample that can be engineered in molecular graphene \cite{Manoharan}. In addition, Zhu {\it et al.} have conceived an alternative method to achieve programmable extreme pseudomagnetic fields with uniform distributions over large areas \cite{ZhuShuze2015} by pulling a nanorribon along one axis (see figure~\ref{Fig_PLL}). They revealed the special shape in which a graphene strip must be cut (like a 2D projection of a musical horn) so that, pulling on its ends yields a uniform pseudomagnetic field.

Experimental confirmation of strain-induced pseudomagnetic fields was reported on scanning tunneling microscopy of graphene nanobubbles \cite{Levy2010}. The scanning
tunneling spectroscopy (STS) spectra measured directly over nanobubbles (strained graphene regions) showed a series of peaks in the same manner as if graphene was subjected to an external magnetic field \cite{Miller2009}. Giant values of the pseudomagnetic fields $\bi{B}_{s}$ of $300\,\mbox{Tesla}$ were determined.  For comparison, (pulsed) magnetic fields of up to 100 Tesla can only be obtained in state-of-the-art experimental facilities.

Subsequently, many experimental confirmations of the pseudomagnetic fields in graphene have been reported under other strain configurations \cite{Yeh2011,Lu2012,Guo2012,Klimov2012,Meng2013,LiSiYu2015,Yan2016,Goergi2016,Jiang2017} and in various graphene-like systems such as molecular graphene \cite{Manoharan}, photonic crystals \cite{Rechtsman2013}, and optical lattices \cite{Tian2015}. Beyond graphene, strain-induced gauge fields have been also characterized on bilayer graphene \cite{Mariani2012,Yan2013,He2014}, borophene \cite{Zabolotskiy2016}, topological insulators \cite{Tang2014}, transition metal dichalcogenides \cite{Roldan2015,Pearce2016,Ochoa2016} and on three-dimensional Dirac and Weyl semimetals \cite{Zubkov2015,Cortijo2016,Cortijo2016b,Pikulin2016,Grushin2016}.

The effects solely due to a position-dependent Fermi velocity tensor on the spinor wavefunction of charge carriers are now discussed. For this purpose, consider the following scenario:
\begin{equation}
\bar{\bi{v}}(x)=v_{F}
\left(\begin{array}{cc}
1 + f(x) & 0\\
0 & 1
\end{array}\right), \ \ \boldsymbol{\Gamma}_{s}=(i f'(x)/2,0),
\end{equation}
i.e. the Fermi velocity varies along the $x$ axis. At the same time, assume $\nabla\times\bi{A}_{s}(\bi{r})=0$ to disregard the effect of the pseudomagnetic field \cite{Oliva2015a,Peres2009,Atanasov2010}. Then, from equation (\ref{HG}) one writes the corresponding time-independent Dirac equation for the spinor wavefunction $\Psi$ as:
\begin{eqnarray}
 \bigl(-i(1+f(x))\partial_{x} - \partial_{y} - i f'(x)/2\bigr)\psi_{2}&=&\varepsilon\psi_{1},\nonumber\\
 \bigl(-i(1+f(x))\partial_{x} + \partial_{y} - i f'(x)/2\bigr)\psi_{1}&=&\varepsilon\psi_{2},\label{LS}
\end{eqnarray}
where $\varepsilon\equiv E/(\hbar v_{F})$ and $E$ is the electron energy. Following the calculation presented in \cite{Oliva2015a}, this system has as solution to the spinor wavefunction:
\begin{equation}\label{sol}
 \Psi(\bi{r})= C \exp\Bigl[i k_{y} y + \int^{x}\frac{i k_{x} - f'(\tilde{x})/2}{1+f(\tilde{x})} d\tilde{x}\Bigr]
 \left(\begin{array}{c}
1 \\
s e^{i\theta}
\end{array}\right),
\end{equation}
where $e^{i\theta}=(k_{x}+ik_{y})/|\varepsilon|$,  $\varepsilon=\pm(k_{x}^{2}+k_{y}^{2})^{1/2}$, $C$ is a normalization constant and $s=\pm1$ denotes the conduction band and valence bands, respectively.

Some remarks follow from equation (\ref{sol}). First:
\begin{equation}\label{mod}
 |\Psi|^{2}\sim(1+f(x))^{-1},
\end{equation}
therefore, a position-dependent Fermi velocity induces an inhomogeneity in the carrier probability density, which was early analysed by using a quantum field theory approach in curved
spaces \cite{deJuan2007}. However, a position-dependent Fermi velocity does not lead to the emergence of band gap structure in the energy spectrum. Besides, given the arbitrariness of the function $f(x)$, equation (\ref{sol}) allows to consider more complex scenarios than those described by Kronig-Penney-like models for the variation of the Fermi velocity along a given spatial direction \cite{Raoux2010,Ratnikov2014,Lima2015b,Lima2016}.

\subsubsection{Time-dependent strain.}

As discussed above, a nonuniform deformation of the graphene lattice can be interpreted as a pseudomagnetic field given by equation (\ref{pseudoB}). Within the same theoretical framework and by analogy with the normal electromagnetic field, a time-dependent deformation must also give rise to a pseudo-electric field given by $\bi{E}_{s}=-\partial_{t}\bi{A}_{s}$ \cite{Oppen2009,Trif2013}, which accelerates electrons and induces an alternating electric current. However, since $\bi{E}_{s}$ couples with opposite signs in the two valleys, it does not generate a net electric current.
Some works have been devoted to reveal observable consequences of this pseudoelectric field, but the predicted effects (a topological electric current \cite{Vaezi2013,Zhang2016} and the modification of the Raman spectrum \cite{Sasaki2014}) are not experimentally confirmed yet.

\begin{figure}[tb]
\centering
\includegraphics[width=\linewidth]{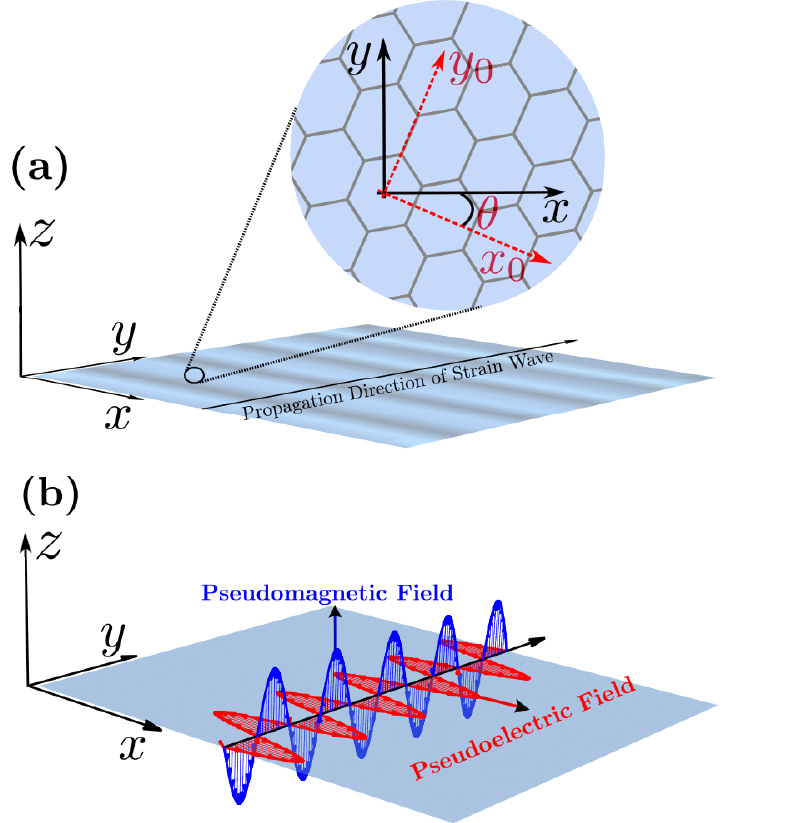}
\caption{(a) A deformation wave propagating in
graphene. The dark regions on the sample represent zones of higher density of carbon atoms. The inset plays the relation between the arbitrary coordinate system $(x,y)$ and
the crystalline coordinate system $(x_{0},y_{0})$. (b) Unstrained graphene under the equivalent  pseudoelectromagnetic wave. The pseudoelectric field lie in graphene plane, whereas the pseudomagnetic field is out-of-plane. Reproduced from \cite{Oliva2016b} with permission.}\label{fig_SW}
\end{figure}

An analogy with a real electromagnetic problem is given by considering a time-dependent deformation field of the graphene lattice $\bi{u}(\bi{r},t)$ of the form \cite{Oliva2016b}:
\begin{equation}\label{def}
\bi{u}=(0,u_{0}\cos(\Lambda y- \omega t)),
\end{equation}
with $u_{0}\ll a\ll 2\pi/\Lambda$, i.e. the atomic displacement $u_{0}$ is much less than the unstrained carbon-carbon distance $a$, while the wavelength $2\pi/\Lambda$ is much greater than $a$. As illustrated in figure \ref{fig_SW}(a), this deformation wave propagates along the $y$ direction, which is rotated by an arbitrary angle $\theta$ with respect to the crystalline coordinate system $(x_{0},y_{0})$ in which the $x_{0}$-axis points along graphene's zigzag direction (the mechanical wave moves along the armchair direction for $\theta=0$, and along the zigzag direction when $\theta=\pi/2$).

A general expression for the pseudovector potential $\bi{A}_{s}$ in the (rotated) frame $(x,y)$ is given by \cite{Zhai2010}:
\begin{eqnarray}\label{StandarA}
A_{s,x}&=&\frac{\beta}{2a}\Bigl(({\epsilon}_{xx} - {\epsilon}_{yy})\cos3\theta - 2{\epsilon}_{xy}\sin3\theta\Bigr),\nonumber \\
A_{s,y}&=&\frac{\beta}{2a}\Bigl(-2{\epsilon}_{xy}\cos3\theta - ({\epsilon}_{xx} - {\epsilon}_{yy})\sin3\theta\Bigr),
\end{eqnarray}
whose periodicity of $2\pi/3$ in $\theta$ reflects the trigonal symmetry of the honeycomb lattice. The following explicit form of the effective gauge field is obtained for the deformation field given in equation (\ref{def}):
\begin{equation}\fl\label{OurA}
 \bi{A}_{s}=\frac{\beta u_{0}\Lambda}{2a}\sin(\Lambda y - \omega t)(\cos3\theta,-\sin3\theta).
\end{equation}

\begin{figure}[tb]
\centering
\includegraphics[width=\linewidth]{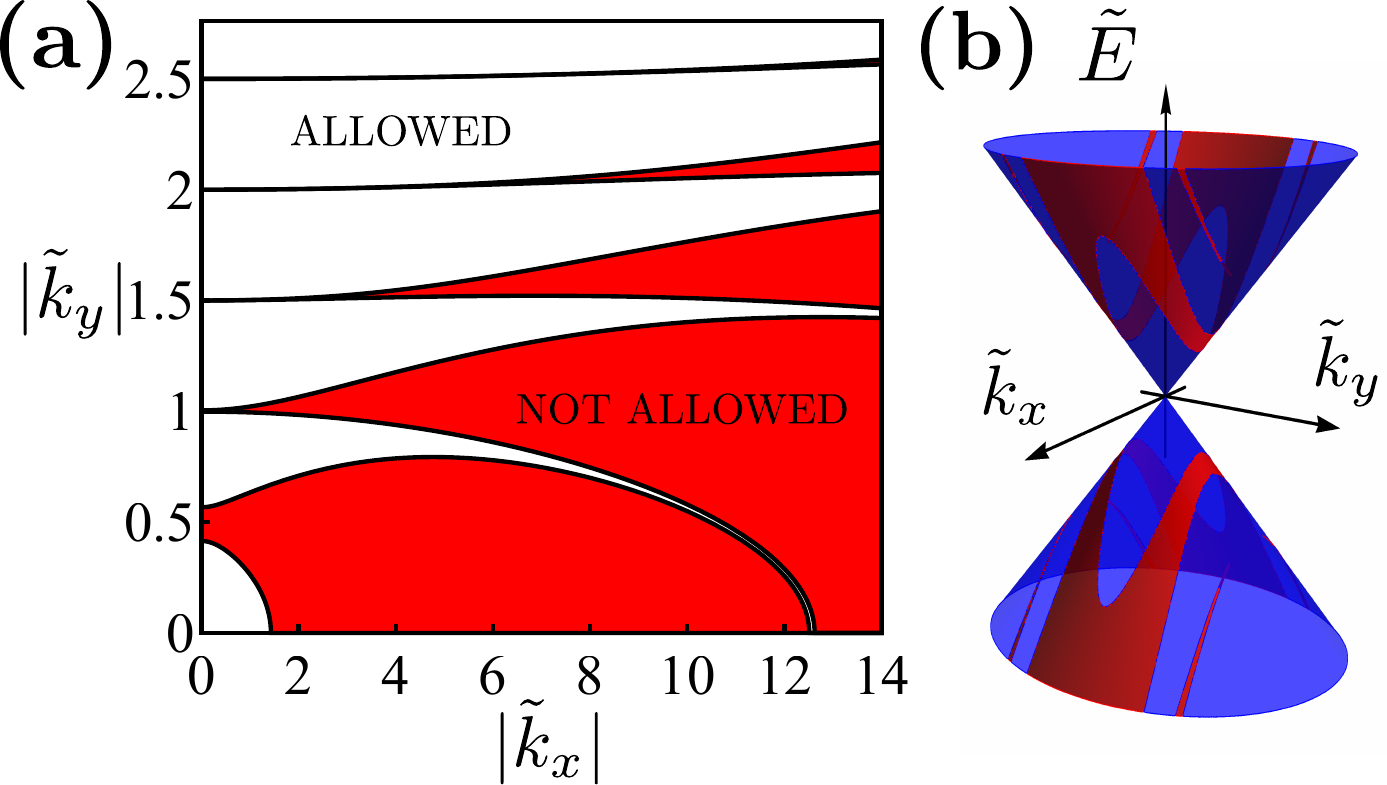}
\caption{(a) Chart of the allowed bands
(white regions) for the quasi momentum $\tilde{\bi{k}}$ (in units of $\Lambda$), with $\tilde{A}_{0}\equiv\beta u_{0}/(2a)=0.15$
and the strain wave propagating along the armchair direction. The diagram is symmetrical with respect to both axes. (b) Associated Dirac cone, whose red (blue) strips correspond to the forbidden
(allowed) values of the
quasi-energy $\tilde{E}(\tilde{k}_{x},\tilde{k}_{y})$ (in units of $\hbar v_{F}\Lambda$) due to the deformation wave. Reproduced from \cite{Oliva2016b} with permission.}\label{fig_Col}
\end{figure}

Hence the deformation wave, equation (\ref{def}), leads to a pseudoelectromagnetic wave that propagates along the $y$-axis with velocity $v_{s}=\omega/\Lambda$, which can be assumed equal to the velocity of sound in graphene \cite{Oliva2016b}. Note that while the pseudomagnetic field $\bi{B}_{s}$ is perpendicular
to the graphene sample (figure~\ref{fig_SW}(b)), the pseudoelectric field $\bi{E}_{s}$ oscillates in the sample plane but, in general, is not perpendicular to the propagation direction of the strain wave. As noted in \cite{Oliva2016b}, the pseudoelectric field oscillates along the propagation direction of the pseudoelectromagnetic wave only  when the deformation wave propagates along the zigzag direction (i.e. when $\theta=\pm\pi/2+2n\pi/3$). In that case, the pseudoelectromagnetic wave behaves as a longitudinal mechanical wave. In contrast, when the deformation wave propagates along the armchair direction, i.e. for $\theta=n\pi/3$, the pseudoelectric field oscillates transversal to the propagation direction of the pseudoelectromagnetic wave, as is the expected behavior of a normal electromagnetic wave.

Including the strain-induced pseudovector potential into the electron dynamics, equation (\ref{OurA}), \textit{via} minimal coupling and disregarding the effect of a position-dependent Fermi velocity and the smaller effect of the electric field, the resulting effective Dirac equation reads:
\begin{equation}\fl\label{DW}
 v_{F}\bsigma\cdot(-i\nabla - \bi{A}_{s})\bPsi=i\partial_{t}\bPsi.
\end{equation}
As solutions of this equation, Oliva-Leyva \textit{et al.} \cite{Oliva2016b} found Volkov-type states \cite{Wolkow1935,Lopez2008} that propagate parallel to the deformation preferably. In addition, they reported the structure of allowed quasi-momentum and quasi-energies shown in figure~\ref{fig_Col}.

The emergent band structure shows the allowed strain waves with respect to the crystalline directions of the graphene lattice. In consequence, a deformation wave might produce a collimation effect over the charge carriers and, accordingly, an alternative mechanism to achieve electron beam collimation beyond other traditional methods \cite{Park2008,Wang2010}.

 It is known that Floquet bands have topological properties \cite{FoaTorres2014} and time-driven strain leads to topological considerations too. Indeed, zero-energy modes appear at the edges of the Floquet zone \cite{Iadecola2013,Delplace2013,Majgraphene2015}
with a complex topological quantum phase diagram \cite{RomanTaboada2017}.

\begin{figure*}[t]
\centering
\includegraphics[width=0.75\linewidth]{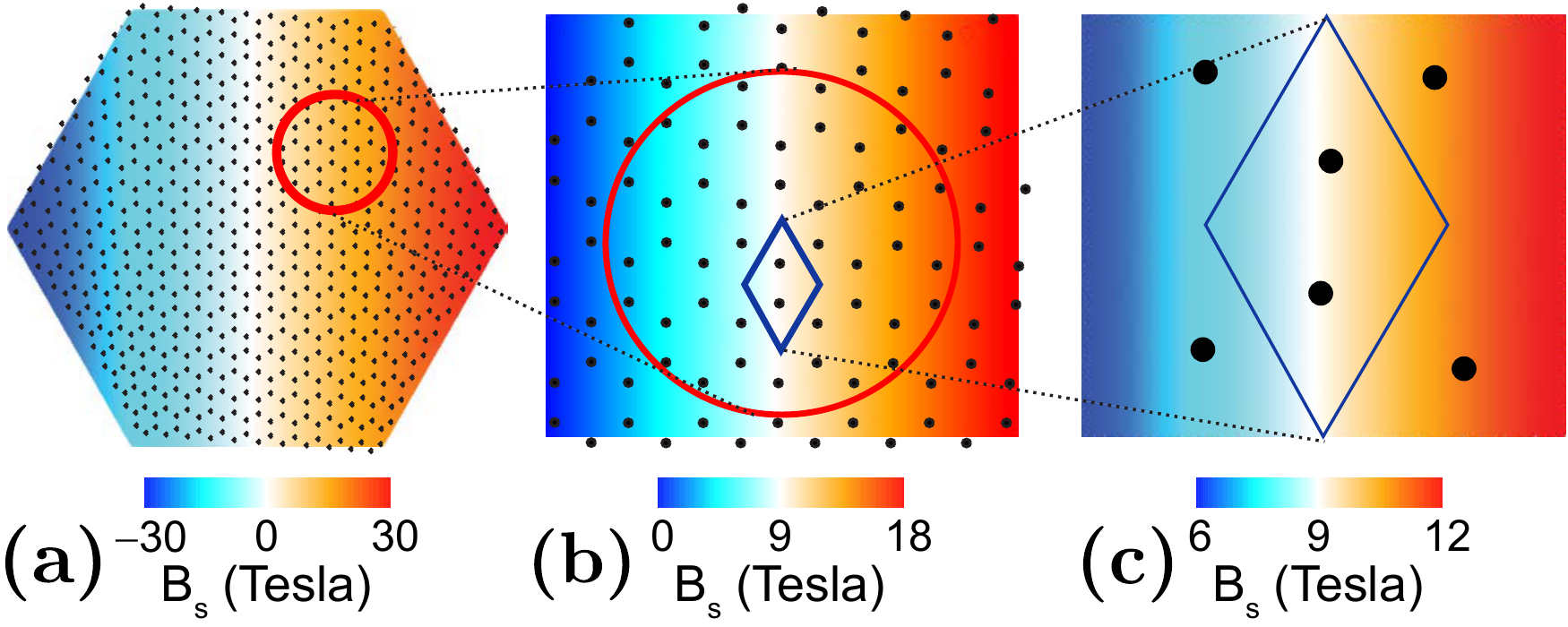}
\caption{In the approach from continuum elasticity, strain-induced fields are defined regardless of spatial scale. A unit cell ``plaquette'' is shown in (b) and (c). One defines the pseudospin Hamiltonian $H_{ps}$ at individual ``plaquettes.'' As a result, strain-induced gauge fields  become discrete themselves. Reproduced from \cite{Barraza2013} with permission.}\label{fig:FSSC_1}
\end{figure*}

\subsection{Strain from the perspective of discrete differential geometry}\label{secdiscrete}

In the approach discussed thus far, and as illustrated in figure \ref{fig:FSSC_1}, continuum deformation fields are superposed to the atomistic lattice \cite{Guinea2012}.

A discrete approach to strain \cite{Sloan2013,Barraza2013,Pacheco2014} represents a consistent route to enhance intuition of the problem. In this approach, the graphene membrane is never described in terms of a continuum deformation field, and it always remains as an atomistic mesh. One  employs geometrical tools that are customarily employed to deal with discrete meshes \cite{DDG} to describe the electronic properties, without any recourse to the continuum deformation field discussed thus far.

Furthermore, given that strain and ripples induce similar changes on the electronic structure (Section \ref{Ripples}), the deployment of a geometry (i.e. a metric (strain) and curvature (ripples)) that always remains faithful to the atomistic mesh makes natural sense in describing arbitrary strained/curved two-dimensional materials. The discrete approach has an unique economy to it, and it provides insights that will be showcased next.

The discussion is divided in four parts: additional motivation statements, showcasing strain-induced gauges in graphene, a discussion of the discrete geometry which captures graphene's shape without recourse to continuum approximations and generalizes to arbitrary 2D materials (Section 7), and a discussion of open, possible developments within this approach.

\subsubsection{Underlying assumptions of the continuum, Dirac approach.} The main assumption of all models describing the effects of mechanical strain in terms of pseudo-electromagnetic gauges is expressed in \cite{Guinea2010} as follows:

``If a mechanical strain varies smoothly on the scale of interatomic distances, it does not break sublattice symmetry but rather deforms the Brillouin zone in such a way that the Dirac cones located in graphene at points $K_+$ and $K_-$ are shifted in opposite directions'' (see \cite{CastroNeto2009,Suzuura2002} as well). Previous statement says that one can understand the effects of mechanical strain on the electronic structure in terms of a semiclassical approach, provided that strain preserves sublattice symmetry.

In this semiclassical approach, a local and strain-induced pseudo-magnetic field $\mathbf{B}_s(\mathbf{r})=\nabla\times {\boldsymbol A}_s(\mathbf{r})$ and a deformation potential $V_s(\mathbf{r})$ are added into a pseudospin Hamiltonian $H_{ps}(\mathbf{q})$ (equation (\ref{HDirac})). The semiclassical approximation is justified if the strain is slowly varying, that is, when it extends over many unit cells and preserves sublattice symmetry \cite{Vozmediano2010,Suzuura2002,Guinea2010}.

Sublattice symmetry is directly linked to the hermiticity of the Dirac Hamiltonian: sublattice symmetry means that phase factors for $A$ and $B$ atoms are complex conjugates, \textit{exactly}, at any given unit cell. From then on, the continuum approach identifies the pseudospin with a spin (i.e. an object that lacks spatial structure) on a continuum media. As illustrated in figure \ref{fig:FSSC_1}(a), this is why a gauge field is customarily drawn to scales smaller than interatomic distances in almost all the literature concerning strain on graphene.

In figure \ref{fig:FSSC_1}(b), one observes a unit cell (that represents one pseudospin) within the diamond shape. That pseudospin has a spatial dimension and a volume proportional to the square of carbon-carbon distances. The main argument of the discrete approach is that gauge fields take a single value within a unit cell, and not the continuum of values observed by the continuum change of color within the unit cell seen in figure \ref{fig:FSSC_1}(c). A brief derivation of the discrete approach to strain on graphene follows.

If atomic positions are explicitly known, one can determine the extent to which phases of atoms $A$ and $B$ cease to be conjugated as the deviation from the origin of a phasor built up by adding $\Delta\boldsymbol{\delta}_i$ for atom A and $\Delta\boldsymbol{\delta'}_i$ for atom B which, as seen in figure \ref{fig:FSSC_2}, are not necessarily equivalent:
\begin{equation}\label{phasor}
\sum_{i=1}^3(\Delta\boldsymbol{\delta}_i+\Delta\boldsymbol{\delta}'_i) \ne \mathbf{0}.
\end{equation}
The deviation of this phasor sum from zero is a direct measure of ``slow-varying strain'' and of ``distortions preserving sublattice symmetry;'' such measure of deviations from zero represents the first insight beyond the continuum theory that is furnished from the discrete approach.

One has to preserve crystal symmetry for reciprocal space to exist, so when crystal symmetry is strongly perturbed, the reciprocal space representation starts to lack physical meaning, presenting a limitation to the semiclassical Dirac hamiltonian theory, including the continuum and discrete ones.  Lack of sublattice symmetry may not allow proper phase conjugation of pseudospin Hamiltonians at unit cells undergoing very large mechanical deformations. Nevertheless, one must realize that this check cannot proceed on a description of the theory within a continuum media, because there is no reference to atoms on a continuum.

In mechanics of continuum media, there is always a suitable spatial scale in which a mechanical distortion appears homogeneous enough. Therefore, it is not possible to assess sublattice symmetry, and proper phase conjugation of pseudospin Hamiltonians $H_{ps}$ is an implicit assumption of that model. Continuum elasticity is based on a pillar known as Cauchy-Born rule, which means that the deformation field is followed at all spatial scales, even within a single unit cell. In reality, Cauchy-Born rule may break down (\textit{c.f.}, figure \ref{Fig_Deformedlattices}(b)) \cite{Zhou,Ericksen2008}.

\subsubsection{Strain-induced gauges.} 
\begin{figure}[t]
\includegraphics[width=\linewidth]{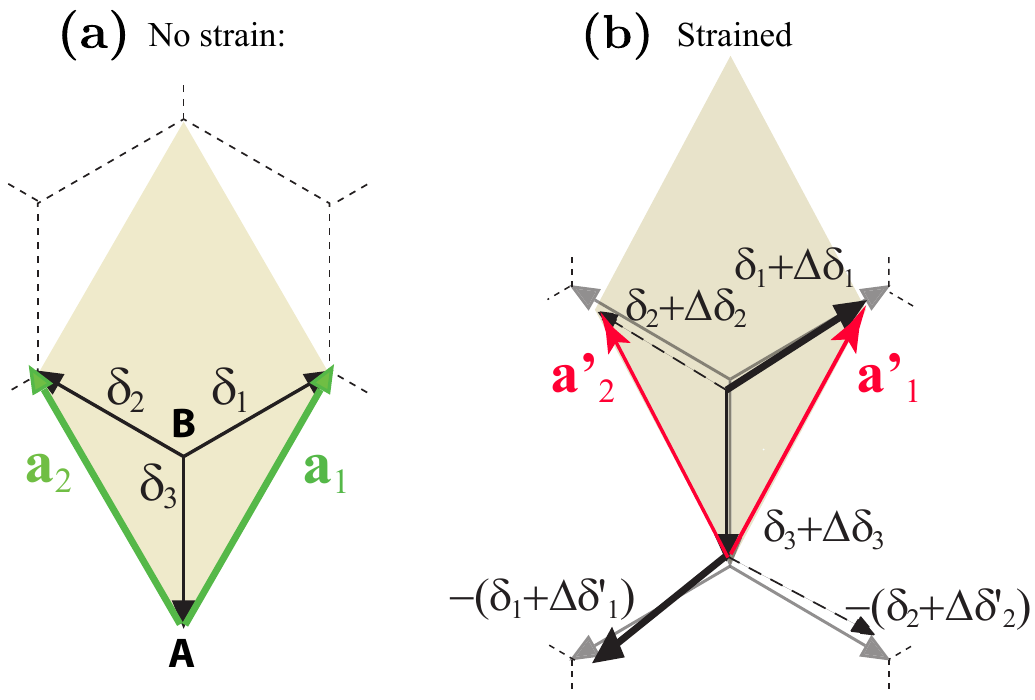}
\caption{(a) Definitions of geometrical parameters in a unit cell. (b) Sublattice symmetry relates to how pairs of nearest-neighbor vectors (either in thick, or dashed lines) are modified due to strain. Reproduced from \cite{Barraza2013} with permission.}\label{fig:FSSC_2}
\end{figure}

Consider the unit cell before (figure~\ref{fig:FSSC_2}(a)) and  after arbitrary strain has been applied (figure~\ref{fig:FSSC_2}(b)). When strain is applied (figure~\ref{fig:FSSC_2}(b)), each local pseudospin Hamiltonian will only have meaning at  unit cells where the phasor in equation (\ref{phasor}) is close to zero:
\begin{equation}\label{eq:applicabilitycondition}
\Delta \boldsymbol{\delta}_j'\simeq\Delta \boldsymbol{\delta}_j\ \ \ \ (j=1,2).
\end{equation}

Equation (\ref{eq:applicabilitycondition}) can be re-expressed in terms of changes of lengths $\Delta L_j$ for pairs of nearest-neighbor vectors $\boldsymbol{\delta}_j$ and $\boldsymbol{\delta}_j'$ (solid and dashed lines are drawn  in figure~\ref{fig:FSSC_2}(b) for better comparison):
\begin{equation}\label{eq:L}
\Delta L_j\equiv |\boldsymbol{\delta}_j+\Delta\boldsymbol{\delta}_j|-|\boldsymbol{\delta}_j+\Delta\boldsymbol{\delta}'_j|.
\end{equation}
Preservation of sublattice symmetry \cite{Guinea2010} requires that $\Delta L_j\simeq 0$  ($j=1,2$).  Forcing  sublattice symmetry to hold amounts to introducing an artificial mechanical constraint on the lattice.

For reciprocal lattice vectors to make sense at each unit cell, equation~(\ref{eq:applicabilitycondition}) must hold (i.e. $\Delta L_j$ should be close to zero). Therefore, only after one knows that the strain distortion is slowly varying from numerical check of equation (\ref{eq:applicabilitycondition}), the next task consists in determining how reciprocal lattice vectors change to first order under mechanical load.  In that case, lattice vectors at any unit cell change as follows:
\begin{equation}
\Delta \mathcal{A}\equiv
\left(
\begin{array}{cc}
\Delta \delta_{1x}-\Delta \delta_{3x}& \Delta \delta_{2x}-\Delta \delta_{3x}\\
\Delta \delta_{1y}-\Delta \delta_{3y}& \Delta \delta_{2y}-\Delta \delta_{3y}
\end{array}
\right).
\end{equation}
This way, to first order on $\Delta \mathcal{A}$, the reciprocal lattice vectors are renormalized by:
\begin{equation}\label{eqreci}
\Delta\mathcal{B}=-2\pi\left(\mathcal{A}^{-1}\Delta\mathcal{A}\mathcal{A}^{-1}\right)^T.
\end{equation}
This additional renormalization of the reciprocal lattice (also given by equation (\ref{UniformGvectors})) was missed in \cite{Kitt2012} (see \cite{Kitt2013,deJuan2013} too). Equation (\ref{eqreci}) can be employed to calculate the shifts of the $K_{\pm}$ points due to strain, and $\mathbf{K}_+$ shifts according to \cite{Barraza2013}:
\begin{eqnarray}\label{eq:discrete}
\Delta\mathbf{K}_+=\nonumber\\
-\frac{4\pi}{3a^2}
\left(\Delta\delta_{1x}-\Delta\delta_{2x},\frac{\Delta \delta_{1x}+\Delta \delta_{2x}-2\Delta \delta_{3x}}{\sqrt{3}}\right),
\end{eqnarray}
where $a$ is the lattice parameter of unstrained graphene and $\delta_{ij}$ is the $j-$component of vector $\boldsymbol{\delta}_i$. Given that $\Delta K_-=-\Delta\mathbf{K}_+$ \cite{Barraza2013}, the $\mathbf{K}_+$ and $\mathbf{K}_-$ points shift in opposite directions \cite{CastroNeto2009,Guinea2010}.

Within a unit cell, local gauge fields can be computed as low energy approximations of a $2\times 2$ pseudospin Hamiltonian that will be soon explicitly expressed. Explicit calculation yields \cite{Barraza2013}:
\begin{equation}\label{eq:cancellation}
\sum_{j=1}^3e^{i\boldsymbol{\delta}_j\cdot\mathbf{K}_+}[1+i(\boldsymbol{\delta}_j\cdot\Delta\mathbf{K}_++\Delta\boldsymbol{\delta}_j\cdot\mathbf{K}_+)]=0.
\end{equation}

The term linear on $\Delta \mathbf{K}_+$ on equation~(\ref{eq:cancellation}) cancels out fictitious $K_+$ point-dependent gauge fields \cite{Kitt2012}, which originated from the term linear on $\Delta \boldsymbol{\delta}_j$ in this same equation: equation (\ref{eq:cancellation}) is a consistency check for the discrete approach.

 In the presence of strain, the low-energy $2\times 2$ pseudospin Hamiltonian takes the following form:
\begin{eqnarray}\label{eq:ps1}
H_{ps}=\nonumber\\
\left(
\begin{array}{cc}
0 & t_0\sum_{j=1}^3ie^{-i\mathbf{K}_+\cdot\boldsymbol{\delta}_j}\boldsymbol{\delta}_j\cdot\mathbf{q}\\
-t_0\sum_{j=1}^3ie^{i\mathbf{K}_+\cdot\boldsymbol{\delta}_j}\boldsymbol{\delta}_j\cdot\mathbf{q} & 0
\end{array}
\right)+\nonumber\\
\left(
\begin{array}{cc}
E_{s,A} & -\sum_{j=1}^3\delta t_{0,j}e^{-i\mathbf{K}_+\cdot\boldsymbol{\delta}_j}\\
-\sum_{j=1}^3\delta t_{0,j}e^{i\mathbf{K}_+\cdot\boldsymbol{\delta}_j} & E_{s,B}
\end{array}
\right),
\end{eqnarray}
 with the first term on the right-hand side reducing to the standard pseudospin Hamiltonian in the absence of strain. The change of the hopping parameter $t_0$ is related to the variation of length, as explained in \cite{Vozmediano2010} and \cite{Suzuura2002} (see equation (\ref{eqGruinesen}) too):
\begin{equation}
\delta t_{0,j}=-\frac{|\beta| t_0}{a^2} \boldsymbol{\delta}_j\cdot\Delta\boldsymbol{\delta}_j.
\end{equation}
 This way, equation~(\ref{eq:ps1}) becomes:
\begin{eqnarray}
\mathcal{H}_{ps}=
\hbar v_F\boldsymbol{\sigma}\cdot \mathbf{q}
+\left(
\begin{array}{cc}
E_{s,A} & f_1^*\\
f_1 & E_{s,B}
\end{array}
\right),
\end{eqnarray}
with $f_1^*=\frac{|\beta|t}{2a^2}
[2\boldsymbol{\delta}_3\cdot\Delta\boldsymbol{\delta}_3
-\boldsymbol{\delta}_1\cdot\Delta\boldsymbol{\delta}_1
-\boldsymbol{\delta}_2\cdot\Delta\boldsymbol{\delta}_2
+\sqrt{3}i(\boldsymbol{\delta}_2\cdot\Delta\boldsymbol{\delta}_2-\boldsymbol{\delta}_1\cdot\Delta\boldsymbol{\delta}_1)]$, and $\hbar v_F\equiv
\frac{\sqrt{3}at_0}{2}$.
The parameter $f_1$ can be expressed in terms of a vector potential: $A_s$ $f_1=-\hbar v_F\frac{eA_s}{\hbar}$. This way:
\begin{eqnarray}\label{eq:Asdiscrete}
A_s&=-\frac{|\beta|\phi_0}{\pi a^3}[
\frac{2\boldsymbol{\delta}_3\cdot\Delta\boldsymbol{\delta}_3
-\boldsymbol{\delta}_1\cdot\Delta\boldsymbol{\delta}_1
-\boldsymbol{\delta}_2\cdot\Delta\boldsymbol{\delta}_2}{\sqrt{3}}\nonumber\\
&-i(
\boldsymbol{\delta}_2\cdot\Delta\boldsymbol{\delta}_2
-\boldsymbol{\delta}_1\cdot\Delta\boldsymbol{\delta}_1)],
\end{eqnarray}
with $\phi_0$ the flux quantum.

In the absence of a full Poisson solver, one may estimate the diagonal entries \cite{Suzuura2002} in equation~(\ref{eq:ps1})  as follows \cite{Barraza2013}:
\begin{equation}\label{eq:EsA}
E_{s,A}=-\frac{0.3 eV}{0.12}\frac{1}{3}\sum_{j=1}^3\frac{|\boldsymbol{\delta}_j-\Delta\boldsymbol{\delta}_j|-a/\sqrt{3}}{a/\sqrt{3}},
\end{equation}
and:
\begin{equation}\label{eq:EsB}
E_{s,B}=-\frac{0.3 eV}{0.12}\frac{1}{3}\sum_{j=1}^3\frac{|\boldsymbol{\delta}_j-\Delta\boldsymbol{\delta}'_j|-a/\sqrt{3}}{a/\sqrt{3}}.
\end{equation}
These entries represent the scalar deformation potential, which are taken to linear order in the average bond increase \cite{Choi2010}. In numerical calculations, the deformation potential given by these entries tends to be asymmetric within the A and B sublattices, which gives rise to a mass term \cite{Barraza2013,Manes2013}. One must keep in mind that the effect of $E_{s,A}$ and $E_{s,B}$ is more complex than a simple shift of the Fermi energy \cite{Barraza2013,Castro2016}: theory that disregards the effects of the deformation potential must include explanation as of why such neglect is physically sensible.

\begin{figure}[t]
\centering
\includegraphics[width=\linewidth]{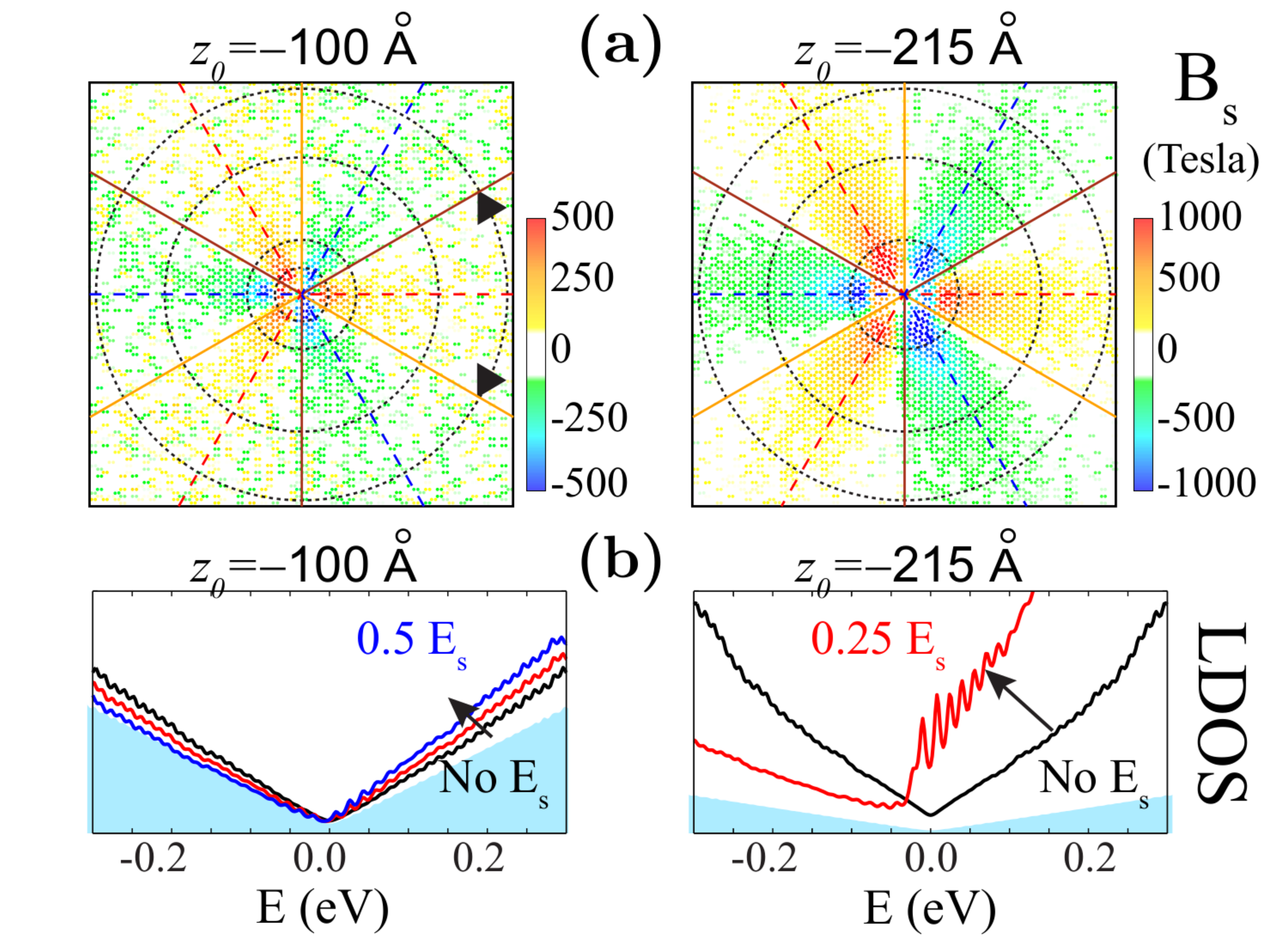}
\caption{(a)~$B_s$ for $z_0=-100$ and $-215$ \AA{} loads. (b)~LDOS with screened values of the deformation potential $E_s$ at $r=0$, for $z_0=-100$ and $-215$ \AA.{} Note the acute effect of $E_s$ on the spectra. Reproduced from \cite{Pacheco2014} with permission. Copyrighted by the American Physical Society.}\label{fig:PRBdiscrete}
\end{figure}

In the absence of significant curvature, the continuum limit is achieved when $\frac{|\Delta\boldsymbol{\delta}_j|}{a}\to 0$ (for $j=1,2,3$). One then has (Cauchy-Born rule):
$\boldsymbol{\delta}_j\cdot \Delta \boldsymbol{\delta}_j\to \boldsymbol{\delta}_j\left(
\begin{array}{cc}
{\epsilon}_{xx}&{\epsilon}_{xy}\\
{\epsilon}_{xy}&{\epsilon}_{yy}
\end{array}\right)\boldsymbol{\delta}_j^T$, where ${\epsilon}_{ij}$ are the entries of the strain tensor.

\begin{figure*}[tb]
\centering
\includegraphics[width=0.85\textwidth]{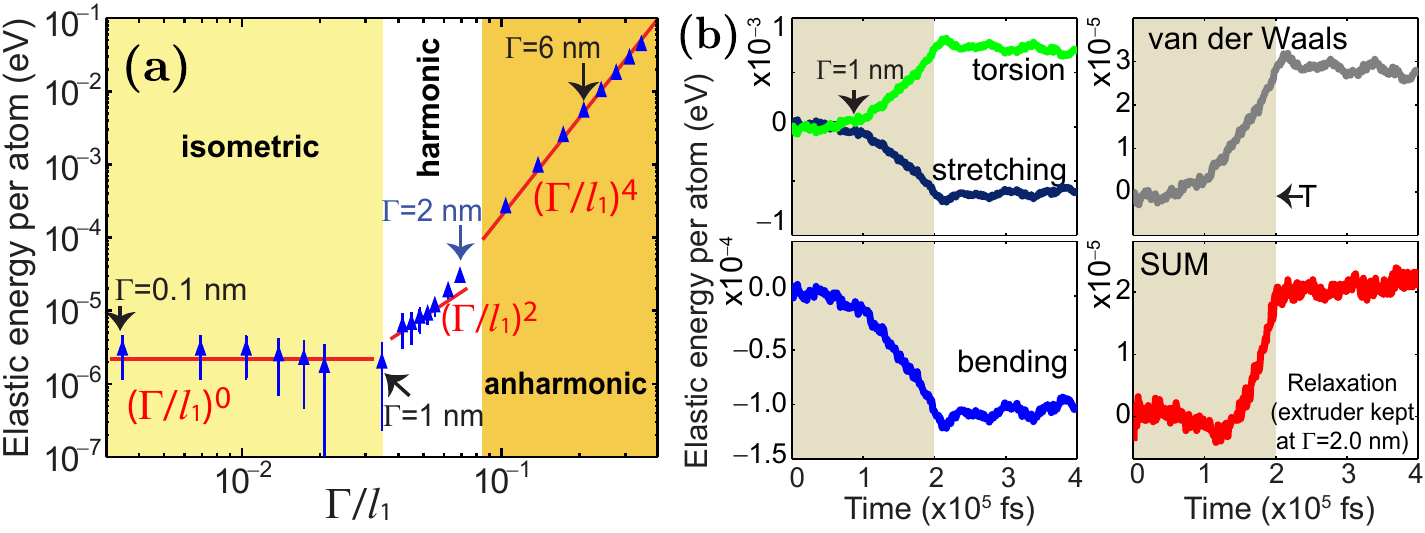}
\caption{(a) The elastic energy versus indentation has three regimes: (i) isometric, where the membrane mostly modifies angles among atoms, but not carbon-carbon distances; (ii) harmonic; and (iii) non-linear, with a force (energy) proportional to the third- (fourth-)power of the indentation, c.f., figure \ref{Fig_Indentation} and equation (\ref{Schwering}). The isometric regime is customarily neglected, but it could exist on plied or corrugated graphene membranes, that must be ``ironed out'' prior to the actual creation of strain. (b) The decomposition of the elastic energy indicates that bending requires much less energy than that needed to stretch carbon bonds. Reproduced from \cite{Sloan2013} with permission. Copyrighted by the American Physical Society.}\label{PRBenergetics}
\end{figure*}

This way equation~(\ref{eq:Asdiscrete}) becomes \cite{Vozmediano2010,Guinea2010}
$A_s\to \frac{|\beta|\phi_0}{2\sqrt{3}\pi a}({\epsilon}_{xx}-{\epsilon}_{yy}-2i{\epsilon}_{xy})$.

The relevant equations from the discrete approach (\ref{eq:Asdiscrete}, \ref{eq:EsA}, \ref{eq:EsB}) take as direct input changes in atomic positions upon strain, never needing fitting onto a continuum. Note that only $N/2$ space-modulated pseudospinor Hamiltonians can be built for a graphene membrane having $N$ atoms.

Figure \ref{fig:PRBdiscrete} \cite{Sloan2013,Barraza2013,Pacheco2014} shows the discrete pseudomagnetic field, resolved over individual unit cells, and a deformation potential $E_s$ that alters the LDOS in a complex manner.

There is a subtle point concerning the nature of the distortions of the graphene lattice: a rippled graphene membrane may require a certain degree of load in order to iron-out ripples and start accumulating actual tensile strain. This point is emphasized within the ``isometric'' regime in figure \ref{PRBenergetics} \cite{Barraza2013} --on the other hand, the harmonic and anharmonic regimes in figure \ref{PRBenergetics} have a dependence consistent with experiments of graphene under load (see Section \ref{elasticcoeffs} and figure \ref{Fig_Indentation}). Additional mechanical and energetic issues concerning graphene suspended on small holes have been addressed by Verbiest and coworkers in the recent past too \cite{Verbiest2016}. Now that the discrete approach has been discussed, a geometry for meshes that finds a niche application in 2D materials is showcased next.

\subsubsection{Further discussion of the discrete geometry.}\label{discgeo}

Strain and ripples induce similar changes on the electronic structure (Sections \ref{Ripples} and \ref{secdiscrete}). This implies that the effects of strained/rippled 2D materials on their electronic properties are essentially geometrical. Previous observation invites a dedicated discussion of the geometry of deformed graphene that is provided in the present Section.

The local geometry of a two-dimensional (2D) surface is determined by four invariants of its metric ($g$) and curvature ($k$) that indicate how much it stretches and curves with respect to a reference non-deformed shape. Suitable choices are the determinant and the trace of $g$, the Gauss curvature $K\equiv\det(k)/\det(g)$, and the mean
curvature $H\equiv Tr(k)/(2 Tr(g))$.

The geometry of 2D materials is commonly studied in terms of a continuous displacement field $\epsilon_{\alpha}(\xi^1,\xi^2)$, where $\xi^1$ and $xi^2$ are local coordinates in the deformed two-dimensional sample. Specifically, the strain tensor is
$\epsilon_{\alpha\beta}=(\partial_{\alpha}\epsilon_{\beta}+\partial_{\beta}\epsilon_{\alpha}+ \partial_{\alpha}\epsilon_{\beta}\partial_{\beta}\epsilon_{\alpha}+ \partial_{\alpha}z\partial_{\beta}z)/2$,
with $z$ an out-of-plane elongation. There, differential geometry and mechanics couple as:
\begin{equation}\label{geocouple}
g_{\alpha\beta}=\delta_{\alpha\beta}+2\epsilon_{\alpha\beta},\qquad k_{\alpha\beta}=\hat{\mathbf{n}}\cdot \frac{\partial \mathbf{g}_{\alpha}}{\partial \xi^{\beta}},
\end{equation}
where $\mathbf{g}_{\alpha}(\xi^1,\xi^2)$ is a tangent vector field, $\delta_{\alpha\beta}$ is the reference metric
and $\hat{\mathbf{n}}=\frac{\mathbf{g}_{\xi^1}\times \mathbf{g}_{\xi^2}}{|\mathbf{g}_{\xi^1}\times \mathbf{g}_{\xi^2}|}$ is the local normal.

\begin{figure}[tb]
\centering
\includegraphics[width=\linewidth]{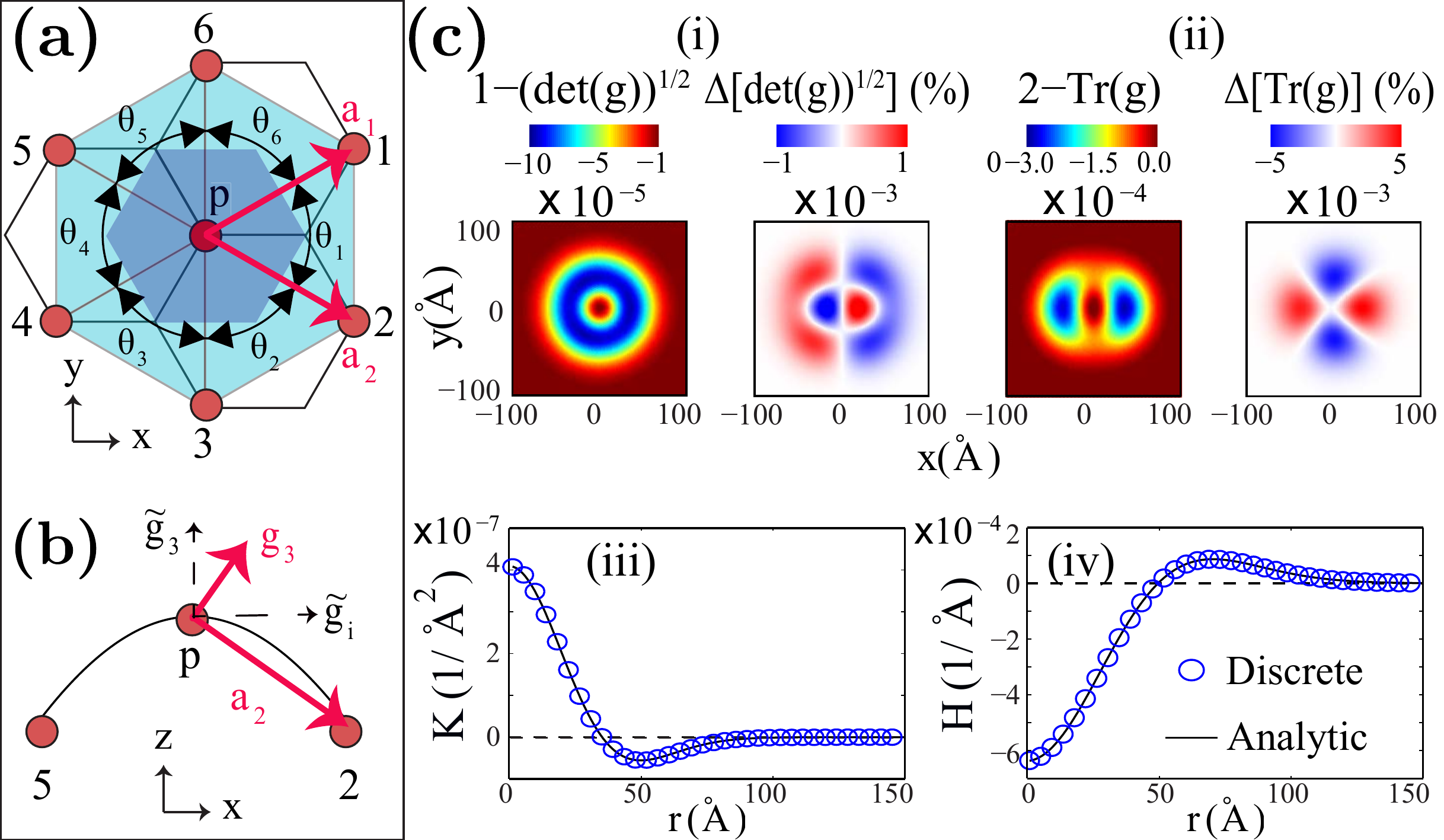}
\caption{(a) Polyhedra used to determine the four geometrical invariants from the metric and curvature.
Circles represent atoms on the A-sublattice. Local lattice vectors are $\mathbf{a}_1$ and $\mathbf{a}_2$; $\theta_i$ are internal angles to edges $\mathbf{e}_i$
and $\mathbf{e}_{i+1}$; and the central shaded hexagon is the Voronoi cell. (b) Side view highlights the differences between continuum and discrete vector fields.
  (c) i: $\sqrt{\det(g)}$, ii: ${Tr}(g)$, iii: $K$ and iv: $H$ for a smooth gaussian bump where discrete and continuum results coincide. Percent differences of $\sqrt{\det(\tilde{g})}-\sqrt{\det(g)}$ and ${Tr}(\tilde{g})-{Tr}(g)$ are also shown. Reproduced from \cite{Pacheco2014} with permission. Copyrighted by the American Physical Society.}\label{fig:nodalpoints}
\end{figure}

But given that a graphene lattice is not a continuum manifold but rather a discrete collection of atoms that are joined by chemical bonds, it is possible to furnish a geometry that does not require transforming the atomistic lattice into a continuum as a necessary and unavoidable preamble to discuss the effects of strain and curvature on material properties. Using the atomistic lattice in figure~\ref{fig:nodalpoints}(a-b), the discrete metric is defined from the local lattice vectors $\mathbf{a}_{\alpha}$ \cite{Pacheco2014,Pacheco2014b}
$g_{\alpha\beta}=\mathbf{a}_{\alpha}\cdot \mathbf{a}_{\beta}$, and the discrete Gauss curvature ($K_D$) originates
from the angle defect $\sum_{i=1}^6\theta_i$ \cite{DDG,Xu2009}:
\begin{equation}\label{eq:DGB}
K_D=(2\pi-\sum_{i=1}^6\theta_i)/A_p.
\end{equation}
Here $\theta_i$ ($i=1,...,6$) are angles between vertices shown in figure \ref{fig:nodalpoints}(a).
The Voronoi tessellation shown in dark blue in figure~\ref{fig:nodalpoints}(a) with an area $A_p$ generalizes the Wigner-Seitz unit cell on conformal 2D geometries. As a consistency check, the angle defect adds up to $2\pi$ on a flat surface, making $K_D=0$.

\begin{figure}[tb]
\centering
\includegraphics[width=\linewidth]{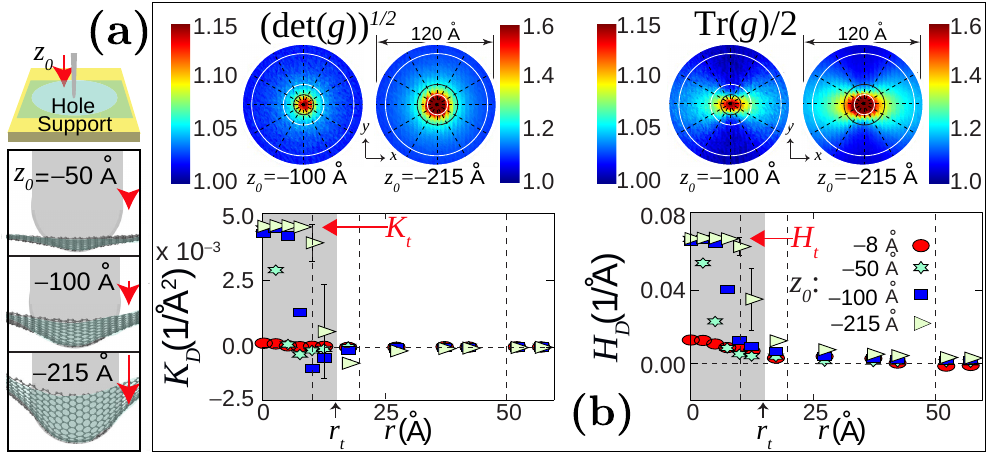}
\caption{(a) Graphene under load by a tip of radius $r_t= 15$ \AA{}. (b) Geometrical invariants as the indentation proceeds: the metric increases unbounded, yet curvatures saturate
to $K_t$ and $H_t$ as graphene conforms to the tip (see flat horizontal lines $K_D=K_t$ and $H_D=H_t$, for $0\le r\lesssim r_t$ at
$z_0=-100$ and $-215$ \AA{}). Reproduced from \cite{Pacheco2014} with permission. Copyrighted by the American Physical Society.}\label{fig:rapidf3}
\end{figure}

The discrete mean curvature $H_D$  is given by:
\begin{equation}\label{eq:Hdiscrete}
H_{D}=\sum_{i=1}^6 \mathbf{e}_i\times(\boldsymbol{\nu}_{i,i+1}-\boldsymbol{\nu}_{i-1,i})\cdot \hat{\mathbf{n}}/4A_p,
\end{equation}
where $\mathbf{v}_i$ is the position of atom $i$ at site $A$, and $\mathbf{e}_i=\mathbf{v}_i-\mathbf{v}_p$ is the edge between points
$p$ and $i$ (note that $\mathbf{a}_{1(2)}=\mathbf{e}_{1(2)}$). $\boldsymbol{\nu}_{i,i+1}$ is the normal to edges $\mathbf{e}_i$ and $\mathbf{e}_{i+1}$ ($i$
is a cyclic index), and  $\hat{\mathbf{n}}=\frac{\sum_{i=1}^6\boldsymbol{\nu}_{i,i+1}A_i}{\sum_{i=1}^6A_i}$ is the area-weighted normal
with $A_i=|\mathbf{e}_i\times \mathbf{e}_{i+1}|/2$ \cite{DDG}. For the purposes of discrete geometry, the metric and curvatures are formally decoupled objects. The discrete geometry is accurate regardless of elastic regime, and the geometrical distortion shown in figure \ref{fig:rapidf3} leads to the strain-derived gauge fields presented in figure \ref{fig:PRBdiscrete}.

\begin{figure}[tb]
\centering
\includegraphics[width=\linewidth]{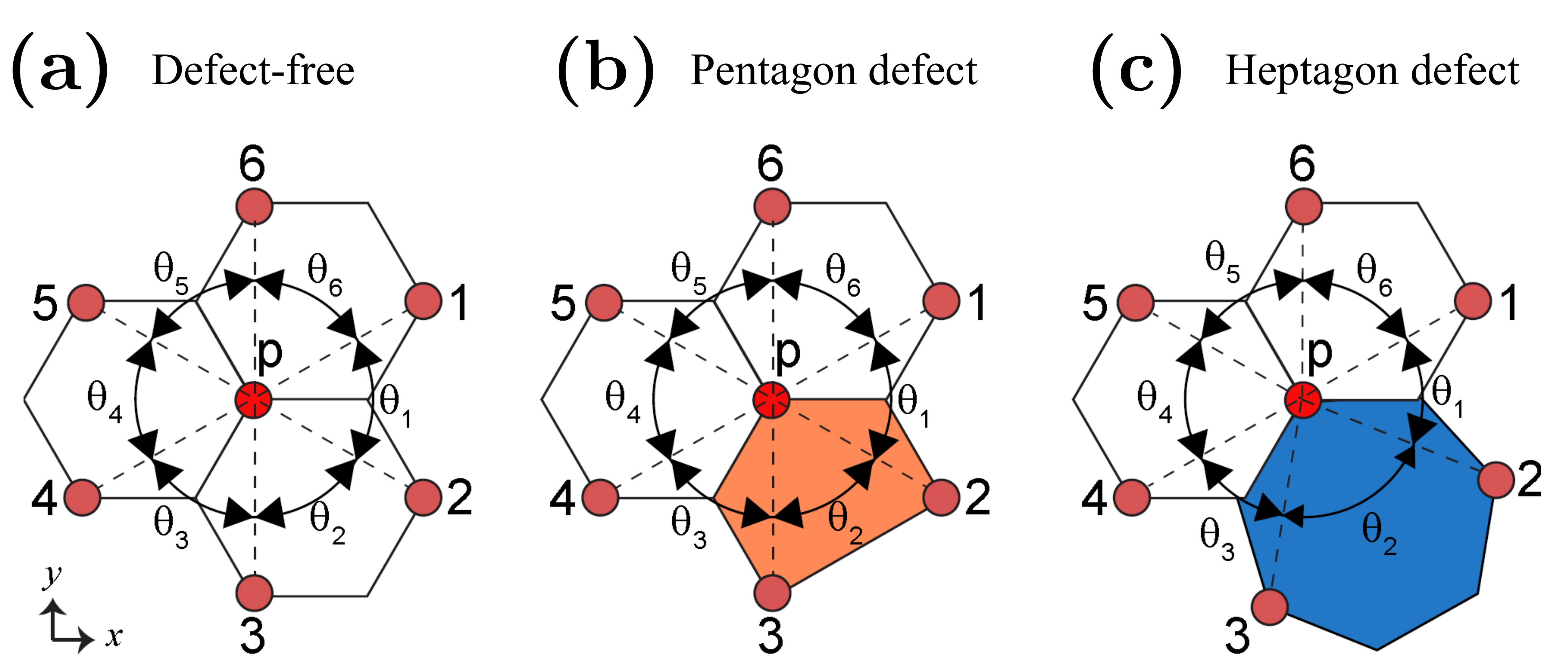}
\caption{(a)
Neighborhood of a graphene atom with no topological defects.  The more common defects are:  (b) a
pentagon defect, (c) a heptagon defect, or a combination of (b) and (c). Reproduced from \cite{Pacheco2014b} with permission. Copyright 2014 American Chemical Society.}\label{defects}
\end{figure}

As discussed in \cite{Pacheco2014b} and figure \ref{defects}, the discrete geometry readily admits generalization to samples with atomistic defects. The discrete metric and curvatures furnish a geometry consistent with crystalline structures, and lead to the faithful characterization of the morphology of 2D materials beyond the effective-continuum paradigm, equation (\ref{geocouple}). As it will be seen when discussing phosphorene, the discrete geometry admits generalization to arbitrary 2D materials.

The discrete approach demonstrates that one can understand the phenomena at play on deformed graphene without necessarily imposing a continuum approximation onto the deformed atomistic lattice, thus providing an increased intuition on the relation among shape and material properties.
The discrete geometry will be employed again, when discussing a reduction of phosphorene's electronic bandgap that arises from curvature later on.

\begin{figure*}[tb]
\centering
\includegraphics[width=0.78\textwidth]{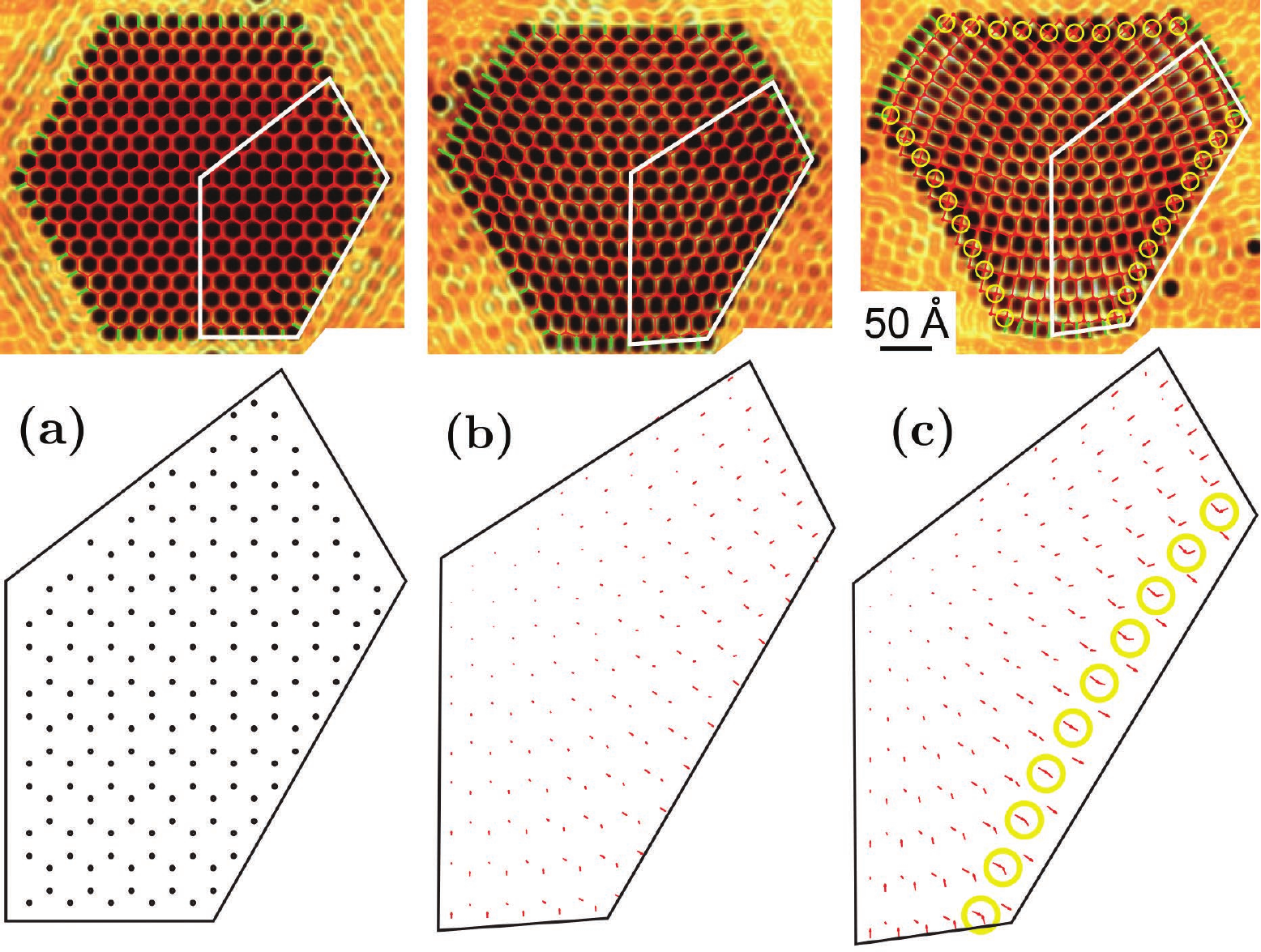}
\caption{Upper row: Evolution of the honeycomb (dual) lattice (``molecular graphene'') as the (direct) triangular lattice is displaced following equation (\ref{dispguinea}) for (a) $s=0$, (b) $s=5$, and (c) $s=10$ $\times 10^{-4}$ \AA$^{-1}$. The dual lattice does not follow the displacement given by equation (\ref{dispguinea}) and red arrows in plots (b) and (c), lower row, show the extent of disagreement. As a result, the pseudomagnetic field is not quite constant. Vectors within yellow circles in (c) highlight the collapse of ``atoms'' positions onto a single point, which means that the number of ``atoms'' is not preserved at molecular graphene's edges. Experimental data adapted by permission from Macmillan Publishers Ltd. from \cite{Manoharan}, copyright (2016).  Dual lattice courtesy of Bradley Klee.}\label{MG_Fig1}
\end{figure*}

\subsubsection{Gaining insight into molecular graphene from its discrete geometry.}

The discrete geometry will be showcased on molecular graphene \cite{Manoharan} next. There, CO molecules are initially arranged as a triangular (direct) lattice on a metallic surface. As seen by a scanning microscope (STM) tip, interstitials among CO molecules create conducting paths, whose intersections conform what mathematicians call a ``dual'' lattice \cite{DDG}. The important point here is that molecular graphene \textit{is} the dual lattice on this problem. Once this identification is set, one can use discrete geometry to understand how ``atomic locations'' of molecular graphene (i.e. the line intersections on the dual lattice) evolve as strain is applied by moving CO molecules on the direct lattice with the aid of an STM tip.

The following equation:
\begin{equation}\label{dispguinea}
\Delta\mathbf{r}=s(2xy,(x^2-y^2)),
\end{equation}
leads to the creation of a constant pseudomagnetic field \cite{Guinea2010} when applied to the graphene lattice and the specific points to demonstrate here are as follows:
\begin{enumerate}
\item{}The direct (CO) lattice was displaced following equation (\ref{dispguinea}) in \cite{Manoharan}.
\item{}As a result, the dual (molecular graphene) lattice is not displaced following the prescription given by equation (\ref{dispguinea}). (Therefore, the pseudomagnetic field that was created is not uniform in the sample, which is probably related to the fact that the $(dI/dV)/dV$ (spectral) data is reported at a single point at the center of the structure on \cite{Manoharan}.)
\item{}For a given magnitude of $s$, unit cells begin to lack hermiticity for large values of $x$ and $y$ away from the center of symmetry, such that the assumption of a deformation preserving sublattice symmetry breaks down. For these large values of $x$ and $y$, Cauchy-Born rule also is violated, as pairs of ``graphene atoms'' move towards one another until they collapse onto a single ``atom.''
\end{enumerate}

Point (i) is demonstrated in figure \ref{MG_Fig1}, upper row. There, the center of mass of the black spots (CO molecules) is recovered for subplot (a). The red dual mesh in \ref{MG_Fig1}(a) results by bisecting lines joining the centers of mass among two consecutive CO molecules and is the graphene lattice. In figures \ref{MG_Fig1}(b) and \ref{MG_Fig1}(c), upper row, the CO lattice is displaced according to equation (\ref{dispguinea}). The red mesh, obtained by bisecting the CO lattice, follows the experiment exactly. Such agreement for both the direct and dual lattices in figure \ref{MG_Fig1}, upper row, permits to affirm that the CO lattice (not the molecular graphene lattice) was displaced according the equation (\ref{dispguinea}) conclusively.

Point (ii) is shown in figure \ref{MG_Fig1}, lower row. The black dots on \ref{MG_Fig1}(a), lower row indicate the intersections of the red dual lattice seen in \ref{MG_Fig1}(a), upper row when $s=0$. The red arrows on figures \ref{MG_Fig1}(b) and \ref{MG_Fig1}(c), lower plots, indicate the discrepancy among the target displacement that would lead to a constant pseudomagnetic field (beginning of arrows) and the experimental location of these crossings (arrows' end).

\begin{figure}[tb]
\centering
\includegraphics[width=0.99\linewidth]{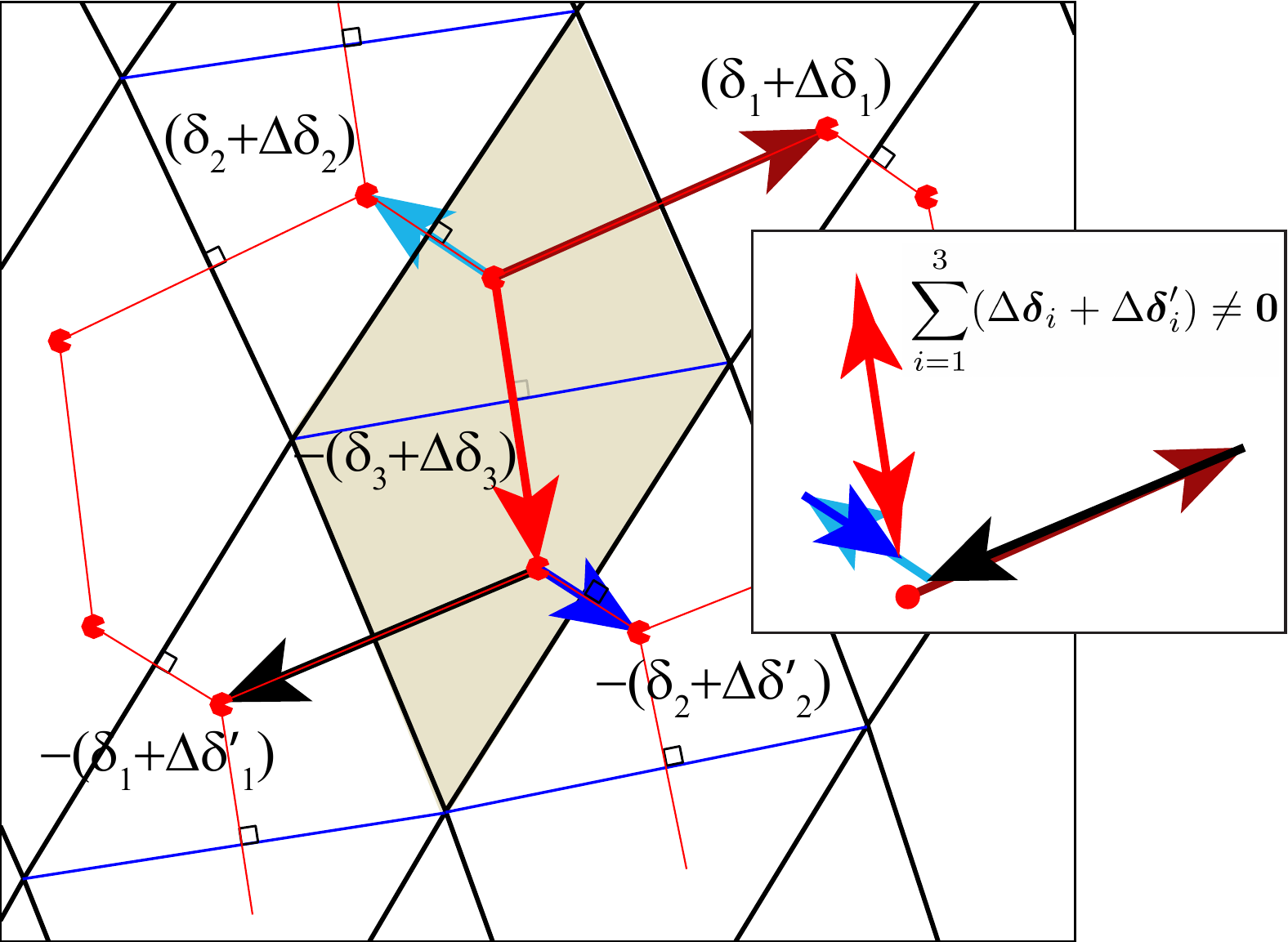}
\caption{Intersections of blue and black lines are positions of CO molecules obtained from \cite{Manoharan}. The red lines bisect the triangular lattice, and lead to the dual molecular graphene lattice. As described in figure \ref{fig:FSSC_2}, the amount to which nearest-neigbouring atoms deviate from one another is related to the extent to which sublattice symmetry is violated at any given unit cell. Inset: equation (\ref{phasor}) is exemplified by the fact that the sum of vectors that originate at the red dot end up elsewhere.}\label{MG_Fig0}
\end{figure}

In figure \ref{MG_Fig0} (a zoom-in from an arbitrary location in figure \ref{MG_Fig1}(c), upper plot), the lack of hermiticity of local pseudospins within a Dirac picture (point (iii)) is demonstrated, along with details for building the dual lattice by bisecting the CO triangular lattice shown in black and blue lines.

\begin{figure*}[tb]
\centering
\includegraphics[width=0.99\textwidth]{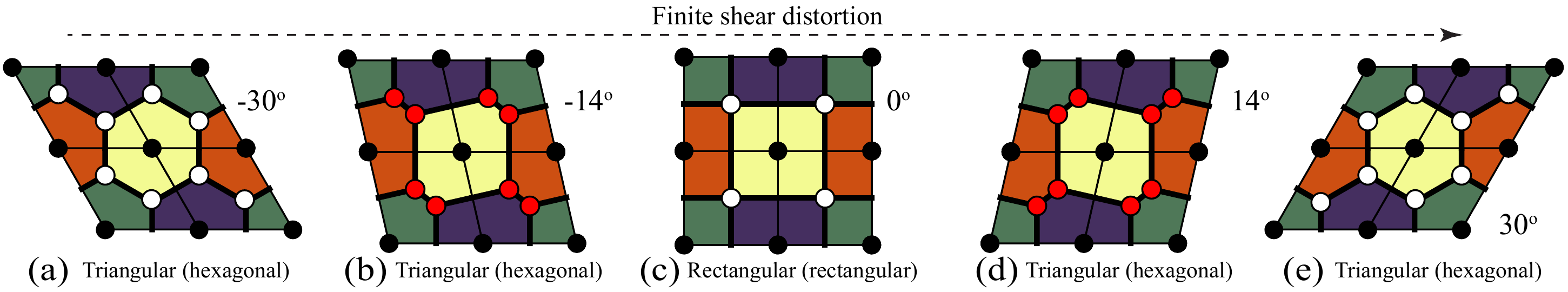}
\caption{A shear distortion converts a triangular lattice onto a rectangular one, and the dual from hexagonal to rectangular: there are less ``atoms'' in the dual lattice in plot (c) when compared to plots (a), (b), (d) and (e). This reduction of the number of atoms in the dual lattice is highlighted by the red circles in (b) and (d). Image courtesy of Bradley Klee.}\label{MG_shear}
\end{figure*}

The last point to prove is violation of Cauchy-Born rule for unit cells away from the center that suffer the largest distortion, as highlighted within yellow circles in figure \ref{MG_Fig1}(c), lower row that show pairs of converging arrows. (This situation is the opposite to the one shown in figure \ref{Fig_Deformedlattices} where the two atoms in the unit cell pull in opposite vertical directions, but both show displacements that go in opposite directions for atoms belonging to complementary sublattices.)

To better visualize this phenomena, figure~\ref{MG_shear} shows the evolution of a triangular lattice and its dual, the honeycomb one, under a horizontal shear strain. The direct (triangular) lattice is shown in black circles and thin black lines, while its dual is shown in white or red dots, and edges appear in thicker black lines. Color was employed to distinguish neighboring honeycombs in the dual lattice.

At the undeformed configuration (subplot \ref{MG_shear}(a)) one notes the existence of two white circles per black circle; these two white circles represent the A and B sublattice sites on the (dual) honeycomb lattice in \cite{Manoharan}.

But in figure~\ref{MG_shear}(b), when shear is introduced onto the triangular CO lattice, a pair of neighbouring intersections of (bold black) bisectors (``molecular graphene atoms'') move in opposite directions in order to come closer together.  As shear reaches the point in which atoms on the (initially triangular) lattice lie along vertical lines (figure~\ref{MG_shear}(c)), the lattice becomes rectangular, and two ``atoms'' on the dual lattice collapse onto one. This means that even the number of ``atoms'' may be ill-defined, so that no pseudospin Hamiltonians (which require two atoms, or sublattices) can be defined at unit cells where the deformation is that extreme. No such discussion exists in the literature, making the discrete approach insightful and relevant for theory and experiments on engineered 2D lattices.

\subsubsection{Open directions within the discrete approach.} It is desirable to generalize these concepts to include back-scattering by incorporating cross-terms among inequivalent $K-$points.

\subsection{Results from density-functional theory}\label{DFT_H}

\textit{Ab initio} or first principles methods are becoming more accessible due to the availability of fast computers. One of the most used \textit{ab initio} methods is based on the Density Functional Theory (DFT) which can be traced to 1964 with a manuscript by Hohenberg and Kohn in which the ground state of an interacting electron gas is obtained through an electron density universal functional \cite{Hohenberg1964}. This work lead to the important theorem which establishes that for any system of interacting particles in an external potential, the external potential is uniquely determined by the ground state density (charge density). This theory was latter generalized by Levy \cite{Levy1979}. Therefore, under the DFT approach, the total energy can be expressed as:
\begin{equation}\label{eq150}
E_{t}(\rho) = E_{k}(\rho) + U(\rho) + E_{ex}(\rho),
\end{equation}
where $E_{k}(\rho)$ is the kinetic energy of a system of noninteracting particles of density $\rho$, $U(\rho)$ is the classical electrostatic energy due to coulombic interactions, and $E_{ex}(\rho)$ is a term that includes the exchange and correlation energies in addition to other many-body contributions. Note that all the terms depend on the charge density.

Approximations are needed to compute the third term of equation (\ref{eq150}). The simplest one is the local density approximation (LDA) which mainly comes from the exchange-correlation energy of a uniform electron gas \cite{Hedin1971,Ceperley1980,Perdew1992}.  The LDA approach can be improved to include inhomogeneous effects of the electron gas by implementing a gradient expansion of the electron density leading to the General Gradient Approximation (GGA) \cite{Perdew1992}. LDA and GGA are the most used exchange-correlation functionals, though there are others which provide more information, but at the expense of the computing time such as hybrid functionals (HS06) \cite{Heyd2003}.

A DFT computation becomes less time consuming and more efficient when using an \textit{ab-initio} pseudopotential to describe the electron-ion interaction. These pseudopotentials are built by considering first, the effect of all the electrons of an atom within the LDA or GGA approximation, and then by separating the obtained valence electron wavefunctions, so the core electrons are not considered in the pseudopotential, therefore, just the outer valence electrons are taken into account \cite{Cohen2016}. In the case of transition metal dichalcogenides (TMDs) heavy atoms such as tungsten, molybdenum, sulphur and selenium are present, thus the spin-orbit coupling needs to be considered in the pseudopotential.

To be able to apply the DFT approach to the lattice dynamics of a system, density functional perturbation theory (DFPT) has been developed: In this method, the second derivative of the total energy with a given perturbation is computed. Depending on the perturbation, different properties can be obtained, for example: If the perturbation is in the ionic positions, phonon dispersions can be computed; if the perturbation is in the unit cell vectors, elastic constants can be calculated \cite{Gonze1992,Baroni2001,Gonze1997}.

Nowadays it is possible to perform DFT calculations on thousands of atoms systems to understand their electronic, mechanical, vibrational and chemical properties and graphene is not the exception. Graphene band structure has been calculated by DFT since the beginning of carbon nanotube research in the 1990's in order to understand not only these amazing 1D tubular structures \cite{Charlier2007}. Moreover, other carbon allotropes such as Fullerenes \cite{Heggie1998} and 3D systems known as Schwarzites \cite{Lenosky1992,Lherbier2014} have also been studied with DFT.

In particular, for DFT calculations of monolayers of 2D materials, such as graphene, hBN, transition metal dichalcogenides (TMD), phosphorene, etc., a 3D unit cell must be used: the layers within the unit cell should be separated by enough vacuum to avoid any interaction, usually more than $15\,\mbox{\AA}$. Therefore, the unit cell must belong to one of the 230 space groups in 3D. It is worth noticing that for DFT calculations knowing the symmetry of the system helps in saving computing time and also allows a better understanding of the physicochemical properties of the structure: In the case of graphene the space group $P6/mmm$ (191) can be used since the stacking is not relevant (one is interested in the properties of just one layer). The associated 1BZ to the 3D unit cell plays a crucial role for determining the electronic band structure and the DOS: The finer the BZ is scanned in the DFT calculation, the better the results obtained.

For graphene, the 3D BZ is hexagonal (an hexagonal prism) \cite{Kittel2005}, thus as it has been seen when graphene is isotropically deformed, the space group symmetry does not change and the BZ symmetry remains the same since the deformation just scales the unit cell up or down. However, for uniaxial strains, along armchair or zigzag directions, the unit cell changes its symmetry and now is centered orthorhombic belonging to the $Cmmm$ (65) space group: The BZ also changes (being now a distorted hexagonal prism), and this is the reason why the notation of the most symmetric points of the BZ, such as the $K$ points, is different when describing strained graphene in a 3D cell for DFT calculations \cite{Aroyo2014,Hinuma2017}. According to the crystallography naming, the $\bi{K}$ ($\bi{K}_+$ in our notation) and $\bi{K}'$ points ($\bi{K}_-$ in our notation) of the $P6/mmm$ (191) space group after uniaxial strain become $\bi{F}_o$ and $\bi{\Delta}_o$ in the $Cmmm$ (65) space group. In addition, when a combination of the uniaxial strains mentioned above is applied to graphene at the same time, the symmetry of the unit cell of the strained structure belongs also to the $Cmmm$ (65) space group, of course, the cell parameters must change depending on the magnitude of the strains as well as the shape of the 1BZ.

\begin{figure}[tb]
\centering
\includegraphics[width=0.95\linewidth]{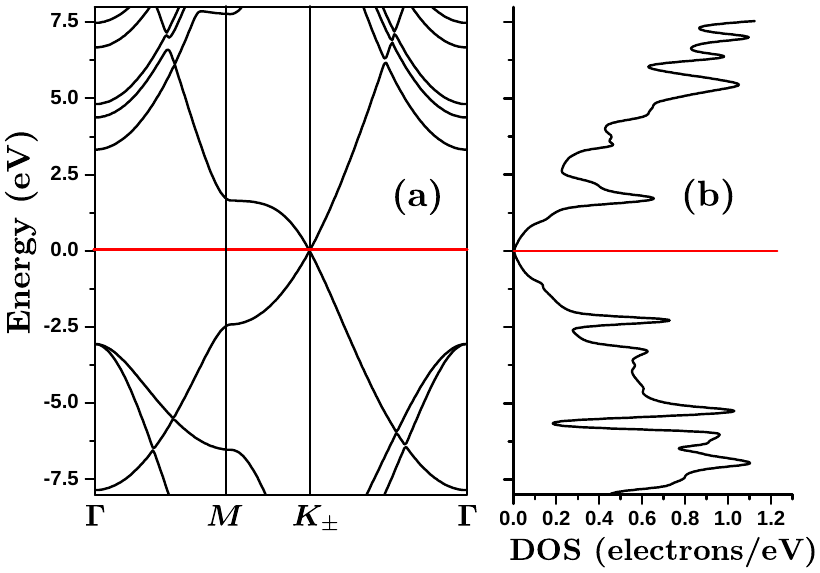}
\caption{(a) Band structure and (b) density of states of graphene calculated using DFT-LDA.}
\label{Fig1_HT}
\end{figure}

From the DFT-LDA graphene band structure and density of states (DOS) it is possible to appreciate their main features: the zero band gap at the $\bi{K}_{\pm}$ points in the Brillouin zone and the linear energy dispersion, closed to the $\bi{K}_{\pm}$ points (see figure \ref{Fig1_HT}).

In 2008, Gui \textit{et al.} \cite{Gui2008} performed DFT-GGA(PW91) calculations \cite{Kresse1996,Kresse1996-2} finding a band gap opening under uniaxial strain of $12.2\,\%$ parallel to the carbon-carbon bonds up to $0.486\,\mbox{eV}$, however, it was pointed later by Farjam and Tabar \cite{Farjam2009} that using the same DFT parameters as Gui \textit{et al.} \cite{Gianozzi2009} a band gap does not open, thus confirming TB results.

\begin{figure}[tb]
\includegraphics[width=\linewidth]{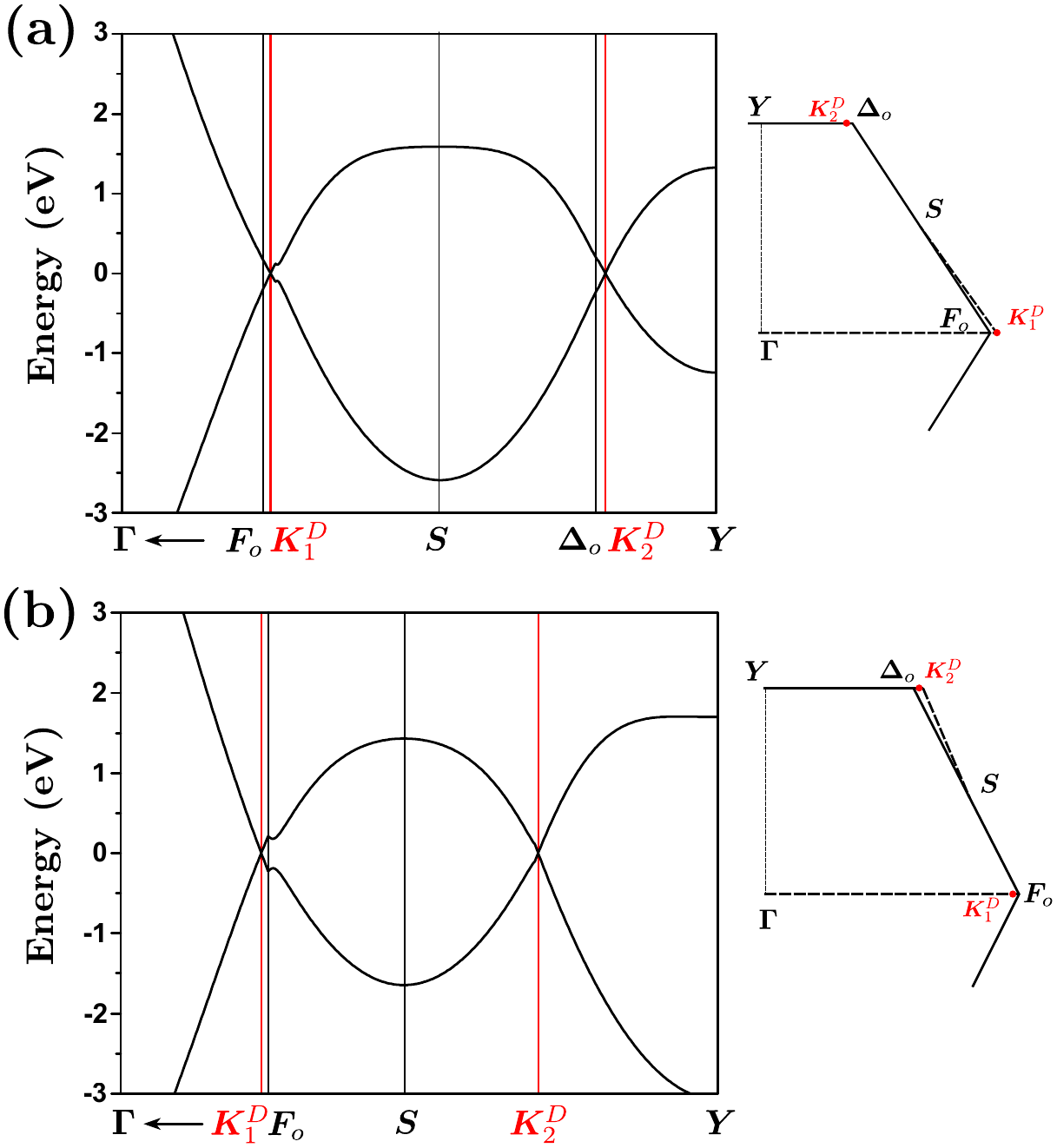}
\caption{DFT-GGA(PW91) Graphene band structure under (a) $7.57\,\%$ tensile strain parallel to the zigzag direction and (b) $11.65\,\%$ tensile strain parallel to the armchair direction. Right panels show the position of the Dirac points $\bi{K}_1^{D}$ and $\bi{K}_2^{D}$ whereas $\bi{F}_o$ and $\bi{\Delta}_o$ are the vertices of the strained 1BZ.}
\label{Fig2_HT_DFT}
\end{figure}

The result obtained by Gui \textit{et al.} was due to a poor scan of the 1BZ, reflected in the low number of $k-$points in the BZ used for the band structure \cite{Gui2009}, which should serve as a warning against predicting materials properties of materials that contain sharp valleys while undersampling $k-$point grids. The robustness of the Dirac cones under uniaxial strain has been confirmed by using the plane-wave code CASTEP \cite{Clark2005, Materials-Studio2016,Castep} under DFT-GGA(PW91) with a plane-wave cutoff of $750\,\mbox{eV}$ with a fine grid of $k-$points when calculating the band structure of the order of 0.001/{\AA} per path segment as shown in figure \ref{Fig2_HT_DFT}: it is shown that under uniaxial strains (parallel to the zigzag edges and parallel to the armchair edges), the Dirac points $\bi{K}_1^{D}$ and $\bi{K}_2^{D}$ are preserved and slightly shifted away from the high symmetry points ($\bi{F}_o$ and $\bi{\Delta}_o$) in the strained 1BZ  (red dots in the 1BZ represent the Dirac points $\bi{K}_1^{D}$ and $\bi{K}_2^{D}$.
This result agrees with TB calculations (figure~\ref{contour1}(b,c))  and with the DFT calculations performed by Choi and colleagues using plane-waves with a cutoff of $400\,\mbox{eV}$ under DFT-GGA-PBE \cite{Choi2010}. The message is clear: DFT calculations of the band structure using a path which just includes the high symmetry points in the 1BZ could miss the Dirac points and a gap may be wrongly reported if the structure is under strain.

In 2015 Kerszberg and Surnarayan performed DFT-LDA and GGA(PZ) \cite{Kerszberg2015} with the ABINIT code \cite{Gonze2002} and considered a combination of uniaxial strains: a tension of $11\,\%$ in the zigzag direction and a compression of $20\,\%$ in armchair direction with a gap opening of $1\,\mbox{eV}$. However, this result needs to be confirmed by other methods.

It is important to bear in mind that the stability of graphene under extreme strains not only depends on energetics of the system; the elastic constants should be obtained and DFPT calculations should also be performed. With DFPT the phonon dispersion can be generated and if the structure is stable there should not be negative frequency values. Another point to consider is that, it is well known that DFT-LDA tends to over bind atoms, so differences depending on different exchange correlation functionals when are expected studying the strain by DFT. In addition, it is also well documented from the beginnings of DFT that it underestimates the band gap \cite{Perdew1985}, thus to correct the gap a many body approach such as GW needs to be used \cite{Aryasetiawan1998}.

\section{Optical properties}\label{Optical}

\begin{figure*}[t]
\centering
\includegraphics[width=\textwidth]{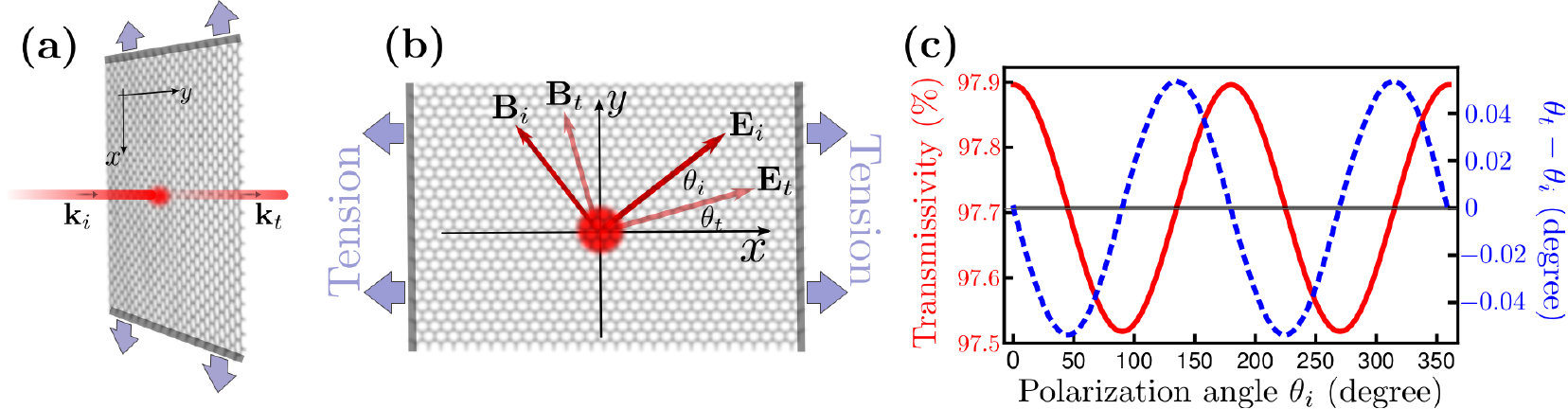}
\caption {(a) Scattering of normal-incident linearly polarized light over strained graphene.
(b) Schematic representation of the dichroism induced by the anisotropic absorption. The electromagnetic fields lie in the graphene plane. (c) Rotation of the transmitted field and transmittance as a function of the incident polarization angle for  uniaxial strain with $\epsilon=0.03$.}
\label{Fig_Modulation}
\end{figure*}

Experiments on the optical response of graphene give an absorbance proportional to the fine structure constant $\alpha$: a more or less constant magnitude $\pi\alpha\approx2.3\,\%$ over a broad range of frequencies from the far-infrared to the visible spectrum \cite{Nair2008,Mak2008}. Thus, absorbance leads to a universal optical conductivity for graphene that is equal to $\sigma_{0}=e^{2}/4\hbar$ at zero chemical potential, which is ultimately a consequence of the behaviour of charge carriers as massless Dirac fermions \cite{Ando2002b,Gusynin2007,Mikhailov2007,Stauber2015,Klimchitskaya2016}.

\subsection{Strain engineering of optical absorption}\label{OpAb}

The strain-induced anisotropy in the electron dynamics results in an anisotropic optical conductivity \cite{Pereira2010,Pellegrino2010,Pellegrino2011,Oliva2014,Oliva2016a,Hernandez2016,Rakheja2016}. Combining the effective Dirac Hamiltonian (\ref{NewH}) and the Kubo formula, the optical conductivity tensor becomes \cite{Oliva2014,Oliva2014C}:
\begin{equation}\label{AC}
\bar{\boldsymbol{\sigma}}(w)\simeq\sigma_{0}(w)\bigl(\bar{\bi{I}} -
2\tilde{\beta}\bar{\boldsymbol{\epsilon}} + \tilde{\beta}\mbox{Tr}(\bar{\boldsymbol{\epsilon}})\bar{\bi{I}}\bigr),
\end{equation}
where $\sigma_{0}(w)$ is the  (isotropic) frequency-dependent conductivity in the absence of strain \cite{Nguyen2016}.

Equation (\ref{AC}) yields $\sigma_{\parallel,\perp}(w)=\sigma_{0}(w)(1\mp\tilde{\beta}\epsilon(1+\nu))$ for  uniaxial strain, where $\sigma_{\parallel}(\sigma_{\perp})$ is the optical conductivity in the  parallel (perpendicular) direction to the applied strain, $\nu$ is the Poisson's ratio and $\epsilon$ is the applied strain \cite{Pereira2010}. Note that the optical conductivity along the direction of the strain decreases while the transverse conductivity increases by the same amount. Thus, an increase of $\sigma_{\perp}$ helps explain the variation of waveguide transmission of hybrid graphene integrated microfibers elongated along their axial direction \cite{Chen2016}.

The strain-induced anisotropy of the optical absorption yields two effects: dichroism, and a modulation of the transmittance as a function of the polarization direction. For normal
incidence of linearly polarized light upon strained graphene, the transmittance is expressed from equation (\ref{AC}) \cite{Oliva2015b} as:
\begin{eqnarray}\label{TR}
 T(\theta_{i})\approx\nonumber\\
 1-\pi\alpha\bigl(1-\tilde{\beta}({\epsilon}_{xx}-{\epsilon}_{yy})\cos2\theta_{i}
 - 2\tilde{\beta}{\epsilon}_{xy}\sin2\theta_{i}\bigr),
\end{eqnarray}
whereas the dichroism is expressed by:
\begin{equation}\label{DiR}
 \theta_{t}-\theta_{i}\approx\alpha\tilde{\beta}\bigl(\frac{{\epsilon}_{yy}-{\epsilon}_{xx}}{2}\sin2\theta_{i} +
 {\epsilon}_{xy}\cos2\theta_{i}\bigr),
\end{equation}
where $\theta_{i}(\theta_{t})$ is the incident (transmitted) polarization angle (see figure \ref{Fig_Modulation}). Equations
(\ref{TR}) and (\ref{DiR}) show modulations with respect to the incident polarization angle $\theta_{i}$ with a period of $\pi$, due to the physical equivalence between $\theta_{i}$ and $\theta_{i} + \pi$ for linearly polarized light at normal incidence.

Equation (\ref{TR}) yields
$T(\theta_{i})=1-\pi\alpha(1-\tilde{\beta}(1+\nu)\epsilon\cos2\theta_{i})$ for uniaxial strain \cite{Pereira2010},
where $\epsilon$ is the magnitude of strain, $\nu$ is
the Poisson ratio and the angle $\theta_{i}$ is measured with respect to the stretching direction.
This periodic modulation $\triangle T$  of the transmittance as a function
of polarization direction has been experimentally confirmed in the visible
light range on strained graphene, and the magnitude $\epsilon$ for uniaxial strain was extracted by means of $\epsilon\approx\triangle T/(2\pi\tilde{\beta}(1+\nu))$ from Raman spectroscopy measurements \cite{Ni2014}.

\subsection{Raman Spectroscopy}

Raman spectroscopy is one of the most important techniques to characterize layered materials because of its non-invasiveness and sensitiveness to defects, number of layers, strain, curvature, doping, \textit{etc.}  In the context of carbon nanostructures, Raman spectroscopy has been used to characterize  fullerenes \cite{Bethune1991}, carbon nanotubes \cite{Dresselhaus2005}, and graphene \cite{Ferrari2006, Ferrari2007, Ferrari2013}.

Materials such as graphite, carbon nanotubes and graphene are characterized by three main Raman features: a signal around 1590 cm$^{-1}$, called the $G$ band, a first order signal due to the in-plane vibrations of the carbon atoms ($E_{2g}$ modes in graphite); another signal at around 1350 cm$^{-1}$, called $D$ band, due to defects in the lattice; and the $2D$ band around 2700 cm$^{-1}$.

\begin{figure}[t]
\centering
\includegraphics[width=0.85\linewidth]{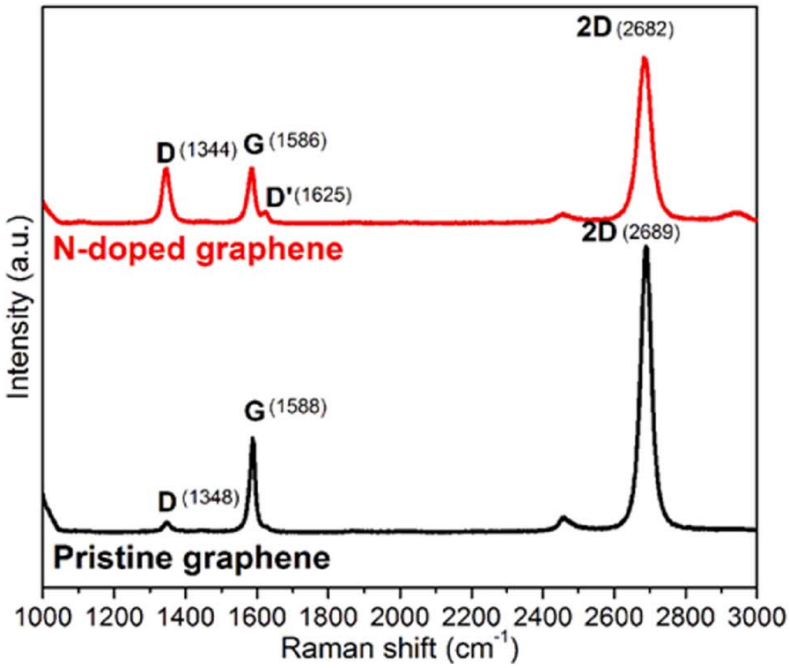}
\caption{Experimental Raman spectra of nitrogen doped graphene and pristine graphene (from \cite{Lv2012}).}
\label{Fig3_HT}
\end{figure}

\begin{figure}[t]
\centering
\includegraphics[width=0.85\linewidth]{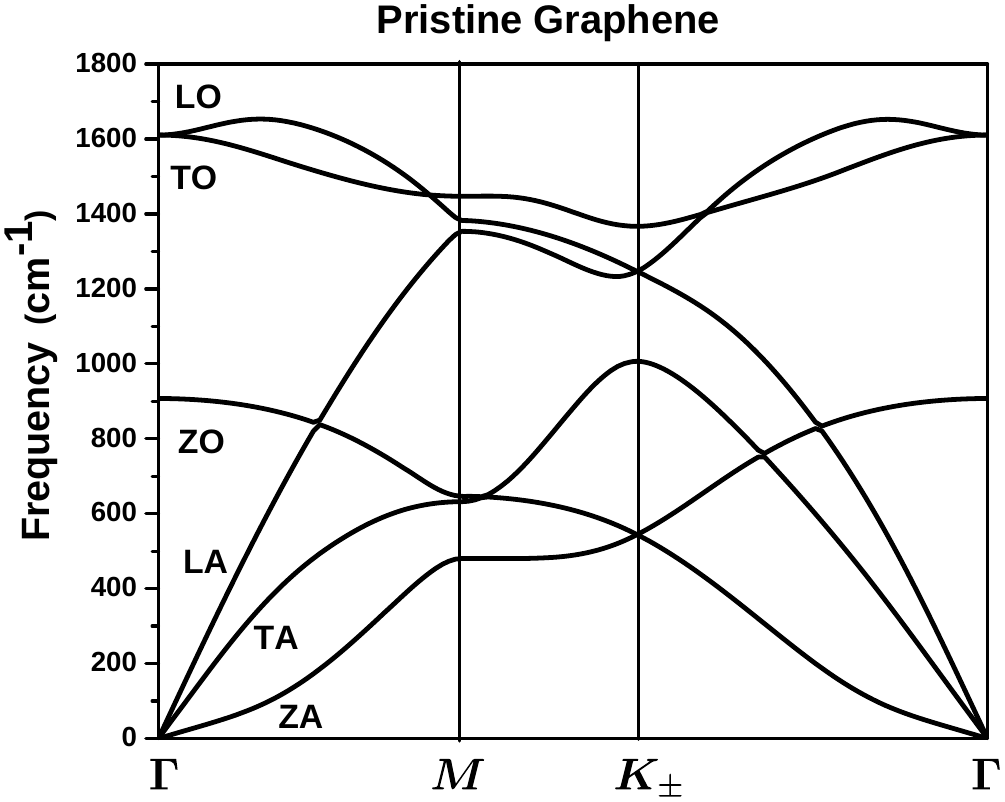}
\caption{Density functional perturbation theory (DFPT) calculation of the phonon dispersion of graphene sowing the transverse, longitudinal and zero acoustic and optical modes.}
\label{Fig4_HT}
\end{figure}

The $D$ and the $2D$ signals involve double resonance processes \cite{Ferrari2006, Ferrari2007, Ferrari2013}: for the $D$ band, after a photon excites an electron, intervalley electron scattering takes place involving a defect and a transverse optical (TO) phonon around the $K_{\pm}$ point. For the $2D$ band, the double resonance does not involve defects, just phonons around the $K_{\pm}$ point. One unique characteristic of graphene is that the $2D$ band exhibits a much higher intensity than the $G$ band.

The Raman signal of pristine graphene is compared to that of nitrogen doped graphene  in figure \ref{Fig3_HT}. Here, the $D'$  signal due to intravalley electron scattering by a defect involving a longitudinal optical (LO) phonon is observed as well. On the other hand, the $D$ and $D'$ signals are strongly reduced in pristine graphene due the small amount of defects. Figure \ref{Fig4_HT} reveals the phonon dispersion calculated with DFPT, in which the longitudinal and acoustic branches of graphene are shown.

\begin{figure}
\centering
\includegraphics[width=0.85\linewidth]{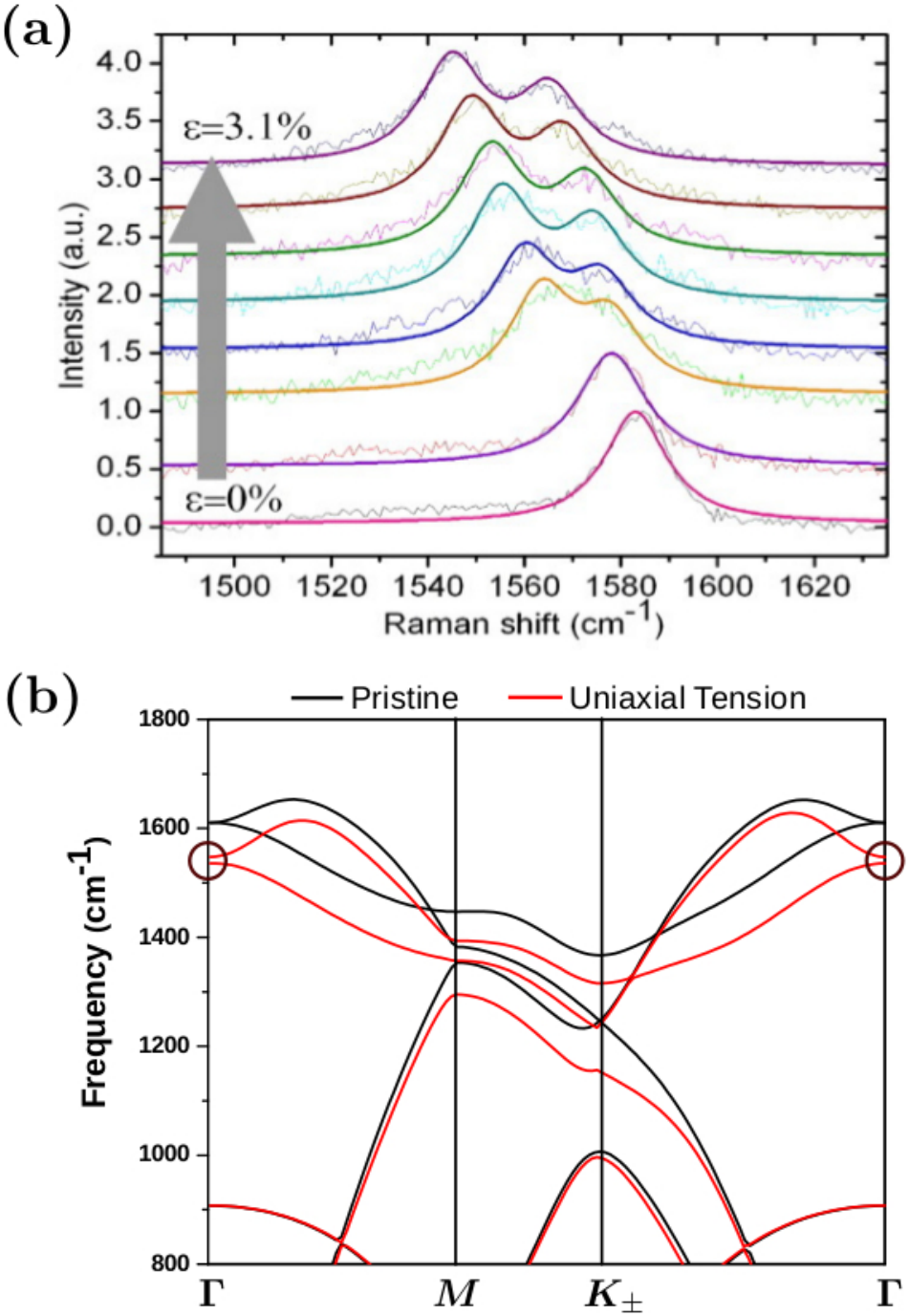}
\caption{Graphene under uniaxial tensile strain. (a) Experimental Raman spectra showing the splitting of the $G$ band (reproduced from \cite{Huangm2009}). (b) DFPT calculation showing the shift and the splitting of graphene phonon bands (emphasized by black circles at the $\Gamma$ point) under $1.63\,\%$ uniaxial strain along the zigzag direction when compared to the phonon dispersion of pristine graphene.}
\label{Fig5_HT}
\end{figure}

Graphene's symmetry is broken when exposed to uniaxial tensile stress. Besides a shift to lower frequencies, a splitting of the LO and TO modes takes place at the $\Gamma$ point, generating the split Raman bands $G^+$ and $G^-$ \cite{Huangm2009, Mohiuddin2009, Huangm2010}. Figure \ref{Fig5_HT}(a) shows a series of experimental graphene's Raman spectra under uniaxial strain \cite{Huangm2009}. In figure \ref{Fig5_HT}(b), a DFPT calculation reproduces the band splitting, thus agreeing with experimental results. The Gr{\"u}neisen parameters that relate the rate of change of phonon frequencies with respect to strain can be measured from the Raman shift with strain \cite{Mohiuddin2009}.

\begin{figure}[t]
\centering
\includegraphics[width=0.85\linewidth]{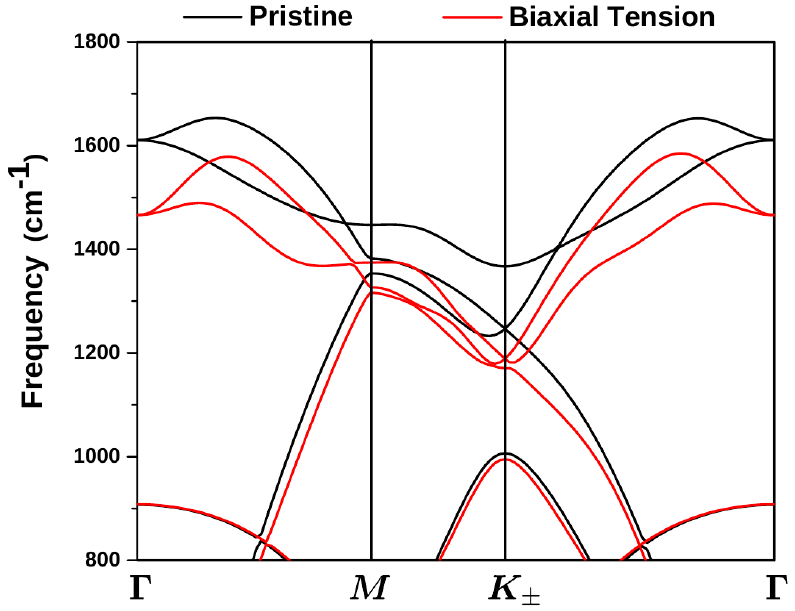}
\caption{Graphene under isotropic biaxial tensile strain: DFPT calculation showing the shift and the splitting of graphene under biaxial tensile strain of $1.3\,\%$ compared to pristine graphene. Note that there are no splitting at the $\Gamma$ point.}
\label{Fig6_HT}
\end{figure}

On the other hand, structural symmetry is preserved under isotropic biaxial strain, figure \ref{Fig6_HT}, and there is no splitting of the $G$ band (although there is a shift towards lower frequencies in the case of tensile stress and to higher frequencies under isotropic compressive biaxial stress \cite{Mohiuddin2009}).

Due to its non-invasiveness, Raman spectroscopy also plays a crucial role in the characterization of other 2D materials such as transition metal dichalcogenides (TMDs) \cite{LV2016, Terrones2014, LiangLiangbo2014} and phosphorene \cite{Zhangshuan2014, LingLiang2015}.

\section{Strain in graphene multilayers}\label{Multilayered}

\begin{figure*}[t]
\centering
\includegraphics[width=0.65\linewidth]{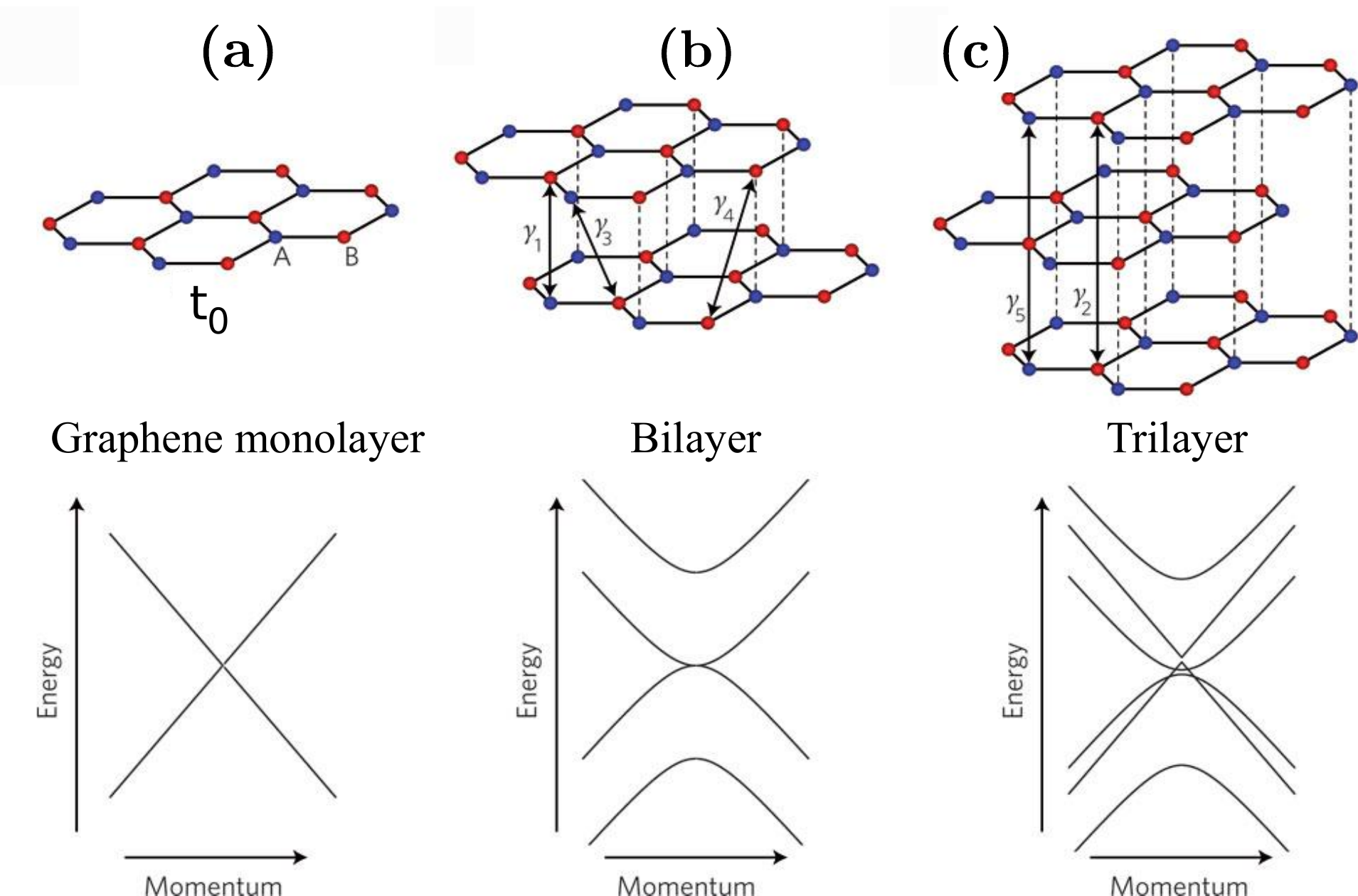}
\caption {Unstrained lattice structure (upper panels) and a sketch of their corresponding energy dispersion (lower panels) for (a) monlayer graphene, (b) Bernal
stacked bilayer graphene and (c) Bernal stacked trilayer graphene. Blue atoms belong to the A bipartite lattice while red atoms belong to the B lattice.
The arrows indicate the different kinds of interactions that appear in a TB calculation,
parametrized by $t_0$ for intra-layer interaction, and $\gamma_i$ with $i=1,...,5$  for inter-layer interactions. The Dirac cone seen in (a) for graphene is replaced by parabolic bands in (b),
while trilayer graphene includes both types of bands.  Reprinted by permission from Macmillan Publishers Ltd: Nature   \cite{Freitag2011}, copyright (2011).}
\label{Multilayerstructure}
\end{figure*}

Even though graphene multilayers are not truly 2D materials, they can be studied with the techniques discussed thus far. Layers can be stacked following different relative placements of the graphene's bipartite lattices A and B in different layers \cite{Kim2016}. Nowadays, it is possible to control the stacking sequences, including the addition of
other different layered materials to build nanocomposites \cite{Lee2015}.
For graphite, the most common  stacking is known as Bernal stacking (see figure \ref{Multilayerstructure}(b)), where $B$ atoms of layer 2 ($B2$) lie directly on top of $A$ atoms of layer 1 ($A1$) while
$B1$ and $A2$ atoms are in the center of the hexagons of the opposing layers \cite{Latil2007,Freitag2011}.
 The structure of an $ABA$ trilayer is presented in figure \ref{Multilayerstructure}(c) as well.

The band structure of graphene multilayers can  be  understood  using  the  Slonczewski-Weiss TB model for graphite \cite{McCann2013}. Within this model, there is a hopping parameter $t_0$ as in monolayer graphene to account for the intra-layer interaction, as indicated in figure \ref{Multilayerstructure}(b) by arrows. For bilayer graphene, there are five hopping parameters $\gamma_i$ with $i=0,...,4$  to account for the different kinds of overlaps of $\pi$-orbitals and four on-site energies $\epsilon_{A1},\epsilon_{B1},\epsilon_{A2},\epsilon_{A2}$ on the four atomic sites. As indicated in figure \ref{Multilayerstructure}(c), these parameters were determined by infrared spectroscopy \cite{Kuzmenko2009}, resulting in $t_{0}=3.16$ eV, $\gamma_1 = 0.381$ eV, $\gamma_3 = 0.38$ eV, $\gamma_4= 0.14$ eV, $\epsilon_{B1} = \epsilon_{A2}=0.022$ eV and $\epsilon_{A1} = \epsilon_{B2}=0$. Two additional parameters ($\gamma_2$ and  $\gamma_5$)  are needed to describe trilayers.

The eigenvalues of the following $4\times4$ Hamiltonian matrix provide the single-particle electronic dispersion of bilayer graphene:
\begin{eqnarray}
 H=\nonumber\\
 \left( \begin{array}{cccc}
\epsilon_{A1} &  -t_{0} f(\boldsymbol{k}) & \gamma_{4}  f(\boldsymbol{k}) & -\gamma_{3}  f^{*}(\boldsymbol{k})\\
 -t_{0} f^{*}(\boldsymbol{k}) & \epsilon_{B1} & \gamma_{1}   & \gamma_{4}  f(\boldsymbol{k})\\
 \gamma_{4} f^{*}(\boldsymbol{k}) & \gamma_{1}   & \epsilon_{A2} & -t_{0}  f(\boldsymbol{k})\\
-\gamma_{3} f(\boldsymbol{k}) & \gamma_{4}  f^{*}(\boldsymbol{k}) & -t_{0}  f^{*}(\boldsymbol{k}) & \epsilon_{B2}\\
\end{array} \right)
\end{eqnarray}
where $f(\boldsymbol{k})$ is given by equation (\ref{finplane}). The four resulting bands are schematically represented in
figure \ref{Multilayerstructure}(b). The red bands are parabolic and touch without a gap within this single-particle picture. Nevertheless, a gap does open at low temperatures
due to electron-electron interaction \cite{Lau2011,Lau2012,Lau2014}. It is also possible to
obtain an effective low-energy Hamiltonian \cite{McCann2013}.

 Trilayer graphene can be treated in a similar way, resulting in the ``mixing''
of bilayer and monolayer energy dispersion features shown in figure \ref{Multilayerstructure}(c).

Strain can be studied in bilayer graphene using the methodology presented for graphene, with the peculiarity of having extra degrees of freedom in the deformation associated with the presence of two layers \cite{Nanda2009,Verberck2012}; band gaps can be opened by in-plane layer distortions or by pulling the layers apart \cite{Verberck2012}.

Also, triaxial strain effects can be described by an effective pseudo-magnetic field in bilayers \cite{Moldovan2016}. Using this methodology, an in-plane strain applied equally to both layers breaks the layer symmetry: at low energy, just one of the layers feels the pseudo-magnetic field while the zero-energy pseudo-Landau level is missing in the other layer. This effect produces a gap between the lowest non-zero levels \cite{Moldovan2016}.

\begin{figure*}[t]
\centering
\includegraphics[width=0.65\linewidth]{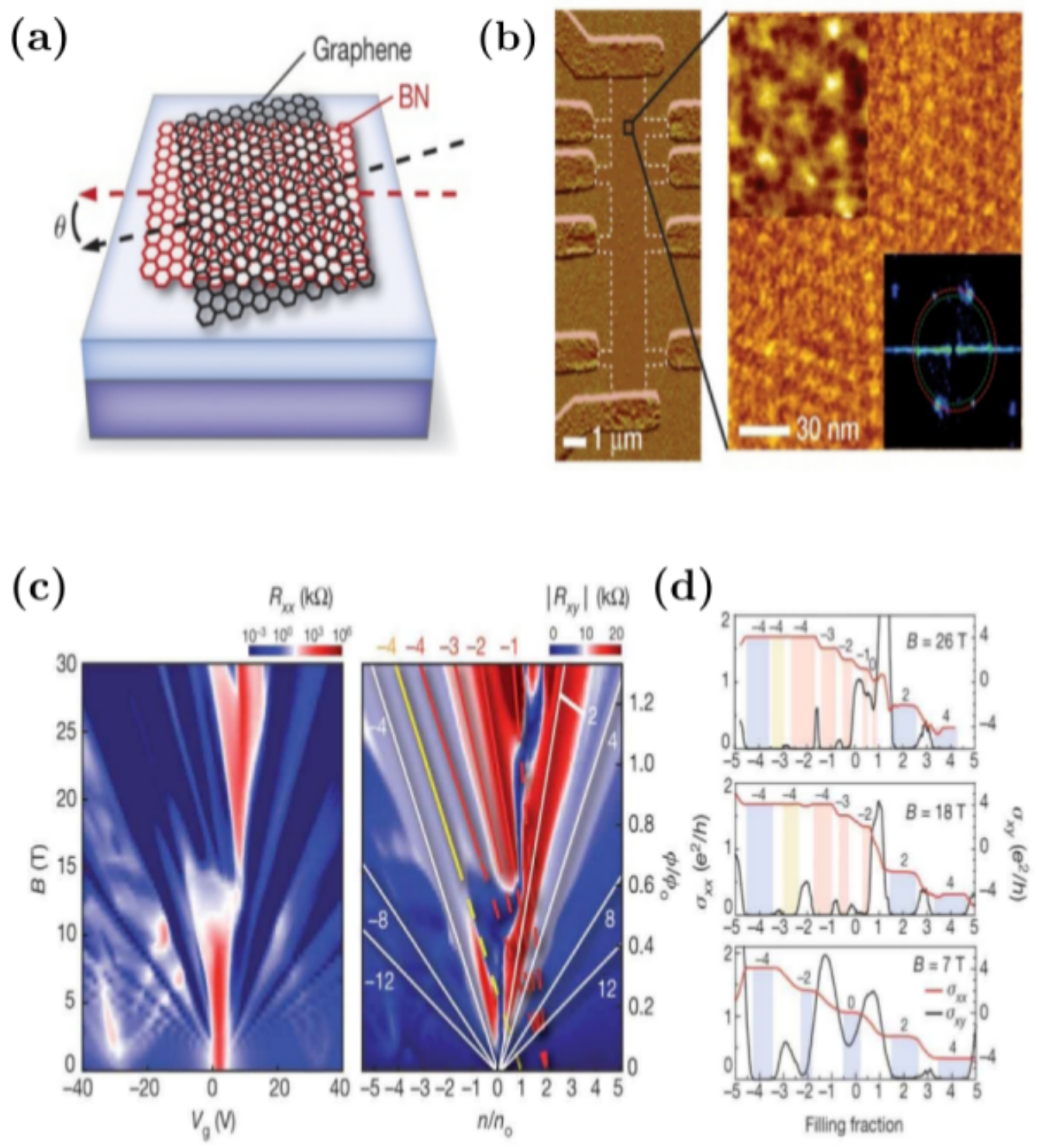}
\caption {Moir\'e superlattice and anomalous quantum Hall states.
(a) Bilayer graphene on hBN. The rotation of graphene by the mismatch angle $\theta$ determines
the wavelength.
(b) Left: AFM image of the multiterminal Hall device. Right: a magnified region of the same device. The
resulting pattern is a triangular lattice.
A fast Fourier confirms the triangular lattice symmetry.
(c)  Diagram showing the longitudinal resistance, $R_{xx}$ (left), and Hall resistance, $R_{xy}$ (right) for the
device presented in (b).  $R_{xx}$ is plotted versus magnetic field on the vertical axis and versus gate voltage $V_{(g)}$.
$R_{xy}$ is plotted as a function of the magnetic flux ratio $\phi/\phi_0$ on the vertical axis and the normalized carrier density
$n/n_{0}$ on the horizontal axis.
QHE states corresponding to the conventional BLG  spectrum
are indicated by white lines. Solid yellow and red lines track the QHE outside
the conventional spectrum, with dashed lines indicating the projected $n/n_{0}$. The slope of each line is shown on the top axis
as well as the intercept. Each pair of parameters are the solution of a Diophantine equation \cite{Dean2013}, characteristic of
topological states in the QHE \cite{Naumis2016}.
(d)  Longitudinal and transversal Hall conductivities corresponding to line cuts at constant magnetic
field (constant $\phi/\phi_0$). The color bars indicate the features with the same color appearing in (c). For magnetic fields of
$12$ T and $26$ T, additional QHE states appear with non-integer Landau level filling fractions. Reprinted by
permission from Macmillan Publishers Ltd: Nature  \cite{Dean2013}, copyright (2013).}
\label{Hofstadter}
\end{figure*}

When two graphene layers are rotated relative to each other by an angle $\theta$ away from Bernal stacking,
a moir\'e pattern is produced \cite{Eva2012}. This  stacking misorientation  mimics the effect
of in-plane pseudomagnetic fields \cite{He2014}.

A very important advance made possible by using rotationally faulted biaxial graphene was the first experimental observation of the Hofstadter butterfly \cite{Dean2013}. This
fractal spectrum was predicted to occur for electrons in a lattice under a constant magnetic field \cite{Hofstadter1976}.
Its importance was paramount since it provided a platform to understand the Quantum Hall effect (QHE) in terms of topological phases.
Originally, the problem of electrons in a constant field was studied by Landau, giving rise to the well
known Landau levels with energy $E=(n+1/2)\hbar\omega$  and $n$ integer. As noted by Hofstadter, the lattice length adds a new
scale in the problem that competes with the magnetic length \cite{Hofstadter1976}. As a result, the spectrum is controlled
by the ratio between the elementary quantum flux ($\phi_0$) and the magnetic flux ($\phi$). This results in a one dimensional effective
problem, where the potential depends on the ratio $\phi/\phi_0$. The corresponding equation is known as the Harper equation, and is quite similar
to the equation dictating the dynamics of uniaxial strained graphene.

As it was discussed with graphene monolayers in previous sections, the spectrum depends upon the ratio $\phi/\phi_0$ and it leads to a periodic or quasiperiodic behavior depending on  whether $\phi/\phi_0$ is rational or irrational. The corresponding spectrum is a complex fractal, known as the
Hofstadter butterfly \cite{Hofstadter1976} with interesting topological properties \cite{Naumis2016}.
For small ratios of the fluxes, Landau levels are recovered. Unfortunately,
for atomic systems this requires the use of magnetic fields well beyond the available sources. However, such fractal
spectrum was measured recently by using bilayer graphene over a hBN sustrate \cite{Dean2013}.

As shown in figure \ref{Hofstadter}(a),
the mismatch angle $\theta$ between both lattices determines  the moir\'e pattern (see Section \ref{Deformations}). In panel (b)
of figure \ref{Hofstadter}, the resulting triangular pattern and the device used to measure the electronic
properties are shown. In figure \ref{Hofstadter}(c), the longitudinal resistance $R_{xx}$ (left), and Hall resistance, $R_{xy}$ (right) measured by Dean \textit{et al.} \cite{Dean2013} are presented. $R_{xx}$ is plotted as a function of the gate voltage $V_g$ and magnetic field $B(T)$,
while $R_{xy}$ is shown as a function of the magnetic flux ratio $\phi/\phi_0$ on the vertical axis
and the normalized carrier density $n/n_{0}$ on the horizontal axis ($n_{0}$ is the carrier density at zero
gate voltage). Both plots reveal fountain-like structures characteristic of the Hofstadter butterfly \cite{Naumis2016} (there are van Hove singularities associated
to the electronic structure).

Moreover, the slope and intercept with the vertical axis of each observed
line in figure \ref{Hofstadter}(c) are the solution of the Diophantine equation, characteristic of topological states in the QHE, and in figure \ref{Hofstadter}(d), the longitudinal and transversal Hall conductivities corresponding to line cuts for
three constant magnetic fields (constant $\phi/\phi_0$) are presented.

\section{Extensions to other strained 2D materials}\label{Other2D}

A brief guide to other strained 2D materials is presented in this Section with the aim of helping readers to pin-point areas of new research in which the previous discussed methods are also applicable. There are some recent reviews and books already covering this subject \cite{Roldan2015,Amorim2016,Deng2016,KolobovBook}, and this short review covers the following materials: silicene and other group-IV two-dimensional materials, phosphorene, transition metal-dichalcogenides, and layered monochalcogenides.

Layered monochalcogenides have seen a recent resurgence in interest \cite{Zhao2014,KaiSCIENCE,MauriSCIENCE}, yet they still lack a dedicated review thus far. For that reason, the present document contains a more detailed description of these materials when compared to the coverage of silicene, phosphorene and transition metal-dichalcogenides, which have been reviewed by other authors already.

Let us remark that when describing strain in Section \ref{DescriptionStrain}, a mention was made to random strain fields, which may arise on crystalline lattices at finite temperature; thermal effects on strain were highlighted in figure \ref{Freestanding} too.

Similarly, it has been argued in Sections \ref{Ripples} and \ref{discgeo} that curvature (ripples) and metric (strain) command similar effects on electronic and optical properties of 2D materials. Therefore, some of the aspects highlighted on this Section concern the effects of temperature and curvature on the properties of 2D materials, which may not be as thoroughly addressed in other reviews. In that sense, the present approach is thus general and original as well.

\begin{figure}[t]
\centering
\includegraphics[width=\linewidth]{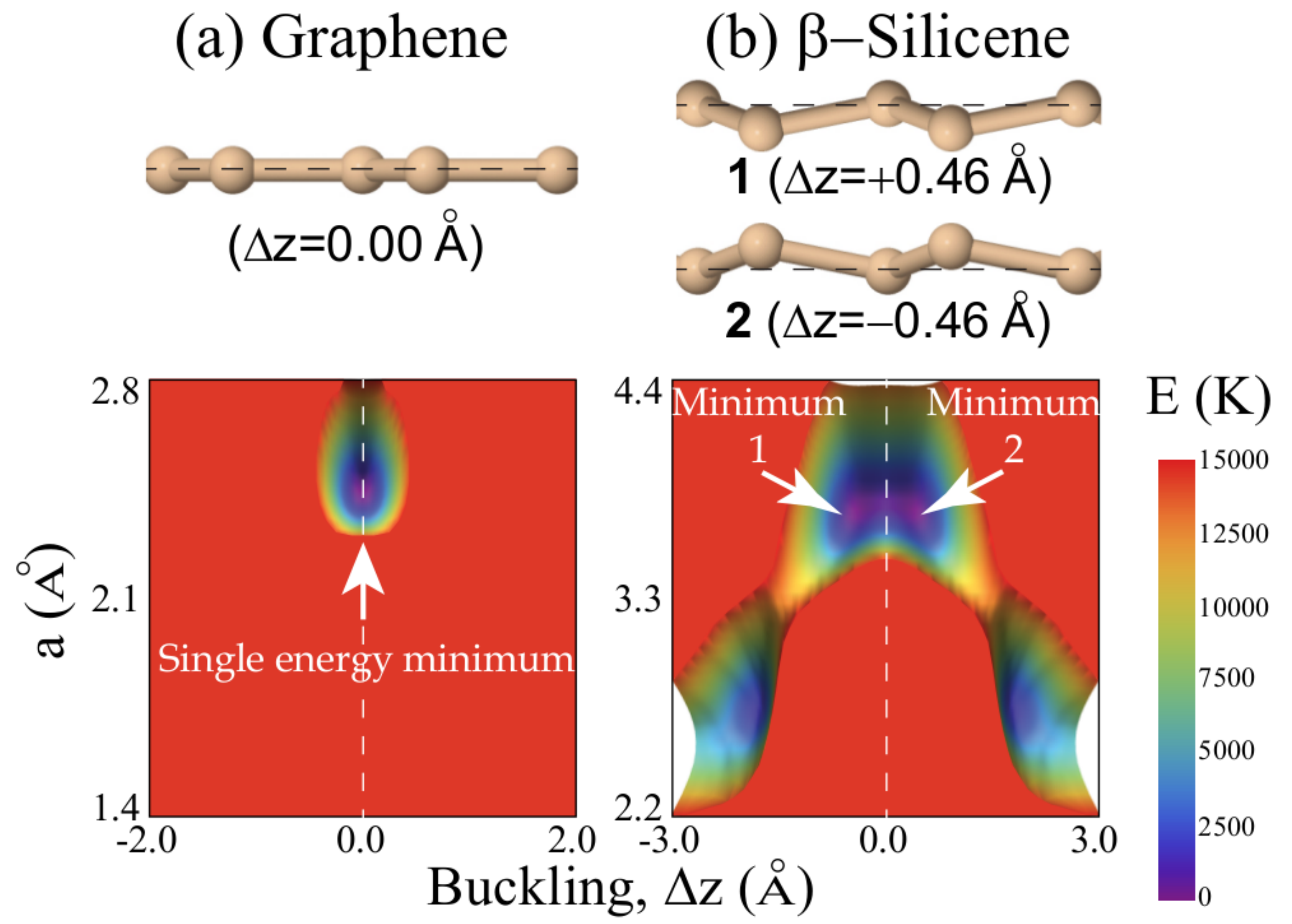}
\caption {Highlighting the planar graphene structure, and the two-fold structural degeneracy of $\beta-$silicene. The atomistic models correspond to the minima of the energy profiles shown in the lower row.}
\label{landscape}
\end{figure}

\subsection{Silicene and other group-IV two-dimensional materials}\label{sec:silicene}

Silicene \cite{Cahangirov2009,Tao2014,Silicene2016,Kara2012} is an interesting material in part due to its compatibility with
current electronic technologies.
If silicene was flat (a structure called $\alpha$-silicene), all of the methods discussed for studying strain on graphene would carry over \cite{Kara2012}.

However, silicene is low-buckled (a structure called $\beta-$silicene that is shown in figure \ref{landscape}). This structure is similar to that of graphene, with the particularity that silicon atoms in the two triangular bipartite sublattices are
vertically displaced by $0.46\,\mbox{\AA}$. The interatomic distance between atoms
is  $2.28\,\mbox{\AA}$, which is larger than the carbon-carbon distance in graphene \cite{Kara2012}.

In figure \ref{landscape}, lower row, one sees in color the energy $E$ needed to displace $A$ and $B$ atoms vertically ($\Delta z$, horizontal axis) and to increase the lattice constant $a$ with the application of in-plane isotropic biaxial strain (vertical axis). The energy profile $E(\Delta z,a)$ arises when changing interatomic distances and therefore it is an elastic energy landscape \cite{WalesBook}.

While graphene has a single minima on $E(\Delta z,a)$ at $E(0,1.42\,\mbox{\AA})$ , $\beta$-silicene displays two such minima, which arise due to the fact that the structure has the same energy regardless of whether the $B$ atom is vertically displaced by $+ 0.46\,\mbox{\AA}$ or by $-0.46\,\mbox{\AA}$ with respect to the $A$ atom. This is to say, $\beta$-silicene has a two-fold degenerate atomistic structure. As a novel research avenue, $\beta$-silicene is yet to be thoroughly studied from the perspective of its structural degeneracies. As it will be shown when discussing layered monochalcogenides, structural degeneracies may be the norm rather than the exception in 2D materials beyond graphene, playing a preponderant role on material effects that are induced from the combined effects of strain and temperature.

In order to properly account for the electronic properties of $\beta$-silicene, one must take into account the coupling
of  $\pi$ and $\sigma$ electrons, which leads to an effective $8\times 8$ matrix Hamiltonian \cite{Guzman2007,Huertas2006}.

Not considering the gap opening due to spin-orbit coupling, DFT calculations indicate that the Dirac point is preserved  for  $\beta$ silicene up to $5\,\%$ of strain\cite{Guzman2007}. Higher strain induces hole doped Dirac states because of weakened bonds \cite{Kaloni2013}. The Gr\"{u}neisen parameter shows a  significant variation,  from $1.64\,\%$ for a strain of $5\,\%$
to $1.42\,\%$ for $25\,\%$ strain \cite{Kaloni2013}.

\begin{figure}
\centering
\includegraphics[width=\linewidth]{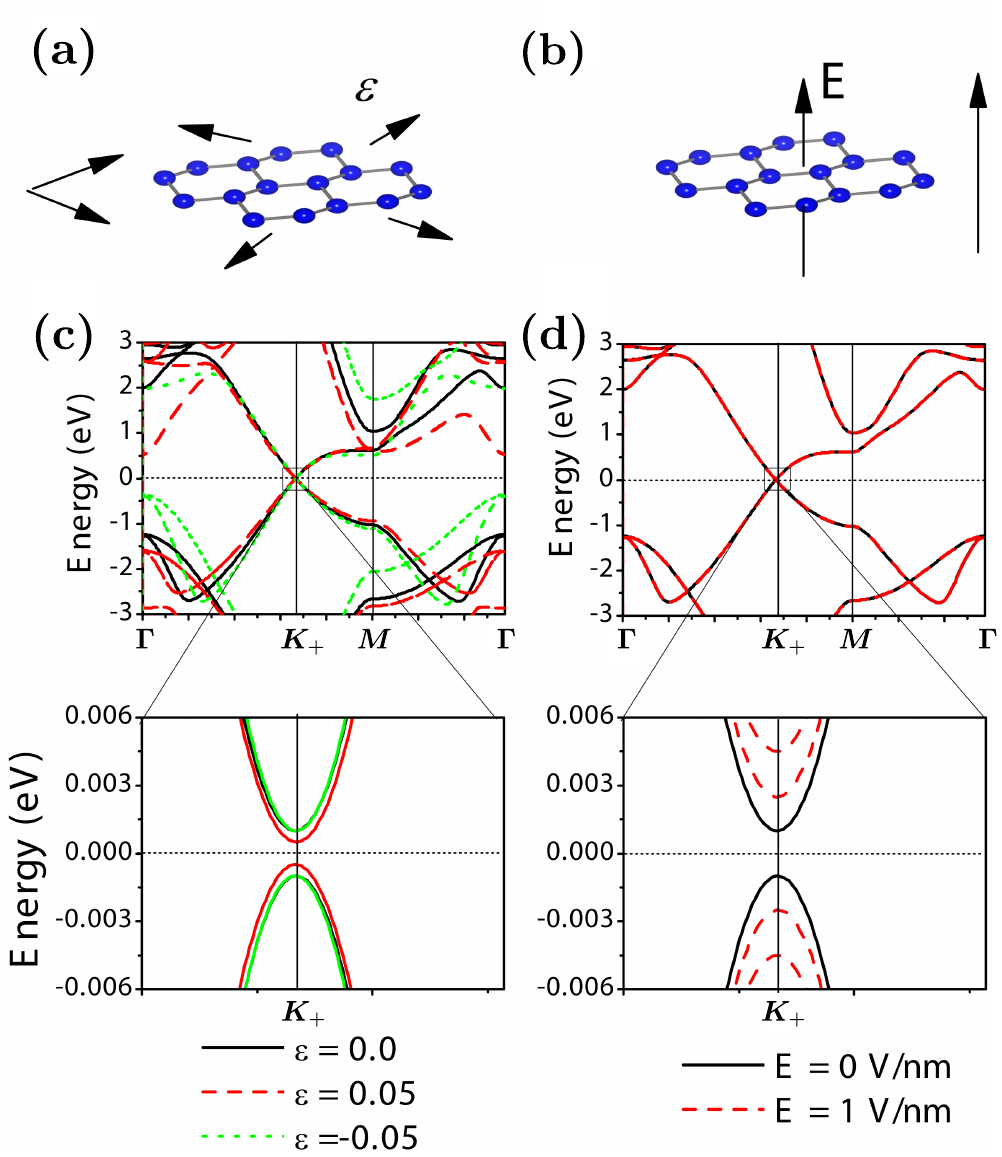}
\caption {The electronic properties of silicene are independently tuned by (a) in-plane biaxial strain $\varepsilon$, and (b) an out-of-plane E-field $E_z$. Subplots (c) and (d) are band dispersions under typical values of $\varepsilon$ or $E_z$, respectively. Insets are zoom-ins of the band dispersion near the $\bi{K}_{+}$ point (the Fermi level is set to zero; note the overlapping bands for $\varepsilon=0.00$ and $0.05$ on (c)). Reproduced from \cite{Yan2015b} with permission.}
\label{APL_Fig1}
\end{figure}

\begin{figure}
\centering
\includegraphics[width=\linewidth]{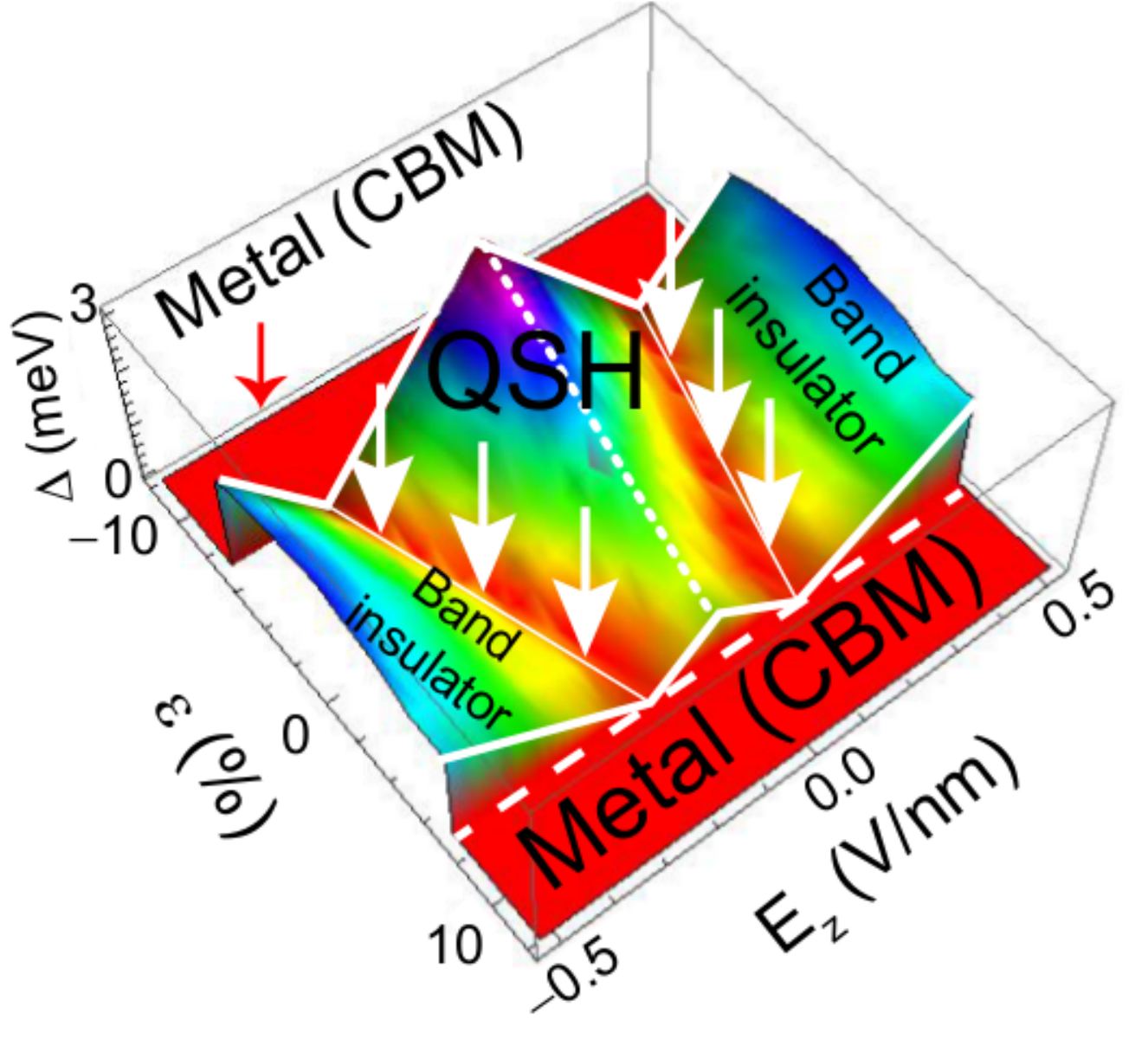}
\caption {Topological quantum phase diagram of $\beta$-silicene with respect to in-plane biaxial strain $\varepsilon$ and out-of-plane $E-$field $E_z$. The vertical axis is the band gap $\Delta$. The critical E-field $E_c$, at which there is a phase transition from a topological insulator into a band insulator have been indicated by vertical white arrows. The metallic state has a zero value of $\Delta$ that is shown in red, and the area marked by QSH represents the topological spin Hall state phase. Reproduced from \cite{Yan2015b} with permission.}
\label{APL_Fig3}
\end{figure}

But due to a larger spin-orbit coupling than that of graphene, $\beta$-silicene is a quantum spin Hall insulator \cite{Kane2005} with a topological quantum phase transition controlled by an out-of-plane electric field \cite{Ezawa1,Ezawa2}, that can be further tuned by the application of an in-plane isotropic biaxial strain $\varepsilon$ owing to the curvature-dependent spin-orbit coupling (SOC) \cite{Huertas2006}. It is a $Z_2 = 1$ topological insulator phase for biaxial strain $|\varepsilon|$ smaller than $0.07$, and the band gap can be tuned from $0.7\,\mbox{meV}$ for $\varepsilon = +0.07$ up to a fourfold $3.0\,\mbox{meV}$ for $\varepsilon = -0.07$.

As seen in figures \ref{APL_Fig1} and \ref{APL_Fig3}, first-principles calculations also show that the critical field strength $E_c$ can be tuned by more than $113\,\%$, with the absolute values nearly 10 times stronger \cite{Yan2015b,Yan2015} than theoretical predictions based on a TB model \cite{Ezawa1,Ezawa2}. Due to the curvature-enhanced SOC, the buckling structure of the honeycomb lattice thus enhances the tunability of both the quantum phase transition and the SOC-induced band gap, which are crucial for the design of field-effect topological field-effect transistors based on 2D materials.

Atomistic defects play an important role in the structural properties of silicene nanoribbons, as the Young modulus exhibits
a complex dependence on the combinations of vacancies and on temperature \cite{ChavezCastillo2015}.

Additional group-IV 2D materials have been proposed, and experimental efforts to synthesize them are well under way. Still within the context of strain, this Subsection is concluded by highlighting relevant energetic considerations that raise the stakes to generate crystalline 2D phases out of group-IV elements.

\begin{figure}[tb]
\includegraphics[width=\linewidth]{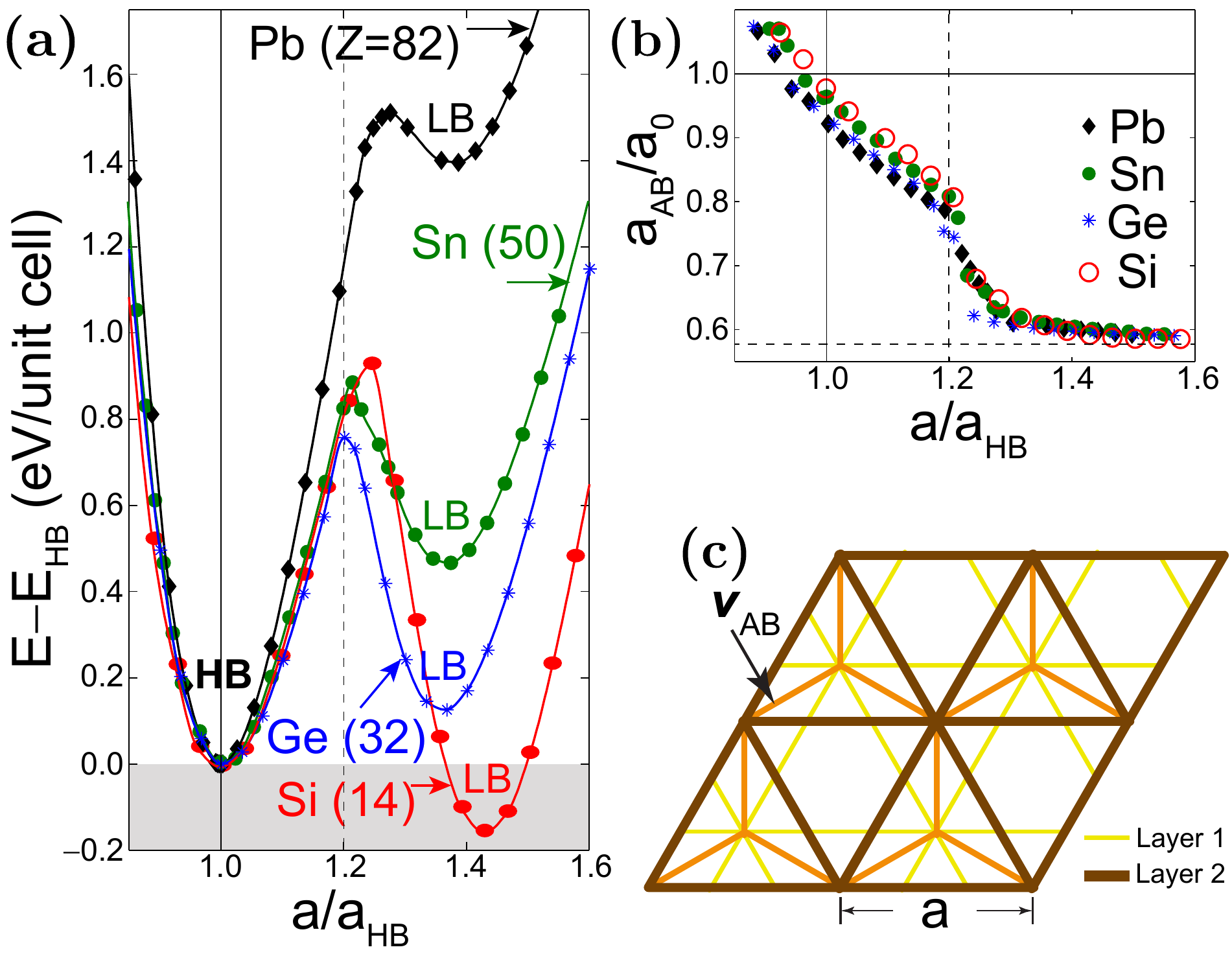}
\caption{(a) In going from silicon down to lead, the high-buckled phase becomes more stable with increasing atomic number of column IV elements.
(b) Nearest-neighbor distances $a_{AB}\equiv |\mathbf{v}_{AB}|$ approach the lattice constant $a$ ($a_{AB}\simeq a$) at the high-buckled
energy minimum; the structure transitions to a low-buckled phase at roughly 1.2$a_{HB}$. (c) The high-buckled structure is a HCP bilayer. Reproduced from \cite{Rivero2014} with permission.  Copyrighted by the American Physical Society.}\label{fig:F1}
\end{figure}

In figure \ref{fig:F1}(a) the total DFT energy as a function of the lattice constant is displayed. Thus, while silicene has a structural minima at the low-buckled, $\beta-$phase, the actual structural minima for two-dimensional materials in group-IV is at the minima labeled HB at the figure, which is a hexagonal-close-packed bilayer with nine-fold coordination shown in figure \ref{fig:F1}(a) \cite{Rivero2014} and whose lattice constant is labeled $a_{HB}$. Within the context of strain, the actual structural minima is not the one proposed in \cite{Ma2012} and latter reproduced in \cite{Xu2013}, which are actually realized only after applying a $\sim40\,\%$ isotropic in-plane biaxial strain to the unstrained (labeled HB) structure.

A possibility for tuning the band gap in low-buckled two-dimensional tin is to add fluorine atoms to the two-dimensional structure \cite{Ma2012}, a result confirmed in \cite{Xu2013}. As it turns out, such fluorinated structure is not the structural minima either \cite{Rivero2014}. These structural considerations constrain the realization of fluorinated 2D tin as a viable 2D topological insulator.

\subsection{Phosphorene}\label{bp}

\begin{figure}[tb]
\includegraphics[width=\linewidth]{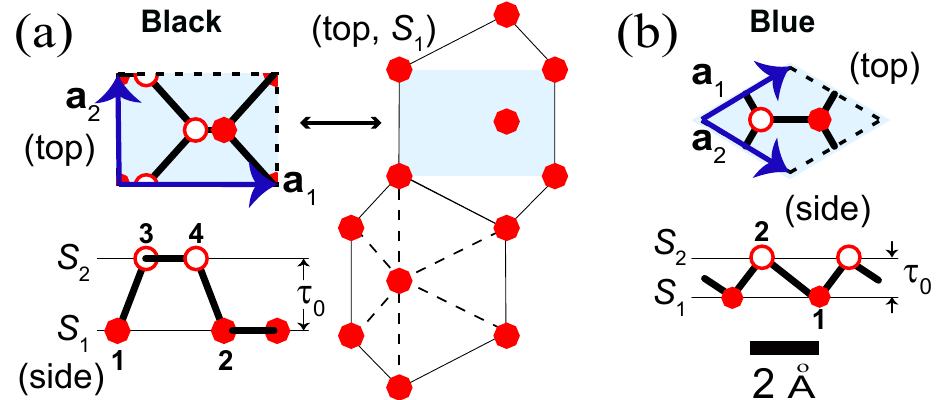}
\caption{The unit cells for (a) black and (b) blue phosphorene monolayers are formed by two sublayers ($S_1$ and $S_2$) separated by a distance $\tau_0$. Reproduced from \cite{Mehrshad2015} with permission.}\label{fig:PNASF1}
\end{figure}

Phosphorene is a single-layer material obtained by exfoliation of layered black phosphorus (BP).   Black phosphorene, figure \ref{fig:PNASF1}(a), has a semiconducting gap that is tunable with the number of layers, and by in-plane strain \cite{Roldan2015,BlackStrain1,BlackStrain2,KatPhos,Cakir2014}.

The  outstanding performance of transistors built using black phosphorene ignited intense research activities \cite{Phosphorene1,Likai2014}. Many groups consider this material as better suited for electronics than graphene due to several factors: the ease of fabrication, a
$0.2\,\mbox{eV}$ direct gap, and a band topology that is not altered by thickness that has an impressive anisotropy \cite{Das2014,Kou2015}.
It possess a reasonable on/off ratio ($10^{4}$-$10^{5}$) with a good carrier mobility (around $1000$ cm$^2$/Vs) suitable for many applications \cite{Kou2015}. Concerning the mechanical properties, it has a highly anisotropic Young modulus and Poisson ratio \cite{Jiang2014}.
All these remarkable features are a result of the stacked  layered structure and weak van der Waals (vdW) interlayer interactions. Its rectangular unit cell leads to a four-fold degenerate structural ground state \cite{Mehboudi2016}.

Phosphorene oxidizes readily \cite{Wood2014,Favron2015} and structural defects provide the only structural mechanism known to dissociate $O_2$ dimers \cite{Utt2015}.

2D semiconductors screen electric fields poorly, and inclusion of many body effects within the context of the GW approximation is essential for the correct description of the electronic bandgap of phosphorene \cite{BlackStrain2,Cakir2014} and its excitonic properties \cite{Xia2015}. Tensile strain on phosphorene enhances electron transport along the zigzag direction, while biaxial strain is able to tune the optical band gap from $0.38\,\mbox{eV}$ (at $0.8\,\%$ strain) to $2.07\,\mbox{eV}$ (at $5.5\,\%$) \cite{BlackStrain2,Cakir2014}.  Another interesting phenomena resulting from \textit{ab initio} calculations is the notable increasing of the electron-phonon interaction by biaxial strain \cite{Ge2015}.

Using a low-energy TB Hamiltonian that includes the spin-orbit interaction for bulk phosphorene, it has been found that a compressive biaxial in-plane strain and a perpendicular tensile strain lead to a topological phase transition \cite{Sisakht2016} with protected edge states. For a width of $100\,\mbox{nm}$, the energy gap is at least three orders of magnitude larger than the thermal energy at room temperature \cite{Sisakht2016}.

Phosphorene \cite{Phosphorene1,Churchill,Dresselhaus} is also predicted to have many allotropes that are either  semiconducting or metallic depending on their two-dimensional atomistic structure \cite{TomanekBlue,allotropes,Tiling,Nebraska}, and the unit cell of blue phosphorene (which is similar to that of $\beta$-silicene) is displayed on figure~\ref{fig:PNASF1}(b).

Considering the existence of reviews on this material \cite{Kou2015,Jiang2014}, this section ends by highlighting a geometrical (strain/curvature induced) effect on the electronic bandgap.

\begin{figure}[tb]
\includegraphics[width=\linewidth]{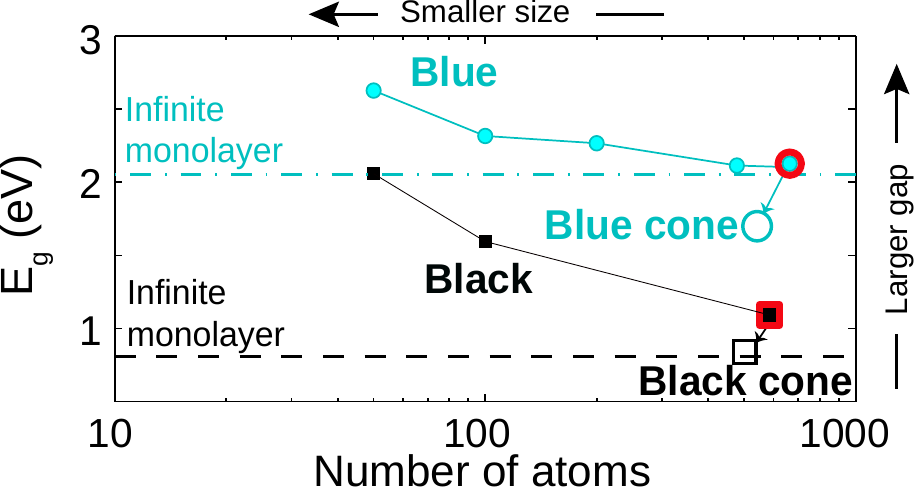}
\caption{The semiconducting electronic bandgap of finite-size planar black and blue phosphorene monolayers increases as the number of atoms decreases due to quantum confinement. In an opposite trend, the gap decreases on a curved structure; such decrease on the bandgap thus has a purely geometrical origin. (Dash and dash-dot lines indicate these gaps when the number of atoms is infinite.)  Reproduced from \cite{Mehrshad2015} with permission.}\label{fig:PNASF3}
\end{figure}

Theoretical studies of defects on planar phosphorene indicate that dislocation lines do not induce localized electronic states  within the electronic bandgap \cite{Jakobson2}. This somehow unexpected result was employed to build semiconducting phosphorene cones that have a finite curvature \cite{Mehrshad2015}. Black and blue phosphorene monolayers are both semiconducting 2-D materials with a direct bandgap, and the semiconducting gap evolves with their shape. As seen in figure \ref{fig:PNASF3}, a reduction of the semiconducting gap can be induced by curvature.

As in graphene, another emerging field is the study of strained black phosphorus multilayers.  It has been found that a periodic stress produces a remarkable shift of the optical absorption band-edge, up to $0.7\,\mbox{eV}$ between the regions under tensile and compressive stress \cite{Quereda2016}. This tunability greatly exceeds the reported value for strained transition metal dichalcogenides.  According to theoretical models, the periodic stress modulation can yield to quantum confinement of carriers at low temperature \cite{Quereda2016}.

\subsection{Transition metal-dichalcogenide monolayers}

Another family of layered materials, analogous to graphene, where layers interact by weak van der Waals forces corresponds to transition metal-dichalcogenide (TMD) monolayers, with chemical composition $\mbox{MX}_2$, where $M$ corresponds to a transition metal and $X$ represents a chalcogen. TMD monolayers are promising materials for solving fundamental scientific and technological challenges \cite{Vogel2015}.

Although the first report on TMD monolayers dates back 30 years \cite{Joensen1986}, a renewed surge of interest occurred after the discovery of graphene and the development of new techniques to deal with ultrathin layered materials. In 2000, it was predicted that TMD monolayers possess a direct band gap \cite{seifert2000}, and this prediction was confirmed only in 2010, when exfoliated MoS$_2$ monolayers exhibited a direct band gap around $1.8\,\mbox{eV}$ \cite{Mak2010}.

\begin{figure}[t]
\centering
\includegraphics[width=0.9\linewidth]{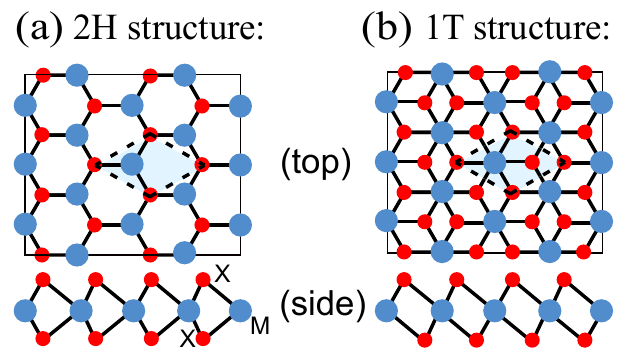}
\caption {Top and side views of the $\mbox{MX}_2$ monolayer crystal structure with phase: (a) 2H  and (b) 1T. Unit cells are shown within dashed lines.}
\label{TMDC-1}
\end{figure}

As illustrated in figure~\ref{TMDC-1}, TMD monolayers are structurally of the form X--M--X, with a hexagonally packed plane of metal atoms sandwiched between two planes of chalcogen atoms. Metal atoms can have either octahedral or trigonal prismatic coordination. For instance, the most stable phase of the Mo and W-based TMD monolayers is a trigonal prismatic phase ($2H$), while Hf and Zr-based TMDs have an octahedral ground state phase ($1T$). The preferred phase adopted by TMD monolayers depends on the $d$-electron count of the transition metal strongly \cite{KolobovBook}.

A key fact that triggered the attention to TMD monolayers was the building of a high-quality field-effect transistor based on $\mbox{MoX}_2$ monolayers \cite{Radisavljevic2011}. TMD monolayers have electronic properties ranging from semiconducting to superconducting depending on their chemical composition. Particularly, group-VI TMD monolayers (e.g. MoS$_2$, WS$_2$, MoSe$_2$ and WSe$_2$) exhibit semiconductor behaviour, with a direct bandgap in the range of $1$ to $2\,\mbox{eV}$,  which provides a wide variety of promising electronic and optoelectronic applications. TMD bilayers and multilayers possess an indirect band gap. Further features of TMD monolayers will be discussed next.

\begin{figure}[t]
\centering
\includegraphics[width=0.8\linewidth]{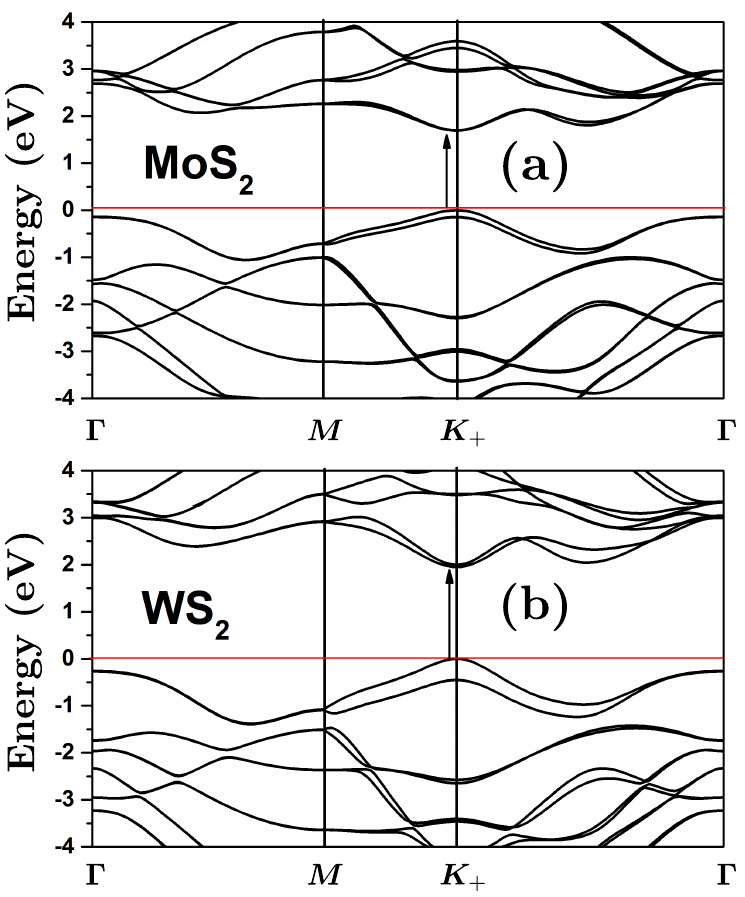}
\caption {DFT-GGA band structucture of TMD monolayers (a) MoS$_2$ and (b) WS$_2$. Note the direct band gap at the $\bi{K}_+$ point (shown by the arrow) and the splitting of the valence band at $\bi{K}_+$ caused by the strong spin-orbit interaction.}
\label{TMD-Humberto_1}
\end{figure}

TMD monolayers with a $2H$ phase (MoS$_2$, WS$_2$, MoSe$_2$ and WSe$_2$) belong to the hexagonal $P\bar6m2$ (187) group, thus having hexagonal symmetry in both real and reciprocal space (similar to graphene). As shown in figure~\ref{TMD-Humberto_1}, TMD monolayers with a $2H$ phase
 exhibit a direct band gap at the $\bi{K}_{\pm}$ point as calculated by DFT. In addition, in the same figure, the splitting in the valence band at the $\bi{K}_{+}$ point for MoS$_2$ and WS$_2$ caused by the strong SOC is shown. The splitting is larger in WS$_2$ ($0.42\,\mbox{eV}$) than in MoS$_2$ ($0.155\,\mbox{eV}$) because W is heavier than Mo. Due to time-inversion symmetry, the spin polarization at the valence band reverts sign in between the $\bi{K}_+$ and the $\bi{K}_-$ valleys.

The strong SOC \cite{Zhu2011} mentioned above is due to the lack of inversion symmetry and the presence of $d$-orbitals associated with the transition-metal atoms. This property makes TMD monolayers potential candidates for spintronic devices \cite{Zibouche2014}.

Applications for TMD monolayers in spintronics, photonics and in a new area that it is just emerging called valleytronics \cite{Xu2014,Yang2012} are being heavily sought. At the same time, semiconducting TMD monolayers posses strong exciton-binding energies that promise a new age of atomic-scale photonics \cite{Cao2015}. Regarding valleytronics, TMD monolayers are ideal systems since they exhibit different spin polarization at the $\bi{K}_{+}$ and $\bi{K}_-$ valleys. Different valleys can be independently addressed by circularly-polarized light having orthogonal polarizations. Analogous to spin, valley polarization becomes an additional degree of freedom which can be used in the future for new devices \cite{zeng2012,Mak2012}.

Moreover, TMD monolayers have no surface dangling bonds, making the production of heteroestructures in the vertical direction without the requirement of lattice matching possible. TMD nanoribbons can have interesting properties due to edges: enhanced catalytic activity \cite{Reyes2014} that is useful for several applications like dye-sensitized solar cells \cite{ZhangZhen2016}, or robust electrocatalysis, which may be useful for hydrogen generation \cite{Wang2016}. Three basic mechanisms explain the active properties due to edges:  quantum confinement,  edge topology, and electronic interaction among edges for very narrow nanoribbons \cite{Reyes2014}.

TMD monolayers command large mechanical strength, flexibility and stretchability. For instance, a MoS$_2$ monolayer has a Young's modulus of $\approx180\,\mbox{N/m}$ (corresponding to a 3D Young's modulus of $\approx270\,\mbox{GPa}$) and a breaking strength which approaches the upper theoretical limit, as measured by nanoindentation experiments \cite{Bertolazzi2011}. At the same time, MoS$_2$ monolayers withstand strain levels greater than $10\,\%$ and can also be folded \cite{Castellanos2014}. Similar mechanical properties are expected for all $\mbox{MX}_2$ monolayers because of their similar atomistic structure. In consequence, TMD monolayers are ideal materials to engineer their electronic and optical properties by means of mechanical deformations.

\begin{figure}[t]
\centering
\includegraphics[width=0.8\linewidth]{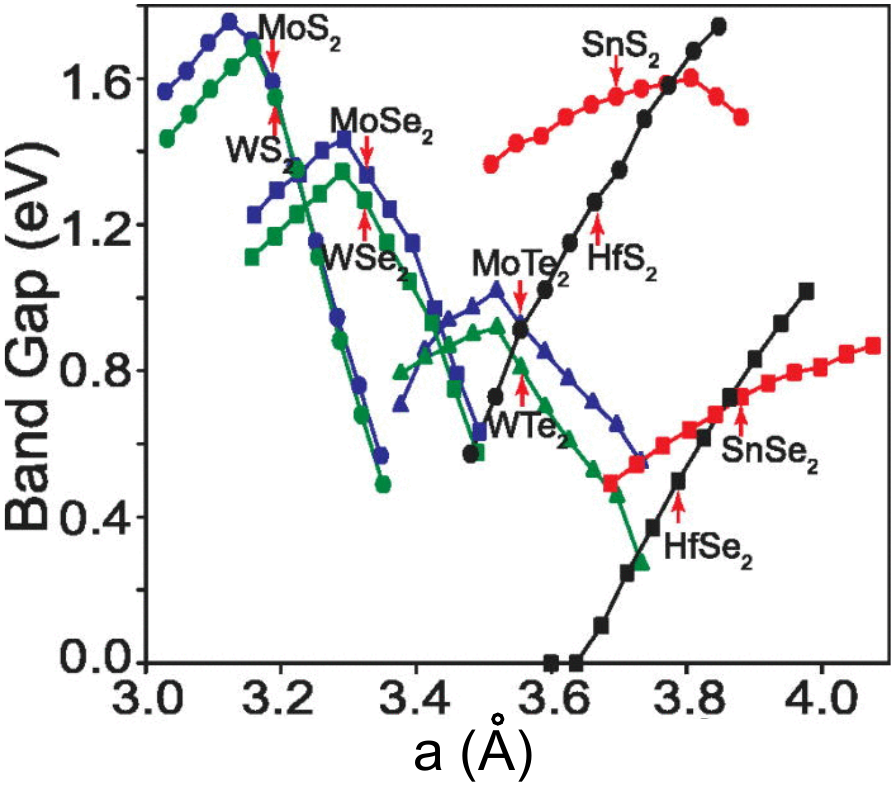}
\caption {Theoretical band gap for different MX$_2$ monolayers as function of the lattice constant $a$ modified by uniform isotropic strain ranging from $-5\,\%$ to $5\,\%$. Red arrow indicates the equilibrium lattice constant for each monolayer. Reprinted with permission from \cite{Guzman2014}. Copyright 2014 AIP Publishing LLC.}
\label{TMDC-2}
\end{figure}

Effects of uniform strain on the electronic and phononic band structures of TMD monolayers have been mostly studied by DFT calculations \cite{Johari2012,Guzman2014,Ghorbani2013,Scalise2014,Zibouche2014,Maniadaki2016,Zhao2017}.
Figure~\ref{TMDC-2} illustrates the theoretical evolution of the band gap for several TMD monolayers under uniform isotropic strain in the range of $-5\,\%$ to $5\,\%$. Whereas 1T structures (HfS$_2$ and HfSe$_2$) show a increase of the band gap under isotropic strain, 2H structures (group-VI TMDs) present a linear reduction of the band gap when the lattice parameter is increased. Eventually the gap disappears for $11\,\%$ of uniform isotropic strain leading to a semiconductor-metal transition \cite{Ghorbani2013}. The results are similar for uniaxial strain, but with a smaller modulation of the band gap \cite{Maniadaki2016}.

From a experimental side, works have been mostly limited to strained MoS$_2$ monolayers. Pioneer experiments  demonstrated that the band gap of MoS$_{2}$ monolayer can be tuned by strain by means of photoluminescence spectroscopy, with a redshift at a rate of $\approx70\,\mbox{ meV}/\%$ strain \cite{He2013,Conley2013}. This scenario has been confirmed in experiments that demonstrate a continuous and reversible tuning of the optical band gap by as much as $500\,\mbox{meV}$ under large biaxial strain \cite{Lloyd2016}.

\begin{figure}[t]
\centering
\includegraphics[width=0.9\linewidth]{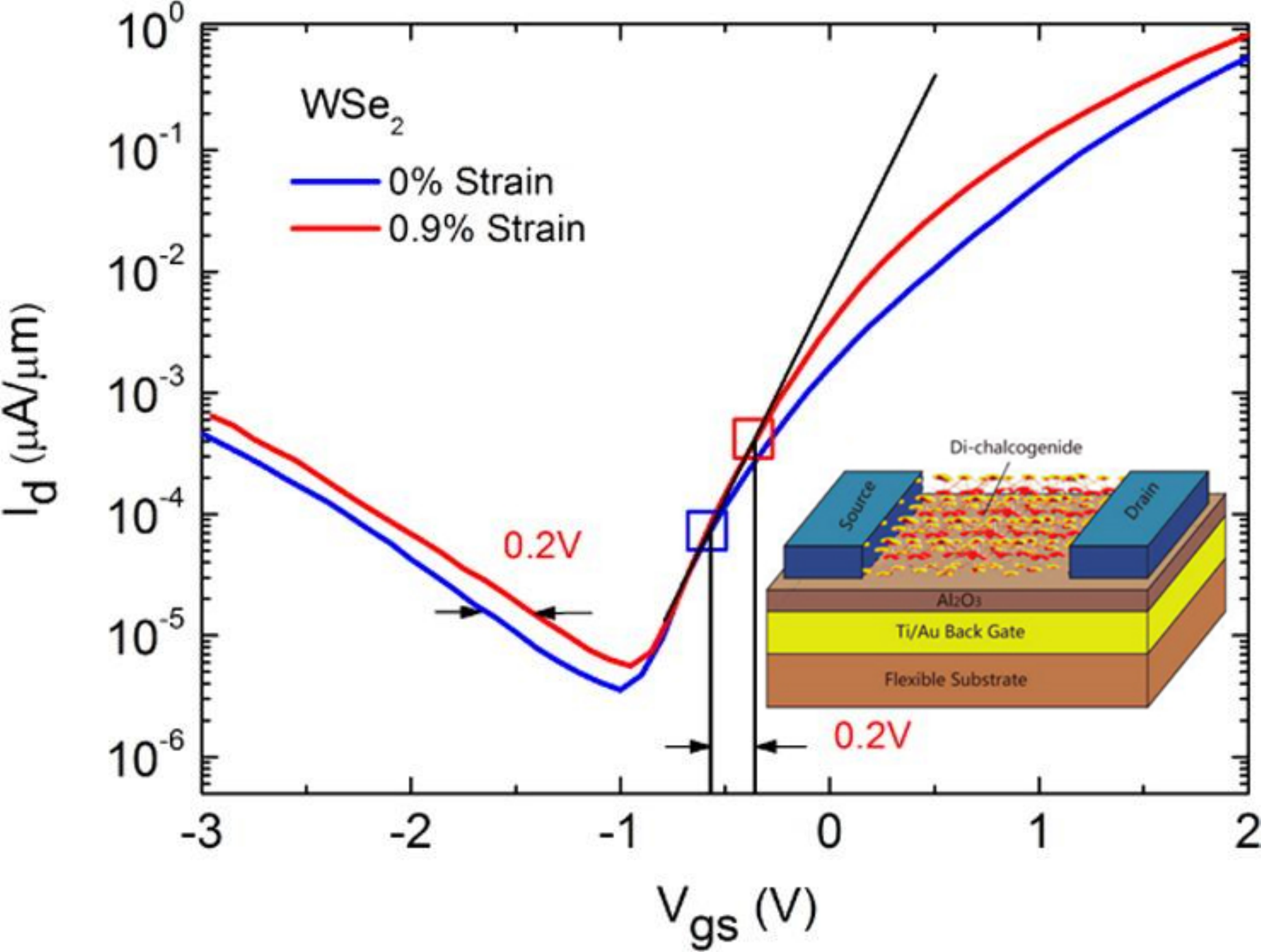}
\caption {Transport characteristics of a WSe$_2$ FET on a flexible polyimide substrate. $V_{gs}$ is the gate voltage while $I_{d}$ is the current. The measurement
were made at room temperature for $0\,\%$ (blue curve) and $0.9\,\%$ (red curve) of tensile strain.
A higher flat band voltage of $0.2\,\mbox{V}$ is observed under strain as indicated by the arrows.
Reprinted with permission from \cite{Shen2016}. Copyright 2016 American Chemical Society.}
\label{Straintransistor}
\end{figure}

Strained MoS$_2$ and WSe$_2$ field-effect transistors (FETs) have been fabricated on a $500\,\mu\mbox{m}$ flexible polyimide substrate \cite{Shen2016}. In figure~\ref{Straintransistor}, their transport characteristics are reported \cite{Shen2016}.
The conclusion is that strain helps in tuning the band gap, just as curvature helped tune phosphorene's band gap \cite{Mehrshad2015}. Figure~\ref{Straintransistor}  \cite{Shen2016} shows a substantial band gap reduction of 100 meV in WSe$_2$ under a modest uniaxial tensile strain smaller than $1\,\%$.

\begin{figure}[t]
\centering
\includegraphics[width=0.83\linewidth]{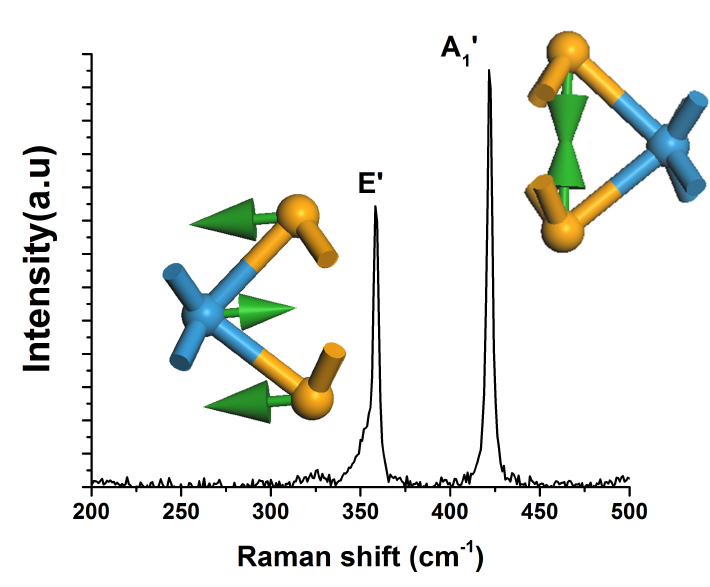}
\caption {Raman spectrum of three layers of WS$_2$ with the excitation wavelength of 488 nm. Atomic models of the in-plane mode $E'$ and out of plane mode $A'_1$ are shown.}
\label{TMD-Humberto_2}
\end{figure}

As in graphene, Raman spectroscopy plays an important role in the characterization of TMD monolayers. There are three main Raman active modes in monolayer TMDs: two in-plane modes called $E'$ and $E''$, and one out-of-plane mode labeled $A'_1$. The $E''$ mode cannot be observed in a backscattering geometry; this is why there are just two main first-order peaks in the Raman spectra shown in figure~\ref{TMD-Humberto_2} \cite{Gutierrez2013,Lin2016}.

The point group associated to TMD monolayers has a $D3h$ symmetry and bilayers have a $D3d$ symmetry. Generalizing, for an odd number of layers the point group will be $D3h$ (no inversion symmetry) and for an even number of layers will be $D3d$ (inversion symmetry). In the case of bulk the point group is $D6h$ (inversion symmetry). It is important to bear this in mind, since Raman spectroscopy uses the character tables' irreducible  representations for its notation. This is why Raman active modes are $A_{1g}$ (out of plane) and $E_{2g}$ (in-plane) in bulk samples.  Raman spectroscopy  can be used to differentiate monolayers from multilayers in a non destructive way \cite{Berkdemir2013,Terrones2014}.

\begin{figure}[t]
\centering
\includegraphics[width=0.85\linewidth]{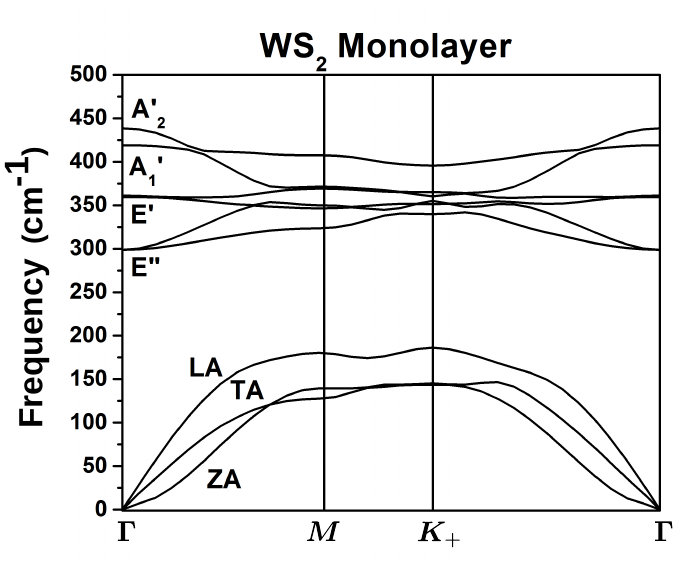}
\caption {DFT-GGA calculated phonon dispersion of monolayer WS$_2$.}
\label{TMD-Humberto_3}
\end{figure}

In figure~\ref{TMD-Humberto_3}, the DFT-LDA phonon dispersion of WS$_2$ shows the $E'$ and $E''$ and $A'_1$ branches. Note that the observed Raman frequencies correspond to the values of the branches $A'_1$ and  $E'$ at the $\Gamma$ point (see figure~\ref{TMD-Humberto_2}).

 Conley and colleagues \cite{Conley2013} have studied MoS$_2$ monolayers and bilayers under uniaxial tensile strains from 0 to $2.2\,\%$ finding that the $E'$ in plane mode splits into two modes and the out of plane mode $A'_1$ does not change. This behavior is similar to the one observed in graphene under uniaxial or non-isotropic biaxial strain. In this context, Rice and colleagues found that uniaxial tensile strain of the order of $0.7\,\%$ causes a small shift of the out of plane mode and a bigger shift of the in-plane mode on monolayers and few-layers of MoS$_2$ \cite{Rice2013}.

As indicated before, Raman spectroscopy is very sensitive to strain and defects, and the quality of TMDs can be determined by this technique.

\subsection{Monochalcogenide monolayers}

This review concludes by highlighting in a dramatic way, the interplay between structure, strain and temperature on two-dimensional materials that possess structural degeneracies. This discussion opens this Review towards new research avenues and new materials in which strain continues to be important, now playing a combined role along with thermal and optical excitations.

\begin{figure}[tb]
\includegraphics[width=\linewidth]{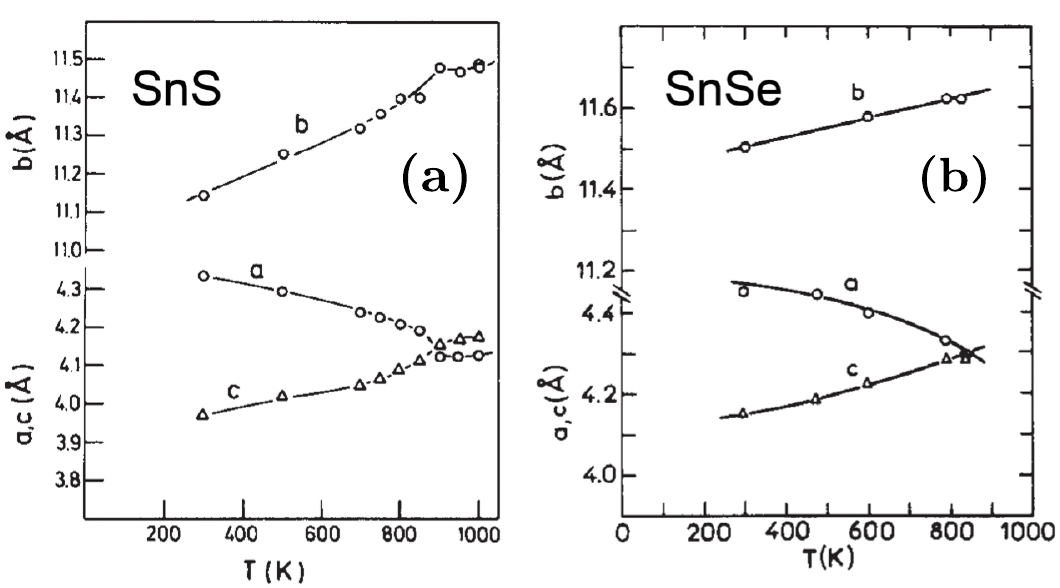}
\caption{The structural phase transition in bulk, layered SnS and SnSe is signaled by the coalescence of in-plane lattice parameters, labeled $a$ and $c$ on this figure. Adapted from \cite{Chattopadhyay1986} with permission.}
\label{fig:bulktransition}
\end{figure}

Group-IV monochalcogenides contain one element from column-IV (Si, Ge, Sn, Pb) and a chalcogen heavier than Oxygen (S, Se, Te). To facilitate the discussion, an average atomic number $\bar{Z}$ is defined as follows \cite{Mehboudi2016}:
\begin{equation}
\bar{Z}=\frac{1}{4}\sum_{i=1}^4Z_i,
\end{equation}
where the sum is over the four atomic elements on a unit cell, each having atomic number $Z_i$ ($i=1,2,3,4$). Materials such as PbSe and PbTe, for which $\bar{Z}>50$ realize a rocksalt bulk structure. They are topological crystalline insulators \cite{Fu2015} and will not be discussed here.

On the other hand, group-IV monochalcogenides with  $\bar{Z}<50$, such as SnSe ($\bar{Z}=42$), SnS ($\bar{Z}=33$), GeSe ($\bar{Z}=33$) and GeS ($\bar{Z}=24$), are known to realize a layered, black-phosphorus-like phase \cite{Lefebvre1998} (see Section \ref{bp} and figure \ref{fig:PNASF1}(a)). The important point for the ensuing discussion is that these layered monochalcogenides undergo strain-like, two-dimensional structural phase transitions at finite temperature and prior to melting, where atomistic coordination turns from three- to five-fold.

The signature of these structural transitions in bulk group-IV monochalcogenides is provided in figure \ref{fig:bulktransition} \cite{Chattopadhyay1986}, where a coalescence of in-plane lattice parameters (labeled $a$ and $c$, where $a\simeq c$) was shown to occur in bulk layered SnS and SnSe. The structural transition turns a $Pnma$ structure onto a $Cmcm$ one, and the temperature-driven evolution of lattice parameters has an effect that is akin to strain, which makes the present discussion appropriate and relevant within the context of the present review.

\begin{figure}[tb]
\includegraphics[width=\linewidth]{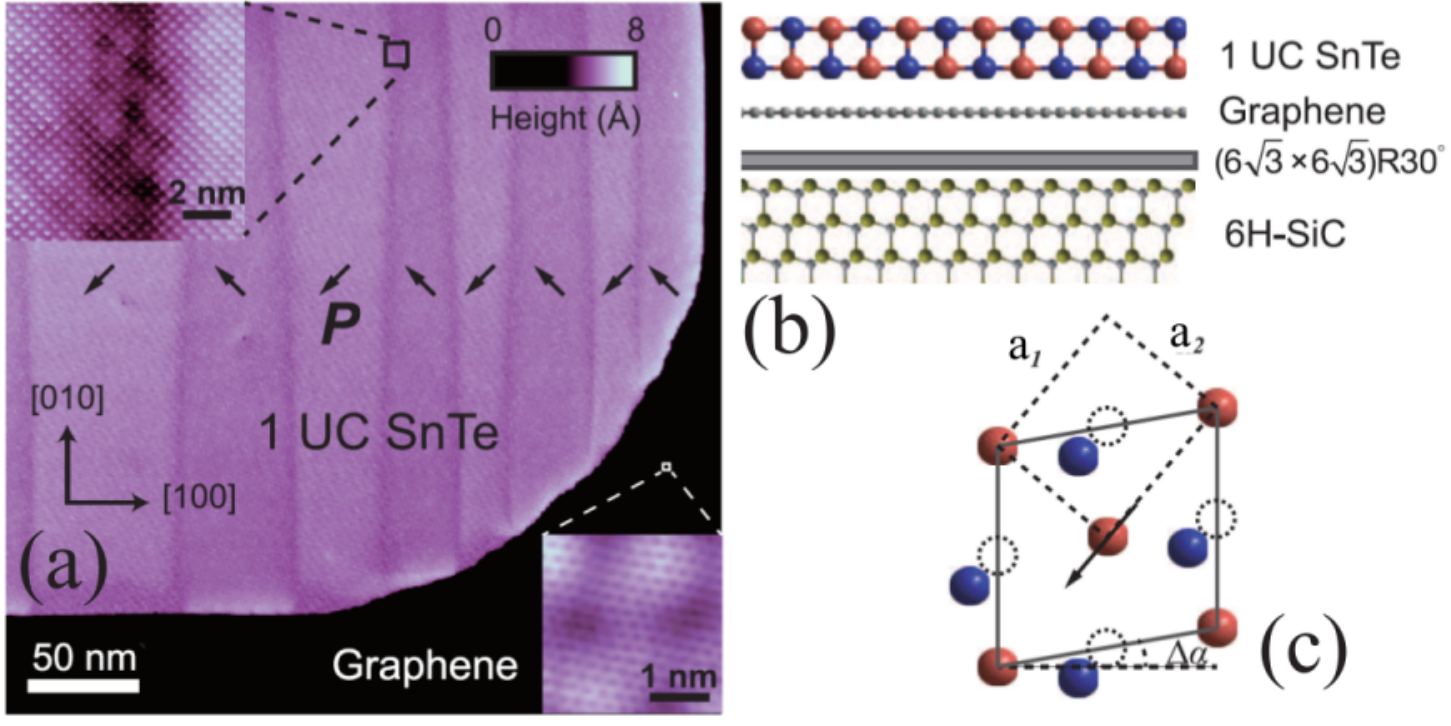}
\caption{(a) Experimental domains in ultrathin SnTe. (b) SnTe and the surface it is grown on. (c) A rhombus, whose diagonals are related to in-plane lattice parameters $a_1$ and $a_2$ and to the angle $\Delta \alpha$. Reproduced with permission from \cite{KaiSCIENCE}.}
\label{Kai}
\end{figure}

This structural phase transition was shown to lead to an unexpectedly large thermal figure of merit $ZT$ in bulk samples \cite{Zhao2014}, which may result in novel thermoelectric applications based on bulk SnSe. The nature of the structural transition was attributed to electronic (orbital) modes \cite{Li2015}.

Turning to the discussion of monochalcogenide monolayers, figure \ref{Kai}(a) displays an experimental domain structure in ultrathin SnTe. In turn, figure \ref{Kai}(b) shows structural details schematically, and figure \ref{Kai}(c) displays a rhombus, whose diagonals are twice the in-plane orthogonal lattice parameters $a_1$ and $a_2$ similar to those seen in figure \ref{fig:PNASF1}(a).

The following equation:
\begin{equation}\label{deltaalpha}
a_1/a_2=(1+\sin(\Delta \alpha))/\cos(\Delta \alpha),
\end{equation}
relates the magnitude of the angle $\Delta\alpha$ shown in figure \ref{Kai}(c) to the ratio of in-plane lattice parameters.

\begin{figure}[t]
\centering
\includegraphics[width=0.9\linewidth]{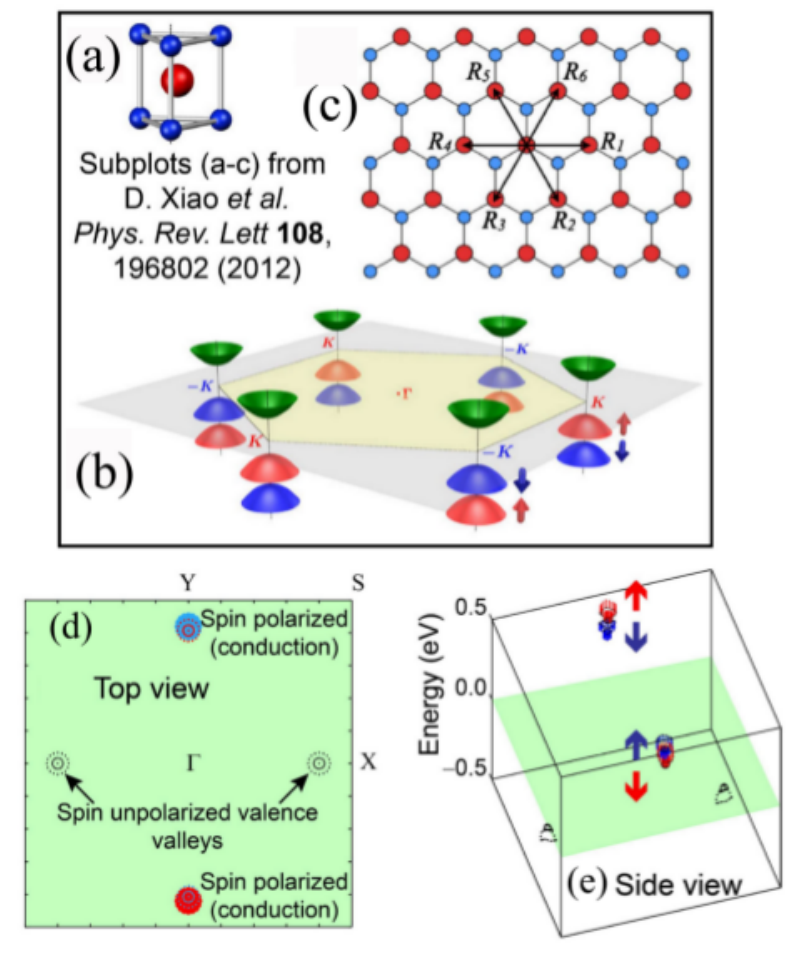}
\caption{Contrasting the spin polarization, number of valleys, and their placement on the first Brillouin zone for MoS2 and SnSe monolayers: (a-c) The MoS$_2$ monolayer displays spin splittings at the valence-band valley, and the conduction and valence band valleys lie at the same location in reciprocal space. (d-e) Spin-polarized conduction-band valley and valence band valleys for SnSe: the electron/valley couplings on subplots (d-e) are unique to IV-VI monolayers and can originate a new platform for valleytronics in 2D. Subplots (a-c) are adapted from \cite{Xiao2012} with permission. Copyrighted by the American Physical Society.}
\label{fig:comparisonmonochalcogenides}
\end{figure}

The salient characteristics of these materials are:
\begin{itemize}
\item{}A rectangular unit cell. As seen in figure \ref{fig:comparisonmonochalcogenides}, this unit cell with a reduced symmetry leads to a unique placement of valleys in the first Brillouin zone \cite{Kaxiras,Hennig0,Hennig1,Gomes1}. Valleys in dichalcogenide monolayers are addressable with circularly-polarized light, but the valleys in monochalcogenide monolayers can be addressed with linearly-polarized light \cite{Rodin2016}. (The extraordinary effect of lower structural symmetry on valley placement is touched upon in Ref~\cite{Rivero2014} already.)
\item{}An inherent piezoectricity, given their diatomic chemical composition, and the placement of such atoms within the unit cell \cite{Carvalho,LiYang,Zhu2015}.
\end{itemize}

\begin{figure}[t]
\centering
\includegraphics[width=\linewidth]{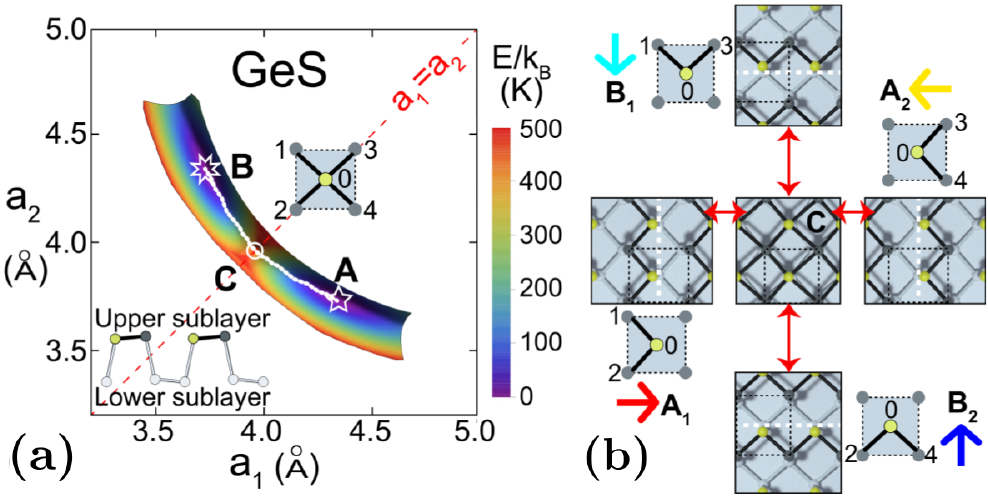}
\caption {(a) The elastic energy landscape $E(a_1,a_2)$ as a function of lattice parameters $a_1$ and $a_2$ as exemplified on a GeS monolayer. A dashed white curve joins points $A$ and $B$ at two degenerate minima ($E_A=E_B=0$). The circle labeled $C$ at $(4.0\,\mbox{\AA},4.0\,\mbox{\AA})$ is a saddle point in which atom 0 forms bonds to four in-plane neighbors, and the elastic energy barrier is defined by $E_C$. (b) Atomistic decorations increase the structural degeneracy at points $A$ and $B$. The four degenerate ground states are named $A_1$, $A_2$, $B_1$ and $B_2$, and assigned in-plane arrows that label them uniquely. Reproduced from \cite{Mehboudi2016} with permission. Copyright (2016) American Chemical Society.}
\label{fig:NL1}
\end{figure}

\begin{figure}[tb]
\centering
\includegraphics[width=\linewidth]{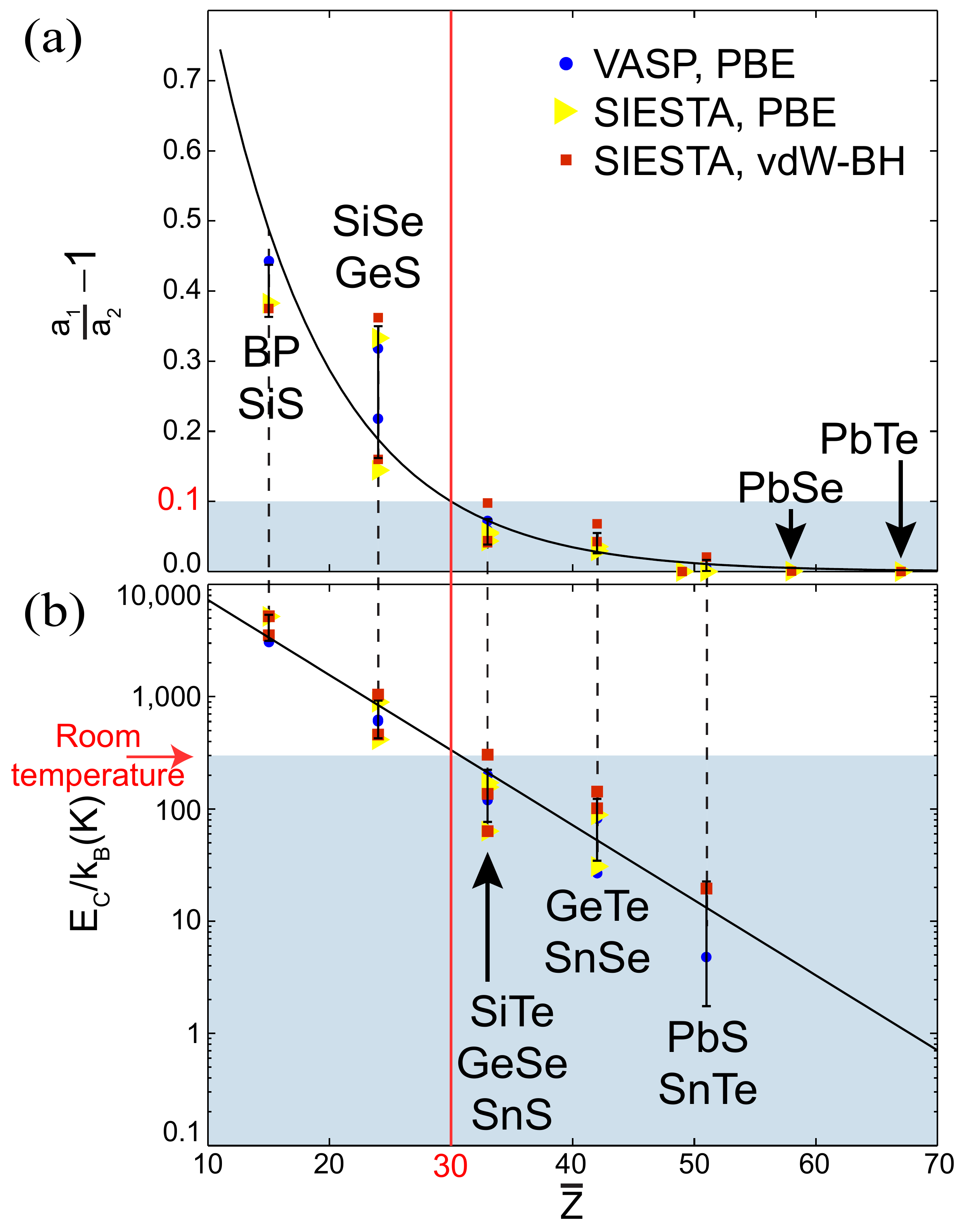}
\caption{(a) The ratio $a_1/a_2$ among orthogonal in-plane lattice constants decreases exponentially with the mean atomic number $\bar{Z}$, and (b) the energy barrier $E_C$ decays exponentially with $\bar{Z}$ as well, as obtained from three different DFT calculations as indicated in the figure. $E_C/k_B<300$ K (and $a_1/a_2\leq 1.1$) for $\mathbf{Z}\geq 30$, so that GeSe, SnS and SnSe monolayers undergo a strain-like 2D phase transition near room temperature. Structures with $a_1\simeq a_2$ display a five-fold-coordinated and non-degenerate ground state with $E_C\simeq 0$. Solid lines are exponential fits.  Reproduced from \cite{Mehboudi2016} with permission. Copyright (2016) American Chemical Society.}
\label{fig:NL2}
\end{figure}

Just as it is the case for silicene (figure \ref{landscape}), 2D materials with rectangular unit cells such as black phosphorene and monochalcogenide monolayers have a natural structural degeneracy, with a high energy barrier $E_C$ (i.e. the energy needed to switch among degenerate structural ground states) for black phosphorene, and a relatively low $E_C$ for monochalcogenide monolayers (figure \ref{fig:NL2}). In the latter case, and as seen in figure \ref{fig:NL1}(a), the degeneracy occurs on monochalcogenide monolayers upon exchange of lattice vectors. Additional degeneracies occur upon reflection of the basis vectors, figure \ref{fig:NL1}(b), yielding a four-fold degenerate structural ground state \cite{Mao_2015}. The energy landscape originally presented in \cite{Mehboudi2016} has been verified in \cite{Wang_2016}, and the fourfold degenerate structural ground states have also been confirmed \cite{Wang_2016,Wu_2016}.

The values of $a_1/a_2$ and $E_C$ displayed in figure \ref{fig:NL2} show an exponential decay with mean atomic number $\bar{Z}$. Light compounds such as black phosphorus monolayer or SiS monolayers ($\bar{Z}=15$) have the largest values of $a_1/a_2$ and $E_C$ and melt directly, without an intermediate two-dimensional structural phase transition.

On the other hand, ultrathin Pb-based monochalcogenides ($\bar{Z}>48$) have a rock-salt structure so that $a_1/a_2=1$ and $E_C=0$ \cite{Fu2015}. All remaining monochalcogenide monolayers (MMs) have values of $a_1/a_2$ and $E_C$ lying somewhere in between, which implies their possibility of displaying two-dimensional phase transitions.

\begin{figure*}[t]
\centering
\includegraphics[width=\textwidth]{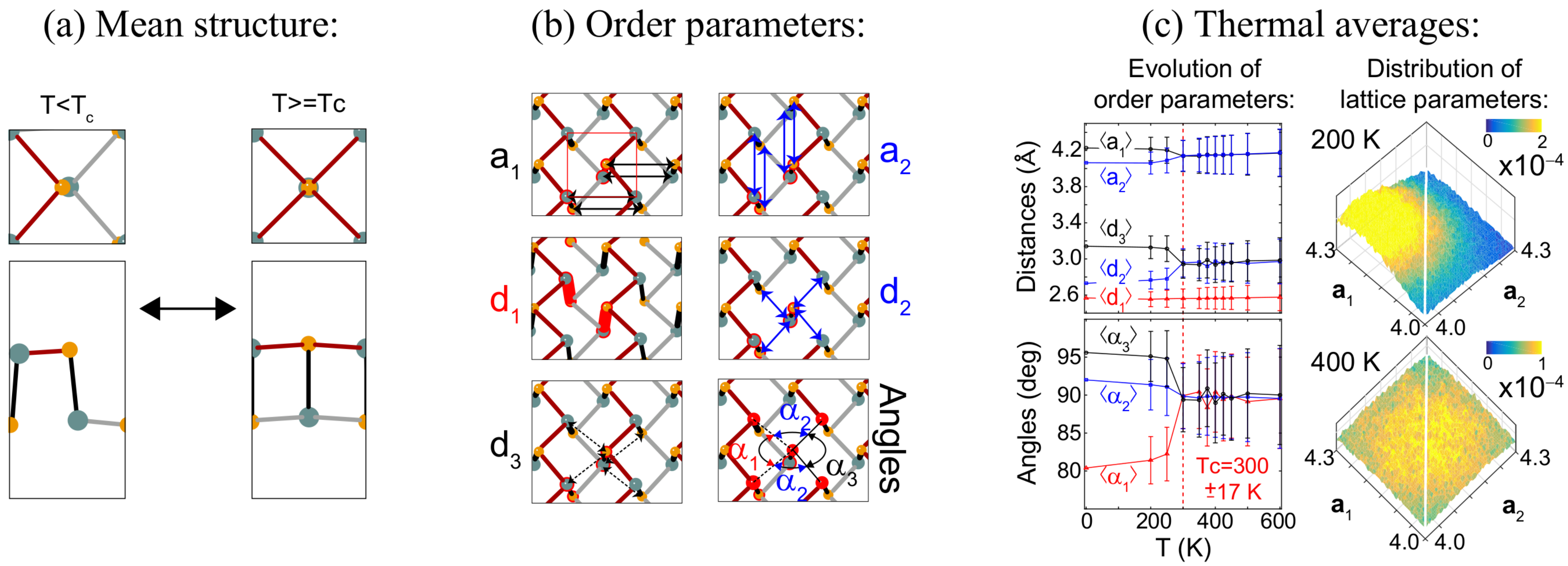}
\caption {(a) Depiction of the structural transition. (b) Order parameters highlighting the transition.
(c) Left: thermal averages for order parameters shown in (b) as a function of $T$ for a GeSe ML. $T_c$ is reached when average values agree ($\langle a_1\rangle= \langle a_2\rangle$, $\langle d_2\rangle =\langle d_3\rangle$, and $\langle\alpha_1\rangle=\langle\alpha_3\rangle$). Right: the distribution of lattice parameters $a_1$ and $a_2$ leads to the error bars on the subplots on the left. The line $a_1=a_2$ is shown in white. Adapted from \cite{Mehboudi2016b} with permission. Copyrighted by the American Physical Society.}
\label{allops}
\end{figure*}

Indeed, the ferroelectric-to-paraelectric transition has been experimentally demonstrated \cite{KaiSCIENCE} and is signaled by a sudden decay of $\Delta \alpha$ (figure \ref{Kai}(c)) to zero. This two-dimensional structural phase transition has been probed via molecular dynamics (MD) in figure \ref{allops} \cite{Mehboudi2016,Mehboudi2016b}. The distribution of lattice parameters seen in figure \ref{allops}(c-f) attest to the large range of fluctuations in these structurally-degenerate two-dimensional structures that are already known to give rise to floppy and anharmonic phonon modes \cite{Li2015}.

\begin{figure}[t]
\centering
\includegraphics[width=\linewidth]{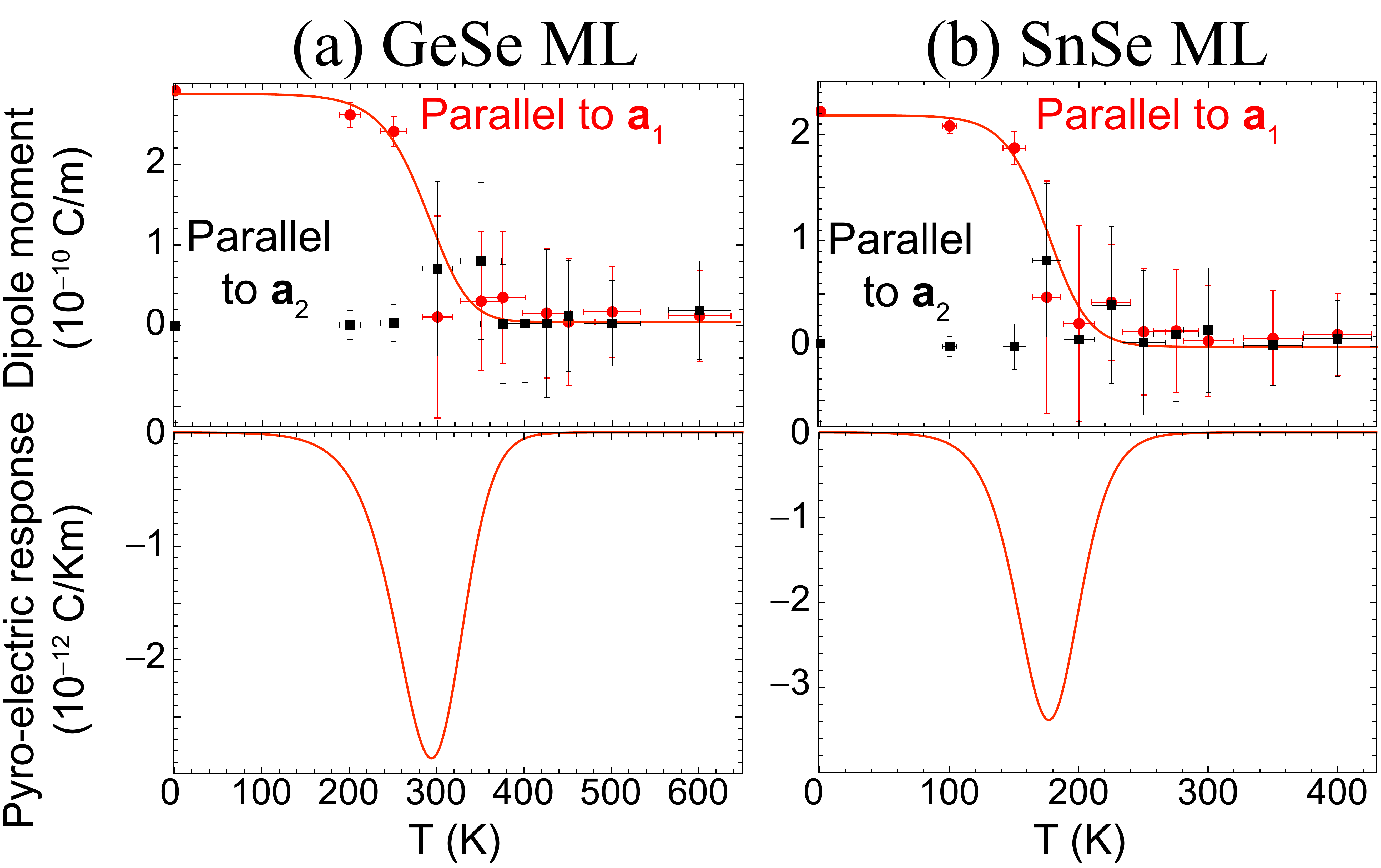}
\caption {The structural transition in monochalcogenide monolayers (MLs) quenches the intrinsic electric dipole and creates a pyroelectric response. Reproduced from \cite{Mehboudi2016b} with permission. Copyrighted by the American Physical Society.}
\label{pyro}
\end{figure}

Equation (\ref{deltaalpha}) indicates that $\Delta \alpha=0$ for $a_1/a_2=1$, so that the experimental observation is captured in calculations shown in figure \ref{allops}(c), and in figure \ref{pyro} concerning the quenching of the dipole moment. Figure \ref{allops} details the nature of the 2D transition, the structural parameters that were tracked as a function of temperature, and the collapse of all order parameters for $T>T_c$, which implies a sudden change of lattice constants akin to strain past the transition temperature.

Unlike the process followed in \cite{Mehboudi2016b} in which lattice constants are let to evolve with temperature, the lattice constants (and hence $\Delta \alpha$) remain fixed throughout the ferroelectric to paraelectric transition discussed in \cite{Copy2016,Rangel2016} (i.e. $a_1/a_2$ and $\Delta \alpha$ are temperature-independent in those works), which implies that these models do not describe experimental conditions. An immediate consequence of fixing lattice parameters at finite temperature is the overestimation of the transition temperature or, within the context of this review, the estimation of the transition temperature on clamped (read strained) structures.

As discussed in \cite{Mehboudi2016b}, some consequences of the structural transition are as follows: no in-gap states develop as the structural transition takes place, so that these phase-change materials remain semiconducting below and above $T_c$. Nevertheless, as the in-plane lattice transforms from a rectangle onto a square at $T_c$ (for a sudden structural change akin to strain), the electronic, spin, optical, and piezo-electric properties dramatically depart from earlier predictions.

Indeed, the $Y-$ and $X-$points in the Brillouin zone become effectively equivalent at $T_c$, leading to a symmetric electronic structure. The spin polarization at the conduction valley edge vanishes, and the hole conductivity must display an anomalous thermal increase at $T_c$. The linear optical absorption band edge must change its polarization as well, making this structural and electronic evolution verifiable optically.  A pyroelectric response of about $3\times 10^{-12}$ $C/K m$ was also predicted for monolayers of GeSe and SnSe (due to their inherent AB-stacking, a net dipole moment only exists in structures with an odd number of layers), and the quenching of the dipole moment was latter confirmed in \cite{Copy2016}. The results in \cite{Mehboudi2016b} thus uncover the fundamental role of temperature as a control knob for the physical properties of few-layer group-IV monochalcogenides, and conclude the review of this novel material family.

It is relevant for the present discussion to note that the ratio of in-plane lattice constants of monochalcogenides increases with the number of layers \cite{Gomes1}. This raises the transition temperatures in going from monolayers to the bulk, which means that $E_c$ is tunable by the number of layers too.

This review concerns the interplay among electronic, optical, and mechanical properties of 2D materials, making it relevant to address a new optomechanical coupling that is predicted to occur in monochalcogenide monolayers that underscores the vibrancy of this field. In brief, linearly polarized light can address individual valleys and create a population of excited carriers that screen the underlying dipole moment characteristic of these materials. In turn, the structure compresses anisotropically \cite{newest}. This finding means that illumination can play an active role in creating non-uniform strain in this barely explored material, showing promising ways ahead concerning the interrelation among strain and material properties in 2D materials.

\section{Conclusions and outlook}

The effect of mechanical (geometrical) deformations in graphene and other 2D materials remains an active field of research, and the possibility of  tailoring very strong magnetic fields via pseudomagnetic potentials produced by substrates is opening new research avenues. Strained/bent 2D materials are a playground for interesting
physical systems, like Luttinger liquids, topological phases, broken chiral symmetry phases due to Kekul\'e ordering, 2D structural phase transitions, and so on. Topological phases, spintronics, direct band gap semiconductors or band gap tunability are among the most important areas of actual research.

Although research in graphene is mature, open questions remain, like the kind of perturbations needed to treat
graphene over substrates, or even more basic questions like the mismatch between energetic and geometry descriptions observed in
uniformly-deformed graphene. Even the existence of gaps in graphene strained using different ways is a highly debated issue
and more work is needed to elucidate this question.

Furthermore, the treatment of
pseudomagnetic potentials needs to be formalized using techniques borrowed from the electromagnetic case using concepts like magnetic groups
and lattices, and can also employ tools like discrete geometry.  Multilayered systems still represent a tantalizing promise to
design at will materials by stacking one-atom thickness surfaces.

Deformations and structural phase transitions in other 2D materials are still in the process of being explored where, there are many new interesting questions and intriguing research directions that deserve attention as well.

\ack

The authors thank Juan Carlos Obeso, Pedro E. Roman-Taboada, Jos\'e Eduardo Barrios-Vargas, Raad E. Haleoot, Ryu Nakae and Manuel Mon for support in revising the manuscript, help with the figures and discussion of the subject. They also thank V. Pereira, S. Roche, A. Kitt, A. Pacheco-Sanjuan, M.A.H. Vozmediano, F. de Juan, M.A Zubkov, E. Harriss, P. Kumar, L. Bellaiche, B. Klee, G.E. Volovik and G. Montambaux for enlightening discussions during several phases of our research in this field. This work was partially supported by DGAPA-UNAM project IN102717 (G.G.N.), by an Early Career Grant from the DOE to S.B.L. (DE-SC0016139), by NSF XSEDE (TG-PHY 090002) and by the Arkansas High Performance Computing Center which is funded through multiple NSF grants and the Arkansas Economic Development Commission. M.O.L. acknowledges a postdoctoral fellowship from DGAPA-UNAM. H.T. acknowledges NSF (EFRI 1433311).

\section*{References}

\begin{thebibliography}{100}
\expandafter\ifx\csname url\endcsname\relax
  \def\url#1{{\tt #1}}\fi
\expandafter\ifx\csname urlprefix\endcsname\relax\def\urlprefix{URL }\fi
\providecommand{\eprint}[2][]{\url{#2}}

\bibitem{Novoselov2004}
Novoselov K~S, Geim A~K, Morozov S~V, Jiang D, Zhang Y, Dubonos S~V, Grigorieva I~V and Firsov A~A 2004 Electric field effect in atomically thin carbon films
  \textit{ Science\/} {\bf 306} 666--669

\bibitem{Novoselov2005}
Novoselov K~S, Jiang D, Schedin F, Booth T~J, Khotkevich V~V, Morozov S~V and
  Geim A~K 2005 Two-dimensional gas of massless Dirac fermions in graphene
  \textit{ Proc. Natl. Acad. Sci. USA\/} {\bf 102} 10451--10453

\bibitem{Roddaro2007}
Roddaro S, Pingue P, Piazza V, Pellegrini V and Beltram F 2007 {The optical
  visibility of graphene: interference colors of ultrathin graphite on SiO$_2$}
  \textit{ Nano Lett.\/} {\bf 7} 2707--2710

\bibitem{Balandin2008}
Balandin A~A, Ghosh S, Bao W, Calizo I, Teweldebrhan D, Miao F and Lau C~N 2008
  Superior thermal conductivity of single-layer graphene \textit{ Nano Lett.\/}
  {\bf 8} 902--907

\bibitem{Lee2008}
Lee C, Wei X, Kysar J~W and Hone J 2008 {Measurement of the elastic properties
  and intrinsic strength of monolayer graphene} \textit{ Science\/} {\bf 321}
  385--388

\bibitem{Vozmediano2010}
Vozmediano M, Katsnelson M and Guinea F 2010 {Gauge fields in graphene}
  \textit{ Physics Reports\/} {\bf 496} 109--148

\bibitem{Novoselov2012}
{Novoselov K S}, {Falko V I}, {Colombo L}, {Gellert P R}, {Schwab M G} and {Kim
  K} 2012 {A roadmap for graphene} \textit{ Nature\/} {\bf 490} 192--200

\bibitem{Bissett2014}
Bissett M~A, Tsuji M and Ago H 2014 {Strain engineering the properties of
  graphene and other two-dimensional crystals} \textit{ Phys. Chem. Chem.
  Phys.\/} {\bf 16} 11124--11138

\bibitem{Roldan2015}
Rold{\'a}n R, Castellanos-Gomez A, Cappelluti E and Guinea F 2015 {Strain
  engineering in semiconducting two-dimensional crystals} \textit{ J. of Phys.:
  Condens. Matter\/} {\bf 27} 313201

\bibitem{BarrazaLopez2015}
Barraza-Lopez S 2015 {Discrete differential geometry and the properties of
  conformal two-dimensional materials} \textit{ Synth. Met.\/} {\bf 210} 32--41

\bibitem{Galiotis2015}
Galiotis C, Frank O, Koukaras E~N and Sfyris D 2015 {Graphene mechanics:
  current status and perspectives} \textit{ Annu. Rev. Chem. Biomol. Eng.\/}
  {\bf 6} 121--140

\bibitem{Jing2015}
Jiang J~W, Wang B~S, Wang J~S and Park H~S 2015 {A review on the flexural mode
  of graphene: lattice dynamics, thermal conduction, thermal expansion,
  elasticity and nanomechanical resonance} \textit{ J. Phys.: Condens.
  Matter\/} {\bf 27} 083001

\bibitem{Amorim2016}
Amorim B, Cortijo A, de~Juan F, Grushin A, Guinea F, Guti{\'e}rrez-Rubio A,
  Ochoa H, Parente V, Rold{\'a}n R, San-Jose P, Schiefele J, Sturla M and
  Vozmediano M 2016 {Novel effects of strains in graphene and other two
  dimensional materials} \textit{ Phys. Rep.\/} {\bf 617} 1--54

\bibitem{Deng2016}
Deng S and Berry V 2016 {Wrinkled, rippled and crumpled graphene: an overview
  of formation mechanism, electronic properties, and applications} \textit{
  Mater. Today\/} {\bf 19} 197--212

\bibitem{Meunier2016}
Meunier V, {Souza Filho} A~G, Barros E~B and Dresselhaus M~S 2016 {Physical
  properties of low-dimensional ${sp}^{2}$-based carbon nanostructures}
  \textit{ Rev. Mod. Phys.\/} {\bf 88} 025005

\bibitem{Pereira2009a}
Pereira V~M, {Castro Neto} A~H and Peres N~M~R 2009 Tight-binding approach to
  uniaxial strain in graphene \textit{ Phys. Rev. B\/} {\bf 80} 045401

\bibitem{Pletikosi2009}
{Pletikosi\ifmmode \acute{c}\else {\'c}\fi{}} I, Kralj M, Pervan P, Brako R,
  Coraux J, N'Diaye A~T, Busse C and Michely T 2009 {Dirac cones and minigaps
  for graphene on Ir(111)} \textit{ Phys. Rev. Lett.\/} {\bf 102} 056808

\bibitem{Yankowitz2012}
Yankowitz M, Xue J, Cormode D, Sanchez-Yamagishi J~D, Watanabe K, Taniguchi T,
  Jarillo-Herrero P, Jacquod P and LeRoy B~J 2012 {Emergence of superlattice
  Dirac points in graphene on hexagonal boron nitride} \textit{ Nat. Phys.\/}
  {\bf 8} 382--386

\bibitem{Ortix2012}
Ortix C, Yang L and van~den Brink J 2012 {Graphene on incommensurate
  substrates: Trigonal warping and emerging Dirac cone replicas with halved
  group velocity} \textit{ Phys. Rev. B\/} {\bf 86} 081405

\bibitem{Ponomarenko2013}
Ponomarenko L A \textit{et al} 2013 {Cloning of Dirac fermions
  in graphene superlattices} \textit{ Nature\/} {\bf 497} 594--597

\bibitem{Dean2013}
Dean C R  \textit{et al} 2013 {Hofstadter's butterfly and the fractal quantum Hall effect in moir{\'e} superlattices} \textit{ Nature\/} {\bf 497} 598--602

\bibitem{Hofstadter1976}
Hofstadter D~R 1976 {Energy levels and wave functions of Bloch electrons in rational and irrational magnetic fields} \textit{ Phys. Rev. B\/} {\bf 14} 2239--2249

\bibitem{Zheng2013}
Liu Z \textit{et al} 2013 {In-plane
  heterostructures of graphene and hexagonal boron nitride with controlled domain sizes} \textit{ Nat. Nanotechnol.\/} {\bf 8} 119--124

\bibitem{NaumisTerrones2009}
Naumis G~G, Terrones M, Terrones H and Gaggero-Sager L~M 2009 {Design of
  graphene electronic devices using nanoribbons of different widths} \textit{
  Appl. Phys. Lett.\/} {\bf 95} 182104

\bibitem{Diniz2016}
Diniz G, Vernek E and Souza F 2017 Graphene-based spin switch device via
  modulated Rashba field and strain \textit{ Physica E Low. Dimens. Syst.
  Nanostruct.\/} {\bf 85} 264--270

\bibitem{Ong2012}
Ong M~T and Reed E~J 2012 {Engineered piezoelectricity in graphene} \textit{
  ACS Nano\/} {\bf 6} 1387--1394

\bibitem{Levy2010}
Levy N, Burke S~A, Meaker K~L, Panlasigui M, Zettl A, Guinea F, {Castro Neto}
  A~H and Crommie M~F {2010} {Strain-induced pseudo-magnetic fields greater
  than 300 Tesla in graphene nanobubbles} \textit{ {Science}\/} {\bf {329}}
  {544--547}

\bibitem{Xiaomu2015}
Xiaomu W, He T, Weiguang X, Yi S, Wen-Tian M, Mohammad A~M, Qian-Yi X, Yi Y,
  Jian-Bin X and Tian-Ling R 2015 {Observation of a giant two-dimensional
  band-piezoelectric effect on biaxial-strained graphene} \textit{ NPG Asia
  Mater.\/} {\bf 7} e154

\bibitem{Carvalho}
Gomes L~C, Carvalho A and Castro~Neto A~H 2015 Enhanced piezoelectricity and
  modified dielectric screening of two-dimensional group-IV monochalcogenides
  \textit{ Phys. Rev. B\/} {\bf 92} 214103

\bibitem{LiYang}
Fei R, Li W, Li J and Yang L 2015 Giant piezoelectricity of monolayer group IV
  monochalcogenides: SnSe, SnS, GeSe, and GeS \textit{ Appl. Phys. Lett.\/}
  {\bf 107} 173104

\bibitem{Hennig}
Blonsky M~N, Zhuang H~L, Singh A~K and Hennig R~G 2015 \textit{Ab initio}
  prediction of piezoelectricity in two-dimensional materials \textit{ ACS
  Nano\/} {\bf 9} 9885--9891

\bibitem{Droth2016}
Droth M, Burkard G and Pereira V~M 2016 Piezoelectricity in planar boron
  nitride via a geometric phase \textit{ Phys. Rev. B\/} {\bf 94} 075404

\bibitem{Thorpe2012}
Kumar A, Wilson M and Thorpe M~F 2012 {Amorphous graphene: a realization of
  Zachariasen{\rq}s glass} \textit{ J. Phys.: Condens. Matter\/} {\bf 24}
  485003

\bibitem{Mehboudi2016}
Mehboudi M, Dorio A~M, Zhu W, van~der Zande A, Churchill H~O~H, Pacheco-Sanjuan
  A~A, Harriss E~O, Kumar P and Barraza-Lopez S 2016 Two-dimensional disorder
  in black phosphorus and monochalcogenide monolayers \textit{ Nano Lett.\/}
  {\bf 16} 1704--1712

\bibitem{Naumis2015Frontiers}
Naumis G~G 2015 {Low-frequency vibrational modes anomalies and rigidity: A key
  to understanding the glass and the electronic properties of flexible
  materials from a Topological Perspective} \textit{ Front. Mater.\/} {\bf 2}
  44

\bibitem{newest}
Haleoot R, Paillard C, Mehboudi M, Xu B, Bellaiche L and Barraza-Lopez S 2017
  {Photostrictive two-dimensional materials in the monochalcogenide family}
  \textit{ Phys. Rev. Lett.\/}  {\bf 118} 227401

\bibitem{Montambaux2009a}
Montambaux G, Pi{\'e}chon F, Fuchs J~N and Goerbig M~O 2009 {Merging of Dirac
  points in a two-dimensional crystal} \textit{ Phys. Rev. B\/} {\bf 80} 153412

\bibitem{Montambaux2009b}
Montambaux G, Pi{\'e}chon F, Fuchs J~N and Goerbig M~O 2009 {A universal
  Hamiltonian for motion and merging of Dirac points in a two-dimensional
  crystal} \textit{ EPJ B\/} {\bf 72} 509--520

\bibitem{Montambaux2012}
de~Gail R, Fuchs J~N, Goerbig M, Pi{\'e}chon F and Montambaux G 2012
  {Manipulation of Dirac points in graphene-like crystals} \textit{ Physica B
  Condens. Matter.\/} {\bf 407} 1948--1952

\bibitem{Montambaux2013}
Bellec M, Kuhl U, Montambaux G and Mortessagne F 2013 {Topological Transition
  of Dirac Points in a Microwave Experiment} \textit{ Phys. Rev. Lett.\/} {\bf
  110} 033902

\bibitem{Roman2015a}
Roman-Taboada P and Naumis G~G 2014 {Spectral butterfly, mixed
  Dirac-Schr{\"o}dinger fermion behavior, and topological states in armchair
  uniaxial strained graphene} \textit{ Phys. Rev. B\/} {\bf 90} 195435

\bibitem{CastroNeto2009}
{Castro Neto} A~H, Guinea F, Peres N~M~R, Novoselov K~S and Geim A~K 2009 {The
  electronic properties of graphene} \textit{ Rev. Mod. Phys.\/} {\bf 81}
  109--162

\bibitem{Naumis1994}
Naumis G~G, Barrio R~A and Wang C 1994 {Effects of frustration and localization
  of states in the Penrose lattice} \textit{ Phys. Rev. B\/} {\bf 50}
  9834--9842

\bibitem{Naumis2002}
Naumis G~G, Wang C and Barrio R~A 2002 {Frustration effects on the electronic
  density of states of a random binary alloy} \textit{ Phys. Rev. B\/} {\bf 65}
  134203

\bibitem{Naumis2007}
Naumis G~G 2007 {Internal mobility edge in doped graphene: Frustration in a
  renormalized lattice} \textit{ Phys. Rev. B\/} {\bf 76} 153403

\bibitem{BarriosVargas2011a}
Barrios-Vargas J~E and Naumis G~G 2011 {Doped graphene: the interplay between
  localization and frustration due to the underlying triangular symmetry}
  \textit{ J. Phys.: Condens. Matter\/} {\bf 23} 375501

\bibitem{BarriosVargas2013}
Barrios-Vargas J~E and Naumis G~G 2013 {Pseudo-gap opening and Dirac point
  confined states in doped graphene} \textit{ Solid State Comm.\/} {\bf 162}
  23--27

\bibitem{Ashcroft}
Ashcroft N and Mermin N 1976 \textit{ Solid State Physics\/} (Fort Worth:
  Saunders College Publishing)

\bibitem{Meyer2007}
Meyer J~C, Geim A~K, Katsnelson M~I, Novoselov K~S, Booth T~J and Roth S 2007
  {The structure of suspended graphene sheets} \textit{ Nature\/} {\bf 446}
  60--63

\bibitem{Malvern}
Malvern L 1969 \textit{ Introduction to the mechanics of a continuous medium\/}
  (Englefford, New Jersey, USA: Prentice Hall)

\bibitem{Lai}
Lai W~M Rubin D and Krempl E 2010 \textit{ Introduction to continuum
  mechanics\/} (Oxford, U.K.: Butterworth-Heinemann)

\bibitem{Landau}
Landau L and Lifshitz E 1976 \textit{ Theory of Elasticity\/} (Oxford, U.K.:
  Pergamon Press)

\bibitem{Kitt2013}
Kitt A~L, Pereira V~M, Swan A~K and Goldberg B~B 2013 {Erratum:
  Lattice-corrected strain-induced vector potentials in graphene [Phys. Rev. B
  \textbf{85} , 115432 (2012)]} \textit{ Phys. Rev. B\/} {\bf 87} 159909

\bibitem{Midtvedt2016}
Midtvedt D, Lewenkopf C~H and Croy A 2016 {Strain--displacement relations for
  strain engineering in single-layer 2D materials} \textit{ 2D Mater.\/} {\bf
  3} 011005

\bibitem{Zhou}
Zhou J and Huang R 2007 Internal lattice relaxation of single-layer graphene
  under in-plane deformation \textit{ J. Mech. Phys. Solids\/} {\bf 56}
  1609--1623

\bibitem{Ericksen2008}
Ericksen J 2008 On the Cauchy—Born rule \textit{ Math. \& Mech. of Solids\/}
  {\bf 13} 199--220

\bibitem{Gomez2016}
Gomez-Arias W~A and Naumis G~G 2016 Analytical calculation of electron group
  velocity surfaces in uniform strained graphene \textit{ Int. J. Mod. Phys.
  B\/} {\bf 30} 1550263

\bibitem{Artaud2016}
{Artaud A}, {Magaud L}, {Le Quang T}, {Guisset V}, {David P}, {Chapelier C} and
  {Coraux J} 2016 {Universal classification of twisted, strained and sheared
  graphene moir{\'e} superlattices} \textit{ Sci. Rep.\/} {\bf 6} 25670

\bibitem{Janot}
Janot C 1995 \textit{ Quasicrystals: A Primer\/} 2nd ed Monographs on the
  Physics and Chemistry of Materials (Oxford University Press, USA)

\bibitem{Naumis1998}
Naumis G~G and Aragon J~L {1998} {The influence of phason disorder on the
  electronic spectrum and eigenstates of Fibonacci lattices} \textit{ {Phys.
  Lett. A}\/} {\bf {244}} {133--138}

\bibitem{NaumisThorpe1999}
Naumis G~G, Wang C~M, Thorpe M~F and Barrio R~A {1999} {Coherency of phason
  dynamics in Fibonacci chains} \textit{ {Phys. Rev. B}\/} {\bf {59}}
  {14302--14312}

\bibitem{NaumisHierarchy2005}
Naumis G~G {2005} {Phason hierarchy and electronic stability of quasicrystals}
  \textit{ {Phys. Rev. B}\/} {\bf {71}}

\bibitem{NaumisAragon2003}
Naumis G~G and Aragon J~L {2003} {Analytic expressions for the vertex
  coordinates of quasiperiodic lattices} \textit{ {Z. Kristallogr.}\/} {\bf
  {218}} {397--402}

\bibitem{Janssen2004}
Janssen T and Radulescu O {2004} {Theory of phasons in aperiodic crystals}
  \textit{ {Ferroelectrics}\/} {\bf {305}} {179--184}

\bibitem{Macia2006}
Maci{\'a} E 2006 {The role of aperiodic order in science and technology}
  \textit{ Rep. Prog. Phys.\/} {\bf 69} 397

\bibitem{Diaye2008}
N'Diaye A~T, Coraux J, Plasa T~N, Busse C and Michely T 2008 {Structure of
  epitaxial graphene on Ir(111)} \textit{ New J. Phys.\/} {\bf 10} 043033

\bibitem{Wang2016}
Wang F, Li Y, Shifa T~A, Liu K, Wang F, Wang Z, Xu P, Wang Q and He J 2016
  {Selenium-enriched nickel selenide nanosheets as a robust electrocatalyst for
  hydrogen generation} \textit{ Angewandte Chemie\/} {\bf 128} 7033--7038

\bibitem{Lubensky1985}
Lubensky T~C, Ramaswamy S and Toner J 1985 Hydrodynamics of icosahedral
  quasicrystals \textit{ Phys. Rev. B\/} {\bf 32} 7444--7452

\bibitem{Colin15}
Daniels C, Horning A, Phillips A, Massote D~V~P, Liang L, Bullard Z, Sumpter
  B~G and Meunier V 2015 {Elastic, plastic, and fracture mechanisms in graphene
  materials} \textit{ J. Phys.: Condens. Matter\/} {\bf 27} 373002

\bibitem{Liu2015}
Liu K and Wu J 2015 {Mechanical properties of two-dimensional materials and
  heterostructures} \textit{ J. Mater. Res.\/} {\bf 31} 832--844

\bibitem{Akinwande2016}
Akinwande D \textit{et al} 2017 A review on mechanics and
  mechanical properties of 2D materials--Graphene and beyond \textit{ Extreme
  Mechanics Letters\/} {\bf 13} 42--77

\bibitem{Bunch2008}
Bunch J~S, Verbridge S~S, Alden J~S, van~der Zande A~M, Parpia J~M, Craighead
  H~G and McEuen P~L 2008 {Impermeable atomic membranes from graphene sheets}
  \textit{ Nano Lett.\/} {\bf 8} 2458--2462

\bibitem{Bao2009}
Bao W, Miao F, Chen Z, Zhang H, Jang W, Dames C and Lau C~N 2009 {Controlled
  ripple texturing of suspended graphene and ultrathin graphite membranes}
  \textit{ Nat. Nanotechnol.\/} {\bf 4} 562--566

\bibitem{Kim2011}
Kim K, Lee Z, Malone B~D, Chan K~T, Alem{\'a}n B, Regan W, Gannett W, Crommie
  M~F, Cohen M~L and Zettl A 2011 {Multiply folded graphene} \textit{ Phys.
  Rev. B\/} {\bf 83} 245433

\bibitem{Bunch2007}
Bunch J~S, van~der Zande A~M, Verbridge S~S, Frank I~W, Tanenbaum D~M, Parpia
  J~M, Craighead H~G and McEuen P~L 2007 {Electromechanical resonators from
  graphene sheets} \textit{ Science\/} {\bf 315} 490--493

\bibitem{Wang2014}
Wang Q and Arash B 2014 {A review on applications of carbon nanotubes and
  graphenes as nano-resonator sensors} \textit{ Comp. Mat. Sci.\/} {\bf 82}
  350--360

\bibitem{Mathew2016}
Mathew J~P, Patel R~N, Borah A, Vijay R and Deshmukh M~M 2016 {Dynamical strong
  coupling and parametric amplification of mechanical modes of graphene drums}
  \textit{ Nat. Nanotechnol.\/} {\bf 11} 747--751

\bibitem{Kim2015}
Kim S~J, Choi K, Lee B, Kim Y and Hong B~H 2015 {Materials for flexible,
  stretchable electronics: graphene and 2D materials} \textit{ Annu. Rev.
  Mater. Res.\/} {\bf 45} 63--84

\bibitem{Jang16}
Jang H, Park Y~J, Chen X, Das T, Kim M~S and Ahn J~H 2016 {Graphene-based
  flexible and stretchable electronics} \textit{ Adv. Mater. (Weinheim,
  Ger.)\/} {\bf 28} 4184--4202

\bibitem{Reddy2006}
Reddy C~D, Rajendran S and Liew K~M 2006 {Equilibrium configuration and
  continuum elastic properties of finite sized graphene} \textit{
  Nanotechnology\/} {\bf 17} 864--870

\bibitem{Scarpa2010}
Scarpa F, Adhikari S, Gil A~J and Remillat C 2010 {The bending of single layer
  graphene sheets: the lattice versus continuum approach} \textit{
  Nanotechnology\/} {\bf 21} 125702

\bibitem{Wei2009}
Wei X, Fragneaud B, Marianetti C~A and Kysar J~W 2009 {Nonlinear elastic
  behavior of graphene: \textit{Ab initio} calculations to continuum
  description} \textit{ Phys. Rev. B\/} {\bf 80} 205407

\bibitem{Colombo2011}
Colombo L and Giordano S 2011 {Nonlinear elasticity in nanostructured
  materials} \textit{ Rep. Prog. Phys.\/} {\bf 74} 116501

\bibitem{Lindahl2012}
Lindahl N, Midtvedt D, Svensson J, Nerushev O~A, Lindvall N, Isacsson A and
  Campbell E~E~B 2012 {Determination of the bending rigidity of graphene via
  electrostatic actuation of buckled membranes} \textit{ Nano Lett.\/} {\bf 12}
  3526--3531

\bibitem{Landau1986}
Landau L~D and Lifshitz E~M 1986 \textit{ Course of Theoretical Physics, Theory
  of Elasticity\/} 3rd ed vol~7 (Pergamon Press Oxford)

\bibitem{Nelson2004}
Nelson D, S W and Piran T 2004 \textit{ Statistical mechanics of membranes and
  surfaces\/} 2nd ed (World Scientific Pub)

\bibitem{Begley2004}
Begley M~R and Mackin T~J 2004 {Spherical indentation of freestanding circular
  thin films in the membrane regime} \textit{ J. Mech. Phys. Solids.\/} {\bf
  52} 2005--2023

\bibitem{Komaragiri2005}
Komaragiri U, Begley M~R and Simmonds J~G 2004 {The mechanical response of
  freestanding circular elastic films under point and pressure loads} \textit{
  J. Appl. Mech.\/} {\bf 72} 203--212

\bibitem{Cadelano2009}
Cadelano E, Palla P~L, Giordano S and Colombo L 2009 {Nonlinear elasticity of
  monolayer graphene} \textit{ Phys. Rev. Lett.\/} {\bf 102} 235502

\bibitem{Huang2011}
Huang M, Pascal T~A, Kim H, Goddard W~A and Greer J~R 2011
  {Electronic-mechanical coupling in graphene from in situ nanoindentation
  experiments and multiscale atomistic simulations} \textit{ Nano Lett.\/} {\bf
  11} 1241--1246

\bibitem{Lee2013}
Lee G~H, Cooper R~C, An S~J, Lee S, van~der Zande A, Petrone N, Hammerberg A~G,
  Lee C, Crawford B, Oliver W, Kysar J~W and Hone J 2013 {High-strength
  chemical-vapor{\textendash}deposited graphene and grain boundaries} \textit{
  Science\/} {\bf 340} 1073--1076

\bibitem{Lopez2015}
Lopez-Polin G, Gomez-Navarro C, Parente V, Guinea F, Katsnelson M~I,
  Perez-Murano F and Gomez-Herrero J 2011 {Increasing the elastic modulus of
  graphene by controlled defect creation} \textit{ Nat. Phys.\/} {\bf 11}
  26--31

\bibitem{Herbert2011}
Herbert E~G, Oliver W~C, de~Boer M~P and Pharr G~M 2011 {Measuring the elastic
  modulus and residual stress of freestanding thin films using nanoindentation
  techniques} \textit{ J. Mater. Res.\/} {\bf 24} 2974--2985

\bibitem{Wong2010}
Wong C~L, Annamalai M, Wang Z~Q and Palaniapan M 2010 {Characterization of
  nanomechanical graphene drum structures} \textit{ J. Micromech. Microeng.\/}
  {\bf 20} 115029

\bibitem{Politano2015}
Politano A and Chiarello G 2015 {Probing the Young's modulus and Poisson's
  ratio in graphene/metal interfaces and graphite: a comparative study}
  \textit{ Nano Res.\/} {\bf 8} 1847--1856

\bibitem{Zakharchenko2009}
Zakharchenko K~V, Katsnelson M~I and Fasolino A 2009 {Finite temperature
  lattice properties of graphene beyond the quasiharmonic approximation}
  \textit{ Phys. Rev. Lett.\/} {\bf 102} 046808

\bibitem{Liu2007}
Liu F, Ming P and Li J 2007 {\textit{Ab initio} calculation of ideal strength
  and phonon instability of graphene under tension} \textit{ Phys. Rev. B\/}
  {\bf 76} 064120

\bibitem{Zhang2011}
Zhang D~B, Akatyeva E and {Dumitric\ifmmode \u{a}\else \u{a}\fi{}} T 2011
  {Bending ultrathin graphene at the margins of continuum mechanics} \textit{
  Phys. Rev. Lett.\/} {\bf 106} 255503

\bibitem{Zhang2015}
Zhang G and Zhang Y~W 2015 Strain effects on thermoelectric properties of
  two-dimensional materials \textit{ Mechanics of Materials\/} {\bf 91, Part 2}
  382 -- 398

\bibitem{BaiKeKe2014}
Bai K~K, Zhou Y, Zheng H, Meng L, Peng H, Liu Z, Nie J~C and He L 2014
  {Creating One-Dimensional Nanoscale Periodic Ripples in a Continuous Mosaic
  Graphene Monolayer} \textit{ Phys. Rev. Lett.\/} {\bf 113} 086102

\bibitem{Fabien2015}
Jean F, Zhou T, Blanc N, Felici R, Coraux J and Renaud G 2015 {Topography of
  the graphene/Ir(111) moir{\'e} studied by surface X-ray diffraction} \textit{
  Phys. Rev. B\/} {\bf 91} 245424

\bibitem{Hattab2012}
Hattab H \textit{et al} 2012 Interplay of Wrinkles, Strain, and Lattice Parameter in Graphene on
  Iridium \textit{ Nano Lett.\/} {\bf 12} 678--682

\bibitem{Ackerman2016}
Ackerman M~L, Kumar P, Neek-Amal M, Thibado P~M, Peeters F~M and Singh S 2016
  {Anomalous Dynamical Behavior of Freestanding Graphene Membranes} \textit{
  Phys. Rev. Lett.\/} {\bf 117} 126801

\bibitem{Pereira2009b}
Pereira V~M and {Castro Neto} A~H 2009 Strain Engineering of Graphene's
  Electronic Structure \textit{ Phys. Rev. Lett.\/} {\bf 103} 046801

\bibitem{Kamien2014}
Castle T, Cho Y, Gong X, Jung E, Sussman D~M, Yang S and Kamien R~D 2014 Making
  the cut: Lattice \textit{Kirigami} rules \textit{ Phys. Rev. Lett.\/} {\bf
  113} 245502

\bibitem{Blees2015}
Blees M, Barnard A~W, Rose P~A, Roberts S~P, McGill K~L, Huang P~Y, Ruyack A~R,
  Kevek J~W, Kobrin B, Muller D~A, Muller D~A and McEuen P~L 2015 Graphene
  kirigami \textit{ Nature\/} {\bf 524} 204--207

\bibitem{Grosso2015}
Grosso B~F and Mele E~J 2015 Bending rules in graphene Kirigami \textit{ Phys.
  Rev. Lett.\/} {\bf 115} 195501

\bibitem{Castlee2016}
Castle T, Sussman D~M, Tanis M and Kamien R~D 2016 Additive lattice kirigami
  \textit{ Sci. Adv.\/} {\bf 2} e1601258

\bibitem{meyersolid07}
Meyer J, Geim A, Katsnelson M, Novoselov K, Obergfell D, Roth S, Girit C and
  Zettl A 2007 {On the roughness of single- and bi-layer graphene membranes}
  \textit{ Solid State Comm.\/} {\bf 143} 101--109

\bibitem{stolyarova07}
Stolyarova E, Rim K~T, Ryu S, Maultzsch J, Kim P, Brus L~E, Heinz T~F,
  Hybertsen M~S and Flynn G~W 2007 {High-resolution scanning tunneling
  microscopy imaging of mesoscopic graphene sheets on an insulating surface}
  \textit{ Proc. Natl. Acad. Sci. U.S.A.\/} {\bf 104} 9209--9212

\bibitem{Vinogradov12}
Vinogradov N~A, Zakharov A~A, Kocevski V, Rusz J, Simonov K~A, Eriksson O,
  Mikkelsen A, Lundgren E, Vinogradov A~S, M{\aa}rtensson N and Preobrajenski
  A~B 2012 {Formation and Structure of Graphene Waves on Fe(110)} \textit{
  Phys. Rev. Lett.\/} {\bf 109} 026101

\bibitem{Sloan2013}
Sloan J~V, Sanjuan A~A~P, Wang Z, Horvath C and Barraza-Lopez S 2013 {Strain
  gauge fields for rippled graphene membranes under central mechanical load: An
  approach beyond first-order continuum elasticity} \textit{ Phys. Rev. B\/}
  {\bf 87} 155436

\bibitem{Barraza2013}
Barraza-Lopez S, Sanjuan A~A~P, Wang Z and {Vanevi\ifmmode \acute{c}\else
  {\'c}\fi{}} 2013 {Strain-engineering of graphene's electronic structure
  beyond continuum elasticity} \textit{ Solid State Comm.\/} {\bf 166} 70--75

\bibitem{Monteverde2015}
Monteverde U, Pal J, Migliorato M, Missous M, Bangert U, Zan R, Kashtiban R and
  Powell D 2015 {Under pressure: Control of strain, phonons and bandgap opening
  in rippled graphene} \textit{ Carbon\/} {\bf 91} 266--274

\bibitem{Geim2007}
{Geim A K} and {Novoselov K S} 2007 {The rise of graphene} \textit{ Nat.
  Mater.\/} {\bf 6} 183--191

\bibitem{Bolotin2008}
Bolotin K~I, Sikes K~J, Hone J, Stormer H~L and Kim P 2008
  {Temperature-dependent transport in suspended graphene} \textit{ Phys. Rev.
  Lett.\/} {\bf 101} 096802

\bibitem{Castro2009}
Castro E~V, Ochoa H, Katsnelson M~I, Gorbachev R~V, Elias D~C, Novoselov K~S,
  Geim A~K and Guinea F 2010 {Limits on charge carrier mobility in suspended
  graphene due to flexural phonons} \textit{ Phys. Rev. Lett.\/} {\bf 105}
  266601

\bibitem{Mayorov2011}
Mayorov A~S \textit{et al} 2011
  {Micrometer-Scale Ballistic Transport in Encapsulated Graphene at Room
  Temperature} \textit{ Nano Lett.\/} {\bf 11} 2396--2399

\bibitem{DasSarma2011}
{Das Sarma} S, Adam S, Hwang E~H and Rossi E 2011 {Electronic transport in
  two-dimensional graphene} \textit{ Rev. Mod. Phys.\/} {\bf 83} 407--470

\bibitem{Roche08}
Cresti A, Nemec N, Biel B, Niebler G, Triozon F, Cuniberti G and Roche S 2008
  {Charge transport in disordered graphene-based low dimensional materials}
  \textit{ Nano Res.\/} {\bf 1} 361--394

\bibitem{DresselhausBook}
Saito R, Dresselhaus G and Dresselhaus M~S 1998 \textit{ Physical Properties of
  Carbon Nanotubes\/} (Singapore: World Scientific)

\bibitem{Oliva2013}
Oliva-Leyva M and Naumis G~G 2013 Understanding electron behavior in strained
  graphene as a reciprocal space distortion \textit{ Phys. Rev. B\/} {\bf 88}
  085430

\bibitem{Volovik2015}
Volovik G and Zubkov M 2015 {Emergent geometry experienced by fermions in
  graphene in the presence of dislocations} \textit{ Ann. Phys. (N.Y.)\/} {\bf
  356} 255--268

\bibitem{Oliva2015a}
Oliva-Leyva M and Naumis G~G 2015 Generalizing the Fermi velocity of strained
  graphene from uniform to nonuniform strain \textit{ Phys. Lett. A\/} {\bf
  379} 2645--2651

\bibitem{Salvador2012}
Xu P, Yang Y, Barber S~D, Ackerman M~L, Schoelz J~K, Qi D, Kornev I~A, Dong L,
  Bellaiche L, Barraza-Lopez S and Thibado P~M 2012 Atomic control of strain in
  freestanding graphene \textit{ Phys. Rev. B\/} {\bf 85} 121406

\bibitem{Evers2008}
Evers F and Mirlin A~D 2008 {Anderson transitions} \textit{ Rev. Mod. Phys.\/}
  {\bf 80} 1355--1417

\bibitem{Hasegawa}
Hasegawa Y, Konno R, Nakano H and Kohmoto M 2006 {Zero modes of tight-binding
  electrons on the honeycomb lattice} \textit{ Phys. Rev. B\/} {\bf 74} 033413

\bibitem{Naumis2014}
Naumis G~G and Roman-Taboada P 2014 {Mapping of strained graphene into
  one-dimensional Hamiltonians: Quasicrystals and modulated crystals} \textit{
  Phys. Rev. B\/} {\bf 89} 241404

\bibitem{Roman2015b}
Roman-Taboada P and Naumis G~G 2015 {Spectral butterfly and electronic
  localization in rippled-graphene nanoribbons: Mapping onto effective
  one-dimensional chains} \textit{ Phys. Rev. B\/} {\bf 92} 035406

\bibitem{deJuan2012}
de~Juan F, Sturla M and Vozmediano M~A~H 2012 {Space dependent Fermi velocity
  in strained graphene} \textit{ Phys. Rev. Lett.\/} {\bf 108} 227205

\bibitem{Volovik2014}
Volovik G~E and Zubkov M~A 2014 {Emergent Horava gravity in graphene} \textit{
  Ann. Phys. (N.Y.)\/} {\bf 340} 352--368

\bibitem{Pacheco2014}
{Pacheco Sanjuan} A~A, Wang Z, Imani H~P, {Vanevi\ifmmode \acute{c}\else
  {\'c}\fi{}} M and Barraza-Lopez S 2014 {Graphene's morphology and electronic
  properties from discrete differential geometry} \textit{ Phys. Rev. B\/} {\bf
  89} 121403

\bibitem{Guinea2010}
Guinea F, Katsnelson M~I and Geim A~K 2010 {Energy gaps and a zero-field
  quantum Hall effect in graphene by strain engineering} \textit{ Nat. Phys.\/}
  {\bf 6} 30--33

\bibitem{Guinea2010b}
Guinea F, Geim A~K, Katsnelson M~I and Novoselov K~S 2010 Generating quantizing
  pseudomagnetic fields by bending graphene ribbons \textit{ Phys. Rev. B\/}
  {\bf 81} 035408

\bibitem{Pereira2010}
Pereira V~M, Ribeiro R~M, Peres N~M~R and Castro-Neto A~H 2010 {Optical
  properties of strained graphene} \textit{ Europhys. Lett.\/} {\bf 92} 67001

\bibitem{Oliva2015b}
Oliva-Leyva M and Naumis G~G 2015 {Tunable dichroism and optical absorption of
  graphene by strain engineering} \textit{ 2D Mater.\/} {\bf 2} 025001

\bibitem{Reich2002}
Reich S, Maultzsch J, Thomsen C and Ordej{\'o}n P 2002 {Tight-binding
  description of graphene} \textit{ Phys. Rev. B\/} {\bf 66} 035412

\bibitem{Foa2014}
Foa-Torres L~E, Roche S and Charlier J~C 2014 \textit{ Introduction to
  Graphene-Based Nanomaterials\/} 1st ed (Cambridge University Press)

\bibitem{BarriosVargas2012}
Barrios-Vargas J~E and Naumis G~G 2012 {Critical wavefunctions in disordered
  graphene} \textit{ J. Phys.: Condens. Matter\/} {\bf 24} 255305

\bibitem{Li2010}
Li Y, Jiang X, Liu Z and Liu Z 2010 {Strain effects in graphene and graphene
  nanoribbons: The underlying mechanism} \textit{ Nano Res.\/} {\bf 3} 545--556

\bibitem{Katsnelson2008}
Katsnelson M~I and Geim A~K 2008 {Electron scattering on microscopic
  corrugations in graphene} \textit{ Philos. Trans. R. Soc. Lond. A\/} {\bf
  366} 195--204

\bibitem{Hatsugai2011}
Hatsugai Y 2011 {Topological aspect of graphene physics} \textit{ J. Phys.:
  Conf. Ser.\/} {\bf 334} 012004

\bibitem{Naumis2016}
Naumis G~G 2016 {Topological map of the Hofstadter butterfly: Fine structure of
  Chern numbers and Van Hove singularities} \textit{ Phys. Lett. A\/} {\bf 380}
  1772--1780

\bibitem{Abergel2010}
Abergel D~S~L, Apalkov V, Berashevich J, Ziegler K and Chakraborty T {2010}
  {Properties of graphene: a theoretical perspective} \textit{ {Adv. Phys.}\/}
  {\bf {59}} {261--482}

\bibitem{Bostwick2009}
Bostwick A, McChesney J~L, Emtsev K~V, Seyller T, Horn K, Kevan S~D and
  Rotenberg E 2009 {Quasiparticle Transformation during a Metal-Insulator
  Transition in Graphene} \textit{ Phys. Rev. Lett.\/} {\bf 103} 056404

\bibitem{BarriosVargas2011b}
Barrios-Vargas J~E and Naumis G~G 2011 {Electrical conductivity and resonant
  states of doped graphene considering next-nearest neighbor interaction}
  \textit{ Philos. Mag.\/} {\bf 91} 3844--3857

\bibitem{Martinazzo2010}
Martinazzo R, Casolo S and Tantardini G~F 2010 The Effect of Atomic-Scale
  Defects and Dopants on Graphene Electronic Structure \textit{ Physics and
  Applications of Graphene - Theory\/} ed Mikhailov D~S (InTech) chap~3

\bibitem{Kirkpatrick1972}
Kirkpatrick S and Eggarter T~P 1972 Localized states of a binary alloy \textit{
  Phys. Rev. B\/} {\bf 6} 3598--3609

\bibitem{Kohmoto1986}
Kohmoto M and Sutherland B 1986 Electronic states on a Penrose lattice \textit{
  Phys. Rev. Lett.\/} {\bf 56} 2740--2743

\bibitem{Bernard2002}
Bernard D and LeClair A 2002 {A classification of 2D random Dirac fermions}
  \textit{ J. Phys. A\/} {\bf 35} 2555

\bibitem{Tikhonenko2009}
Tikhonenko F~V, Kozikov A~A, Savchenko A~K and Gorbachev R~V 2009 {Transition
  between Electron Localization and Antilocalization in Graphene} \textit{
  Phys. Rev. Lett.\/} {\bf 103} 226801

\bibitem{Katsnelson2012}
Katsnelson M 2012 \textit{ {Graphene: Carbon in two dimensions}\/} 1st ed
  (Cambridge University Press)

\bibitem{deJuan2013}
de~Juan F, Ma{\~n}es J~L and Vozmediano M~A~H 2013 {Gauge fields from strain in
  graphene} \textit{ Phys. Rev. B\/} {\bf 87} 165131

\bibitem{Thorpe1985}
Phillips J~C and Thorpe M~F 1985 {Constraint theory, vector percolation and
  glass formation} \textit{ Solid State Comm.\/} {\bf 53} 699--702

\bibitem{Flores2010}
Flores-Ruiz H~M, Naumis G~G and Phillips J~C 2010 Heating through the glass
  transition: A rigidity approach to the boson peak \textit{ Phys. Rev. B\/}
  {\bf 82} 214201

\bibitem{Phillips1979}
Phillips J 1979 {Topology of covalent non-crystalline solids I: Short-range
  order in chalcogenide alloys} \textit{ J. Non-Cryst. Solids\/} {\bf 34}
  153–181

\bibitem{Suzuura2002}
Suzuura H and Ando T 2002 {Phonons and electron-phonon scattering in carbon
  nanotubes} \textit{ Phys. Rev. B\/} {\bf 65} 235412

\bibitem{KimNeto2008}
Kim E~A and Neto A~H~C 2008 {Graphene as an electronic membrane} \textit{
  Europhys. Lett.\/} {\bf 84} 57007

\bibitem{Ribeiro}
Ribeiro R~M, Pereira V~M, Peres N~M~R, Briddon P~R and {Castro Neto} A~H 2009
  {Strained graphene: tight-binding and density functional calculations}
  \textit{ New J. Phys.\/} {\bf 11} 115002

\bibitem{Mohiuddin2009}
Mohiuddin T~M~G \textit{et al} 2009 {Uniaxial strain in graphene by Raman spectroscopy: $G$ peak splitting,
  Gr{\"u}neisen parameters, and sample orientation} \textit{ Phys. Rev. B\/}
  {\bf 79} 205433

\bibitem{Ding2010}
Ding F, Ji H, Chen Y, Herklotz A, D{\"o}rr K, Mei Y, Rastelli A and Schmidt O~G
  2010 {Stretchable Graphene: A Close Look at Fundamental Parameters through
  Biaxial Straining} \textit{ Nano Lett.\/} {\bf 10} 3453--3458

\bibitem{Cheng2011}
Cheng Y~C, Zhu Z~Y, Huang G~S and Schwingenschl{\"o}gl U 2011 {Gr{\"u}neisen
  parameter of the $G$ mode of strained monolayer graphene} \textit{ Phys. Rev.
  B\/} {\bf 83} 115449

\bibitem{Oliva2016a}
Oliva-Leyva M and Naumis G~G 2016 Effective Dirac Hamiltonian for anisotropic
  honeycomb lattices: Optical properties \textit{ Phys. Rev. B\/} {\bf 93}
  035439

\bibitem{Naumov2011}
Naumov I~I and Bratkovsky A~M 2011 {Gap opening in graphene by simple periodic
  inhomogeneous strain} \textit{ Phys. Rev. B\/} {\bf 84} 245444

\bibitem{Kerszberg2015}
Kerszberg N and Suryanarayana P 2015 {Ab initio strain engineering of graphene:
  opening bandgaps up to 1 eV} \textit{ RSC Adv.\/} {\bf 5} 43810--43814

\bibitem{Zubkov2015}
Zubkov M 2015 {Emergent gravity and chiral anomaly in Dirac semimetals in the
  presence of dislocations} \textit{ Ann. Phys. (N.Y.)\/} {\bf 360} 655--678

\bibitem{Goerbig2008}
Goerbig M~O, Fuchs J~N, Montambaux G and Pi{\'e}chon F 2008 {Tilted anisotropic
  Dirac cones in quinoid-type graphene and
  $\ensuremath{\alpha}\mbox{-}(\mbox{BEDT-TTF})_{2}\mbox{I}_{3}$} \textit{
  Phys. Rev. B\/} {\bf 78} 045415

\bibitem{Choi2010}
Choi S~M, Jhi S~H and Son Y~W 2010 {Effects of strain on electronic properties
  of graphene} \textit{ Phys. Rev. B\/} {\bf 81} 081407

\bibitem{Ni2008}
Ni Z~H, Yu T, Lu Y~H, Wang Y~Y, Feng Y~P and Shen Z~X 2008 {Uniaxial strain on
  graphene: Raman spectroscopy study and band-gap opening} \textit{ ACS Nano\/}
  {\bf 2} 2301--2305

\bibitem{Ni2009}
{Ni Zhen Hua}, {Yu Ting}, {Lu Yun Hao}, {Wang Ying Ying}, {Feng Yuan Ping} and
  {Shen Ze Xiang} 2009 {Uniaxial strain on graphene: Raman spectroscopy study
  and band-gap opening} \textit{ ACS Nano\/} {\bf 3} 483--483

\bibitem{Fradkin}
Fradkin E 2013 \textit{ Field Theories of Condensed Matter Physics\/} 2nd ed
  (Cambridge University Press)

\bibitem{Ando2002b}
Ando T, Zheng Y and Suzuura H 2002 {Dynamical conductivity and zero-mode
  anomaly in honeycomb lattices} \textit{ J. Phys. Soc. Jpn.\/} {\bf 71}
  1318--1324

\bibitem{Wang2015}
Wang J, Zhang G, Ye F and Wang X 2015 {Mechanical manipulations on electronic
  transport of graphene nanoribbons} \textit{ J. Phys.: Condens. Matter\/} {\bf
  27} 225305

\bibitem{Fasolino}
Fasolino A, Los J~H and Katsnelson M~I 2007 {Intrinsic ripples in graphene}
  \textit{ Nat. Matter.\/} {\bf 6} 858--861

\bibitem{Dumitrica}
Levente~Tapaszt{\'o} L, Dumitrica T, Kim S~J, Nemes-Incze P, Hwang C and
  Bir{\'o} L~P 2012 {Breakdown of continuum mechanics for nanometre-wavelength
  rippling of graphene} \textit{ Nat. Phys.\/} {\bf 8} 858--861

\bibitem{Huertas2006}
Huertas-Hernando D, Guinea F and Brataas A 2006 Spin-orbit coupling in curved
  graphene, fullerenes, nanotubes, and nanotube caps \textit{ Phys. Rev. B\/}
  {\bf 74} 155426

\bibitem{Manes2013}
Ma\~nes J~L, de~Juan F, Sturla M and Vozmediano M~A~H 2013 Generalized
  effective Hamiltonian for graphene under nonuniform strain \textit{ Phys.
  Rev. B\/} {\bf 88} 155405

\bibitem{Carrillo2016}
Carrillo-Bastos R, Le{\'o}n C, Faria D, Latg{\'e} A, Andrei E~Y and Sandler N
  2016 {Strained fold-assisted transport in graphene systems} \textit{ Phys.
  Rev. B\/} {\bf 94} 125422

\bibitem{Dean2010}
Dean C~R, Young A~F, Meric I, Lee C, Wang L, Sorgenfrei S, Watanabe K,
  Taniguchi T, Kim P, Shepard K~L and Hone J 2010 {Boron nitride substrates for
  high-quality graphene electronics} \textit{ Nat. Nanotechnol.\/} {\bf 5}
  722--726

\bibitem{Kindermann2012}
Kindermann M, Uchoa B and Miller D~L 2012 Zero-energy modes and gate-tunable
  gap in graphene on hexagonal boron nitride \textit{ Phys. Rev. B\/} {\bf 86}
  115415

\bibitem{Amet2012}
Amet F, Williams J~R, Garcia A~G~F, Yankowitz M, Watanabe K, Taniguchi T and
  Goldhaber-Gordon D 2012 Tunneling spectroscopy of graphene-boron-nitride
  heterostructures \textit{ Phys. Rev. B\/} {\bf 85} 073405

\bibitem{Haas2007}
Hass J, Feng R, Mill{\'a}n-Otoya J~E, Li X, Sprinkle M, First P~N, de~Heer W~A,
  Conrad E~H and Berger C 2007 {Structural properties of the multilayer
  graphene/$4H$-$\mathrm{Si}\mathrm{C}(000\overline{1})$ system as determined
  by surface x-ray diffraction} \textit{ Phys. Rev. B\/} {\bf 75} 214109

\bibitem{Satija2013}
Satija I~I and Naumis G~G 2013 {Chern and Majorana modes of quasiperiodic
  systems} \textit{ Phys. Rev. B\/} {\bf 88} 054204

\bibitem{NaumisTraceMap1999}
Naumis G~G 1999 {Use of the trace map for evaluating localization properties}
  \textit{ Phys. Rev. B\/} {\bf 59} 11315--11321

\bibitem{Nemec2006}
Nemec N and Cuniberti G 2006 Hofstadter butterflies of carbon nanotubes:
  Pseudofractality of the magnetoelectronic spectrum \textit{ Phys. Rev. B\/}
  {\bf 74} 165411

\bibitem{Skomski2014}
Skomski R, Dowben P~A, {Sky Driver} M and Kelber J~A 2014 {Sublattice-induced
  symmetry breaking and band-gap formation in graphene} \textit{ Mater.
  Horiz.\/} {\bf 1} 563--571

\bibitem{Jeil2015}
Jeil J, DaSilva A~M, MacDonald A~H and Shaffique A 2015 {Origin of band gaps in
  graphene on hexagonal boron nitride} \textit{ Nat. Commun.\/} {\bf 6} 6308

\bibitem{Eryin2016}
Wang E \textit{et al} 2016 {Gaps induced by inversion
  symmetry breaking and second-generation Dirac cones in graphene/hexagonal
  boron nitride} \textit{ Nat. Phys.\/} {\bf 12} 1111--1115

\bibitem{Gutierrez2016}
Gutierrez C, Kim C~J, Brown L, Schiros T, Nordlund D, Lochocki E~B, Shen K~M,
  Park J and Pasupathy A~N 2016 {Imaging chiral symmetry breaking from Kekule
  bond order in graphene} \textit{ Nat. Phys.\/} {\bf 12} 950--958

\bibitem{Summerfield2016}
Summerfield A \textit{et al}
  2016 {Strain-engineered graphene grown on hexagonal boron nitride by
  molecular beam epitaxy} \textit{ Sci. Rep.\/} {\bf 6} 22440

\bibitem{Jarillo2011}
Xue J, Sanchez-Yamagishi J, Bulmash D, Jacquod P, Deshpande A, Watanabe K,
  Taniguchi T, Jarillo-Herrero P and LeRoy B~J 2011 {Scanning tunnelling
  microscopy and spectroscopy of ultra-flat graphene on hexagonal boron
  nitride} \textit{ Nat. Mater.\/} {\bf 10} 282--285

\bibitem{SanJose2014}
San-Jose P, Guti{\'e}rrez-Rubio A, Sturla M and Guinea F 2014 {Electronic
  structure of spontaneously strained graphene on hexagonal boron nitride}
  \textit{ Phys. Rev. B\/} {\bf 90} 115152

\bibitem{Wang2012}
Wang Z~F, Liu F and Chou M~Y 2012 Fractal Landau-Level Spectra in Twisted
  Bilayer Graphene \textit{ Nano Lett.\/} {\bf 12} 3833--3838

\bibitem{Zhou2007}
{Zhou S Y}, {Gweon G-H}, {Fedorov A V}, {First P N}, A d~H~W, {Lee D-H},
  {Guinea F}, {Castro Neto A H} and {Lanzara A} 2007 {Substrate-induced bandgap
  opening in epitaxial graphene} \textit{ Nat. Mater.\/} {\bf 6} 770--775

\bibitem{Zhou2008}
{Zhou SY}, {Siegel DA}, {Fedorov AV}, {Gabaly FEl}, {Schmid AK}, {Neto AH
  Castro}, {Lee D-H} and {Lanzara A} 2008 {Origin of the energy bandgap in
  epitaxial graphene} \textit{ Nat. Mater.\/} {\bf 7} 259--260

\bibitem{Bostwick2007}
{Bostwick Aaron}, {Ohta Taisuke}, {Seyller Thomas}, {Horn Karsten} and
  {Rotenberg Eli} 2007 {Quasiparticle dynamics in graphene} \textit{ Nat.
  Phys.\/} {\bf 3} 36--40

\bibitem{Kong2010}
Kong L \textit{et al} 2010 {Graphene/Substrate Charge Transfer
  Characterized by Inverse Photoelectron Spectroscopy} \textit{ J. Phys. Chem.
  C\/} {\bf 114} 21618--21624

\bibitem{Gaddam2011}
Gaddam S, Bjelkevig C, Ge S, Fukutani K, Dowben P~A and Kelber J~A 2011 {Direct
  graphene growth on MgO: origin of the band gap} \textit{ J. Phys.: Condens.
  Matter\/} {\bf 23} 072204

\bibitem{Cervantes2016}
Garc{\'i}a-Cervantes H, Gaggero-Sager L~M, Sotolongo-Costa O, Naumis G~G and
  Rodr{\'i}guez-Vargas I 2016 {Angle-dependent bandgap engineering in gated
  graphene superlattices} \textit{ AIP Adv.\/} {\bf 6} 035309

\bibitem{Sattari2016}
Sattari F 2016 {Spin transport in graphene superlattice under strain} \textit{
  J. Magn. Magn. Mater\/} {\bf 414} 19--24

\bibitem{Liang2011}
Qi X~L and Zhang S~C 2011 {Topological insulators and superconductors} \textit{
  Rev. Mod. Phys.\/} {\bf 83} 1057--1110

\bibitem{ZhuShuze2015}
Zhu S, Stroscio J~A and Li T 2015 {Programmable extreme pseudomagnetic fields
  in graphene by a uniaxial stretch} \textit{ Phys. Rev. Lett.\/} {\bf 115}
  245501

\bibitem{Manoharan}
Gomes K~K, Mar W, Ko W, Guinea F and Manoharan H~C 2012 {Designer Dirac
  fermions and topological phases in molecular graphene} \textit{ Nature\/}
  {\bf 483} 306--310

\bibitem{Manoharan2013}
Polini M, Guinea F, Lewenstein M, Manoharan H~C and Pellegrini V 2013
  {Artificial honeycomb lattices for electrons, atoms and photons} \textit{
  Nat. Nanotechnol.\/} {\bf 8} 625--633

\bibitem{Liu2014}
{Liu Y}, {Li Y Y}, {Rajput S}, {Gilks D}, {Lari L}, {Galindo P L}, {Weinert M},
  {Lazarov V K} and {Li L} 2014 {Tuning Dirac states by strain in the
  topological insulator Bi$_2$Se$_3$} \textit{ Nat. Phys.\/} {\bf 10} 294--299

\bibitem{Guassi2015}
Guassi M~R, Diniz G~S, Sandler N and Qu F 2015 {Zero-field and
  time-reserval-symmetry-broken topological phase transitions in graphene}
  \textit{ Phys. Rev. B\/} {\bf 92} 075426

\bibitem{Su1979}
Su W~P, Schrieffer J~R and Heeger A~J 1979 {Solitons in polyacetylene} \textit{
  Phys. Rev. Lett.\/} {\bf 42} 1698--1701

\bibitem{Pershoguba2012}
Pershoguba S~S and Yakovenko V~M 2012 {Shockley model description of surface
  states in topological insulators} \textit{ Phys. Rev. B\/} {\bf 86} 075304

\bibitem{DasSarma2013}
Ganeshan S, Sun K and Das~Sarma S 2013 Topological Zero-Energy Modes in Gapless
  Commensurate Aubry-Andr\'e-Harper Models \textit{ Phys. Rev. Lett.\/} {\bf
  110} 180403

\bibitem{Volovik2011}
Heikkil{\"a} T~T, Kopnin N~B and Volovik G~E 2011 Flat bands in topological
  media \textit{ JETP Letters\/} {\bf 94} 233--239

\bibitem{Volovik2013}
Volovik G~E 2013 Flat Band in Topological Matter \textit{ J. Supercond. Nov.
  Magn.\/} {\bf 26} 2887--2890

\bibitem{Iadecola2013}
Iadecola T, Campbell D, Chamon C, Hou C~Y, Jackiw R, Pi S~Y and Kusminskiy S~V
  2013 {Materials design from nonequilibrium steady states: Driven graphene as
  a tunable semiconductor with topological properties} \textit{ Phys. Rev.
  Lett.\/} {\bf 110} 176603

\bibitem{Delplace2013}
Delplace P, G\'omez-Le\'on A and Platero G 2013 Merging of Dirac points and
  Floquet topological transitions in ac-driven graphene \textit{ Phys. Rev.
  B\/} {\bf 88} 245422

\bibitem{Majgraphene2015}
San-Jose P, Lado J~L, Aguado R, Guinea F and Fern\'andez-Rossier J 2015
  Majorana Zero Modes in Graphene \textit{ Phys. Rev. X\/} {\bf 5} 041042

\bibitem{RomanTaboada2017}
Roman-Taboada P and Naumis G~G 2017 Topological flat bands in time-periodically
  driven uniaxial strained graphene nanoribbons \textit{ Phys. Rev. B\/} {\bf
  95} 115440

\bibitem{Wallace1947}
Wallace P~R 1947 The band theory of graphite \textit{ Phys. Rev.\/} {\bf 71}
  622--634

\bibitem{Slonczewski1958}
Slonczewski J~C and Weiss P~R 1958 Band Structure of Graphite \textit{ Phys.
  Rev.\/} {\bf 109} 272--279

\bibitem{Semedoff1984}
Semenoff G~W 1984 Condensed-matter simulation of a three-dimensional anomaly
  \textit{ Phys. Rev. Lett.\/} {\bf 53} 2449--2452

\bibitem{Bjorken}
Bjorken J~D and Drell S~D 1964 \textit{ Relativistic Quantum Mechanics\/} (New
  York: McGraw-Hill)

\bibitem{Katsnelson2007}
Katsnelson M~I and Novoselov K~S 2007 Graphene: New bridge between condensed
  matter physics and quantum electrodynamics \textit{ Solid State Comm.\/} {\bf
  143} 3--13

\bibitem{Cooper2012}
Cooper D~R, D'Anjou B, Ghattamaneni N, Harack B, Hilke M, Horth A, Majlis N,
  Massicotte M, Vandsburger L and Yu V 2012 Experimental review of graphene
  \textit{ ISRN Condensed Matter Physics\/} {\bf 2012} 501686

\bibitem{Pellegrino2011}
Pellegrino F~M~D, Angilella G~G~N and Pucci R 2011 {Linear response correlation
  functions in strained graphene} \textit{ Phys. Rev. B\/} {\bf 84} 195407

\bibitem{Oliva2017}
Oliva-Leyva M and Wang C 2017 Low-energy theory for strained graphene: an
  approach up to second-order in the strain tensor \textit{ Journal of Physics:
  Condensed Matter\/} {\bf 29} 165301

\bibitem{Betancur2015}
Betancur-Ocampo Y, Cifuentes-Quintal M, Cordourier-Maruri G and de~Coss R 2015
  Landau levels in uniaxially strained graphene: A geometrical approach
  \textit{ Ann. Phys. (N.Y.)\/} {\bf 359} 243 -- 251

\bibitem{Ramezani2013}
Masir M~R, Moldovan D and Peeters F 2013 {Pseudo magnetic field in strained
  graphene: Revisited} \textit{ Solid State Comm.\/} {\bf 175--176} 76--82

\bibitem{Shallcross2016b}
Ray N, Rost F, Weckbecker D, Vogl M, Sharma S, Gupta R, Pankratov O and
  Shallcross S 2016 {Going beyond $k\cdot p$ theory: a general method for
  obtaining effective Hamiltonians in both high and low symmetry situations} arXiv:1607.00920

\bibitem{Oliva2014}
Oliva-Leyva M and Naumis G~G 2014 {Anisotropic AC conductivity of strained
  graphene} \textit{ J. Phys.: Condens. Matter\/} {\bf 26} 125302

\bibitem{Oliva2014C}
Oliva-Leyva M and Naumis G~G 2014 {Corrigendum: Anisotropic AC conductivity of
  strained graphene (2014 J. Phys.: Condens. Matter 26 125302)} \textit{ J.
  Phys.: Condens. Matter\/} {\bf 26} 279501

\bibitem{Jang2014}
Jang W~J, Kim H, Shin Y~R, Wang M, Jang S~K, Kim M, Lee S, Kim S~W, Song Y~J
  and Kahng S~J 2014 {Observation of spatially-varying Fermi velocity in
  strained-graphene directly grown on hexagonal boron nitride} \textit{
  Carbon\/} {\bf 74} 139--145

\bibitem{Goerbig2011}
Goerbig M~O 2011 Electronic properties of graphene in a strong magnetic field
  \textit{ Rev. Mod. Phys.\/} {\bf 83} 1193--1243

\bibitem{Storz2016}
Storz O, Cortijo A, Wilfert S, Kokh K~A, Tereshchenko O~E, Vozmediano M~A~H,
  Bode M, Guinea F and Sessi P 2016 Mapping the effect of defect-induced strain
  disorder on the Dirac states of topological insulators \textit{ Phys. Rev.
  B\/} {\bf 94}(12) 121301

\bibitem{Khaidukov2016}
Khaidukov Z~V and Zubkov M~A 2016 Landau levels in graphene in the presence of
  emergent gravity \textit{ Eur. Phys. J. B\/} {\bf 89} 213

\bibitem{Zhang1994}
Zhang Y 1994 {Motion of electrons in semiconductors under inhomogeneous strain
  with application to laterally confined quantum wells} \textit{ Phys. Rev.
  B\/} {\bf 49} 14352--14366

\bibitem{Linnik2012}
Linnik T~L 2012 {Effective Hamiltonian of strained graphene} \textit{ J. Phys.:
  Condens. Matter\/} {\bf 24} 205302

\bibitem{Yang2011}
Yang H~T 2011 {Strain induced shift of Dirac points and the pseudo-magnetic
  field in graphene} \textit{ J. Phys.: Condens. Matter\/} {\bf 23} 505502

\bibitem{Kitt2012}
Kitt A~L, Pereira V~M, Swan A~K and Goldberg B~B 2012 Lattice-corrected
  strain-induced vector potentials in graphene \textit{ Phys. Rev. B\/} {\bf
  85} 115432

\bibitem{Sasaki2008}
Sasaki K and Saito R 2008 Pseudospin and Deformation-Induced Gauge Field in
  Graphene \textit{ Progress of Theoretical Physics Supplement\/} {\bf 176}
  253--278

\bibitem{Cabra2013}
Cabra D~C, Grandi N~E, Silva G~A and Sturla M~B 2013 Low-energy electron-phonon
  effective action from symmetry analysis \textit{ Phys. Rev. B\/} {\bf 88}
  045126

\bibitem{Yang2015}
Yang B 2015 Dirac cone metric and the origin of the spin connections in
  monolayer graphene \textit{ Phys. Rev. B\/} {\bf 91} 241403

\bibitem{Iorio2015}
Iorio A and Pais P 2015 Revisiting the gauge fields of strained graphene
  \textit{ Phys. Rev. D\/} {\bf 92} 125005
  
\bibitem{Arias2015}
Arias E, Hern\'andez A~R and Lewenkopf C 2015 Gauge fields in graphene with
  nonuniform elastic deformations: A quantum field theory approach \textit{
  Phys. Rev. B\/} {\bf 92} 245110
  
\bibitem{Pavel2017}
Castro-Villarreal P and Ruiz-S\'anchez R 2017 Pseudomagnetic field in curved graphene
  \textit{ Phys. Rev. B\/} {\bf 95} 125432

\bibitem{Cortijo2016}
Cortijo A and Zubkov M~A 2016 {Emergent gravity in the cubic tight-binding
  model of Weyl semimetal in the presence of elastic deformations} \textit{
  Ann. Phys. (N.Y.)\/} {\bf 366} 45--56

\bibitem{Froggatt1991}
Froggatt C~D and Nielsen H~B 1991 \textit{ {Origin of Symmetry}\/} (Singapore:
  World Scientific)

\bibitem{Volovik2003}
Volovik G~E 2003 \textit{ The universe in a helium droplet\/} (Oxford:
  Clarendon Press)

\bibitem{Horava2005}
{Ho\ifmmode \check{r}\else \v{r}\fi{}ava} P 2005 {Stability of Fermi surfaces
  and $K$ theory} \textit{ Phys. Rev. Lett.\/} {\bf 95} 016405

\bibitem{Falko2013}
Gradinar D~A, Mucha-Kruczy\ifmmode~\acute{n}\else \'{n}\fi{}ski M, Schomerus H
  and Fal'ko V~I 2013 Transport Signatures of Pseudomagnetic Landau Levels in
  Strained Graphene Ribbons \textit{ Phys. Rev. Lett.\/} {\bf 110} 266801

\bibitem{Roy2013}
Roy B, Hu Z~X and Yang K 2013 Theory of unconventional quantum Hall effect in
  strained graphene \textit{ Phys. Rev. B\/} {\bf 87} 121408

\bibitem{Cosma2014}
Cosma D~A, Mucha-Kruczy\ifmmode~\acute{n}\else \'{n}\fi{}ski M, Schomerus H and
  Fal'ko V~I 2014 Strain-induced modifications of transport in gated graphene
  nanoribbons \textit{ Phys. Rev. B\/} {\bf 90} 245409

\bibitem{Roy2014}
Roy B and Juri\ifmmode \check{c}\else \v{c}\fi{}i\ifmmode~\acute{c}\else
  \'{c}\fi{} V 2014 Strain-induced time-reversal odd superconductivity in
  graphene \textit{ Phys. Rev. B\/} {\bf 90} 041413

\bibitem{Bahamon2015}
Bahamon D~A, Qi Z, Park H~S, Pereira V~M and Campbell D~K 2015 Conductance
  signatures of electron confinement induced by strained nanobubbles in
  graphene \textit{ Nanoscale\/} {\bf 7} 15300--15309

\bibitem{Burgos2015}
Burgos R, Warnes J, Lima L~R~F and Lewenkopf C 2015 Effects of a random gauge
  field on the conductivity of graphene sheets with disordered ripples \textit{
  Phys. Rev. B\/} {\bf 91} 115403

\bibitem{Stegmann2016}
Stegmann T and Szpak N 2016 Current flow paths in deformed graphene: from
  quantum transport to classical trajectories in curved space \textit{ New J.
  Phys.\/} {\bf 18} 053016

\bibitem{Venderbos2016}
Venderbos J~W~F and Fu L 2016 Interacting Dirac fermions under a spatially
  alternating pseudomagnetic field: Realization of spontaneous quantum Hall
  effect \textit{ Phys. Rev. B\/} {\bf 93} 195126

\bibitem{Kauppila2016}
Kauppila V~J, Aikebaier F and Heikkil\"a T~T 2016 Flat-band superconductivity
  in strained Dirac materials \textit{ Phys. Rev. B\/} {\bf 93} 214505

\bibitem{Yeh2016}
Yeh N~C, Hsu C~C, Teague M~L, Wang J~Q, Boyd D~A and Chen C~C 2016 Nanoscale
  strain engineering of graphene and graphene-based devices \textit{ Acta
  Mechanica Sinica\/} {\bf 32} 497--509

\bibitem{Mikkel2016b}
Settnes M, Leconte N, Barrios-Vargas J~E, Jauho A~P and Roche S 2016 Quantum
  transport in graphene in presence of strain-induced pseudo-Landau levels
  \textit{ 2D Mater.\/} {\bf 3} 034005

\bibitem{Mikkel2016c}
Settnes M, Power S~R, Brandbyge M and Jauho A~P 2016 Graphene nanobubbles as
  valley filters and beam splitters \textit{ Phys. Rev. Lett.\/} {\bf 117}
  276801

\bibitem{Miller2009}
Miller D~L, Kubista K~D, Rutter G~M, Ruan M, de~Heer W~A, First P~N and
  Stroscio J~A 2009 Observing the Quantization of Zero Mass Carriers in
  Graphene \textit{ Science\/} {\bf 324} 924--927

\bibitem{Yeh2011}
Yeh N~C, Teague M~L, Yeom S, Standley B, Wu R~P, Boyd D and Bockrath M 2011
  Strain-induced pseudo-magnetic fields and charging effects on CVD-grown
  graphene \textit{ Surf. Sci.\/} {\bf 605} 1649--1656

\bibitem{Lu2012}
Lu J, Neto A~C and Loh K~P 2012 Transforming moir\'{e} blisters into geometric
  graphene nano-bubbles \textit{ Nat. Commun.\/} {\bf 3} 823

\bibitem{Guo2012}
Donghui G, Takahiro K, Takahiro M, Keigo I, Susumu O and Junji N 2012
  Observation of Landau levels in potassium-intercalated graphite under a zero
  magnetic field \textit{ Nat. Commun.\/} {\bf 3} 1068

\bibitem{Klimov2012}
Klimov N~N, Jung S, Zhu S, Li T, Wright C~A, Solares S~D, Newell D~B, Zhitenev
  N~B and Stroscio J~A 2012 Electromechanical properties of graphene drumheads
  \textit{ Science\/} {\bf 336} 1557--1561

\bibitem{Meng2013}
Meng L, He W~Y, Zheng H, Liu M, Yan H, Yan W, Chu Z~D, Bai K, Dou R~F, Zhang Y,
  Liu Z, Nie J~C and He L 2013 Strain-induced one-dimensional Landau level
  quantization in corrugated graphene \textit{ Phys. Rev. B\/} {\bf 87} 205405

\bibitem{LiSiYu2015}
Li S~Y, Bai K~K, Yin L~J, Qiao J~B, Wang W~X and He L 2015 Observation of
  unconventional splitting of Landau levels in strained graphene \textit{ Phys.
  Rev. B\/} {\bf 92} 245302

\bibitem{Yan2016}
Yan W, Li S~Y, Yin L~J, Qiao J~B, Nie J~C and He L 2016 Spatially resolving
  unconventional interface Landau quantization in a graphene monolayer-bilayer
  planar junction \textit{ Phys. Rev. B\/} {\bf 93} 195408

\bibitem{Goergi2016}
Georgi A \textit{et al} 2017 Tuning the pseudospin polarization of graphene by a
  pseudomagnetic field \textit{ Nano Lett.\/} {\bf 17} 2240-2245

\bibitem{Jiang2017}
Jiang Y, Mao J, Duan J, Lai X, Watanabe K, Taniguchi T and Andrei E~Y 2017
  Visualizing Strain-induced Pseudomagnetic Fields in Graphene through an hBN
  Magnifying Glass \textit{ Nano Lett.\/} {\bf 17} 2839-2843

\bibitem{Rechtsman2013}
Rechtsman M~C, Zeuner J~M, Tunnermann A, Nolte S, Segev M and Szameit A 2013
  Strain-induced pseudomagnetic field and photonic Landau levels in dielectric
  structures \textit{ Nat. Photon.\/} {\bf 7} 153–158

\bibitem{Tian2015}
Tian B, Endres M and Pekker D 2015 Landau Levels in Strained Optical Lattices
  \textit{ Phys. Rev. Lett.\/} {\bf 115} 236803

\bibitem{Mariani2012}
Mariani E, Pearce A~J and von Oppen F 2012 Fictitious gauge fields in bilayer
  graphene \textit{ Phys. Rev. B\/} {\bf 86} 165448

\bibitem{Yan2013}
Yan W, He W~Y, Chu Z~D, Liu M, Meng L, Dou R~F, Zhang Y, Liu Z, Nie J~C and He
  L 2013 Strain and curvature induced evolution of electronic band structures
  in twisted graphene bilayer \textit{ Nat. Commun.\/} {\bf 4} 2159

\bibitem{He2014}
He W~Y, Su Y, Yang M and He L 2014 {Creating in-plane pseudomagnetic fields in
  excess of 1000 T by misoriented stacking in a graphene bilayer} \textit{
  Phys. Rev. B\/} {\bf 89} 125418

\bibitem{Zabolotskiy2016}
Zabolotskiy A~D and Lozovik Y~E 2016 Strain-induced pseudomagnetic field in the
  Dirac semimetal borophene \textit{ Phys. Rev. B\/} {\bf 94} 165403

\bibitem{Tang2014}
Tang E and Fu L 2014 Strain-induced partially flat band, helical snake states
  and interface superconductivity in topological crystalline insulators
  \textit{ Nat. Phys.\/} {\bf 10} 964

\bibitem{Pearce2016}
Pearce A~J, Mariani E and Burkard G 2016 Tight-binding approach to strain and
  curvature in monolayer transition-metal dichalcogenides \textit{ Phys. Rev.
  B\/} {\bf 94} 155416

\bibitem{Ochoa2016}
Ochoa H, Zarzuela R and Tserkovnyak Y 2017 Emergent gauge fields from curvature
  in single layers of transition-metal dichalcogenides \textit{ Phys. Rev.
  Lett.\/} {\bf 118} 026801

\bibitem{Cortijo2016b}
Cortijo A, Ferreir\'os Y, Landsteiner K and Vozmediano M~A~H 2015 Elastic Gauge
  Fields in Weyl Semimetals \textit{ Phys. Rev. Lett.\/} {\bf 115} 177202

\bibitem{Pikulin2016}
Pikulin D~I, Chen A and Franz M 2016 Chiral Anomaly from Strain-Induced Gauge
  Fields in Dirac and Weyl Semimetals \textit{ Phys. Rev. X\/} {\bf 6} 041021

\bibitem{Grushin2016}
Grushin A~G, Venderbos J~W~F, Vishwanath A and Ilan R 2016 Inhomogeneous Weyl
  and Dirac semimetals: transport in axial magnetic fields and Fermi arc
  surface states from pseudo-Landau levels \textit{ Phys. Rev. X\/} {\bf 6}
  041046

\bibitem{Peres2009}
Peres N~M~R 2009 Scattering in one-dimensional heterostructures described by
  the Dirac equation \textit{ J. Phys.: Condens. Matter\/} {\bf 21} 095501

\bibitem{Atanasov2010}
Atanasov V and Saxena A 2010 Tuning the electronic properties of corrugated
  graphene: Confinement, curvature, and band-gap opening \textit{ Phys. Rev.
  B\/} {\bf 81} 205409

\bibitem{deJuan2007}
de~Juan F, Cortijo A and Vozmediano M~A~H 2007 Charge inhomogeneities due to
  smooth ripples in graphene sheets \textit{ Phys. Rev. B\/} {\bf 76} 165409

\bibitem{Raoux2010}
Raoux A, Polini M, Asgari R, Hamilton A~R, Fazio R and MacDonald A~H 2010
  Velocity-modulation control of electron-wave propagation in graphene \textit{
  Phys. Rev. B\/} {\bf 81} 073407

\bibitem{Ratnikov2014}
Ratnikov P~V and Silin A~P 2014 Novel type of superlattices based on gapless
  graphene with the alternating Fermi velocity \textit{ JETP Letters\/} {\bf
  100} 311--318

\bibitem{Lima2015b}
Lima J~R and Moraes F 2015 Indirect band gap in graphene from modulation of the
  Fermi velocity \textit{ Solid State Comm.\/} {\bf 201} 82 -- 87

\bibitem{Lima2016}
Lima J~R~F, Pereira L~F~C and Bezerra C~G 2016 Controlling resonant tunneling
  in graphene via Fermi velocity engineering \textit{ J. Appl. Phys.\/} {\bf
  119} 244301

\bibitem{Oppen2009}
von Oppen F, Guinea F and Mariani E 2009 Synthetic electric fields and phonon
  damping in carbon nanotubes and graphene \textit{ Phys. Rev. B\/} {\bf 80}
  075420

\bibitem{Trif2013}
Trif M, Upadhyaya P and Tserkovnyak Y 2013 Theory of electromechanical coupling
  in dynamical graphene \textit{ Phys. Rev. B\/} {\bf 88} 245423

\bibitem{Vaezi2013}
Vaezi A, Abedpour N, Asgari R, Cortijo A and Vozmediano M~A~H 2013 Topological
  electric current from time-dependent elastic deformations in graphene
  \textit{ Phys. Rev. B\/} {\bf 88} 125406

\bibitem{Zhang2016}
Zhang K, Zhang E, Chen H and Zhang S 2016 Odd-parity currents induced by
  dynamic deformations in graphene-like systems \textit{ J. Phys.: Condens.
  Matter\/} {\bf 28} 455301

\bibitem{Sasaki2014}
Sasaki K~i, Gotoh H and Tokura Y 2014 Valley-antisymmetric potential in
  graphene under dynamical deformation \textit{ Phys. Rev. B\/} {\bf 90} 205402

\bibitem{Oliva2016b}
Oliva-Leyva M and Naumis G~G 2016 Sound waves induce Volkov-like states, band
  structure and collimation effect in graphene \textit{ J. Phys.: Condens.
  Matter\/} {\bf 28} 025301

\bibitem{Zhai2010}
Zhai F, Zhao X, Chang K and Xu H~Q 2010 Magnetic barrier on strained graphene:
  A possible valley filter \textit{ Phys. Rev. B\/} {\bf 82} 115442

\bibitem{Wolkow1935}
Wolkow D~M 1935 {\"U}ber eine klasse von l{\"o}sungen der Diracschen gleichung
  \textit{ Z. Phys.\/} {\bf 94} 250--260

\bibitem{Lopez2008}
L\'opez-Rodr\'{\i}guez F~J and Naumis G~G 2008 Analytic solution for electrons
  and holes in graphene under electromagnetic waves: Gap appearance and
  nonlinear effects \textit{ Phys. Rev. B\/} {\bf 78} 201406

\bibitem{Park2008}
Park C~H, Son Y~W, Yang L, Cohen M~L and Louie S~G 2008 Electron Beam
  Supercollimation in Graphene Superlattices \textit{ Nano Lett.\/} {\bf 8}
  2920--2924

\bibitem{Wang2010}
Wang Z and Liu F 2010 Manipulation of Electron Beam Propagation by
  Hetero-Dimensional Graphene Junctions \textit{ ACS Nano\/} {\bf 4} 2459--2465

\bibitem{FoaTorres2014}
Usaj G, Perez-Piskunow P~M, {Foa Torres} L~E~F and Balseiro C~A 2014
  {Irradiated graphene as a tunable Floquet topological insulator} \textit{
  Phys. Rev. B\/} {\bf 90} 115423

\bibitem{Guinea2012}
Guinea F 2012 Strain engineering in graphene \textit{ Solid State Comm.\/} {\bf
  152} 1437--1441

\bibitem{DDG}
Bobenko A~I, Schr{\"o}der P, Sullivan J~M and Ziegler G~M (eds) 2008 \textit{
  Discrete differential geometry\/} 1st ed (\textit{ Oberwolfach Seminars\/}
  vol~38) (Birkh{\"a}user, Basel)

\bibitem{Castro2016}
Castro E~V, Cazalilla M~A and Vozmediano M~A~H 2016 Raise and collapse of
  strain-induced pseudo-Landau levels in graphene arXiv:1610.08988

\bibitem{Verbiest2016}
Verbiest G~J, Stampfer C, Huber S~E, Andersen M and Reuter K 2016 Interplay
  between nanometer-scale strain variations and externally applied strain in
  graphene \textit{ Phys. Rev. B\/} {\bf 93} 195438

\bibitem{Pacheco2014b}
Pacheco~Sanjuan A~A, Mehboudi M, Harriss E~O, Terrones H and Barraza-Lopez S
  2014 Quantitative chemistry and the discrete geometry of conformal atom-thin
  crystals \textit{ ACS Nano\/} {\bf 8} 1136--1146

\bibitem{Xu2009}
Xu Z and Xu G 2009 Discrete schemes for Gaussian curvature and their
  convergence \textit{ Comput. Math. Appl.\/} {\bf 57} 1187--1195

\bibitem{Hohenberg1964}
Hohenberg P and Kohn W 1964 Inhomogeneous electron gas \textit{ Phys. Rev.\/}
  {\bf 136} B864--B871

\bibitem{Levy1979}
Levy M 1979 Universal variational functionals of electron densities,
  first-order density matrices, and natural spin-orbitals and solution of the
  v-representability problem \textit{ Proc. Natl. Acad. Sci. U.S.A.\/} {\bf 76}
  6062--6065

\bibitem{Hedin1971}
Hedin L and Lundqvist B~I 1971 Explicit local exchange-correlation potentials
  \textit{ J. Phys. C\/} {\bf 4} 2064

\bibitem{Ceperley1980}
Ceperley D~M and Alder B~J 1980 Ground state of the electron gas by a
  stochastic method \textit{ Phys. Rev. Lett.\/} {\bf 45} 566--569

\bibitem{Perdew1992}
Perdew J~P and Wang Y 1992 Accurate and simple analytic representation of the
  electron-gas correlation energy \textit{ Phys. Rev. B\/} {\bf 45}
  13244--13249

\bibitem{Heyd2003}
Heyd J, Scuseria G~E and Ernzerhof M 2003 Hybrid functionals based on a
  screened Coulomb potential \textit{ J. Chem. Phys.\/} {\bf 118} 8207--8215

\bibitem{Cohen2016}
Cohen M~L and Louie G 2016 \textit{ Fundamentals of Condensed Matter Physics\/}
  1st ed (Cambridge, Cambridge University Press)

\bibitem{Gonze1992}
Gonze X, Allan D~C and Teter M~P 1992 Dielectric tensor, effective charges, and
  phonons in \ensuremath{\alpha}-quartz by variational density-functional
  perturbation theory \textit{ Phys. Rev. Lett.\/} {\bf 68} 3603--3606

\bibitem{Baroni2001}
Baroni S, de~Gironcoli S, Dal~Corso A and Giannozzi P 2001 Phonons and related
  crystal properties from density-functional perturbation theory \textit{ Rev.
  Mod. Phys.\/} {\bf 73} 515--562

\bibitem{Gonze1997}
Gonze X 1997 First-principles responses of solids to atomic displacements and
  homogeneous electric fields: Implementation of a conjugate-gradient algorithm
  \textit{ Phys. Rev. B\/} {\bf 55} 10337--10354

\bibitem{Charlier2007}
Charlier J~C, Blase X and Roche S 2007 Electronic and transport properties of
  nanotubes \textit{ Rev. Mod. Phys.\/} {\bf 79} 677--732

\bibitem{Heggie1998}
Heggie M~I, Terrones M, Eggen B~R, Jungnickel G, Jones R, Latham C~D, Briddon
  P~R and Terrones H 1998 Quantitative density-functional study of nested
  fullerenes \textit{ Phys. Rev. B\/} {\bf 57} 13339--13342

\bibitem{Lenosky1992}
Lenosky T, Gonze X, Teter M and Elser V 1992 Energetics of negatively curved
  graphitic carbon \textit{ Nature\/} {\bf 355} 333--335

\bibitem{Lherbier2014}
Lherbier A, Terrones H and Charlier J~C 2014 Three-dimensional massless Dirac
  fermions in carbon schwarzites \textit{ Phys. Rev. B\/} {\bf 90} 125434

\bibitem{Kittel2005}
Kittel C 2005 \textit{ {Introduction to solid state Physics}\/} 8th ed (United
  States, Wiley \& Sons)

\bibitem{Aroyo2014}
Aroyo M~I, Orobengoa D, de~la Flor G, Tasci E~S, Perez-Mato J~M and
  Wondratschek H 2014 Brillouin-zone database on the Bilbao crystallographic
  server \textit{ Acta Cryst. A\/} {\bf 70} 126--137

\bibitem{Hinuma2017}
Hinuma Y, Pizzi G, Kumagai Y, Oba F and Tanaka I 2017 Band structure diagram
  paths based on crystallography \textit{ Comp. Mater. Sci.\/} {\bf 128}
  140--184

\bibitem{Gui2008}
Gui G, Li J and Zhong J 2008 Band structure engineering of graphene by strain:
  First-principles calculations \textit{ Phys. Rev. B\/} {\bf 78} 075435

\bibitem{Kresse1996}
Kresse G and Furthm{\"u}ller J 1996 Efficiency of ab-initio total energy
  calculations for metals and semiconductors using a plane-wave basis set
  \textit{ Comp. Mater. Sci.\/} {\bf 6} 15--50

\bibitem{Kresse1996-2}
Kresse G and Furthm{\"u}ller J 1996 Efficient iterative schemes for ab initio
  total-energy calculations using a plane-wave basis set \textit{ Phys. Rev.
  B\/} {\bf 54} 11169--11186

\bibitem{Farjam2009}
Farjam M and Rafii-Tabar H 2009 Comment on ``Band structure engineering of
  graphene by strain: First-principles calculations'' \textit{ Phys. Rev. B\/}
  {\bf 80} 167401

\bibitem{Gianozzi2009}
Giannozzi P \textit{et al} 2009 QUANTUM ESPRESSO: a modular and open-source software
  project for quantum simulations of materials \textit{ J. Phys.: Condens.
  Matter\/} {\bf 21} 395502

\bibitem{Gui2009}
Gui G, Li J and Zhong J 2009 Reply to ``Comment on `Band structure engineering
  of graphene by strain: First-principles calculations' '' \textit{ Phys. Rev.
  B\/} {\bf 80} 167402

\bibitem{Clark2005}
Clark S~J, Segall M~D, Pickard C~J, Hasnip P~J, Probert M~J, Refson K and Payne
  M~C 2005 First principles methods using CASTEP \textit{ Z. Kristallogr.\/}
  {\bf 220} 567--570

\bibitem{Materials-Studio2016}
Milman V \textit{et al} 2010 Electron and vibrational spectroscopies using DFT, plane waves and pseudopotentials: CASTEP implementation \textit{ J. Mol. Struct.\/} {\bf 954} 22-35

\bibitem{Castep}
 2016 \textit{ On line Castep theory manual, Materials Studio 2016, Biovia
  Dassault Syst\`{e}mes\/}

\bibitem{Gonze2002}
Gonze X \textit{et al} 2002 First-principles computation of material properties: the
  \textit{ABINIT} software project \textit{ Comp. Mater. Sci.\/} {\bf 25}
  478--492

\bibitem{Perdew1985}
Perdew J~P 1985 Density functional theory and the band gap problem \textit{
  Int. J. Quantum Chem.\/} {\bf 28} 497--523

\bibitem{Aryasetiawan1998}
Aryasetiawan F and Gunnarsson O 1998 The GW method \textit{ Rep. Prog. Phys.\/}
  {\bf 61} 237--312

\bibitem{Nair2008}
Nair R~R, Blake P, Grigorenko A~N, Novoselov K~S, Booth T~J, Stauber T, Peres
  N~M~R and Geim A~K 2008 {Fine structure constant defines visual transparency
  of graphene} \textit{ Science\/} {\bf 320} 1308

\bibitem{Mak2008}
Mak K~F, Sfeir M~Y, Wu Y, Lui C~H, Misewich J~A and Heinz T~F 2008 {Measurement
  of the optical conductivity of graphene} \textit{ Phys. Rev. Lett.\/} {\bf
  101} 196405

\bibitem{Gusynin2007}
Gusynin V~P, Sharapov S~G and Carbotte J~P 2007 {AC conductivity of graphene:
  from tight-binding model to 2+1-dimensional quantum electrodynamics} \textit{
  Int. J. Mod. Phys. B\/} {\bf 21} 4611--4658

\bibitem{Mikhailov2007}
Mikhailov S~A and Ziegler K 2007 {New electromagnetic mode in graphene}
  \textit{ Phys. Rev. Lett.\/} {\bf 99} 016803

\bibitem{Stauber2015}
Stauber T, Noriega-P{\'e}rez D and Schliemann J 2015 {Universal absorption of
  two-dimensional systems} \textit{ Phys. Rev. B\/} {\bf 91} 115407

\bibitem{Klimchitskaya2016}
Klimchitskaya G~L and Mostepanenko V~M 2016 {Conductivity of pure graphene:
  Theoretical approach using the polarization tensor} \textit{ Phys. Rev. B\/}
  {\bf 93} 245419

\bibitem{Pellegrino2010}
Pellegrino F~M~D, Angilella G~G~N and Pucci R 2010 {Strain effect on the
  optical conductivity of graphene} \textit{ Phys. Rev. B\/} {\bf 81} 035411

\bibitem{Hernandez2016}
Hern{\'a}ndez-Ortiz S, Valenzuela D, Raya A and S{\'a}nchez-Madrigal S 2016
  {Light absorption in distorted graphene} \textit{ Int J. Mod. Phys. B\/} {\bf
  30} 1650084

\bibitem{Rakheja2016}
Rakheja S and Sengupta P 2016 The tuning of light-matter coupling and dichroism
  in graphene for enhanced absorption: Implications for graphene-based optical
  absorption devices \textit{ J. Phys. D: Appl. Phys.\/} {\bf 49} 115106

\bibitem{Nguyen2016}
Nguyen H~V and Nguyen V~H 2016 {Comment on ``Orientation dependence of the
  optical spectra in graphene at high frequencies''} \textit{ Phys. Rev. B\/}
  {\bf 94} 117401

\bibitem{Chen2016}
Chen J~H, Luo W, Chen Z~X, Yan S~C, Xu F and Lu Y~Q 2016 {Mechanical modulation
  of a hybrid graphene--microfiber structure} \textit{ Adv. Opt. Mater.\/} {\bf
  4} 853--857

\bibitem{Ni2014}
Ni G~X, Yang H~Z, Ji W, Baeck S~J, Toh C~T, Ahn J~H, Pereira V~M and Ozyilmaz B
  2014 {Tuning optical conductivity of large-scale CVD graphene by strain
  engineering} \textit{ Adv. Mater. (Weinheim, Ger.)\/} {\bf 26} 1081--1086

\bibitem{Lv2012}
Lv R \textit{et al} {2012} {Nitrogen-doped
  graphene: beyond single substitution and enhanced molecular sensing} \textit{
  {Sci. Rep.}\/} {\bf {2}} 586

\bibitem{Bethune1991}
Bethune D~S, Meijer G, Tang W~C, Rosen H~J, Golden W~G, Seki H, Brown C~A and
  Devries M~S {1991} {Vibrational Raman and infrared-spectra of
  chromatographically separated C$_{60}$ and C$_{70}$ fullerene clusters}
  \textit{ {Chem. Phys. Lett.}\/} {\bf {179}} {181--186}

\bibitem{Dresselhaus2005}
Dresselhaus M~S, Dresselhaus G, Saito R and Jorio A {2005} {Raman spectroscopy
  of carbon nanotubes} \textit{ Phys. Rep.\/} {\bf {409}} 47--99

\bibitem{Ferrari2006}
Ferrari A~C, Meyer J~C, Scardaci V, Casiraghi C, Lazzeri M, Mauri F, Piscanec
  S, Jiang D, Novoselov K~S, Roth S and Geim A~K {2006} {Raman spectrum of
  graphene and graphene layers} \textit{ {Phys. Rev. Lett.}\/} {\bf {97}} 187401

\bibitem{Ferrari2007}
Ferrari A~C {2007} {Raman spectroscopy of graphene and graphite: Disorder,
  electron-phonon coupling, doping and nonadiabatic effects} \textit{ {Solid
  State Comm.}\/} {\bf {143}} {47--57}

\bibitem{Ferrari2013}
Ferrari A~C and Basko D~M {2013} {Raman spectroscopy as a versatile tool for
  studying the properties of graphene} \textit{ {Nat. Nanotechnol.}\/} {\bf
  {8}} {235--246}

\bibitem{Huangm2009}
Huang M, Yan H, Chen C, Song D, Heinz T~F and Hone J 2009 Phonon softening and
  crystallographic orientation of strained graphene studied by Raman
  spectroscopy \textit{ Proc. Natl. Acad. Sci. U.S.A.\/} {\bf 106} 7304--7308

\bibitem{Huangm2010}
Huang M, Yan H, Heinz T~F and Hone J {2010} {Probing strain-induced electronic
  structure change in graphene by Raman spectroscopy} \textit{ {Nano Lett.}\/}
  {\bf {10}} {4074--4079}

\bibitem{LV2016}
Lv R, Terrones H, Elias A~L, Perea-Lopez N, Gutierrez H~R, Cruz-Silva E,
  Rajukumar L~P, Dresselhaus M~S and Terrones M {2015} {Two-dimensional
  transition metal dichalcogenides: Clusters, ribbons, sheets and more}
  \textit{ {Nano Today}\/} {\bf {10}} {559--592}

\bibitem{Terrones2014}
Terrones H \textit{et al} {2014} {New first order Raman-active modes in few layered
  transition metal dichalcogenides} \textit{ {Sci. Rep.}\/} {\bf {4}} 4215

\bibitem{LiangLiangbo2014}
Liang L and Meunier V {2014} {First-principles Raman spectra of MoS$_2$, WS$_2$
  and their heterostructures} \textit{ {Nanoscale}\/} {\bf {6}} {5394--5401}

\bibitem{Zhangshuan2014}
Zhang S, Yang J, Xu R, Wang F, Li W, Ghufran M, Zhang Y~W, Yu Z, Zhang G, Qin Q
  and Lu Y {2014} {Extraordinary photoluminescence and strong
  temperature/angle-dependent Raman responses in few-layer phosphorene}
  \textit{ {ACS Nano}\/} {\bf {8}} {9590--9596}

\bibitem{LingLiang2015}
Ling X, Liang L, Huang S, Puretzky A~A, Geohegan D~B, Sumpter B~G, Kong J,
  Meunier V and Dresselhaus M~S {2015} {Low-frequency interlayer breathing
  modes in few-layer black phosphorus} \textit{ {Nano Lett.}\/} {\bf {15}}
  {4080--4088}

\bibitem{Kim2016}
Kim C~J, S\'anchez-Castillo A, Ziegler Z, Ogawa Y, Noguez C and Park J 2016
  {Chiral atomically thin films} \textit{ Nat. Nanotech.\/} {\bf 11} 520--524

\bibitem{Lee2015}
Lee T, Min S~H, Gu M, Jung Y~K, Lee W, Lee J~U, Seong D~G and Kim B~S 2015
  {Layer-by-Layer Assembly for Graphene-Based Multilayer Nanocomposites:
  Synthesis and Applications} \textit{ Chem. Mater.\/} {\bf 27} 3785--3796

\bibitem{Latil2007}
Latil S, Meunier V and Henrard L 2007 Massless fermions in multilayer graphitic
  systems with misoriented layers: \textit{Ab initio} calculations and
  experimental fingerprints \textit{ Phys. Rev. B\/} {\bf 76} 201402

\bibitem{Freitag2011}
{Freitag, M} 2011 {Graphene: Trilayers unravelled} \textit{ Nat. Phys.\/} {\bf
  7} 596--597

\bibitem{McCann2013}
McCann E and Koshino M 2013 {The electronic properties of bilayer graphene}
  \textit{ Rep. Prog. Phys.\/} {\bf 76} 056503

\bibitem{Kuzmenko2009}
Kuzmenko A~B, Crassee I, van~der Marel D, Blake P and Novoselov K~S 2009
  {Determination of the gate-tunable band gap and tight-binding parameters in
  bilayer graphene using infrared spectroscopy} \textit{ Phys. Rev. B\/} {\bf
  80} 165406

\bibitem{Lau2011}
{Bao W}, {Jing L}, {Velasco J }, {Lee Y}, {Liu G}, {Tran D}, {Standley B},
  {Aykol M}, {Cronin S B}, {Smirnov D}, {Koshino M}, {McCann E}, {Bockrath M}
  and {Lau C N} 2011 {Stacking-dependent band gap and quantum transport in
  trilayer graphene} \textit{ Nat. Phys.\/} {\bf 7} 948--952

\bibitem{Lau2012}
Velasco J, Jing L, Bao W, Lee Y, Kratz P, Aji V, Bockrath M, Lau N~C, Varma C,
  Stillwell R, Smirnov D, Zhang F, Jung J and MacDonald A~H 2012 {Transport
  spectroscopy of symmetry-broken insulating states in bilayer graphene}
  \textit{ Nat. Nanotechnol.\/} {\bf 7} 156--160

\bibitem{Lau2014}
Velasco J, Lee Y, Zhang F, Myhro K, Tran D, Deo M, Smirnov D, MacDonald A~H and
  Lau C~N 2014 {Competing ordered states with filling factor two in bilayer
  graphene} \textit{ Nat. Commun.\/} {\bf 5} 4550

\bibitem{Nanda2009}
Nanda B~R~K and Satpathy S 2009 {Strain and electric field modulation of the
  electronic structure of bilayer graphene} \textit{ Phys. Rev. B\/} {\bf 80}
  165430

\bibitem{Verberck2012}
Verberck B, Partoens B, Peeters F~M and Trauzettel B 2012 {Strain-induced band
  gaps in bilayer graphene} \textit{ Phys. Rev. B\/} {\bf 85} 125403

\bibitem{Moldovan2016}
Moldovan D and Peeters F~M 2016 {Strain engineering of the electronic
  properties of bilayer graphene quantum dots} \textit{ Phys. Status Solidi
  RRL\/} {\bf 10} 39--45

\bibitem{Eva2012}
Andrei E~Y, Li G and Du X 2012 {Electronic properties of graphene: a
  perspective from scanning tunneling microscopy and magnetotransport} \textit{
  Rep. Prog. Phys.\/} {\bf 75} 056501

\bibitem{Zhao2014}
Zhao L~D, Lo S~H, Zhang Y, Sun H, Tan G, Uher C, Wolverton C, Dravid V~P and
  Kanatzidis M~G 2014 Ultralow thermal conductivity and high thermoelectric
  figure of merit in SnSe crystals \textit{ Nature\/} {\bf 508} 373--377

\bibitem{KaiSCIENCE}
Chang K \textit{et al} 2016 Discovery of robust in-plane
  ferroelectricity in atomic-thick SnTe \textit{ Science\/} {\bf 353} 274--278

\bibitem{MauriSCIENCE}
Sessi P \textit{et al}
  2016 Robust spin-polarized midgap states at step edges of topological
  crystalline insulators \textit{ Science\/} {\bf 354} 1269--1273

\bibitem{Cahangirov2009}
Cahangirov S, Topsakal M, Akt\"urk E, \ifmmode~\mbox{\c{S}}\else \c{S}\fi{}ahin
  H and Ciraci S 2009 Two- and One-Dimensional Honeycomb Structures of Silicon
  and Germanium \textit{ Phys. Rev. Lett.\/} {\bf 102} 236804

\bibitem{Tao2014}
Tao L, Cinquanta E, Chiappe D, Grazianetti C, Fanciulli M, Dubey M, Moll A and
  Akinwande D 2014 Silicene field-effect transistors operating at room
  temperature \textit{ Nat. Nanotechnol.\/} {\bf 10} 227--231

\bibitem{Silicene2016}
Chowdhury S and Jana D 2016 A theoretical review on electronic, magnetic and
  optical properties of silicene \textit{ Rep. Prog. Phys.\/} {\bf 79} 126501

\bibitem{Kara2012}
Kara A, Enriquez H, Seitsonen A~P, Voon L~L~Y, Vizzini S, Aufray B and
  Oughaddou H 2012 {A review on silicene --- New candidate for electronics}
  \textit{ ‎Surf. Sci. Rep.\/} {\bf 67} 1--18

\bibitem{WalesBook}
Wales D 2001 \textit{ Energy Landscapes: Applications to Clusters, Biomolecules
  and Glasses\/} (UK: Cambridge U. Press)

\bibitem{Guzman2007}
Guzm{\'a}n-Verri G~G and {Lew Yan Voon} L~C 2007 {Electronic structure of
  silicon-based nanostructures} \textit{ Phys. Rev. B\/} {\bf 76} 075131

\bibitem{Kaloni2013}
Kaloni T~P, Cheng Y~C and Schwingenschl{\"o}gl U 2013 {Hole doped Dirac states
  in silicene by biaxial tensile strain} \textit{ J. Appl. Phys.\/} {\bf 113}
  104305

\bibitem{Yan2015b}
Yan J~A, Cruz M~A~D, Barraza-Lopez S and Yang L 2015 Strain-tunable topological
  quantum phase transition in buckled honeycomb lattices \textit{ Appl. Phys.
  Lett.\/} {\bf 106} 183107

\bibitem{Kane2005}
Kane C~L and Mele E~J 2005 Quantum spin Hall effect in graphene \textit{ Phys.
  Rev. Lett.\/} {\bf 95} 226801

\bibitem{Ezawa1}
Ezawa M 2012 A topological insulator and helical zero mode in silicene under an
  inhomogeneous electric field \textit{ New J. Phys.\/} {\bf 14} 033003

\bibitem{Ezawa2}
Ezawa M 2012 Valley-polarized metals and quantum anomalous Hall effect in
  silicene \textit{ Phys. Rev. Lett.\/} {\bf 109} 055502

\bibitem{Yan2015}
Yan J~A, Gao S~P, Stein R and Coard G 2015 {Tuning the electronic structure of
  silicene and germanene by biaxial strain and electric field} \textit{ Phys.
  Rev. B\/} {\bf 91} 245403

\bibitem{ChavezCastillo2015}
Chavez-Castillo M~R, Rodriguez-Meza M~A and Meza-Montes L 2015 {Size, vacancy
  and temperature effects on Young's modulus of silicene nanoribbons} \textit{
  RSC Adv.\/} {\bf 5} 96052--96061

\bibitem{Rivero2014}
Rivero P, Yan J~A, Garc\'{\i}a-Su\'arez V~M, Ferrer J and Barraza-Lopez S 2014
  Stability and properties of high-buckled two-dimensional tin and lead
  \textit{ Phys. Rev. B\/} {\bf 90} 241408

\bibitem{Ma2012}
Ma Y, Dai Y, Guo M, Niu C and Huang B 2012 Intriguing Behavior of Halogenated
  Two-Dimensional Tin \textit{ J. Phys. Chem. C\/} {\bf 116} 12977--12981

\bibitem{Xu2013}
Xu Y, Yan B, Zhang H~J, Wang J, Xu G, Tang P, Duan W and Zhang S~C 2013
  Large-gap quantum spin Hall insulators in tin films \textit{ Phys. Rev.
  Lett.\/} {\bf 111} 136804

\bibitem{BlackStrain1}
Rodin A~S, Carvalho A and Castro~Neto A~H 2014 Strain-induced gap modification
  in black phosphorus \textit{ Phys. Rev. Lett.\/} {\bf 112} 176801

\bibitem{BlackStrain2}
Fei R and Yang L 2014 Strain-engineering the anisotropic electrical conductance
  of few-layer black phosphorus \textit{ Nano Lett.\/} {\bf 14} 2884--2889

\bibitem{KatPhos}
Rudenko A~N and Katsnelson M~I 2014 Quasiparticle band structure and
  tight-binding model for single- and bilayer black phosphorus \textit{ Phys.
  Rev. B\/} {\bf 89} 201408

\bibitem{Cakir2014}
{\ifmmode \mbox\c{C}\else \c{C}\fi{}ak\ifmmode \imath \else \i \fi{}r} D, Sahin
  H and Peeters F~m~c~M 2014 {Tuning of the electronic and optical properties
  of single-layer black phosphorus by strain} \textit{ Phys. Rev. B\/} {\bf 90}
  205421

\bibitem{Phosphorene1}
Liu H, Neal A~T, Zhu Z, Luo Z, Xu X, Tománek D and Ye P~D 2014 Phosphorene: An
  Unexplored 2D Semiconductor with a High Hole Mobility \textit{ ACS Nano\/}
  {\bf 8} 4033--4041

\bibitem{Likai2014}
{Li Likai}, {Yu Yijun}, {Ye Guo Jun}, {Ge Qingqin}, {Ou Xuedong}, {Wu Hua},
  {Feng Donglai}, {Chen Xian Hui} and {Zhang Yuanbo} 2014 {Black phosphorus
  field-effect transistors} \textit{ Nat. Nanotechnol.\/} {\bf 9} 372--377

\bibitem{Das2014}
Saptarshi~Das M~D and Roelofs A 2014 Ambipolar phosphorene field effect
  transistor \textit{ ACS Nano\/} {\bf 8} 11730--11738

\bibitem{Kou2015}
Liangzhi~Kou C~C and Smith S~C 2015 Phosphorene: fabrication, properties, and
  applications \textit{ J. Phys. Chem. Letters\/} {\bf 6} 2794--2805

\bibitem{Jiang2014}
Jiang J~W and Park H~S 2014 {Mechanical properties of single-layer black
  phosphorus} \textit{ J. Phys. D: Appl. Phys.\/} {\bf 47} 385304

\bibitem{Mehrshad2015}
Mehboudi M, Utt K, Terrones H, Harriss E~O, Pacheco~SanJuan A~A and
  Barraza-Lopez S 2015 {Strain and the optoelectronic properties of nonplanar
  phosphorene monolayers} \textit{ Proc. Natl. Acad. Sci. U.S.A.\/} {\bf 112}
  5888--5892

\bibitem{Wood2014}
Wood J~D, Wells S~A, Jariwala D, Chen K~S, Cho E, Sangwan V~K, Liu X, Lauhon
  L~J, Marks T~J and Hersam M~C 2014 Effective passivation of exfoliated black
  phosphorus transistors against ambient degradation \textit{ Nano Lett.\/}
  {\bf 14} 6964--6970

\bibitem{Favron2015}
Favron A, Gaufr{\`e}s E, Fossard F, Phaneuf-L'Heureux A~L, Tang N~Y~W, Levesque
  P~L, Loiseau A, Leonelli R, Francoeur S and Martel R 2015 Photooxidation and
  quantum confinement effects in exfoliated black phosphorus \textit{ Nat.
  Mater.\/} {\bf 14} 826--832

\bibitem{Utt2015}
Utt K~L, Rivero P, Mehboudi M, Harriss E~O, Borunda M~F, Pacheco~SanJuan A~A
  and Barraza-Lopez S 2015 Intrinsic Defects, Fluctuations of the Local Shape,
  and the Photo-Oxidation of Black Phosphorus \textit{ ACS Cent. Sci.\/} {\bf
  1} 320--327

\bibitem{Xia2015}
Wang X, Jones A~M, Seyler K~L, Tran V, Jia Y, Zhao H, Wang H, Yang L, Xu X and
  Xia F 2015 {Highly anisotropic and robust excitons in monolayer black
  phosphorus} \textit{ Nat. Nanotech.\/} {\bf 10} 517--521

\bibitem{Ge2015}
Ge Y, Wan W, Yang F and Yao Y 2015 {The strain effect on superconductivity in
  phosphorene: a first-principles prediction} \textit{ New J. Phys.\/} {\bf 17}
  035008

\bibitem{Sisakht2016}
{Taghizadeh Sisakht} E, Fazileh F, Zare M~H, Zarenia M and Peeters F~M 2016
  {Strain-induced topological phase transition in phosphorene and in
  phosphorene nanoribbons} \textit{ Phys. Rev. B\/} {\bf 94} 085417

\bibitem{Churchill}
Churchill H~O~H and Jarillo-Herrero P 2014 Two-dimensional crystals: Phosphorus
  joins the family \textit{ Nat. Nano\/} {\bf 9} 330--331

\bibitem{Dresselhaus}
Ling X, Wang H, Huang S, Xia F and Dresselhaus M~S 2015 The renaissance of
  black phosphorus \textit{ Proc. Natl. Acad. Sci. U.S.A.\/} {\bf 112}
  4523--4530

\bibitem{TomanekBlue}
Zhu Z and Tom\'anek D 2014 Semiconducting layered blue phosphorus: A
  computational study \textit{ Phys. Rev. Lett.\/} {\bf 112} 176802

\bibitem{allotropes}
Guan J, Zhu Z and Tom\'anek D 2014 Phase coexistence and metal-insulator
  transition in few-layer phosphorene: A computational study \textit{ Phys.
  Rev. Lett.\/} {\bf 113} 046804

\bibitem{Tiling}
Guan J, Zhu Z and Tom{\'a}nek D 2014 Tiling phosphorene \textit{ ACS Nano\/}
  {\bf 8} 12763--12768

\bibitem{Nebraska}
Wu M, Fu H, Zhou L, Yao K and Zeng X~C 2015 Nine new phosphorene polymorphs
  with non-honeycomb structures: A much extended family \textit{ Nano Lett.\/}
  {\bf 15} 3557--3562

\bibitem{Jakobson2}
Liu Y, Xu F, Zhang Z, Penev E~S and Yakobson B~I 2014 Two-dimensional
  mono-elemental semiconductor with electronically inactive defects: The case
  of phosphorus \textit{ Nano Lett.\/} {\bf 14} 6782--6786

\bibitem{Quereda2016}
Quereda J, San-Jose P, Parente V, Vaquero-Garzon L, Molina-Mendoza A~J, Agraït
  N, Rubio-Bollinger G, Guinea F, Rold{\'a}n R and Castellanos-Gomez A 2016
  {Strong modulation of optical properties in black phosphorus through
  strain-engineered rippling} \textit{ Nano Lett.\/} {\bf 16} 2931--2937

\bibitem{Vogel2015}
Vogel E~M and Robinson J~A 2015 {Two-dimensional layered transition-metal
  dichalcogenides for versatile properties and applications} \textit{ MRS
  Bull.\/} {\bf 40} 558--563

\bibitem{Joensen1986}
Joensen P, Frindt R and Morrison S 1986 Single-layer MoS$_2$ \textit{ Mat. Res.
  Bull.\/} {\bf 21} 457--461

\bibitem{seifert2000}
Seifert G, Terrones H, Terrones M, Jungnickel G and Frauenheim T 2000 Structure
  and electronic properties of ${\mathrm{MoS}}_{2}$ nanotubes \textit{ Phys.
  Rev. Lett.\/} {\bf 85} 146--149

\bibitem{Mak2010}
Mak K~F, Lee C, Hone J, Shan J and Heinz T~F 2010 Atomically thin
  ${\mathrm{MoS}}_{2}$: A new direct-gap semiconductor \textit{ Phys. Rev.
  Lett.\/} {\bf 105} 136805

\bibitem{KolobovBook}
Alexander V~Kolobov J~T 2016 \textit{ Two-dimensional transition-metal
  dichalcogenides\/} 1st ed Springer Series in Materials Science 239 (Springer
  International Publishing)

\bibitem{Radisavljevic2011}
Radisavljevic B, Radenovic A, Brivio J, Giacometti V and Kis A 2011
  {Single-layer MoS$_2$ transistors} \textit{ Nat. Nanotechnol.\/} {\bf 6}
  147--150

\bibitem{Zhu2011}
Zhu Z~Y, Cheng Y~C and Schwingenschl\"ogl U 2011 Giant spin-orbit-induced spin
  splitting in two-dimensional transition-metal dichalcogenide semiconductors
  \textit{ Phys. Rev. B\/} {\bf 84} 153402

\bibitem{Zibouche2014}
Zibouche N, Kuc A, Musfeldt J and Heine T 2014 Transition-metal dichalcogenides
  for spintronic applications \textit{ Annalen der Physik\/} {\bf 526} 395--401

\bibitem{Xu2014}
Xu X, Yao W, Xiao D and Heinz T~F {2014} {Spin and pseudospins in layered
  transition metal dichalcogenides} \textit{ {Nat. Phys.}\/} {\bf {10}}
  {343--350}

\bibitem{Yang2012}
Yang S, Wang L, Tian X, Xu Z, Wang W, Bai X and Wang E {2012} {The piezotronic
  effect of zinc oxide nanowires studied by in situ TEM} \textit{ {Adv.
  Mat.}\/} {\bf {24}} {4676--4682}

\bibitem{Cao2015}
Cao L 2015 {Two-dimensional transition-metal dichalcogenide materials: Toward
  an age of atomic-scale photonics} \textit{ MRS Bull.\/} {\bf 40} 592--599

\bibitem{zeng2012}
Zeng H, Dai J, Yao W, Xiao D and Cui X {2012} {Valley polarization in MoS$_2$
  monolayers by optical pumping} \textit{ {Nat. Nanotechnol.}\/} {\bf {7}}
  {490--493}

\bibitem{Mak2012}
Mak K~F, He K, Shan J and Heinz T~F {2012} {Control of valley polarization in
  monolayer MoS$_2$ by optical helicity} \textit{ {Nat. Nanotechnol.}\/} {\bf
  {7}} {494--498}

\bibitem{Reyes2014}
Reyes-Retana J~A, Naumis G~G and Cervantes-Sodi F 2014 {Centered honeycomb
  NiSe$_2$ nanoribbons: Structure and electronic properties} \textit{ J. Phys.
  Chem. C\/} {\bf 118} 3295--3304

\bibitem{ZhangZhen2016}
Zhang X, Zhen M, Bai J, Jin S and Liu L 2016 {Efficient
  NiSe-Ni$_3$Se$_2$/graphene electrocatalyst in dye-sensitized solar cells: The
  role of hollow hybrid nanostructure} \textit{ ACS Appl. Mater. Interfaces\/}
  {\bf 8} 17187--17193

\bibitem{Bertolazzi2011}
Bertolazzi S, Brivio J and Kis A 2011 Stretching and Breaking of Ultrathin MoS2
  \textit{ ACS Nano\/} {\bf 5} 9703--9709

\bibitem{Castellanos2014}
Castellanos-Gomez A, van~der Zant H~S~J and Steele G~A 2014 Folded MoS2 layers
  with reduced interlayer coupling \textit{ Nano Res.\/} {\bf 7} 572--578

\bibitem{Guzman2014}
Guzman D~M and Strachan A 2014 Role of strain on electronic and mechanical
  response of semiconducting transition-metal dichalcogenide monolayers: An
  ab-initio study \textit{ J. Appl. Phys.\/} {\bf 115} 243701

\bibitem{Johari2012}
Johari P and Shenoy V~B 2012 Tuning the Electronic Properties of Semiconducting
  Transition Metal Dichalcogenides by Applying Mechanical Strains \textit{ ACS
  Nano\/} {\bf 6} 5449--5456

\bibitem{Ghorbani2013}
Ghorbani-Asl M, Borini S, Kuc A and Heine T 2013 {Strain-dependent modulation
  of conductivity in single-layer transition-metal dichalcogenides} \textit{
  Phys. Rev. B\/} {\bf 87} 235434

\bibitem{Scalise2014}
Scalise E, Houssa M, Pourtois G, Afanas′ev V and Stesmans A 2014
  First-principles study of strained 2D MoS$_2$ \textit{ Physica E Low Dimens.
  Syst. Nanostruct.\/} {\bf 56} 416--421

\bibitem{Maniadaki2016}
Maniadaki A~E, Kopidakis G and Remediakis I~N 2016 Strain engineering of
  electronic properties of transition metal dichalcogenide monolayers \textit{
  Solid State Comm.\/} {\bf 227} 33--39

\bibitem{Zhao2017}
Zhao Y, Zhang Z and Ouyang G 2017 Lattice Strain Effect on the Band Offset in
  Single-Layer MoS2: An Atomic-Bond-Relaxation Approach \textit{ J. Phys. Chem.
  C\/} {\bf 121} 5366--5371

\bibitem{He2013}
He K, Poole C, Mak K~F and Shan J 2013 {Experimental demonstration of
  continuous electronic structure tuning via strain in atomically thin MoS$_2$}
  \textit{ Nano Lett.\/} {\bf 13} 2931--2936

\bibitem{Conley2013}
Conley H~J, Wang B, Ziegler J~I, Haglund R~F, Pantelides S~T and Bolotin K~I
  2013 Bandgap engineering of strained monolayer and bilayer MoS$_2$ \textit{
  Nano Lett.\/} {\bf 13} 3626--3630

\bibitem{Lloyd2016}
Lloyd D, Liu X, Christopher J~W, Cantley L, Wadehra A, Kim B~L, Goldberg B~B,
  Swan A~K and Bunch J~S 2016 Band gap engineering with ultralarge biaxial
  strains in suspended monolayer MoS$_2$ \textit{ Nano Lett.\/} {\bf 16}
  5836--5841

\bibitem{Shen2016}
Shen T, Penumatcha A~V and Appenzeller J 2016 {Strain engineering for
  transition metal dichalcogenides based field effect transistors} \textit{ ACS
  Nano\/} {\bf 10} 4712--4718

\bibitem{Gutierrez2013}
Guti{\'e}rrez H~R, Perea-L{\'o}pez N, El{\'i}as A~L, Berkdemir A, Wang B, Lv R,
  L{\'o}pez-Ur{\'\i}as F, Crespi V~H, Terrones H and Terrones M 2013
  Extraordinary room-temperature photoluminescence in triangular WS$_2$
  monolayers \textit{ Nano Lett.\/} {\bf 13} 3447--3454

\bibitem{Lin2016}
Lin Z, Carvalho B~R, Kahn E, Lv R, Rao R, Terrones H, Pimenta M~A and Terrones
  M 2016 Defect engineering of two-dimensional transition metal dichalcogenides
  \textit{ 2D Mater.\/} {\bf 3} 022002

\bibitem{Berkdemir2013}
Berkdemir A, Gutierrez H~R, Botello-Mendez A~R, Perea-Lopez N, Elias A~L, Chia
  C~I, Wang B, Crespi V~H, Lopez-Urias F, Charlier J~C, Terrones H and Terrones
  M {2013} {Identification of individual and few layers of WS$_2$ using Raman
  Spectroscopy} \textit{ {Sci. Rep.}\/} {\bf {3}} {1755}

\bibitem{Rice2013}
Rice C, Young R~J, Zan R, Bangert U, Wolverson D, Georgiou T, Jalil R and
  Novoselov K~S 2013 Raman-scattering measurements and first-principles
  calculations of strain-induced phonon shifts in monolayer MoS${}_{2}$
  \textit{ Phys. Rev. B\/} {\bf 87} 081307

\bibitem{Chattopadhyay1986}
Chattopadhyay T, Pannetier J and {Von Schnering} H 1986 Neutron diffraction
  study of the structural phase transition in SnS and SnSe \textit{ J. Phys.
  Chem. Solids\/} {\bf 47} 879--885

\bibitem{Fu2015}
Ando Y and Fu L 2015 Topological crystalline insulators and topological
  superconductors: from concepts to materials \textit{ Annu. Rev. Condens.
  Matter Phys.\/} {\bf 6} 361--381

\bibitem{Lefebvre1998}
Lefebvre I, Szymanski M~A, Olivier-Fourcade J and Jumas J~C 1998 Electronic
  structure of tin monochalcogenides from SnO to SnTe \textit{ Phys. Rev. B\/}
  {\bf 58} 1896--1906

\bibitem{Li2015}
Li C, Hong J, May A, Bansal D, Chi S, Hong T, Ehlers G and Delaire O 2015
  Orbitally driven giant phonon anharmonicity in SnSe \textit{ Nat. Phys.\/}
  {\bf 11} 1063--1069

\bibitem{Xiao2012}
Xiao D, Liu G~B, Feng W, Xu X and Yao W 2012 Coupled spin and valley physics in
  monolayers of ${\mathrm{MoS}}_{2}$ and other group-VI dichalcogenides
  \textit{ Phys. Rev. Lett.\/} {\bf 108} 196802

\bibitem{Kaxiras}
Tritsaris G~A, Malone B~D and Kaxiras E 2013 Optoelectronic properties of
  single-layer, double-layer, and bulk tin sulfide: A theoretical study
  \textit{ J. Appl. Phys.\/} {\bf 113} 233507

\bibitem{Hennig0}
Singh A~K and Hennig R~G 2014 Computational prediction of two-dimensional
  group-IV mono-chalcogenides \textit{ Appl. Phys. Lett.\/} {\bf 105} 042103

\bibitem{Hennig1}
Zhuang H~L and Hennig R~G 2014 Computational discovery, characterization, and
  design of single-layer materials \textit{ JOM\/} {\bf 66} 366--374

\bibitem{Gomes1}
Gomes L~C and Carvalho A 2015 Phosphorene analogues: Isoelectronic
  two-dimensional group-IV monochalcogenides with orthorhombic structure
  \textit{ Phys. Rev. B\/} {\bf 92} 085406

\bibitem{Rodin2016}
Rodin A~S, Gomes L~C, Carvalho A and Castro~Neto A~H 2016 Valley physics in tin
  (II) sulfide \textit{ Phys. Rev. B\/} {\bf 93} 045431

\bibitem{Zhu2015}
Zhu Z, Guan J, Liu D and Tománek D 2015 Designing Isoelectronic Counterparts
  to Layered Group V Semiconductors \textit{ ACS Nano\/} {\bf 9} 8284--8290

\bibitem{Mao_2015}
Mao X, Souslov A, Mendoza C~I and Lubensky T 2015 Mechanical instability at
  finite temperature \textit{ Nat. Commun.\/} {\bf 6} 5968

\bibitem{Wang_2016}
Wang H and Qian X 2017 Two-dimensional multiferroics in monolayer group IV
  monochalcogenides \textit{ 2D Mater.\/} {\bf 4} 015042

\bibitem{Wu_2016}
Wu M and Zeng X~C 2016 Intrinsic ferroelasticity and/or multiferroicity in
  two-dimensional phosphorene and phosphorene analogues \textit{ Nano Lett.\/}
  {\bf 16} 3236--3241

\bibitem{Mehboudi2016b}
Mehboudi M, Fregoso B~M, Yang Y, Zhu W, van~der Zande A, Ferrer J, Bellaiche L,
  Kumar P and Barraza-Lopez S 2016 Structural phase transition and material
  properties of few-layer monochalcogenides \textit{ Phys. Rev. Lett.\/} {\bf
  117} 246802

\bibitem{Copy2016}
Fei R, Kang W and Yang L 2016 Ferroelectricity and phase transitions in
  monolayer group-IV monochalcogenides \textit{ Phys. Rev. Lett.\/} {\bf 117}
  097601

\bibitem{Rangel2016}
Rangel T, Fregoso B~M, Mendoza B~S, Morimoto T, Moore J~E and Neaton J~B 2016
  Giant bulk photovoltaic effect and spontaneous polarization of single-layer
  monochalcogenides arXiv:1610.06589

\end{thebibliography}

\providecommand{\newblock}{}

\end{document}